\documentclass{article}[11pt]

\usepackage{times}
\usepackage{fullpage}
\usepackage{cite,color,xspace}
\usepackage{latexsym}
\usepackage{todonotes}
\usepackage{amsmath}
\usepackage{amsfonts}

\usepackage{bibunits}

\usepackage{placeins,float}

\usepackage{subfig}

\usepackage{tikz,tikz-qtree,tikz-qtree-compat,tikz-3dplot}

\usepackage{amsthm}
\usepackage{cite}
\usepackage{url}
\usepackage{graphicx}
\usepackage{algorithm}
\usepackage[noend]{algpseudocode}
\usepackage[matha]{mathabx}
\usepackage{txfonts} % for semijoin                                             

\usepackage{appendix}

\newcommand{\rev}[1]{#1}

\newcommand{\polylog}{\mathrm{polylog}}

\newcommand{\agm}{\textsf{AGM}\xspace}
\newcommand{\support}{\textsf{support}\xspace}
\newcommand{\nprr}{\textsf{NPRR}\xspace}

\newcommand{\firstsplit}{\text{\sf Split-First-Thick-Dimension}\xspace}
\newcommand{\bcp}{{\sf BCP}\xspace}

\newcommand{\ms}{\text{Minesweeper}\xspace}
\newcommand{\cds}{\text{CDS}\xspace}

\newcommand{\outspace}{\mathcal O}

\newcommand{\Coverwidth}{\text{Cover-width}\xspace}
\newcommand{\coverwidth}{\text{cover-width}\xspace}
\newcommand{\cw}{\text{\sf cw}\xspace}
\newcommand{\gbresolution}{\text{gap box resolution}\xspace}
\newcommand{\gbresolutions}{\text{gap box resolutions}\xspace}
\newcommand{\gbresolvent}{\text{gap box resolvent}\xspace}
\newcommand{\gbresolvents}{\text{gap box resolvents}\xspace}
\newcommand{\obresolution}{\text{output box resolution}\xspace}
\newcommand{\obresolutions}{\text{output box resolutions}\xspace}
\newcommand{\obresolvent}{\text{output resolvent}\xspace}

\newcommand{\atoms}{\text{atoms}}

\newcommand{\cert}{\mathcal C}
\newcommand{\dnfcert}{\mathcal C_{\text{dnf}}}
\newcommand{\gaocert}{\mathcal C_<^{\text{gao}}}
\newcommand{\gaoboxcert}{\mathcal C_{\Box}^{\text{gao}}}
\newcommand{\boxcert}{\mathcal C_{\Box}}
\newcommand{\btree}{\text{B-tree}}

\newcommand{\ar}[1]{\todo[inline, color=green]{#1 \hfill --Atri}}

\newcommand{\tw}{\mathsf{tw}}

\newcommand{\fhtw}{\text{fhtw}}

\newcommand{\calH}{\mathcal H}

\newcommand{\calR}{\mathcal R}
\newcommand{\calA}{\mathcal A}
\newcommand{\calG}{\mathcal G}
\newcommand{\calB}{\mathcal B}
\newcommand{\calN}{\mathcal N}
\newcommand{\calD}{\mathcal D}
\newcommand{\calP}{\mathcal P}
\newcommand{\calV}{\mathcal V}

\newcommand{\calE}{\mathcal E}
\newcommand{\calF}{\mathcal F}

\newcommand{\odd}{\mathcal{O}}
\newcommand{\even}{\mathcal{E}}

\newcommand{\eat}[1]{}

\newcommand{\eps}{\epsilon}

\newcommand{\true}{\textnormal{\sc true}}
\newcommand{\false}{\textnormal{\sc false}}

\newcommand{\la}{\leftarrow}

\newcommand{\be}{\begin{enumerate}}
\newcommand{\ee}{\end{enumerate}}
\newcommand{\bi}{\begin{itemize}}
\newcommand{\ei}{\end{itemize}}
\newcommand{\beq}{\begin{equation}}
\newcommand{\eeq}{\end{equation}}

\newcommand{\bp}{\begin{proof}}
\newcommand{\ep}{\end{proof}}
\newcommand{\bcor}{\begin{cor}}
\newcommand{\ecor}{\end{cor}}
\newcommand{\bthm}{\begin{thm}}
\newcommand{\ethm}{\end{thm}}
\newcommand{\blmm}{\begin{lmm}}
\newcommand{\elmm}{\end{lmm}}
\newcommand{\bdefn}{\begin{defn}}
\newcommand{\edefn}{\end{defn}}
\newcommand{\bprop}{\begin{prop}}
\newcommand{\eprop}{\end{prop}}
\newcommand{\bconj}{\begin{conj}}
\newcommand{\econj}{\end{conj}}
\newcommand{\bopm}{\begin{opm}}
\newcommand{\eopm}{\end{opm}}
\newcommand{\brmk}{\begin{rmk}}
\newcommand{\ermk}{\end{rmk}}

\newcommand{\suchthat}{\ | \ }
\newcommand{\inner}[1]{\langle #1 \rangle}
\newcommand{\dbox}[1]{\langle #1 \rangle}

\newcommand{\vars}{\textnormal{vars}}

\newcommand{\mv}[1]{\mathbf{#1}}

\theoremstyle{plain}                   % default
\newtheorem{thm}{Theorem}[section]
\newtheorem{lmm}[thm]{Lemma}
\newtheorem{prop}[thm]{Proposition}
\newtheorem{cor}[thm]{Corollary}

\theoremstyle{definition}              % Examples and all
\newtheorem{example}[thm]{Example}

\newtheorem{opm}[thm]{Open Problem}
\newtheorem{conj}[thm]{Conjecture}

\newtheorem{defn}[thm]{Definition}

\newtheorem{rmk}[thm]{Remark}
\newtheorem{claim}{Claim}

\newcommand{\bbox}{
\begin{center}
\begin{tabular}{|c|}
\hline
}
\newcommand{\ebox}{
\\
\hline
\end{tabular}
\end{center}
}

\newlength{\toppush}
\algrenewcommand\algorithmicrequire{\textbf{Input:}}
\algrenewcommand\algorithmicensure{\textbf{Output:}}
\algrenewcommand\algorithmicwhile{\textbf{While}}
\algrenewcommand\algorithmicfor{\textbf{For}}
\algrenewcommand\algorithmicreturn{\textbf{Return}}
\algrenewcommand\algorithmicif{\textbf{If}}

\newcommand{\D}{\mathbf{D}}

\newcommand{\wgets}{\quad\gets\quad}
\newcommand{\wsuchthat}{\quad|\quad}
\newcommand{\weq}{\quad=\quad}

\newcommand{\wsubseteq}{\quad\subseteq\quad}
\newcommand{\wc}{\;\;,\;\;}

\newcommand{\pair}[2]{(#1,#2)}
\newcommand{\tO}{\tilde{O}}
\newcommand{\abs}[1]{\left|#1\right|}

\newcommand{\prefix}[1]{\mathrm{prefix}\left(#1\right)}
\newcommand{\st}[1]{\mathcal{#1}}

\newcommand{\UB}{\inner{\lambda,\ldots,\lambda}}
\newcommand{\tetris}{\text{\sf Tetris}\xspace}
\newcommand{\tetrisreloaded}{\text{\sf Tetris-Reloaded}\xspace}
\newcommand{\tetrispreloaded}{\text{\sf Tetris-Preloaded}\xspace}
\newcommand{\tetrispreloadedas}{\text{\sf Tetris-Preloaded-LB}\xspace}
\newcommand{\tetrisreloadedas}{\text{\sf Tetris-Reloaded-LB}\xspace}
\newcommand{\tetrisskeleton}{\text{\sf TetrisSkeleton}\xspace}
\newcommand{\tetrisnewskeleton}{\text{\sf TetrisSkeleton2}\xspace}
\newcommand{\Resolve}{\text{\sf Resolve}\xspace}

\newcommand{\nocache}{{\textsc{Tree Ordered Geometric Resolution}}}
\newcommand{\ordered}{{\textsc{Ordered Geometric Resolution}}}
\newcommand{\geo}{{\textsc{Geometric Resolution}}}

\definecolor{ao}{rgb}{0,0.5,0}

\newcommand{\rao}{splitting attribute order}
\newcommand{\RAO}{SAO\xspace}
\newcommand{\Initialize}{\mathsf{Initialize}}

\newcommand{\sqrtC}{\sqrt{\abs{\st C}}}
\newcommand{\sqrthatC}{\sqrt{\abs{\hat{\st C}}}}
\newcommand{\prefixes}[1]{\mathrm{prefixes}\left(#1\right)}
\newcommand{\lbm}{\text{\sf Balance}}
\newcommand{\updatelbm}{\text{\sf Update-Balance}}
\newcommand{\balance}{\mathsf{Balance}}

\newcommand{\kb}{knowledge base\xspace}
\newcommand{\DT}[1]{\hat{\st T}(#1)}

\newcommand{\gap}{\mathcal{G}}

\usepackage{hyperref}

\newtheorem*{example*}{Example}
\newcommand{\continued}[1]{{\bf \ref{#1} Continued}}

\begin{document}
%\title{A Geometric Resolution-based Framework for Joins:\\Worst-case and Beyond}
\title{Joins via Geometric Resolutions: Worst-case and Beyond}

\author{Mahmoud Abo Khamis$^{\footnotemark[1]\;\;\;\footnotemark[3]}$ \and Hung Q. Ngo$^{\footnotemark[1]\;\;\;\footnotemark[3]}$ \and Christopher R\'e$^{\footnotemark[2]}$ \and Atri Rudra$^{\footnotemark[1]}$}
\date{\footnotemark[1]~~ Department of Computer Science and Engineering\\
University at Buffalo, SUNY\\
{\tt \{mabokham,hungngo,atri\}@buffalo.edu}\\
\vspace*{3mm}
\footnotemark[2]~~ Department of Computer Science\\
Stanford University\\
{\tt {chrismre@cs.stanford.edu}}\\
\vspace*{3mm}
\footnotemark[3]~~ LogicBlox Inc.\\
{\tt \{mahmoud.abokhamis,hung.ngo\}@logicblox.com}
}

\maketitle

%\fontsize{10pt}{10.2pt}
%\selectfont

%\begin{bibunit}[acm]
\begin{abstract}
We present a simple geometric framework for the relational join. Using this framework, we design an algorithm that achieves the fractional hypertree-width bound, which generalizes classical and recent worst-case algorithmic results on computing joins. In addition, we use our framework and the same algorithm to show a series of what are colloquially known as beyond worst-case results. The framework allows us to prove results for data stored in Btrees, multidimensional data structures, and even multiple indices per table. A key idea in our framework is formalizing the inference one does with an index as a type of geometric resolution; transforming the algorithmic problem of computing joins to a geometric problem. Our notion of geometric resolution can be viewed as a geometric analog of logical resolution. In addition to the geometry and logic connections, our algorithm can also be thought of as backtracking search with memoization. 
%Some of our results can be obtained even without the memoization
%component.
\end{abstract}

\newpage

%%% Temp Intro

%!TEX root = main.tex

\section{Introduction}
\label{sec:intro}

Efficient processing of the natural join operation is a key problem in
database management systems \cite{DBLP:books/aw/AbiteboulHV95,
  DBLP:books/cs/Maier83,DBLP:books/cs/Ullman89}.  A large number of
algorithms and heuristics for computing joins have been proposed and
implemented in database systems, including Block-Nested loop join,
Hash-Join, Grace, Sort-merge, index-nested, double pipelined, PRISM,
etc.  \cite{graefe93, Blanas:2011:DEM:1989323.1989328,
  Kim:2009:SVH:1687553.1687564, Chaudhuri:1998:OQO:275487.275492}.  In
addition to their role in database management, joins (or variants) are powerful
enough to capture many fundamental problems in logic and constraint
satisfaction
\cite{DBLP:journals/jcss/KolaitisV00,
%  DBLP:conf/stoc/ChandraM77, 
%DBLP:journals/jacm/Fagin83,
%  DBLP:conf/stoc/Vardi82, 
%DBLP:journals/jcss/GottlobLS02,
%  DBLP:journals/jacm/GottlobMS09, DBLP:journals/tcs/ChekuriR00,
  DBLP:conf/pods/PapadimitriouY97}, 
or subgraph listing problems
\cite{skew, NPRR} which are central in social
\cite{DBLP:conf/www/SuriV11, Tsourakakis:2008:FCT:1510528.1511415} and
biological network analysis \cite{alon-network-motifs, 
%Alon02,
DBLP:journals/bioinformatics/PrzuljCJ04}.

Not surprisingly, there has been a great deal of work on joins in
various settings. A celebrated result is Yannakakis' algorithm, which
shows that acyclic join queries can be computed in linear
time~\cite{DBLP:conf/vldb/Yannakakis81} in data complexity (modulo a
$\log$ factor). Over the years, this result was generalized to
successively larger classes of queries based on various notions of
widths: from {\em treewidth} 
(tw)~\cite{DBLP:conf/aaai/DechterP88,MR855559}, 
{\em degree of acyclicity}~\cite{DBLP:journals/ai/GyssensJC94,
  DBLP:conf/adbt/GyssensP82}, 
{\em query width} (qw)~\cite{DBLP:journals/tcs/ChekuriR00}, to
{\em generalized hypertree width}
(ghw)~\cite{DBLP:journals/sigmod/Scarcello05,
  DBLP:journals/jcss/GottlobLS03}.
From the bound of Atserias, Grohe and Marx~\cite{GM06, AGM08} (\agm bound), 
and its algorithmic proof~\cite{NPRR}, we recently know that there is
a class of join algorithms that are optimal in the worst case,
in the sense that for each join query the algorithm runs in time
linear in the size of the worst-case output~\cite{NPRR,leapfrog,skew}. 
%In fact, one of these algorithms, the LeapFrog TrieJoin algorithm is already
%implemented in a commercial database engine~\cite{leapfrog}. 
Combining a worst-case optimal join algorithm with Yannakakis' algorithm
yields an algorithm running in time \rev{$O(\log N\cdot (N^{\text{fhtw}}+Z))$}, where
fhtw stands for {\em fractional hypertree width}~\cite{GM06}, a more
general notion than the widths mentioned above, and $Z$ is the output size.
%(Recent and even more general notions of 
%{\em adaptive width}~\cite{DBLP:journals/mst/Marx11}
%and {\em submodular width}~\cite{DBLP:journals/jacm/Marx13} are 
%beyond the scope of our results in this paper.)

However, worst-case can be pathological. For example, input relations are 
typically already pre-processed and stored in sophisticated indices to
facilitate fast query answering (in even {\em sub-linear} time). Motivated
by this, recent work has gone beyond worst-case analysis to notions
that are closer to {\it instance or pointwise} optimality. These
beyond worst-case results have as their starting point the work of Demaine et
al.~\cite{DBLP:conf/soda/DemaineLM00} and Barbay and
Kenyon~\cite{DBLP:conf/soda/BarbayK02,DBLP:journals/talg/BarbayK08},
who designed beyond worst-case algorithms for set intersection and
union problems, which were recently extended to join
processing~\cite{nnrr}.\footnote{This algorithm has been implemented
in a commercial database system, LogicBlox, with promising but
initial results. In our preliminary experimental results, the new
algorithm on some queries on real social network data showed up to
two, even three orders of magnitude speedup over several existing commercial 
database engines~\cite{Nguyen:2015:JPG:2764947.2764948}.}
  
As one might expect, the algorithms that achieve
the above varied results are themselves varied; they make a wide range
of seemingly incompatible assumptions: data are indexed or not; the
measures are worst-case or instance-based; they may rely on (or
ignore) detailed structural information about the query or cardinality
information about the underlying tables. With all this variety, our
first result may be surprising: {\it we recover all of the above
mentioned results with a single, simple algorithm.}\footnote{\rev{Here we are referring to the above mentioned results on join algorithms: worst-case and beyond. See Table~\ref{tab:ub} for more details.} We are unable 
to recover the more recent notions of widths, in particular, the notion of 
{\em submodular width}~\cite{DBLP:journals/jacm/Marx13}. See 
Section~\ref{sec:related} for a more detailed discussion.}

Our central algorithmic idea is to cast the problem of evaluating a
join over data in indices as a geometric problem; specifically, we
reduce the join problem to a problem (defined below) in which one
covers a rectangular region of a multidimensional space (with
dimension equal to the number of attributes of the join) with a set of
rectangular boxes. These boxes represent regions in the space in which
we know output tuples are {\em not} present. Such rectangles
are a succinct way to represent the information conveyed by these data
structures. We illustrate these ideas by an example.

\begin{figure}[!htp]
\centering
\subfloat[\rev{Tuples from $R(A,B)$}]{
\begin{tikzpicture}[scale=0.42]
\draw [<->] (0, 9) -- (0, 0) -- (9, 0);
\draw[help lines] (0,0) grid (8,8);
\foreach \i in {0,...,7}
   \node [below] at (\i+0.5, 0) {\scriptsize \i};
\foreach \i in {0,...,7}
   \node [left] at (0, \i+0.5) {\scriptsize \i};
\foreach \i in {1,3,5,7}
   \draw[rounded corners, fill=blue, fill opacity=1] (\i,3) rectangle (\i+1,4);
\foreach \i in {1,5,7}
   \draw[rounded corners, fill=blue, fill opacity=1] (3,\i) rectangle (4,\i+1);
\node [black] at (9.5,0) {$A$};
\node [black] at (0, 9.5) {$B$};
\label{fig:S-rel}
    \end{tikzpicture}
}~~
\subfloat[\rev{Gap boxes from $(A,B)$-ordered Btree}]{
\begin{tikzpicture}[scale=0.42]
\draw [<->] (0, 9) -- (0, 0) -- (9, 0);
\draw[help lines] (0,0) grid (8,8);
\foreach \i in {0,...,7}
   \node [below] at (\i+0.5, 0) {\scriptsize \i};
\foreach \i in {0,...,7}
   \node [left] at (0, \i+0.5) {\scriptsize \i};
\foreach \i in {1,3,5,7}
   \draw[rounded corners, fill=blue, fill opacity=1] (\i,3) rectangle (\i+1,4);
\foreach \i in {1,5,7}
   \draw[rounded corners, fill=blue, fill opacity=1] (3,\i) rectangle (4,\i+1);
\foreach \i in{1, 5, 7}
{
   \draw[rounded corners, fill=orange, fill opacity=.6] (\i,0) rectangle (\i+1,3);
   \draw[rounded corners, fill=orange, fill opacity=.6] (\i,4) rectangle (\i+1,8);
}
\foreach \i in{0, 2, 4, 6}
   \draw[rounded corners, fill=orange, fill opacity=.6] (3,\i) rectangle (4,\i+1);
\foreach \i in{0, 2, 4, 6}
   \draw[rounded corners, fill=orange, fill opacity=.3] (\i,0) rectangle (\i+1,8);
\node [black] at (9.5,0) {$A$};
\node [black] at (0, 9.5) {$B$};
\label{fig:S-AB}
\end{tikzpicture}
}
\caption{\rev{A relation and the corresponding gap boxes from sorted order $(A,B)$.}}
\label{fig:rel-to-boxes}
\end{figure}
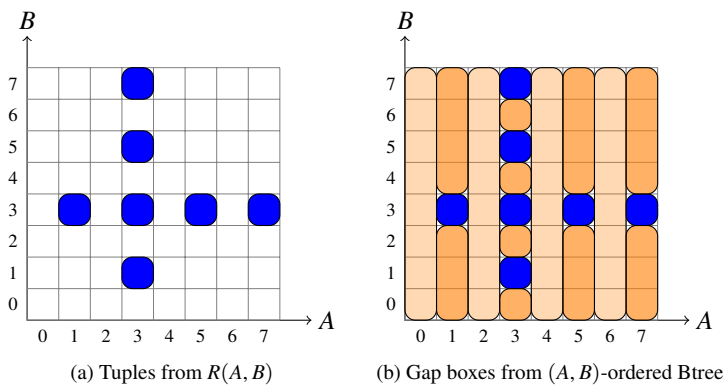

\begin{example}
\rev{
Consider the relation $R(A,B)=\{3\}\times\{1,3,5,7\} \cup \{1,3,5,7\}\times
\{3\}$, which is illustrated in Figure~\ref{fig:S-rel}. For now assume that $R$
is stored in a B-tree with attribute order $(A,B)$. Any two consecutive tuples $(a,b_1)$ and $(a,b_2)$ in $R$ with $b_2>b_1+1$ give rise to a \rev{``tuple-free''} box whose $A$ side contains the single value $\{a\}$ and whose other side spans the values $\{b_1+1, \ldots,  b_2-1\}$.
We call such a box that does not contain any tuples a \emph{gap box}.
Consecutive tuples $(a_1,b_1)$ and $(a_2,b_2)$ for $a_2> a_1+1$ give rise to bigger gap boxes. (Namely, the $A$ side will span the values $\{a_1+1, \ldots, a_2-1\}$, and the $B$ side will span the whole range of $B$.) Figure~\ref{fig:S-AB} illustrates all the gap boxes generated from this $R$.}
Suppose we want to compute the join $R(A, B) \Join S(B, C)$ for some other
relation $S$. Then, the gap boxes from $R$ will  span all
values in the $C$-dimension. Similarly, the gap boxes for $S$ span all values
in the $A$-dimension. And the output tuples are precisely the tuples
$(a,b,c)$ which do not fall into any gap boxes, from both indices of $R$
and $S$.
\end{example}

\rev{It might appear counterintuitive that in this work we chose to compute the join by processing gap boxes rather than tuples of relations (i.e. we are processing the ``absence'' of the data rather than the data directly).
Computing the join over tuples requires first taking the union of tuples of each relation and then taking the intersection over all relations.
In contrast, joining over gap boxes is simpler since we only need to take the union of gap boxes from all relations.
}

\noindent
Throughout this paper we will think of the data as integers for
convenience, but our results assume only that the domains of
attributes are discrete and ordered. For technical reasons,
 %(see
%Appendix~\ref{app:subsec:the case for dyadic}),
 we will
assume that the boxes are {\it dyadic boxes}, i.e., rectangles whose
 endpoints and side lengths can be encoded as powers of
$2$. Importantly, a dyadic interval can be thought of as a 
bitstring, which allows many geometric operations such as containment
and intersection to be reduced to string operations that take time
linear in the length of strings and so logarithmic in the size of the
data. This encoding does increase the number of gap boxes, but by only a
polylogarithmic factor in the input data size.
\footnote{\rev{The exponent in the polylogarithmic factor can be up to the number of attributes $n$. 
However, there are many cases in which it is much smaller. For example, the exponent is $1$ in case of GAO-consistent indices (See~\cite{nnrr} and Definition~\ref{defn:GAO-consistent-box}), and it is $w+1$ in case of queries with treewidth $w$.}}
With this idea, the
central problem in this work is the {\it box cover problem},
which informally is defined as follows.
(Formal definition is in Section~\ref{sec:prelims}.)

\begin{quote}
Given 
%as input a dyadic box $\mv b$, i.e.,
%a multidimensional rectangle, and
 a set of dyadic boxes $\cal A$, i.e.
the gaps from the data,\footnote{We note that our algorithms assume an oracle access to $\cal A$ and some of our algorithms essentially minimize the number of accesses to the oracle.} our goal is to list all the points that are not covered 
by any box in $\cal A$.
%whether (1) $\mv b$
%is covered by some union of input boxes or (2) to find a point which
%is uncovered by $\cal A$.
\end{quote}

The core of our algorithm solves essentially the boolean version of the box 
cover problem, where in addition to the set of boxes $\cal A$, one is also 
given a target box $\mv b$ and the goal is to check if $\mv b$ is covered by 
the union of boxes in $\cal A$.
Our algorithm for the boolean box cover problem is recursive, with the
following steps: We first check if any box $\mv a\in\calA$ contains
$\mv b$. If such $\mv a$ exists, we return it as a witness that $\mv b$ is covered. If not, then we split the box $\mv b$ into two halves $\mv
b_1$ and $\mv b_2$ and recurse. In the recursive steps, we either find
a point in the target box that is not covered (in which case we return
it as a witness that the target box is not covered), or we discover two boxes 
$\mv w_1, \mv w_2\in\calA$ that contain $\mv
b_1$ and $\mv b_2$ respectively. We then construct a single box $\mv w$
by combining $\mv w_1$ and $\mv w_2$ such that $\mv w$ contains $\mv b$, we add $\mv w$ to $\cal A$
\footnote{$\cal A$ is assumed to be a global variable: all levels of the recursion access the same $\cal A$. Check Section~\ref{subsec:the-algo:the-core} for more details.}, and we return it as a witness that $\mv b$ is covered.
%\footnote{We note that the boxes $\mv w_1$ and $\mv w_2$ returned by the recursive calls to $\mv b_1$ and $\mv b_2$ respectively could have been generated by this combination process in the recursive calls.}  
This algorithm needs to answer three
questions:
\begin{itemize}
\item {\it How to find a box $\mv a \in \cal A$ containing the
        target box $\mv b$ if such a box exists?}
  This search procedure should be efficient, ideally in 
  polylogarithmic time in the data size. Dyadic encoding of gap boxes makes this
  goal possible. Our algorithm stores boxes in $\cal A$ in a (multilevel) dyadic
  tree data structure.

\item {\it How to split the input box $\mv b$ into $\mv b_1$ 
and $\mv b_2$?} A first natural scheme is to go in a fixed attribute 
order. We show that this scheme is sufficient to recover all the results 
mentioned earlier in this section. However, we also show that this approach
is fundamentally limited. In particular, we show a novel alternate
scheme that is able to achieve much stronger per-instance guarantees.
%\yell{Vague, need forward pointer?}

\item {\it How to combine $\mv w_1$ and $\mv w_2$ to form $\mv w$?} 
  The combine operation has two competing goals: it should be {\it
  complete} in that it can infer $\mv b$ (or a box that contains $\mv b$) and
  it should be {\it efficient} in that it should take at
  most polylogarithmic time in the data. For that purpose, we introduce a notion
  called {\em geometric resolution} (Figure~\ref{fig:main:resolution}). We
  show that this framework is complete and the
  resolution operation can be implemented efficiently as a simple
  operation on bitstrings.
  % \yell{as shown in Appendix~\ref{app:complete and efficient}}. 
  In conjunction with the efficient search procedure, this implies that the running 
  time of the algorithm is the number of 
  such resolutions (up to polylogarithmic factors in the data size). Thus,
  we can reason about the geometry of these covers instead of the
  algorithmic steps.\footnote{This should be contrasted with
  traditional logical resolution that can potentially require $\Omega(N)$ 
  time for a single step due to large clauses. At a high level, logical resolution is resolving {\em combinatorial} rectangles while our notion resolves geometric rectangles.
  See Appendix~\ref{app:paul} for more discussion and examples.}
\end{itemize}

\begin{table}[ht!]
\begin{center}
\begin{tabular}{|c|c|c|}

\hline

Join Query Type & Run time &  Recovers(R)/Subsumes(S)\\
\hline 
\hline
\multicolumn{3}{|c|}{\textsc{Worst-case Results}}\\
\hline
$\alpha$-acyclic & $N+Z$ &  Yannakakis \cite{DBLP:conf/vldb/Yannakakis81} (R)\\
Arbitrary & $N+\agm$ &  \cite{NPRR, leapfrog} (R)\\
%treewidth  $w$ & $N^{w+1}+Z$ &  \cite{DBLP:conf/aaai/DechterP88,MR855559} (R)\\
%degree of acyclicity $a$  $w$ & $N^{a}+Z$ & None & \cite{DBLP:journals/ai/GyssensJC94, DBLP:conf/adbt/GyssensP82} (R)\\
%qw $w$ $w^*$ & $N^{w}+Z$ & None & \cite{DBLP:journals/tcs/ChekuriR00} (R)\\
%ghw $w$ $w^*$ & $N^{w}+Z$ & None & \cite{DBLP:journals/sigmod/Scarcello05, DBLP:journals/jcss/GottlobLS03} (R)\\
bounded width \rev{$\fhtw$} & \rev{$N^{\fhtw}+Z$} &  \cite{DBLP:conf/vldb/Yannakakis81,NPRR,leapfrog,skew,OZ14,DBLP:conf/aaai/DechterP88,DBLP:journals/tcs/ChekuriR00,DBLP:journals/jcss/GottlobLS03} (S)\\
%$w^* = $ fhtw $\leq$ ghw $\leq$ qw $\leq$ tw $+1$ & & \\
\hline
\hline
\multicolumn{3}{|c|}{\textsc{Certificate Based Results}}\\
\hline
treewidth $w$ & $|\boxcert|^{w+1}+Z$ &  New and \cite{nnrr} (S)\\
%binary relations, acyclic query & $|\boxcert|+Z$ &  New\\
treewidth $1$ & $|\boxcert|+Z$ &  New\\
%% \coverwidth $\text{cw}$ & $|\gaocert|^{\text{cw}}+Z$ & GAO-consistent &
%% New and \cite{nnrr} (S)\\
%% $k$-cycle (Boolean query) & $|\gaocert|^{2-\frac{1}{\lceil k/2\rceil}}$ & GAO-consistent & AYZ
%% \cite{AYZ97} (S)\\
%$k$-clique & $C^{k/2}+Z$ & GAO & New\\
%``knrr"w, arity $\le 2$ $\rho$ &$C^{\rho}+Z$ & Rectangular &New\\
\hline

\end{tabular}
\end{center}
\caption{Overview of our upper bounds achieved by the {\em same} algorithm
called \tetris. The runtimes are up to poly-logarithmic factors and either 
in terms of the total input size $N$ or in the size of the optimal certificate 
$\boxcert$ as well as the output size $Z$. 
%No restrictions on 
%indexes means that the input is presented in a format such that it can 
%be listed in $\tilde{O}(N)$ time; rectangular indexes means the index returns 
%gaps as arbitrary boxes; GAO implies that all the relations are sorted according 
%to a consistent ordering of vertices.
% and NEO is GAO where the consistent 
%ordering is the nested elimination order.
\rev{In the above, the bounded width $\fhtw\leq \text{ghw}\leq\text{qw} \leq \text{tw}+1$.}
Our worst-case result on fractional hypertree width \rev{(fhtw)} implies the other 
worst-case results on various notions of widths.
Our result for treewidth $w$ 
queries subsumes that of \cite{nnrr} since the latter only works for 
indices with mutually consistent sort orders. \agm denotes the AGM-bound
for the query~\cite{AGM08}.}
% GAOs and our result for $k$-cycles subsumes that of AYZ since the 
%latter was only a worst-case result.} % (and explicitly stated for $Z=0$.}
\label{tab:ub}
\end{table}

\noindent
In the first contribution, we show that our algorithm -- named \tetris -- 
is able to recover the worst-case algorithmic results shown in the top-half of 
Table~\ref{tab:ub}, and the recent beyond-worst-case results of~\cite{nnrr}.

Our second contribution is to use these insights to go beyond known results.
In previous work on beyond-worst-case analysis, one made an
assumption that indexes were consistent with a single global
ordering of attributes; a constraint that is not often met in practice. Our
first results remove this restriction, which we believe argues for the
power of the above framework. In particular, we reason
about multiple Btrees on the same relation, multidimensional index
structures like KD-trees and RTrees, and even sophisticated dyadic
trees. In turn, this allows us to extend beyond-worst case
analysis to a larger set of indexing schemes and, conceptually, this
brings us closer to a theory of how indexing and join processing
impact one another. 

The idea of beyond worst-case complexity is captured by a natural
notion of geometric certificate. In particular,
a minimum-sized subset $\boxcert \subseteq \cal A$ whose union is the same as 
the union of all input gap boxes in $\cal A$ is called a 
{\em gap box certificate} for the join problem. 
For beyond worst-case results, $\abs{\boxcert}$ is the analogous quantity 
to input size $N$ that is used in the worst-case results.

There are several reasons for our current certificate framework to use only
`gap' boxes and not input tuples (or more generally boxes that contain the input tuples).
First, gap boxes directly generalize the results from \cite{nnrr}, where
it was shown that $|\boxcert|$ is in the same order as the minimum number of 
comparisons that a comparison-based join algorithm has to perform in order to 
be certain that the output is correct.
Second, we expect the input data to be very sparse in the ambient space.
%reside. (A relation on integers contains finitely many integral points, for
%example. The gap boxes span most of the ``vacuum''.) 
 In particular, we show in this paper that $|\boxcert| = O(N)$ and there are
classes of input instances for which $|\boxcert| = o(N)$ (or even $O(1)$).
Third, gap boxes in some sense capture differences between different input
indices. The same relation indexed in different ways \rev{gives} different sets of gap
boxes which can all be used in evaluating the join.
Last but not least, our move to use gaps rather than the input tuples themselves has a strong parallel with using proof by contradiction to prove logical statements. In hindsight, this parallel is precisely what results in the strong connection between our framework and resolution (indeed resolution is a specific form of proof by contradiction).
%it might be possible to use {\em both} the gap boxes and the
%``input boxes,'' and we leave this question as a future work.

We show that for queries with treewidth $1$ (i.e. query graphs are forests), 
we can compute them in $\tO(\abs{\boxcert}+Z)$ time\footnote{In this paper $\tO$ will hide poly-$\log{N}$ factors as well as factors that just depend on the query size, which is assumed to be a constant.}, where $Z$ is the output size.
%, where $\boxcert$ is the smallest sized certificate that proves that the output is as claimed. 
%%Since $|\boxcert|$ can be a lot smaller that $|\cal A|$, runtime measured
%%based on $|\boxcert|$ is a strictly finer notion of runtime complexity
%%than worst-case runtimes. (Note that $|\boxcert|$ is instance-dependent.)
For general treewidth $w$ join queries, we obtain a weaker runtime of 
$\tO(\abs{\boxcert}^{w+1}+Z)$.

We also develop a new and intriguing result, where we obtain
a runtime of $\tO(\abs{\boxcert}^{n/2}+Z)$ for a query with
$n$ attributes. This subsumes and greatly extends the results on
$3$-cliques from previous work to all queries on $n$ attributes (including
$n$-cliques).
%Moreover, it shows that in some sense the clique queries remain 
%the hardest for beyond worst-case analysis as well.
Our geometric framework plays a crucial role in this result, both in 
the analysis and the design of our algorithm.

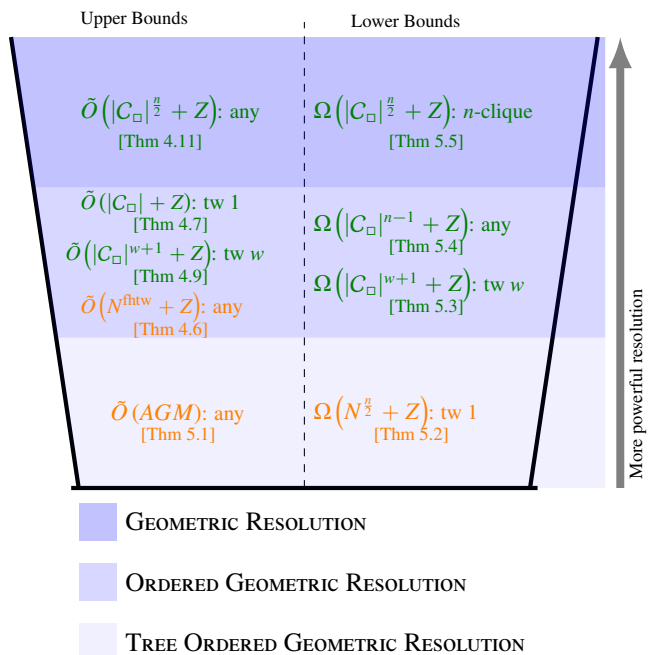
\begin{figure}[!th]
\begin{center}
\begin{tikzpicture}%[scale=0.9]

%%%% The basic structure

\draw [ultra thick] (-3.1,0) -- (3.1,0);
\draw [ultra thick] (-3,0) -- (-3.9,6);
\draw [ultra thick] (3,0) -- (3.9,6);

\draw [dashed] (0,0) -- (0,6.2);

\node [above right] at (-3.1,6) {\textcolor{black}{\scriptsize Upper Bounds}};
\node [above right] at (0.5,6) {\textcolor{black}{\scriptsize Lower Bounds}};

%%%% Resolution types

\path [fill=blue, opacity=.06] (-3,0) -- (-3.3,2) -- (4,2) -- (4,0) -- (-3,0);

\path [fill=blue, opacity=.16] (-3.3,2) -- (-3.6,4) -- (4,4) -- (4,2) -- (-3.3,2);

\path [fill=blue, opacity=.24] (-3.6,4) -- (-3.9,6) -- (4,6) -- (4,4) -- (-3.6,4);

\fill[blue,opacity=.24] (-3,-.2) rectangle (-2.5,-.7);
\node [right] at (-2.5,-.45) {$\geo$};
\fill[blue,opacity=.16] (-3,-1) rectangle (-2.5,-1.5);
\node [right] at (-2.5,-1.25) {$\ordered$};
\fill[blue,opacity=.06] (-3,-1.8) rectangle (-2.5,-2.3);
\node [right] at (-2.5,-2.05) {$\nocache$};

%%% Entries for no cache

\node [right] at (-2.7,1) {\textcolor{orange}{{\small $\tilde{O}\left(AGM\right)$}{\footnotesize : any}}};
\node [right] at (-2.3,.7) {\textcolor{orange}{{\scriptsize [Thm~\ref{thm:app agm}]}}};

\node [right] at (0,1) {\textcolor{orange}{{\small $\Omega\left(N^{\frac{n}{2}}+Z\right)$}{\footnotesize: tw $1$}}};
\node [right] at (0.8,.7) {\textcolor{orange}{{\scriptsize [Thm~\ref{thm:Y-tree}]}}};
%\node [right] at (0,0.5) {\textcolor{ao}{{\small $\omega\left(|\cert|+Z\right)$}{\footnotesize: acyclic}}};

%%%% Entries for ordered

\node [right] at (-3.1,2.4) {\textcolor{orange}{{\footnotesize $\tilde{O}\left(N^{\mathrm{fhtw}}+Z\right)$}{\footnotesize: any}}};
\node [right] at (-2.4,2.1) {\textcolor{orange}{{\scriptsize [Thm~\ref{thm:main fhtw}]}}};
\node [right] at (-3.1,3.8) {\textcolor{ao}{{\footnotesize $\tilde{O}\left(|\boxcert|+Z\right)$}{\footnotesize: tw $1$}}};
\node [right] at (-2.4,3.5) {\textcolor{ao}{{\scriptsize [Thm~\ref{thm:main acyclic arity<=2}]}}};
\node [right] at (-3.3,3.1) {\textcolor{ao}{{\footnotesize $\tilde{O}\left(|\boxcert|^{w+1}+Z\right)$}{\footnotesize: tw $w$}}};
\node [right] at (-2.4,2.8) {\textcolor{ao}{{\scriptsize [Thm~\ref{thm:main:C^{w+1}+Z}]}}};

\node [right] at (0,3.5) {\textcolor{ao}{{\small $\Omega\left(|\boxcert|^{n-1}+Z\right)$}{\footnotesize: any}}};
\node [right] at (1,3.2) {\textcolor{ao}{{\scriptsize [Thm~\ref{thm:cert^n-1-lb}]}}};
\node [right] at (0,2.7) {\textcolor{ao}{{\small $\Omega\left(|\boxcert|^{w+1}+Z\right)$}{\footnotesize: tw $w$}}};
\node [right] at (1,2.4) {\textcolor{ao}{{\scriptsize [Thm~\ref{thm:cert^w+1-lb}]}}};

%%%% Entries for general

\node [right] at (-3.1,5) {\textcolor{ao}{{\small $\tilde{O}\left(|\boxcert|^{\frac{n}{2}}+Z\right)$}{\footnotesize: any}}};
\node [right] at (-2.6,4.6) {\textcolor{ao}{{\scriptsize [Thm~\ref{thm offline as}]}}};

\node [right] at (0,5) {\textcolor{ao}{{\small $\Omega\left(|\boxcert|^{\frac{n}{2}}+Z\right)$}{\footnotesize: $n$-clique}}};
\node [right] at (1,4.6) {\textcolor{ao}{{\scriptsize [Thm~\ref{thm:geo-lb-n-clique}]}}};

%%%% Comparitive powers

\draw [gray, line width=3, -latex] (4.2,0) -- (4.2,6);
\node [rotate=90, below right] at (4.2,0) {\textcolor{black}{\scriptsize More powerful resolution}};

%\node [rotate=90] at (6,2) {{\large $\subsetneq$}};
%\node [rotate=90] at (6,4) {{\large $\subsetneq$}};

\end{tikzpicture}
\end{center}
\caption{An overview of our results and the resolution framework. Bounds for worst-case complexity are denoted by \textcolor{orange}{orange} (where $N$ is the size of the largest relation) and certificate-based results are presented in \textcolor{ao}{green} (where \rev{$\boxcert$} is the optimal certificate). $AGM$ denotes the bound on the output due to AGM and $Z$ denotes the size of the output (per-instance basis). The bounds are presented in format time:query, where {\em any} denotes an arbitrary query on $n$ attributes, %{\em acyclic} denotes a binary acyclic query on $n$ attributes,
 {\em tw} $w$ denotes a query on $n$ attributes with treewidth $w$ ($1<w<n-1$) and $n$-{\em clique} denotes the $n$-variable clique query.} %\ar{ALL: Should we put in explicit pointers to the Theorem numbers next to the results?}}
\label{fig:res}
\end{figure}

Finally, we also use our framework to provide lower bounds on the
number of geometric resolutions that any algorithm needs. In particular, we 
consider three variants of geometric resolution in this paper. 
The most general kind 
(which is not as powerful as general logical resolution) resolves geometric 
boxes, which we call \geo. We are able to recover all of the results in
Table~\ref{tab:ub} with 
a weaker form of geometric resolution called \ordered, which 
corresponds to geometric resolution but when we only combine boxes in a fixed 
attribute order. We also consider an even more special case \nocache, which 
corresponds to ordered geometric resolution when we do {\em not} cache the 
outcome of any resolution. Figure~\ref{fig:res} summarizes where our upper and 
lower bounds fit in these classes of resolution.
%!TEX root = main.tex

%\section{Overview Figure}

There is an intriguing connection between our framework and DPLL with
clause learning used for \#SAT. We address this further in 
Section~\ref{sec:dpll-tetris}.

\section{Related Work} 
%!TEX root = main.tex
\label{sec:related}
\paragraph*{\rev{Acyclic queries and width notions}}
In a seminal work \cite{DBLP:conf/vldb/Yannakakis81}, Yannakakis showed
that if the query is acyclic (or more precisely $\alpha$-acyclic in
Fagin's terminology~\cite{DBLP:journals/jacm/Fagin83}) then it can be
evaluated in time $\tO(N+Z)$, where $N$ is the input size (in terms of
data complexity), and $Z$ is the output size. Researchers have
expanded the classes of tractable queries using an increasingly finer
structural measure called the `width' of the query, measuring how
`far' from being acyclic a query is.  If the query `width' is
bounded by a constant, then the problem is tractable.  The width notion progressed from {\em treewidth}
(tw)~\cite{DBLP:conf/aaai/DechterP88,MR855559}, {\em degree of
acyclicity}~\cite{DBLP:journals/ai/GyssensJC94,
DBLP:conf/adbt/GyssensP82}, {\em query width}
(qw)~\cite{DBLP:journals/tcs/ChekuriR00}, {\em hypertree width} and
{\em generalized hypertree
width}~\cite{DBLP:journals/sigmod/Scarcello05,
DBLP:journals/jcss/GottlobLS03}.

\paragraph*{Worst-case optimal join algorithms}
Atserias, Grohe, and Marx (\agm henceforth)
\cite{GM06, AGM08} %helped bridge the gap between the
%two approaches by 
 derived a bound on the output size (the number of
tuples in the output) using {\em both} the structural information
about the query {\em and} the input relation sizes. Their bound (see
Appendix~\ref{sec prelim}) is a function of the input relation sizes
and a fractional edge cover of the hypergraph representing the
query. By solving a linear program, we can obtain the best possible
bound for the output size. We refer to this best bound as the
{\em \agm bound}. \agm also showed that their bound is tight (in data
complexity) by constructing a family of instances for which the output
size is in the order of the bound. Similar but slightly weaker bounds were
proven by Alon \cite{MR599482} and Friedgut and Kahn \cite{MR1639767}.
All of these results were proved via entropy-based
arguments.

An algorithm whose worst-case runtime matches the \agm bound would be
{\em worst-case optimal}.  Such an algorithm was derived by Ngo, Porat,
R\'e, and Rudra (\nprr henceforth). Soon after, the Leapfrog Triejoin
algorithm \cite{leapfrog} was shown to run within the \agm
bound. An even simpler but generic skeleton of a
class of join algorithms which generalized both \nprr and Leapfrog
Triejoin was shown to run within \rev{the} \agm bound~\cite{skew}. 
%Our first
%contribution is to reprove this result. 

\rev{Combining a worst-case optimal join algorithm with Yannakakis' algorithm
yields an algorithm running in time $\tO(N^{\text{fhtw}}+Z)$, where
fhtw stands for {\em fractional hypertree width}~\cite{GM06}, a more
general notion than the widths mentioned above, and $Z$ is the output size.
Alternatively, the runtime $\tO(N^{\text{fhtw}}+Z)$ can be achieved through \emph{factorized representation}~\cite{OZ14}
or the \tetris algorithm from this paper.}
Marx introduced yet another more general notion of width called {\em
adaptive width} \cite{DBLP:journals/mst/Marx11}, which is equivalent
to {\em submodular width}
\cite{DBLP:conf/stoc/Marx10}, and we are unable to recover this tighter notion of width using the results of this paper.

\paragraph*{Beyond worst-case for joins with \ms}
Beyond worst-case analysis in databases was formalized by Fagin et al.'s
algorithm~\cite{Fagin:2001:OAA:375551.375567} for searching scored 
items in a database. These per instance guarantees are desirable, though they are very hard to achieve:
there have been relatively few such results \cite{geometric-io}. %, self-improving,adaptive-sort}. 
More relevantly, for the sorted set intersection problem,
Demaine, L\'opez-Ortiz, and Munro~\cite{DBLP:conf/soda/DemaineLM00},
with  followups by Barbay and Kenyon \cite{DBLP:conf/soda/BarbayK02,
DBLP:journals/talg/BarbayK08} devised the notion of a {\em certificate}
or a {\em proof}, which is a set of comparisons necessary to certify
that the output is correct. An algorithm running in time proportional
to the minimum certificate size (up to a $\log$-factor and in
data complexity)  can be considered instance-optimal among
comparison-based algorithms.\footnote{As was observed in~\cite{RoughgardenCS264,nnrr}, the $\log$ factor loss is necessary when dealing with comparison-based algorithms.}

The work of Demaine et al. and Barbay et al. was extended to general
join queries~\cite{nnrr} by defining the notion of a {\em comparison
certificate} for a join problem, which roughly speaking is a set of
propositional comparison statements about the input, such that two
inputs satisfy the same set of propositional statements if and only if
they have the same output.  Intuitively, the minimum size of a
comparison certificate is the minimum amount of work a comparison-based join algorithm
has to do to correctly compute the output. A major technical
assumption needed in prior work~\cite{nnrr} was that all relations are
indexed by BTrees according to a single {\em global attribute order}
(GAO) index.  
For example, if the GAO is $A, B, C, D$ (attributes
participating in the query), and $R(A, C)$ is an input relation, then
the BTree/trie for $R$ has to branch on $A$ before $C$. In this
work, we are able to handle more general indexes (KD-trees, dyadic
trees and multiple indices per relation) and do not require this assumption. 
To the best of our knowledge, the current work and that in~\cite{nnrr} are the only two instances that present (near) instance optimal results for a large class of problems.

The analysis from~\cite{nnrr} implies that we can use a `box
certificate' in place of a `comparison certificate' because a box
certificate has size at most the size of a comparison certificate
(see Appendix~\ref{sec:cert}).
This result inspired our investigation into the world of geometric certificates
in this paper. Indeed we were able to generalize the results from \cite{nnrr}
because the box certificates we considered in this paper are more general than
the GAO-consistent boxes in \cite{nnrr}.

\rev{A recent work studied querying big data by accessing only a small part of the data~\cite{Fan:2015:QBD:2745754.2745771}. The notion of certificate used in this work is essentially the smallest part of the data that is sufficient to answer the query.}

\paragraph*{Connections to DPLL} 
As we will see in Section~\ref{sec:dpll-tetris},
%earlier our algorithm \tetris is based on
%resolution. However, the connection of \tetris to resolution-based
%systems runs much deeper. 
 \tetris is essentially a version of the DPLL
algorithm. % (see Appendix~\ref{sec:cnf-dnf} for more on this).
 We would
like to stress that the novelty of our work is to (i) adapt this
well-known framework to a geometric view of joins and (ii) prove sharp
bounds on the run time of \tetris.

\paragraph*{Klee's measure problem in computational geometry}
A variant of the box cover problem (Definition~\ref{defn:bcp-boxcert}) is the \emph{Boolean} box cover problem (Definition~\ref{defn:boolean-bcp}): given a set $\calB$ of $n$-dimensional boxes, determine whether their union covers the entire space.
The Boolean box cover problem is a special case of \emph{Klee's measure problem}: given a set $\calB$ of $n$-dimensional boxes, compute the measure of their union. Klee's measure problem was solved by Overmars and Yap \cite{doi:10.1137/0220065} in time $O(\abs{\calB}^{n/2}\log(\abs{\calB}))$, and later by Chan \cite{6686177} in time $O(\abs{\calB}^{n/2})$. One corollary of this paper (Corollary~\ref{cor:klee-boxcert^{n/2}}) shows that Klee's measure problem over the Boolean semiring can be solved in time $\tO(\abs{\boxcert}^{n/2})$, where $\boxcert$ is any box certificate for $\calB$. (By Definition~\ref{defn:bcp-boxcert}, $\abs{\boxcert}\leq\abs{\calB}$ and there are instances where $\abs{\boxcert}$ is unboundedly smaller than $\abs{\calB}$.) We also present tighter upper bounds for the box cover problem (and hence for Klee's measure problem over the Boolean semiring) in multiple special cases that are common in database joins (e.g. bounded tree-width, acyclicity, GAO-consistency\ldots). Moreover, while the upper bound of $O(\abs{\calB}^{n/2})$ has not been shown to be tight for Klee's measure problem for $n\geq 3$, we show that our upper bounds (including $\tO(\abs{\boxcert}^{n/2})$) are tight for all algorithms that are based on \geo\ (see Corollary~\ref{cor:klee-boxcert^{n/2}-lb}).

\section{Preliminaries} 

%!TEX root = main.tex

\label{sec:prelims}

%\cmr{We have to explain resolution (that it is
%  efficient), state the problem formally, and probably state the
%  results formally? We have to point to an example here if people get
%  stuck. Also, we need to point to the appendix for the argument that
%  the problem can be offline. After this point, only resolution-based
%  arguments!}

\rev{We review the definition of a join query in Section~\ref{sec:join}.}
We give an overview of the strong connection between indices and gap boxes in Section~\ref{sec:gap-boxes} and then move on to our geometric notion of certificates in Section~\ref{sec:boxcert}. We formally define our main geometric problem $\bcp$ in Section~\ref{sec:bcp}.

\rev{
\subsection{Join query}
\label{sec:join}
Let $\mathcal{A}$ be a set of attribute names,
where an attribute $A\in\mathcal{A}$ is a variable over a finite and
discrete domain $\D(A)$.
Let $\mathcal R$ be a set of {\em relation symbols}.
A {\em relational schema} for the symbol $R \in \mathcal R$ of arity $k$ is a 
tuple $\vars(R) = (A_{i_1}, \dots, A_{i_k})$ of distinct attributes that 
defines the attributes of the relation. 
A relational database schema is a set of relational symbols
and associated schemas denoted by $R(\vars(R)), R\in \mathcal R$.
A relational instance for $R(A_{i_1},\dots,A_{i_k})$ is a
subset of $\D(A_{i_1}) \times \dots \times \D(A_{i_k})$. 
A relational database $\calD$ is a collection of instances, one for each 
relational symbol in the schema, denoted by $R^{\calD}$. 
Often the database is clear from context and we drop the superscript $\calD$
from the relation symbols.

A {\em natural join} query (or simply join query) $Q$ is
specified by a finite subset of relational symbols $\atoms(Q) \subseteq
\mathcal R$, denoted by $\Join_{R \in \atoms(Q)} R$. Let $\vars(Q)$ denote
the set of all attributes that appear in some relation in $Q$, that is
\[ \vars(Q) = \{ A \suchthat A \in \vars(R) \text{ for some } R \in \atoms(Q)\}.
\]  
%Given a tuple $\mv t$ we will write $\mv t_{\bar A}$ to emphasize that its support
%is the attribute set $\bar A$. Further, 
%For any $\bar S\subset \bar A$
%we let $\mv t_{\bar S}$ denote $\mv t$ restricted to $\bar S$.
Given a database instance $\calD$, the
output of the query $Q$ on the database instance $\calD$ is denoted 
$Q(\calD)$ and is defined as
\[ Q(\calD) \stackrel{\mathrm{def}}{=} \left\{ \mv t \in \D^{\vars(Q)} 
\suchthat 
\pi_{\vars(R)}(\mv t) \in R^{\calD} \text{ for each } R \in \atoms(Q)\right\} \]
where $\D^{\vars(Q)}$ is a shorthand for 
$\times_{A \in \vars(Q)} \D(A)$, and $\pi$ is the projection operator.
When the instance is clear from the 
context we will refer to $Q(\calD)$ by just $Q$.

For example, in the following so-called {\em triangle query}
\[ Q_\triangle = R(A, B) \Join S(B, C) \Join T(A, C). \]
we have $\vars(Q_\triangle) = \{A, B, C\}$, $\vars(R) = \{A, B\}$,
$\vars(S) = \{B, C\}$, and $\vars(T) = \{A, C\}$.
}

\subsection{Gap boxes and indices} 
\label{sec:gap-boxes}

We informally describe the idea of gap boxes that capture database
indices. 
The set of gap boxes depends intimately on the 
indices that store the relations. For example, for the relation in 
Figure~\ref{fig:S-rel}, Figure~\ref{fig:S-AB} shows the gap boxes generated 
by a BTree that uses the sort order $(A,B)$.
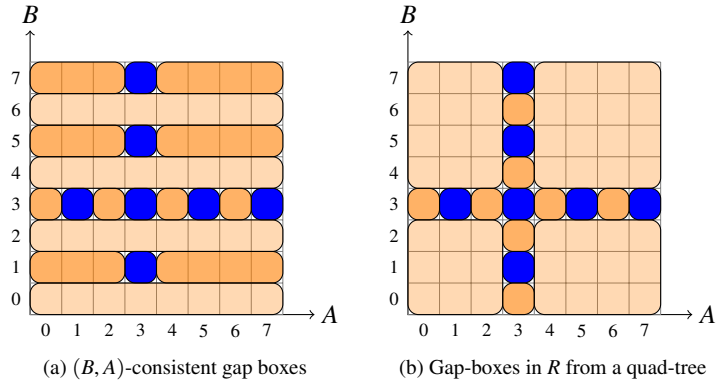
\begin{figure}[!htp]
\centering
\subfloat[\rev{$(B,A)$-consistent gap boxes}]{
\begin{tikzpicture}[scale=0.42]
\draw [<->] (0, 9) -- (0, 0) -- (9, 0);
\draw[help lines] (0,0) grid (8,8);
\foreach \i in {0,...,7}
   \node [below] at (\i+0.5, 0) {\scriptsize \i};
\foreach \i in {0,...,7}
   \node [left] at (0, \i+0.5) {\scriptsize \i};
\foreach \i in {1,3,5,7}
   \draw[rounded corners, fill=blue, fill opacity=1] (\i,3) rectangle (\i+1,4);
\foreach \i in {1,5,7}
   \draw[rounded corners, fill=blue, fill opacity=1] (3,\i) rectangle (4,\i+1);
\foreach \i in{1, 5, 7}
{
   \draw[rounded corners, fill=orange, fill opacity=.6] (0,\i) rectangle (3,\i+1);
   \draw[rounded corners, fill=orange, fill opacity=.6] (4,\i) rectangle (8,\i+1);
}
\foreach \i in{0, 2, 4, 6}
   \draw[rounded corners, fill=orange, fill opacity=.6] (\i,3) rectangle (\i+1,4);
\foreach \i in{0, 2, 4, 6}
   \draw[rounded corners, fill=orange, fill opacity=.3] (0,\i) rectangle (8,\i+1);
\node [black] at (9.5,0) {$A$};
\node [black] at (0, 9.5) {$B$};
\label{fig:S-BA}
\end{tikzpicture}
}~~
\subfloat[\rev{Gap-boxes in $R$ from a quad-tree}] {
\begin{tikzpicture}[scale=0.42]
\draw [<->] (0, 9) -- (0, 0) -- (9, 0);
\draw[help lines] (0,0) grid (8,8);
\foreach \i in {0,...,7}
   \node [below] at (\i+0.5, 0) {\scriptsize \i};
\foreach \i in {0,...,7}
   \node [left] at (0, \i+0.5) {\scriptsize \i};
\foreach \i in {1,3,5,7}
   \draw[rounded corners, fill=blue, fill opacity=1] (\i,3) rectangle (\i+1,4);
\foreach \i in {1,5,7}
   \draw[rounded corners, fill=blue, fill opacity=1] (3,\i) rectangle (4,\i+1);
\foreach \i in{0, 2, 4, 6}
{
   \draw[rounded corners, fill=orange, fill opacity=.6] (\i,3) rectangle (\i+1,4);
   \draw[rounded corners, fill=orange, fill opacity=.6] (3,\i) rectangle (4,\i+1);
}
\draw[rounded corners, fill=orange, fill opacity=.3] (0,0) rectangle (3,3);
\draw[rounded corners, fill=orange, fill opacity=.3] (0,4) rectangle (3,8);
\draw[rounded corners, fill=orange, fill opacity=.3] (4,0) rectangle (8,3);
\draw[rounded corners, fill=orange, fill opacity=.3] (4,4) rectangle (8,8);
\node [black] at (9.5,0) {$A$};
\node [black] at (0, 9.5) {$B$};
\label{fig:S-dyadic}
\end{tikzpicture}
}
\caption{\rev{The gap boxes for the relation in Figure~\ref{fig:S-rel} from sorted order $(B,A)$ and from a quad-tree type index.}}
\label{fig:gap-boxes}
\end{figure}
Figure~\ref{fig:S-BA} shows the gap boxes for the same relation when stored in a
BTree with sort order $(B,A)$. Note that the different sort order manifests
itself in a completely different set of boxes. Finally,
Figure~\ref{fig:S-dyadic} represents the boxes for the same relation when stored
in a quad-tree. In addition to a completely different set of boxes from those in
Figures~\ref{fig:S-AB} and~\ref{fig:S-BA}, the number of boxes is also much
smaller. We will see another example for three attributes soon.

%\hqn{I think in the quad-tree we still need to add the tiny gaps between points,
%vertical and horizontal segments}
%\ar{ALL: Should we keep the figs at the end of the paper in the appendix and put in a forward pointer for an example where the output is not empty?}

\subsection{Geometric Certificates}
\label{sec:boxcert}

For any attribute $A$, $\D(A)$ denotes its domain.
For any join query $Q$, 
let $\atoms(Q)$ denote the set of constituting relations; in other words,
we can write $Q$ as $Q = \ \Join_{R\in \atoms(Q)} R$.
For any relation $R$ ($Q$ included),
$\vars(R)$ denotes the set of its attributes.

We assume each input relation $R$ is already indexed using some data 
structure that 
satisfies the following property. The data structure stores a collection 
$\calB(R)$ of gap boxes whose union contains all points in
$\prod_{A \in \vars(R)} \D(A)$ which are {\em not} tuples in $R$.
Note that there can be multiple indices per relation. Gap boxes from all 
those indices contribute to $\calB(R)$.
By filling out the coordinates not in $\vars(R)$ with ``wild cards''
(i.e. each one of those coordinates spans the entire dimension), we can 
without loss of generality view $\calB(R)$ as a collection of gap boxes in 
the output space $\prod_{A\in \vars(Q)} \D(A)$.

We begin with the notion of certificate. The size of the smallest such 
certificate will replace the input size as the measure of complexity of 
an instance in our beyond worst-case results.

\bdefn[Box certificate]
\rev{A {\em box certificate} for $Q$ is a set of gap boxes that are included in the gap boxes from
$\bigcup_{R\in\atoms(Q)} \calB(R)$ and cover every tuple not in the output.
We use $\boxcert(Q)$ (or just $\boxcert$ if $Q$ is clear from the context) to denote a box certificate of minimum size for the 
instance.}
\label{defn:main:boxcert}
\edefn

We would like to stress the point above that the size of the smallest box certificate is intimately tied to the kind of index being used. In particular, for certain instances the certificate sizes might be much smaller for more powerful kinds of indices. This should be contrasted with the worst-case results of~\cite{NPRR,leapfrog} where BTrees with a single sort order are enough to obtain the optimal worst-case results (and using more powerful indices like quad-trees does not improve the results). Further, our algorithms do not assume the knowledge of $\boxcert$ though they implicitly compute a box certificate $\cert$ such that $|\cert|=\tO(\boxcert)$.

For a more thorough discussion of indices, gap boxes, various notions of 
certificates and how they relate to box certificates, see 
Appendix~\ref{sec:cert}.
In particular, we can show that the notion of a box certificate is 
finer than the notion of comparison-based certificate used in~\cite{nnrr}.
%(see Proposition~\ref{prop:C<<Cgao}).

%It turns out that we will only consider special kinds of boxes in this paper, which we define next.
\paragraph*{Dyadic boxes}
%Consider any join query $Q$ with $\vars(Q) = \{A_1,\dots,A_n\}$.
For simplicity, but without any loss of generality, let us assume
the domain of each attribute is the set of all binary strings of length $d$, 
i.e.
$\D(A) = \{0,1\}^d$, for every $A \in \vars(Q).$
This is equivalent to saying that the domain of each attribute is the set of all  
integers from $0$ to $2^d-1$.
Since $d$ is the number of bits needed to encode a data value of the input, 
$d$ is logarithmic in the input size.

\bdefn[Dyadic interval]
A {\em dyadic interval} is a binary string $x$ of length $|x| \leq d$.
%We use $|x|$ to denote its length. 
This interval represents all the  
binary strings $y$ such that $|y| = d$ and $x$ is a prefix of 
$y$. \rev{Translating to the integral domain, let $i$ be the integer corresponding to the string $x$. The dyadic interval represents 
all integers in the range
$[i2^{d-|x|}, (i+1)2^{d-|x|}-1].$}
The empty string $x = \lambda$ is %(integral value $0$) is
a dyadic interval consisting of all possible values in the domain.
(This serves as a wild-card.)
If $|x| = d$, then it is called a {\em unit dyadic interval}, which
represents a point in the domain.
\edefn

\bdefn[Dyadic box] 
Let $\vars(Q) = \{A_1,\dots,A_n\}$. A {\em dyadic box} is an $n$-tuple of 
dyadic intervals: $\mv b = \dbox{x_1, \dots, x_n}$.
If all components of $\mv b$ are unit dyadic intervals, then $\mv b$
represents a {\em point} in the output space.
The dyadic box is the set of all tuples 
$\mv t = (t_1,\dots,t_n) \in \prod_{i=1}^n\mv D(A_i)$
such that $t_i$ belongs to the dyadic interval $x_i$, for all $i\in[n]$.
\edefn

Note again that some dyadic intervals can be $\lambda$, matching
arbitrary domain values; also,
a dyadic box $\mv b$ contains a dyadic box $\mv b'$ if each of
$\mv b$'s components is a prefix of the corresponding component in $\mv b'$.
The set of all dyadic boxes forms a partially ordered set (poset) under
this containment. % relation.

\begin{figure}[!htp]
\centering
\subfloat[\rev{Gap boxes from $(A,B)$-ordered Btree}]{
\begin{tikzpicture}[scale=.8]
\draw [<->] (0, 4.5) -- (0, 0) -- (4.5, 0);
\draw[help lines] (0,0) grid (4,4);
\foreach \i in {0,...,3}
   \node [below] at (\i+0.5, 0) {\scriptsize \i};
\foreach \i in {0,...,3}
   \node [left] at (0, \i+0.5) {\scriptsize \i};
\node [black] at (5,0) {$A$};
\node [black] at (0, 5) {$B$};
\draw[opacity=0] (-1,-1);

\draw[rounded corners=6pt, fill=blue, fill opacity=1] (0, 3) rectangle (1,4);
\draw[rounded corners=6pt, fill=orange, fill opacity=.6, very thick] (0,0) rectangle (1,3);
\draw[rounded corners=6pt, fill=orange, fill opacity=.3, very thick] (1,0) rectangle (4,4);
\label{fig:dyadic-gaps-1}
\end{tikzpicture}
}~~
\subfloat[\rev{Dyadic gap boxes}] {
\begin{tikzpicture}[scale=.8]
\draw [<->] (0, 4.5) -- (0, 0) -- (4.5, 0);
\draw[help lines] (0,0) grid (4,4);

\draw [gray, |-|] (0,-0.2) -- (4,-0.2); 
\node [below] at (2,-.08) {\tiny{$\lambda$}};
\draw [gray, |-|] (0,-0.5) -- (2,-0.5); 
\node [below] at (1,-.37) {\tiny{$0$}};
\draw [gray, |-|] (2,-0.5) -- (4,-0.5); 
\node [below] at (3,-.37) {\tiny{$1$}};
\draw [gray, |-|] (0,-0.8) -- (1,-0.8); 
\node [below] at (0.5,-.75) {\tiny{$00$}};
\draw [gray, |-|] (1,-0.8) -- (2,-0.8); 
\node [below] at (1.5,-.75) {\tiny{$01$}};
\draw [gray, |-|] (2,-0.8) -- (3,-0.8); 
\node [below] at (2.5,-.75) {\tiny{$10$}};
\draw [gray, |-|] (3,-0.8) -- (4,-0.8); 
\node [below] at (3.5,-.75) {\tiny{$11$}};

\begin{scope}[rotate=-90, shift={(-4, -.1)}]
\draw [gray, |-|] (0,-0.2) -- (4,-0.2); 
\node [left] at (2,-.08) {\tiny{$\lambda$}};
\draw [gray, |-|] (0,-0.5) -- (2,-0.5); 
\node [left] at (1,-.37) {\tiny{$1$}};
\draw [gray, |-|] (2,-0.5) -- (4,-0.5); 
\node [left] at (3,-.37) {\tiny{$0$}};
\draw [gray, |-|] (0,-0.8) -- (1,-0.8); 
\node [left] at (0.5,-.75) {\tiny{$11$}};
\draw [gray, |-|] (1,-0.8) -- (2,-0.8); 
\node [left] at (1.5,-.75) {\tiny{$10$}};
\draw [gray, |-|] (2,-0.8) -- (3,-0.8); 
\node [left] at (2.5,-.75) {\tiny{$01$}};
\draw [gray, |-|] (3,-0.8) -- (4,-0.8); 
\node [left] at (3.5,-.75) {\tiny{$00$}};
\end{scope}

\draw[rounded corners=6pt, fill=blue, fill opacity=1] (0, 3) rectangle (1,4);
\draw[rounded corners=6pt, fill=orange, fill opacity=.6, very thick] (0,0) rectangle (1,2);
\draw[rounded corners=6pt, fill=orange, fill opacity=.6, very thick] (0,2) rectangle (1,3);
\draw[rounded corners=6pt, fill=orange, fill opacity=.3, very thick] (1,0) rectangle (2,4);
\draw[rounded corners=6pt, fill=orange, fill opacity=.3, very thick] (2,0) rectangle (4,4);

\node [black] at (5,0) {$A$};
\node [black] at (0, 5) {$B$};
\label{fig:dyadic-gaps-2}
\end{tikzpicture}
}
\caption{\rev{A relation $R(A, B)$ with a single tuple $(0,3)$, its gap boxes, and corresponding dyadic gap boxes.}}
\label{fig:dyadic-gaps}
\end{figure}
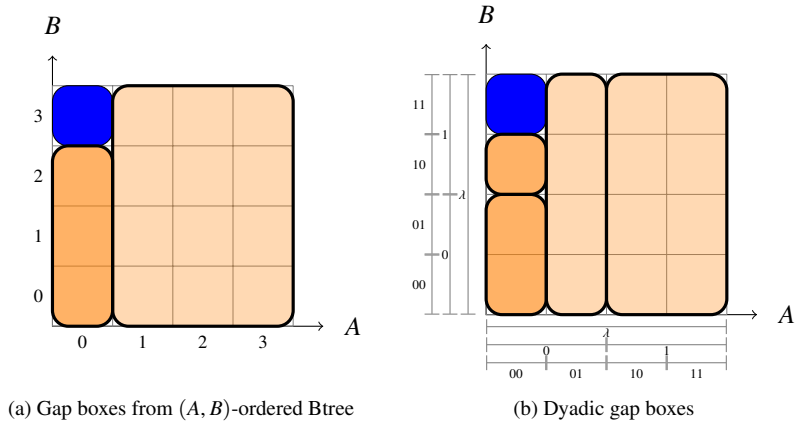

It is straightforward to show that every (not necessarily dyadic) box in $n$
dimensions can be decomposed into a disjoint union of at most $(2d)^n = 
\tO(1)$ dyadic boxes. In particular, 
for every box certificate, there is a dyadic box certificate of size at most 
a factor of $\tO(1)$ larger.
\rev{(See Figure~\ref{fig:dyadic-gaps} for an example.)}
Henceforth, we will assume that all boxes are dyadic boxes. This assumption
is also crucial for the discovery of an optimal box certificate \rev{(i.e. one having minimal size)}.
In particular, the number of dyadic boxes containing a given tuple 
is always at most $\tO(1)$.
(See Appendix~\ref{app:subsec:the case for dyadic} for the details.)

\subsection{The Box Cover Problem}
\label{sec:bcp}

%!TEX root = main.tex

\begin{figure*}[!tbp]
\centering
\subfloat[Gap boxes for $R$]{ %Gap boxes for $R(A,B)=[N/2]\times [N]\setminus [N/2] \cup [N]\setminus [N/2]\times [N/2]$]{
\tdplotsetmaincoords{60}{125}
\begin{tikzpicture}
		[tdplot_main_coords,
			cube/.style={very thick,black},
			grid/.style={very thin,gray},
			axis/.style={->,gray}, scale=.45]

\pgfmathsetmacro{\axesl}{5}
\pgfmathsetmacro{\len}{2}
\pgfmathsetmacro{\wdt}{2}
\pgfmathsetmacro{\hgt}{4}
\pgfmathsetmacro{\shfta}{2}
\pgfmathsetmacro{\shftb}{2}

%%%% Bouding box

\draw[dotted, gray] (4,0,0) -- (4,0,4) -- (0,0,4);
\draw[dotted, gray] (4,0,0) -- (4,0,4) -- (4,4,4) -- (4,4,0) --cycle;
\draw[dotted, gray] (4,4,4) -- (0,4,4) -- (0,4,0) -- (4,4,0) -- cycle;
\draw[dotted, gray] (0,4,4) -- (0,0,4);

\draw (4,2,.1)--(4,2,-.1);
\node[below] at (4,2,-.1) {{\tiny \textcolor{gray}{$2^{d-1}$}}};
\draw (2,3.9,0)--(2,4.1,0);
\node[right] at (2,4.1,0) {{\tiny \textcolor{gray}{$2^{d-1}$}}};

%%%%% Boxes for R(A,B) 

%%Axes

	\draw[axis] (0,0,0) -- (\axesl,0,0) node[anchor=west]{{\footnotesize $A$}};
	\draw[axis] (0,0,0) -- (0,\axesl,0) node[anchor=west]{{\footnotesize $B$}};
	\draw[axis] (0,0,0) -- (0,0,\axesl) node[anchor=west]{{\footnotesize $C$}};

%%Back sides
\fill[opacity=.2, blue] (0,0,0) -- (2,0,0) -- (2,0,4) -- (0,0,4) -- cycle;
\fill[opacity=.2, blue] (0,0,0) -- (0,2,0) -- (0,2,4) -- (0,0,4) -- cycle;

%%Top and bottom
\fill[opacity=.2, blue] (0,0,0) -- (2,0,0) -- (2,2,0) -- (0,2,0) -- cycle;
\fill[opacity=.2, blue] (0,0,4) -- (2,0,4) -- (2,2,4) -- (0,2,4) -- cycle;

%%Front sides
\fill[opacity=.1, blue] (2,0,0) -- (2,0,4) -- (2,2,4) -- (2,2,0)-- cycle;

\begin{scope}[shift={(2,2,0)}]
%%Back sides
\fill[opacity=.2, blue] (0,0,0) -- (2,0,0) -- (2,0,4) -- (0,0,4) -- cycle;
\fill[opacity=.2, blue] (0,0,0) -- (0,2,0) -- (0,2,4) -- (0,0,4) -- cycle;

%%Top and bottom
\fill[opacity=.3, blue] (0,0,0) -- (2,0,0) -- (2,2,0) -- (0,2,0) -- cycle;
\fill[opacity=.2, blue] (0,0,4) -- (2,0,4) -- (2,2,4) -- (0,2,4) -- cycle;

%%Front sides
\fill[opacity=.1, blue] (2,0,0) -- (2,0,4) -- (2,2,4) -- (2,2,0)-- cycle;
\end{scope}

\end{tikzpicture}
\label{fig:bcp-R}
}
\subfloat[Gap boxes for $S$]{
\tdplotsetmaincoords{60}{125}
\begin{tikzpicture}
		[tdplot_main_coords,
			cube/.style={very thick,black},
			grid/.style={very thin,gray},
			axis/.style={->,gray}, scale=.45]

\pgfmathsetmacro{\axesl}{5}
\pgfmathsetmacro{\len}{2}
\pgfmathsetmacro{\wdt}{2}
\pgfmathsetmacro{\hgt}{4}
\pgfmathsetmacro{\shfta}{2}
\pgfmathsetmacro{\shftb}{2}

%%%% Bouding box

\draw[dotted, gray] (4,0,0) -- (4,0,4) -- (0,0,4);
\draw[dotted, gray] (4,0,0) -- (4,0,4) -- (4,4,4) -- (4,4,0) --cycle;
\draw[dotted, gray] (4,4,4) -- (0,4,4) -- (0,4,0) -- (4,4,0) -- cycle;
\draw[dotted, gray] (0,4,4) -- (0,0,4);

%%%%% Boxes for S(B,C) 

%%Axes

	\draw[axis] (0,0,0) -- (\axesl,0,0) node[anchor=west]{{\footnotesize $A$}};
	\draw[axis] (0,0,0) -- (0,\axesl,0) node[anchor=west]{{\footnotesize $B$}};
	\draw[axis] (0,0,0) -- (0,0,\axesl) node[anchor=west]{{\footnotesize $C$}};

\draw (4,2,.1)--(4,2,-.1);
\node[below] at (4,2,-.1) {{\tiny \textcolor{gray}{$2^{d-1}$}}};
\draw (4,0.1,2)--(4,-.1,2);
\node[left] at (4,-.1,2) {{\tiny \textcolor{gray}{$2^{d-1}$}}};

%%Back sides
\fill[opacity=.1, orange] (0,0,0) -- (0,0,2) -- (0,2,2) -- (0,2,0) -- cycle;
\fill[opacity=.2, orange] (0,0,0) -- (4,0,0) -- (4,0,2) -- (0,0,2) -- cycle;

%%Top and bottom
\fill[opacity=.2, orange] (4,0,2) -- (0,0,2) -- (0,2,2) -- (4,2,2) -- cycle;
\fill[opacity=.1, orange] (4,2,2) -- (0,2,2) -- (0,2,0) -- (4,2,0) -- cycle; 

%%Front sides
\fill[opacity=.5, orange] (4,2,2) -- (4,2,0) -- (4,0,0) -- (4,0,2) --cycle;

\begin{scope}[shift={(0,2,2)}]
%%Back sides
\fill[opacity=.4, orange] (0,0,0) -- (0,0,2) -- (0,2,2) -- (0,2,0) -- cycle;
\fill[opacity=.1, orange] (0,0,0) -- (4,0,0) -- (4,0,2) -- (0,0,2) -- cycle;

%%Top and bottom
\fill[opacity=.4, orange] (4,0,2) -- (0,0,2) -- (0,2,2) -- (4,2,2) -- cycle;
\fill[opacity=.4, orange] (4,2,2) -- (0,2,2) -- (0,2,0) -- (4,2,0) -- cycle; 

%%Front sides
\fill[opacity=.7, orange] (4,2,2) -- (4,2,0) -- (4,0,0) -- (4,0,2) --cycle;

\end{scope}

\end{tikzpicture}
%\caption{Gap boxes for $S(B,C)=[N/2]\times [N]\setminus [N/2] \cup [N]\setminus [N/2]\times [N/2]$}
\label{fig:bcp-S}
}
\subfloat[Gap boxes for $T$]{
\tdplotsetmaincoords{60}{125}
\begin{tikzpicture}
		[tdplot_main_coords,
			cube/.style={very thick,black},
			grid/.style={very thin,gray},
			axis/.style={->,gray}, scale=.45]

\pgfmathsetmacro{\axesl}{5}
\pgfmathsetmacro{\len}{2}
\pgfmathsetmacro{\wdt}{2}
\pgfmathsetmacro{\hgt}{4}
\pgfmathsetmacro{\shfta}{2}
\pgfmathsetmacro{\shftb}{2}

%%%% Bouding box

\draw[dotted, gray] (4,0,0) -- (4,0,4) -- (0,0,4);
\draw[dotted, gray] (4,0,0) -- (4,0,4) -- (4,4,4) -- (4,4,0) --cycle;
\draw[dotted, gray] (4,4,4) -- (0,4,4) -- (0,4,0) -- (4,4,0) -- cycle;
\draw[dotted, gray] (0,4,4) -- (0,0,4);

%%%%% Boxes for T(A,C) 

%%Axes

	\draw[axis] (0,0,0) -- (\axesl,0,0) node[anchor=west]{{\footnotesize $A$}};
	\draw[axis] (0,0,0) -- (0,\axesl,0) node[anchor=west]{{\footnotesize $B$}};
	\draw[axis] (0,0,0) -- (0,0,\axesl) node[anchor=west]{{\footnotesize $C$}};

\draw (0,4.1,2)--(0,3.9,2);
\node[right] at (0,4.1,2) {{\tiny \textcolor{gray}{$2^{d-1}$}}};
\draw (2,4.1,0)--(2,3.9,0);
\node[right] at (2,4.1,0) {{\tiny \textcolor{gray}{$2^{d-1}$}}};

%%Back sides
\fill[opacity=.1, green] (0,0,0) -- (0,0,2) -- (2,0,2) -- (2,0,0) -- cycle;
\fill[opacity=.1, green] (0,0,0) -- (0,0,2) -- (0,4,2) -- (0,4,0) -- cycle;

%%Top and bottom
\fill[opacity=.1, green] (0,0,2) -- (0,4,2) -- (2,4,2) -- (2,0,2) --  cycle;

%%Front sides
\fill[opacity=.2, green] (2,4,2) -- (2,0,2) --  (2,0,0) -- (2,4,0) -- cycle;
\fill[opacity=.3, green] (2,4,2) -- (2,4,0) -- (0,4,0) -- (0,4,2) -- cycle;

\begin{scope}[shift={(2,0,2)}]
%%Back sides
\fill[opacity=.3, green] (0,0,0) -- (0,0,2) -- (2,0,2) -- (2,0,0) -- cycle;
\fill[opacity=.1, green] (0,0,0) -- (0,0,2) -- (0,4,2) -- (0,4,0) -- cycle;

%%Top and bottom
\fill[opacity=.4, green] (0,0,2) -- (0,4,2) -- (2,4,2) -- (2,0,2) --  cycle;

%%Front sides
\fill[opacity=.4, green] (2,4,2) -- (2,0,2) --  (2,0,0) -- (2,4,0) -- cycle;
\fill[opacity=.3, green] (2,4,2) -- (2,4,0) -- (0,4,0) -- (0,4,2) -- cycle;

\end{scope}

\end{tikzpicture}
%\caption{Gap boxes for $T(A,C)=[N/2]\times [N]\setminus [N/2] \cup [N]\setminus [N/2]\times [N/2]$}
\label{fig:bcp-T}
}
\subfloat[Union of all gap boxes]{
\tdplotsetmaincoords{60}{125}
\begin{tikzpicture}
		[tdplot_main_coords,
			cube/.style={very thick,black},
			grid/.style={very thin,gray},
			axis/.style={->,gray}, scale=.45]

\pgfmathsetmacro{\axesl}{5}
\pgfmathsetmacro{\len}{2}
\pgfmathsetmacro{\wdt}{2}
\pgfmathsetmacro{\hgt}{4}
\pgfmathsetmacro{\shfta}{2}
\pgfmathsetmacro{\shftb}{2}

%%%% Bouding box

\draw[dotted, gray] (4,0,0) -- (4,0,4) -- (0,0,4);
\draw[dotted, gray] (4,0,0) -- (4,0,4) -- (4,4,4) -- (4,4,0) --cycle;
\draw[dotted, gray] (4,4,4) -- (0,4,4) -- (0,4,0) -- (4,4,0) -- cycle;
\draw[dotted, gray] (0,4,4) -- (0,0,4);

%%Axes

	\draw[axis] (0,0,0) -- (\axesl,0,0) node[anchor=west]{{\footnotesize $A$}};
	\draw[axis] (0,0,0) -- (0,\axesl,0) node[anchor=west]{{\footnotesize $B$}};
	\draw[axis] (0,0,0) -- (0,0,\axesl) node[anchor=west]{{\footnotesize $C$}};

%%%%% Boxes for R(A,B) 

%%Back sides
\fill[opacity=.2, blue] (0,0,0) -- (2,0,0) -- (2,0,4) -- (0,0,4) -- cycle;
\fill[opacity=.2, blue] (0,0,0) -- (0,2,0) -- (0,2,4) -- (0,0,4) -- cycle;

%%Top and bottom
\fill[opacity=.2, blue] (0,0,0) -- (2,0,0) -- (2,2,0) -- (0,2,0) -- cycle;
\fill[opacity=.2, blue] (0,0,4) -- (2,0,4) -- (2,2,4) -- (0,2,4) -- cycle;

%%Front sides
\fill[opacity=.1, blue] (2,0,0) -- (2,0,4) -- (2,2,4) -- (2,2,0)-- cycle;

\begin{scope}[shift={(2,2,0)}]
%%Back sides
\fill[opacity=.2, blue] (0,0,0) -- (2,0,0) -- (2,0,4) -- (0,0,4) -- cycle;
\fill[opacity=.2, blue] (0,0,0) -- (0,2,0) -- (0,2,4) -- (0,0,4) -- cycle;

%%Top and bottom
\fill[opacity=.3, blue] (0,0,0) -- (2,0,0) -- (2,2,0) -- (0,2,0) -- cycle;
\fill[opacity=.2, blue] (0,0,4) -- (2,0,4) -- (2,2,4) -- (0,2,4) -- cycle;

%%Front sides
\fill[opacity=.1, blue] (2,0,0) -- (2,0,4) -- (2,2,4) -- (2,2,0)-- cycle;
\end{scope}

%%%%% Boxes for S(B,C) 

%%Back sides
\fill[opacity=.1, orange] (0,0,0) -- (0,0,2) -- (0,2,2) -- (0,2,0) -- cycle;
\fill[opacity=.2, orange] (0,0,0) -- (4,0,0) -- (4,0,2) -- (0,0,2) -- cycle;

%%Top and bottom
\fill[opacity=.2, orange] (4,0,2) -- (0,0,2) -- (0,2,2) -- (4,2,2) -- cycle;
\fill[opacity=.1, orange] (4,2,2) -- (0,2,2) -- (0,2,0) -- (4,2,0) -- cycle; 

%%Front sides
\fill[opacity=.5, orange] (4,2,2) -- (4,2,0) -- (4,0,0) -- (4,0,2) --cycle;

\begin{scope}[shift={(0,2,2)}]
%%Back sides
\fill[opacity=.4, orange] (0,0,0) -- (0,0,2) -- (0,2,2) -- (0,2,0) -- cycle;
\fill[opacity=.1, orange] (0,0,0) -- (4,0,0) -- (4,0,2) -- (0,0,2) -- cycle;

%%Top and bottom
\fill[opacity=.4, orange] (4,0,2) -- (0,0,2) -- (0,2,2) -- (4,2,2) -- cycle;
\fill[opacity=.4, orange] (4,2,2) -- (0,2,2) -- (0,2,0) -- (4,2,0) -- cycle; 

%%Front sides
\fill[opacity=.7, orange] (4,2,2) -- (4,2,0) -- (4,0,0) -- (4,0,2) --cycle;

\end{scope}

%%%%% Boxes for T(A,C)

%%Back sides
\fill[opacity=.1, green] (0,0,0) -- (0,0,2) -- (2,0,2) -- (2,0,0) -- cycle;
\fill[opacity=.1, green] (0,0,0) -- (0,0,2) -- (0,4,2) -- (0,4,0) -- cycle;

%%Top and bottom
\fill[opacity=.1, green] (0,0,2) -- (0,4,2) -- (2,4,2) -- (2,0,2) --  cycle;

%%Front sides
\fill[opacity=.2, green] (2,4,2) -- (2,0,2) --  (2,0,0) -- (2,4,0) -- cycle;
\fill[opacity=.3, green] (2,4,2) -- (2,4,0) -- (0,4,0) -- (0,4,2) -- cycle;

\begin{scope}[shift={(2,0,2)}]
%%Back sides
\fill[opacity=.3, green] (0,0,0) -- (0,0,2) -- (2,0,2) -- (2,0,0) -- cycle;
\fill[opacity=.1, green] (0,0,0) -- (0,0,2) -- (0,4,2) -- (0,4,0) -- cycle;

%%Top and bottom
\fill[opacity=.4, green] (0,0,2) -- (0,4,2) -- (2,4,2) -- (2,0,2) --  cycle;

%%Front sides
\fill[opacity=.4, green] (2,4,2) -- (2,0,2) --  (2,0,0) -- (2,4,0) -- cycle;
\fill[opacity=.3, green] (2,4,2) -- (2,4,0) -- (0,4,0) -- (0,4,2) -- cycle;

\end{scope}

\end{tikzpicture}
%\caption{$R\Join S\Join T$}
\label{fig:bcp-triang}
}
\caption{\rev{The first three figures show the gap boxes for three relations $R(A,B), S(B,C)$ and $T(A,C)$. Relation $R(A, B)$ contains all tuples $(a,b)\in\{0,1\}^d\times \{0,1\}^d$ such that the first bits (or MSBs) of $a$ and $b$ are complements of each other. Hence, the gap boxes of $R$ are $\dbox{0, 0}$ and $\dbox{1, 1}$, which become $\dbox{0, 0, \lambda}$ and $\dbox{1, 1, \lambda}$ after extending them along the $C$ attribute. Relations $S$ and $T$ are identical to $R$ but with different attributes.}
The last figure shows the union of all the gap boxes. Since the boxes cover all of the bounding box (which is denoted by the dashed gray box), the output is empty.}
\label{fig:triang-bcp}
\end{figure*}
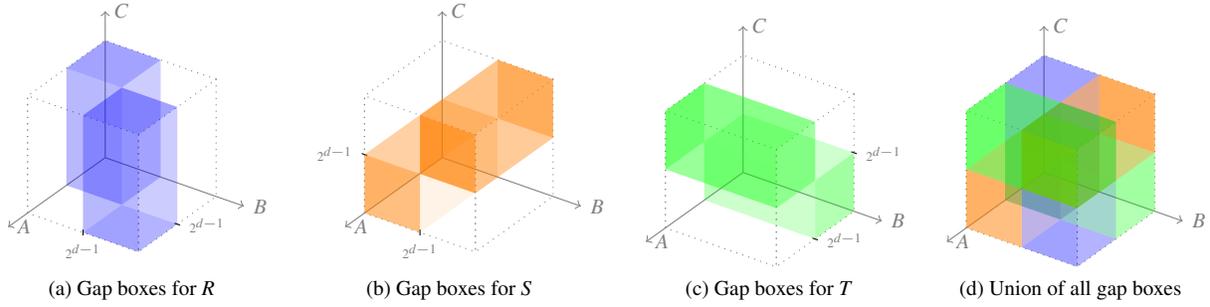

We assume the input index data structure(s) for an input
relation $R$ can return in $\tO(1)$-time the set of all dyadic gap boxes 
in $\calB(R)$ containing a given tuple in $\prod_{A\in\vars(R)}\D(A)$. 
This assumption holds for most of the common indices in relational database 
management systems such as BTree or trie.
The objective of a general join algorithm is to list the set of all output tuples.
Our join algorithm will attempt to take full advantage of the gaps stored in
the input indices: it tries to compute/infer a collection of dyadic
boxes whose union contains all tuples in $\prod_{A\in\vars(Q)} \D(A)$ 
except the output tuples.
(The smallest such collection is called a (dyadic) box certificate as defined
in Definition~\ref{defn:main:boxcert}.)
Recall that an output tuple {\em is also} a (unit) dyadic box.
Hence, the output dyadic boxes and the gap boxes together
fill the entire output space.
Consequently, we can think of a join algorithm as an algorithm that tries
as fast as possible to fill up the entire output space with dyadic boxes
of various shapes and sizes. %This is where the name ``Tetris'' comes from.

Abstracting away from the above idea, we first define a problem called
the {\em box-cover problem} (or $\bcp$). 
%Then, we will explain how 
%\tetris solves $\bcp$, and later in the paper how we can use \tetris to solve the 
%join evaluation problem.

\bdefn[Box Cover Problem]
Given a set $\calA$ of (dyadic) boxes, list all tuples {\em not} covered
by any box in $\calA$, i.e. list all tuples $\mv t$ such that
$\mv t \notin \mv b$ for every $\mv b \in \calA$.
\rev{Define the (box) {\em certificate} for the instance $\calA$ of \bcp, denoted by $\boxcert(\calA)$ (or just $\boxcert$ if $\calA$ is clear from the context), 
to be the smallest subset of $\calA$ such that 
$\bigcup_{\mv b\in \boxcert(\calA)} \mv b=\bigcup_{\mv b\in\calA} \mv b$.} %We will also use $Z$ to denote the number of tuples in the output of the $\bcp$ instance.
\label{defn:bcp-boxcert}
\edefn

\bdefn[Boolean Box Cover Problem]
Given a set $\calA$ of (dyadic) boxes, determine whether their union covers the entire output space, i.e. $\bigcup_{\mv b\in\calA}\mv b=\UB.$
\label{defn:boolean-bcp}
\edefn

Given a join query $Q$ \rev{(as defined in Section~\ref{sec:join})}, $\calB(Q)$ denotes the set of all gap boxes 
from the input indices, i.e. 
$\calB(Q) = \bigcup_{R\in \atoms(Q)} \calB(R).$
The following is straightforwardly true.
\bprop
\label{prop:bcp=join}
On input $\calA = \calB(Q)$, the output of \bcp is exactly the same as the 
output of the join query $Q$. 
And, $|\boxcert(\calB(Q))| = |\boxcert(Q)|$.
\eprop

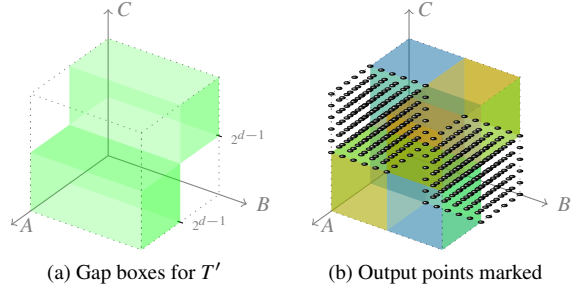
\begin{figure}[!htp]
\begin{center}
\subfloat[Gap boxes for $T'$]{
\tdplotsetmaincoords{60}{125}
\begin{tikzpicture}
		[tdplot_main_coords,
			cube/.style={very thick,black},
			grid/.style={very thin,gray},
			axis/.style={->,gray}, scale=.45]

\pgfmathsetmacro{\axesl}{5}
\pgfmathsetmacro{\len}{2}
\pgfmathsetmacro{\wdt}{2}
\pgfmathsetmacro{\hgt}{4}
\pgfmathsetmacro{\shfta}{2}
\pgfmathsetmacro{\shftb}{2}

%%%% Bouding box

\draw[dotted, gray] (4,0,0) -- (4,0,4) -- (0,0,4);
\draw[dotted, gray] (4,0,0) -- (4,0,4) -- (4,4,4) -- (4,4,0) --cycle;
\draw[dotted, gray] (4,4,4) -- (0,4,4) -- (0,4,0) -- (4,4,0) -- cycle;
\draw[dotted, gray] (0,4,4) -- (0,0,4);

%%%%% Boxes for T(A,C) 

%%Axes

	\draw[axis] (0,0,0) -- (\axesl,0,0) node[anchor=west]{{\footnotesize $A$}};
	\draw[axis] (0,0,0) -- (0,\axesl,0) node[anchor=west]{{\footnotesize $B$}};
	\draw[axis] (0,0,0) -- (0,0,\axesl) node[anchor=west]{{\footnotesize $C$}};

\draw (0,4.1,2)--(0,3.9,2);
\node[right] at (0,4.1,2) {{\tiny \textcolor{gray}{$2^{d-1}$}}};
\draw (2,4.1,0)--(2,3.9,0);
\node[right] at (2,4.1,0) {{\tiny \textcolor{gray}{$2^{d-1}$}}};

%%Back sides
\fill[opacity=.1, green] (0,0,2) -- (0,0,4) -- (2,0,4) -- (2,0,2) -- cycle;
\fill[opacity=.1, green] (0,0,2) -- (0,0,4) -- (0,4,4) -- (0,4,2) -- cycle;

%%Top and bottom
\fill[opacity=.1, green] (0,0,4) -- (0,4,4) -- (2,4,4) -- (2,0,4) --  cycle;

%%Front sides
\fill[opacity=.2, green] (2,4,4) -- (2,0,4) --  (2,0,2) -- (2,4,2) -- cycle;
\fill[opacity=.3, green] (2,4,4) -- (2,4,2) -- (0,4,2) -- (0,4,4) -- cycle;

\begin{scope}[shift={(2,0,-2)}]

%%Back sides
\fill[opacity=.1, green] (0,0,2) -- (0,0,4) -- (2,0,4) -- (2,0,2) -- cycle;
\fill[opacity=.1, green] (0,0,2) -- (0,0,4) -- (0,4,4) -- (0,4,2) -- cycle;

%%Top and bottom
\fill[opacity=.3, green] (0,0,4) -- (0,4,4) -- (2,4,4) -- (2,0,4) --  cycle;

%%Front sides
\fill[opacity=.2, green] (2,4,4) -- (2,0,4) --  (2,0,2) -- (2,4,2) -- cycle;
\fill[opacity=.4, green] (2,4,4) -- (2,4,2) -- (0,4,2) -- (0,4,4) -- cycle;

\end{scope}

\end{tikzpicture}
%\end{center}
%\caption{Gap boxes for $T'(A,C)=[N/2]\times [N/2]\cup [N]\setminus [N/2] \times [N]\setminus [N/2]$}
\label{fig:bcp-T'}
%\end{figure*}
}
\subfloat[Output points marked]{
%\begin{figure*}[!htp]
%\begin{center}
\tdplotsetmaincoords{60}{125}
\begin{tikzpicture}
		[tdplot_main_coords,
			cube/.style={very thick,black},
			grid/.style={very thin,gray},
			axis/.style={->,gray}, scale=.45]

\pgfmathsetmacro{\axesl}{5}
\pgfmathsetmacro{\len}{2}
\pgfmathsetmacro{\wdt}{2}
\pgfmathsetmacro{\hgt}{4}
\pgfmathsetmacro{\shfta}{2}
\pgfmathsetmacro{\shftb}{2}

%%%% Bouding box

\draw[dotted, gray] (4,0,0) -- (4,0,4) -- (0,0,4);
\draw[dotted, gray] (4,0,0) -- (4,0,4) -- (4,4,4) -- (4,4,0) --cycle;
\draw[dotted, gray] (4,4,4) -- (0,4,4) -- (0,4,0) -- (4,4,0) -- cycle;
\draw[dotted, gray] (0,4,4) -- (0,0,4);

%%Axes

	\draw[axis] (0,0,0) -- (\axesl,0,0) node[anchor=west]{{\footnotesize $A$}};
	\draw[axis] (0,0,0) -- (0,\axesl,0) node[anchor=west]{{\footnotesize $B$}};
	\draw[axis] (0,0,0) -- (0,0,\axesl) node[anchor=west]{{\footnotesize $C$}};

%%%%% Boxes for R(A,B) 

%%Back sides
\fill[opacity=.2, blue] (0,0,0) -- (2,0,0) -- (2,0,4) -- (0,0,4) -- cycle;
\fill[opacity=.2, blue] (0,0,0) -- (0,2,0) -- (0,2,4) -- (0,0,4) -- cycle;

%%Top and bottom
\fill[opacity=.2, blue] (0,0,0) -- (2,0,0) -- (2,2,0) -- (0,2,0) -- cycle;
\fill[opacity=.2, blue] (0,0,4) -- (2,0,4) -- (2,2,4) -- (0,2,4) -- cycle;

%%Front sides
\fill[opacity=.1, blue] (2,0,0) -- (2,0,4) -- (2,2,4) -- (2,2,0)-- cycle;

\begin{scope}[shift={(2,2,0)}]
%%Back sides
\fill[opacity=.2, blue] (0,0,0) -- (2,0,0) -- (2,0,4) -- (0,0,4) -- cycle;
\fill[opacity=.2, blue] (0,0,0) -- (0,2,0) -- (0,2,4) -- (0,0,4) -- cycle;

%%Top and bottom
\fill[opacity=.3, blue] (0,0,0) -- (2,0,0) -- (2,2,0) -- (0,2,0) -- cycle;
\fill[opacity=.2, blue] (0,0,4) -- (2,0,4) -- (2,2,4) -- (0,2,4) -- cycle;

%%Front sides
\fill[opacity=.1, blue] (2,0,0) -- (2,0,4) -- (2,2,4) -- (2,2,0)-- cycle;
\end{scope}

%%%%% Boxes for S(B,C) 

%%Back sides
\fill[opacity=.1, orange] (0,0,0) -- (0,0,2) -- (0,2,2) -- (0,2,0) -- cycle;
\fill[opacity=.2, orange] (0,0,0) -- (4,0,0) -- (4,0,2) -- (0,0,2) -- cycle;

%%Top and bottom
\fill[opacity=.2, orange] (4,0,2) -- (0,0,2) -- (0,2,2) -- (4,2,2) -- cycle;
\fill[opacity=.1, orange] (4,2,2) -- (0,2,2) -- (0,2,0) -- (4,2,0) -- cycle; 

%%Front sides
\fill[opacity=.5, orange] (4,2,2) -- (4,2,0) -- (4,0,0) -- (4,0,2) --cycle;

\begin{scope}[shift={(0,2,2)}]
%%Back sides
\fill[opacity=.4, orange] (0,0,0) -- (0,0,2) -- (0,2,2) -- (0,2,0) -- cycle;
\fill[opacity=.1, orange] (0,0,0) -- (4,0,0) -- (4,0,2) -- (0,0,2) -- cycle;

%%Top and bottom
\fill[opacity=.4, orange] (4,0,2) -- (0,0,2) -- (0,2,2) -- (4,2,2) -- cycle;
\fill[opacity=.4, orange] (4,2,2) -- (0,2,2) -- (0,2,0) -- (4,2,0) -- cycle; 

%%Front sides
\fill[opacity=.7, orange] (4,2,2) -- (4,2,0) -- (4,0,0) -- (4,0,2) --cycle;

\end{scope}

%%%%% Boxes for T(A,C)

%%Back sides
\fill[opacity=.1, green] (0,0,2) -- (0,0,4) -- (2,0,4) -- (2,0,2) -- cycle;
\fill[opacity=.1, green] (0,0,2) -- (0,0,4) -- (0,4,4) -- (0,4,2) -- cycle;

%%Top and bottom
\fill[opacity=.1, green] (0,0,4) -- (0,4,4) -- (2,4,4) -- (2,0,4) --  cycle;

%%Front sides
\fill[opacity=.2, green] (2,4,4) -- (2,0,4) --  (2,0,2) -- (2,4,2) -- cycle;
\fill[opacity=.3, green] (2,4,4) -- (2,4,2) -- (0,4,2) -- (0,4,4) -- cycle;

\begin{scope}[shift={(2,0,-2)}]

%%Back sides
\fill[opacity=.1, green] (0,0,2) -- (0,0,4) -- (2,0,4) -- (2,0,2) -- cycle;
\fill[opacity=.1, green] (0,0,2) -- (0,0,4) -- (0,4,4) -- (0,4,2) -- cycle;

%%Top and bottom
\fill[opacity=.3, green] (0,0,4) -- (0,4,4) -- (2,4,4) -- (2,0,4) --  cycle;

%%Front sides
\fill[opacity=.2, green] (2,4,4) -- (2,0,4) --  (2,0,2) -- (2,4,2) -- cycle;
\fill[opacity=.4, green] (2,4,4) -- (2,4,2) -- (0,4,2) -- (0,4,4) -- cycle;

\end{scope}

%%%% Show the grid points in the output

\foreach \x in {2,2.4,...,4}
   \foreach \y in {0,0.4,...,2}
     \foreach \z in {2,2.4,...,4}
        \shade[shading=ball, ball color=black] (\x,\y,\z) circle (.1);

\foreach \x in {0,0.4,...,2}
   \foreach \y in {2,2.4,...,4}
     \foreach \z in {0,0.4,...,2}
        \shade[shading=ball, ball color=black] (\x,\y,\z) circle (.1);

\end{tikzpicture}
\label{fig:bcp-triangnon-empty-gridpts}
}
\end{center}
\caption{Boxes for \bcp\ instance corresponding to $R\Join S\Join T'$ with $R$ and $S$  as in Figure~\ref{fig:triang-bcp} \rev{and $T'(A, C)$ contains all tuples $(a,c)\in\{0,1\}^d\times \{0,1\}^d$ such that the first bits (or MSBs) of $a$ and $c$ are the same (hence the gap boxes for $T'$ are $\dbox{0, \lambda, 1}$ and $\dbox{1, \lambda, 0}$, which are shown in the left figure).} The union of the boxes along  with the output tuples are shown in the right figure.}
\label{fig:another-bcp-eg}
\end{figure}

We illustrate the connection between 
%See Figure~\ref{fig:triang-bcp} for the correspondence between an instance 
the triangle query $Q_\Delta = R(A,B)\Join S(B,C)\Join T(A,C)$ and 
the corresponding instance for the \bcp. Consider the instance
for $Q_{\Delta}$ in which $R$ has pairs $(a,b)\in\{0,1\}^d\times \{0,1\}^d$ such that the first bits (or MSBs) of $a$ and $b$ are complements of each other. In this case, the gaps in $R$ can be represented in a dyadic tree as depicted in Figure~\ref{fig:bcp-R}: there are two gap boxes corresponding to all triples $(a,b,c)$ such that the first bits of $a$ and $b$ are $0$ and $1$ respectively.\footnote{By contrast one gap box in Figure~\ref{fig:bcp-R} would correspond to roughly $2^{d-1}$ gap boxes if $R$ was stored in a BTree.} Further, let $(b,c)\in S$ ($(a,c)\in T$ resp.) if and only if the first bits of $b$ and $c$ ($a$ and $c$ resp.)  are different. The corresponding gap boxes are depicted in Figures~\ref{fig:bcp-S} and~\ref{fig:bcp-T}. Then the \bcp\ instance corresponds to the six gap boxes the union of which covers the entire output space (as depicted in Figure~\ref{fig:bcp-triang}), since the output of $Q_{\Delta}$ is empty for the given instance.
See Figure~\ref{fig:another-bcp-eg} for 
another instance for the same join query when the output is non-empty.

\bdefn[Support of a dyadic box]
Let $\mv b = \dbox{x_1,\dots,x_n}$ be a dyadic box. Its {\em support},
denoted by $\support(\mv b)$, is the
set of all attributes $A_i$ for which $x_i \neq \lambda$.
It follows that, if $\mv b \in \calB(R)$ for some relation $R$,
then the $\support(\mv b) \subseteq \vars(R)$.
\edefn

\bdefn[Supporting hypergraph of a set of boxes]
Let $\calA$ be a collection of dyadic boxes. The {\em supporting hypergraph of
$\calA$}, denoted by $\calH(\calA)$, is the hypergraph whose vertex set is
the set $\calV$ of all attributes participating in boxes of $\calA$,
and whose edge set is the set of all $\support(\mv b)$, $\mv b \in \calA$.
\edefn

\rev{
\begin{example}
Consider the relations depicted in Figure~\ref{fig:triang-bcp}.
The gap boxes for $R(A, B)$ are $\dbox{0,0,\lambda}$ and $\dbox{1,1,\lambda}$, the support of each is $\{A, B\}\subseteq \vars(R)$. Similarly, gap boxes of $S$ and $T$ have supports $\{B,C\}$ and $\{A,C\}$ respectively.
The supporting hypergraph of gap boxes from all three relations has vertices $\calV=\{A, B, C\}$ and hyperedges $\calE=\left\{\{A,B\},\{B,C\},\{A,C\}\right\}$.
\end{example}
}

\bprop\label{prop:tw box vs tw Q}
Let $Q$ be any join query, and $\tw$ denote {\em
tree-width}, then $\tw(\calH(\calB(Q))) \leq \tw(Q)$.
\eprop
\bp
For every box $\mv b \in \calB(Q)$, we have $\support(\mv b) \subseteq \vars(R)$
for some $R \in \atoms(Q)$. Thus, every edge of the hypergraph $\calH(\calB(Q))$
is a subset of some edge of the hypergraph of $Q$. 
This means every tree decomposition of the hypergraph of $Q$ is a tree
decomposition of the hypergraph $\calH(\calB(Q))$. The proposition follows
trivially.
\ep

A dyadic segment $x$ is {\em non-trivial} if $x \neq \lambda$ and $x$ is
not a unit segment.
Let $\calA$ be a set of dyadic boxes on attribute set 
$\calV = \{A_1,\dots,A_n\}$.
A {\em global attribute order} (GAO) is an ordering $\sigma$
of attributes in $\calV$. 

\bdefn[GAO-consistent boxes]
\label{defn:GAO-consistent-box}
Let $\calA$ be a set of dyadic boxes on $\calV$ and $\sigma$ be a GAO
on $\calV$. Then, $\calA$ is said to be {\em $\sigma$-consistent}
if the following conditions are met:
%\bi
 (a) For every box $\mv b  = \dbox{x_1,\dots,x_n} \in \calA$, 
     there is at most one $x_i$ for which $x_i$ is non-trivial.
 (b) For every box $\mv b  = \dbox{x_1,\dots,x_n} \in \calA$, 
     if $x_i$ is non-trivial, then $x_j = \lambda$ for all $j$ such
     that $A_j$ comes after $A_i$ in $\sigma$.
%\ei
\edefn
Note that
if $Q$ is a join query whose input relations are indexed consistently with
a GAO $\sigma$,
then $\calB(Q)$ is $\sigma$-consistent. (See \cite{nnrr} for the definition
of GAO-consistent indices. In short, the search tree for each relation
is indexed using an attribute order consistent with the GAO.)

\section{Upper Bounds} 

%!TEX root = main.tex

%\cmr{Here, we show the worst-case results in one
%  subsection. Then, we show the beyond worst case results. The
%  lowerbounds may be in this section or the next?}

%\ar{Still have to make my pass here.}

We formally define the notion of geometric resolution in Section~\ref{sec:geo-res}. Our main algorithm \tetris\ is presented in Section~\ref{sec:the-algo}. We present rederivations of existing results using \tetris\ in Section~\ref{SEC:WORST-CASE} (worst-case results) and in Section~\ref{SEC:BEYOND-WC} (beyond-worst case results, which recover and generalize results from~\cite{nnrr}). Finally, we present our new beyond worst-case result that works for arbitrary queries in Section~\ref{SEC:CERT-N/2}.

\subsection{Geometric Resolution}
\label{sec:geo-res}

Our algorithm uses the framework of {\em geometric resolution}, which is a 
special case of logical resolution. 
The two input {\em clauses} to geometric resolution 
are two dyadic boxes, say,
\[\mv w_1=\dbox{y_1,\dots,y_n} \text{ and } \mv w_2 =\dbox{z_1,\dots,z_n}\]
that have to satisfy the following two properties:
%\begin{enumerate}
(1) There exists a position $\ell\in [n]$ and a string $x$ such that $y_{\ell}=x0$ and $z_{\ell}=x1$ (where $x$ can be $\lambda$ and $xb$ denotes the concatenation of string $x$ and bit $b$); and
(2) For every other $j\in[n]\setminus \{\ell\}$, either $y_j$ is a prefix of $z_j$ or $z_j$ is a prefix of $y_j$.
%\end{enumerate}

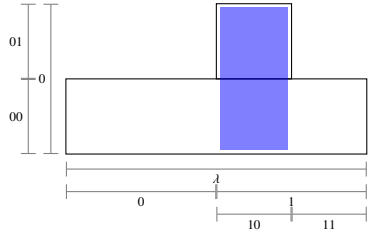
\begin{figure}[!ht]
\begin{center}
\begin{tikzpicture}

%%%% The dyadic boxes

\draw (0,0) rectangle (4,1);
\draw (2,1) rectangle (3,2);
\fill[blue, opacity=.5] (2.05,0.05) rectangle (2.95,1.95);

%%%%%% The dyadic intervals
\draw [gray, |-|] (0,-0.2) -- (4,-0.2); 
\node [below] at (2,-.15) {\tiny{$\lambda$}};
\draw [gray, |-|] (0,-0.5) -- (2,-0.5); 
\node [below] at (1,-.45) {\tiny{$0$}};
\draw [gray, |-|] (2,-0.5) -- (4,-0.5); 
\node [below] at (3,-.45) {\tiny{$1$}};
\draw [gray, |-|] (2,-0.8) -- (3,-0.8); 
\node [below] at (2.5,-.75) {\tiny{$10$}};
\draw [gray, |-|] (3,-0.8) -- (4,-0.8); 
\node [below] at (3.5,-.75) {\tiny{$11$}};

\draw [gray, |-|] (-0.2,0) -- (-0.2,2);
\node [left] at (-.15, 1) {\tiny{$0$}};
\draw [gray, |-|] (-0.5,0) -- (-0.5,1);
\node [left] at (-.45, .5) {\tiny{$00$}};
\draw [gray, |-|] (-0.5,1) -- (-0.5,2);
\node [left] at (-.45, 1.5) {\tiny{$01$}};
\end{tikzpicture}
\end{center}
\caption{Geometric resolution on the vertical axis between two dyadic rectangles $\dbox{\lambda,00}$ (bottom box) and $\dbox{10,01}$ (top box). The resolution result ($\dbox{10,0}$) is highlighted and is drawn slightly smaller than its correct size for illustration purposes. }
\label{fig:main:resolution}
\end{figure}

The result of the geometric resolution or the {\em resolvent} is the dyadic box
\[\mv w=\dbox{y_1\cap z_1,\dots,y_{\ell-1}\cap z_{\ell-1},x,y_{\ell+1}\cap z_{\ell+1},\dots,y_n\cap z_n},\]
where we use $y_i \cap z_i$ to denote the {\em longer} of the two
strings $y_i$, $z_i$. Geometrically, $\mv w_1$ and $\mv w_2$ are adjacent in
the $\ell$th dimension, and in the other dimensions we are taking the intersection
of those two dyadic segments which are contained in one another. 
For the rest of the paper, unless we explicitly mention otherwise, whenever 
we say resolution we mean geometric resolution.
Pictorially this can be visualized for $n=2$ as in Figure~\ref{fig:main:resolution}.

\rev{We briefly explain the name `resolution'. In propositional logic, the 
resolution of two clauses $D_1$ and $D_2$ is a clause $D$ such that every 
truth assignment satisfying both $D_1$ and $D_2$ must satisfy $D$ (and $D$ has the minimal number of literals possible). 
The geometric resolution of two boxes $\mv w_1$ and $\mv w_2$ is a 
box $\mv w$ such that every point covered by neither $\mv w_1$
nor $\mv w_2$ must not be covered by $\mv w$ (and $\mv w$ is maximal).}

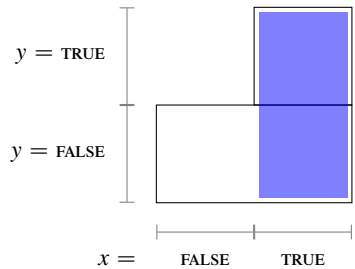
\begin{figure}[!ht]
\begin{center}
\begin{tikzpicture}[scale =1.3, every node/.style={transform shape}]
%%%% The dyadic boxes

\draw (0,0) rectangle (2,1);
\draw (1,1) rectangle (2,2);
\fill[blue, opacity=.5] (1.05,0.05) rectangle (1.95,1.95);

%%%%%% The dyadic intervals
\node [anchor=base, below] at (-.4,-.4) {\scriptsize{$x=$}};
\draw [gray, |-|] (0,-0.3) -- (1,-0.3); 
\node [anchor=base, below] at (0.5,-.4) {\scriptsize{$\false$}};
\draw [gray, |-|] (1,-0.3) -- (2,-0.3); 
\node [anchor=base, below] at (1.5,-.4) {\scriptsize{$\true$}};

\draw [gray, |-|] (-0.3,0) -- (-0.3,1);
\node [left] at (-.4, .5) {\scriptsize{$y=\false$}};
\draw [gray, |-|] (-0.3,1) -- (-0.3,2);
\node [left] at (-.4, 1.5) {\scriptsize{$y=\true$}};
\end{tikzpicture}
\end{center}
\caption{\rev{Geometric interpretation of the logical resolution of two clauses $D_1=(y)$ and $D_2=(\bar x \vee \bar y)$.
The negation of $D_1$ is the conjunction $C_1=(\bar y)$, corresponding to the bottom rectangle. 
The negation of $D_2$ is the conjunction $C_2=(x \wedge y)$, corresponding to the top rectangle.
The geometric resolution of the two rectangles, which is highlighted, corresponds to the conjunction $C=(x)$, whose negation is the clause $D=(\bar x)$, which is exactly the logical resolution of the two clauses $D_1$ and $D_2$.
(Compare to Figure~\ref{fig:main:resolution}.)}}
\label{fig:main:log-resolution}
\end{figure}

\rev{In particular, logical resolution has a geometric interpretation. The negation of a (disjunctive) clause is a conjunction corresponding to a box in the Boolean cube. The logical resolution of two clauses is a clause corresponding to a box which is the geometric resolution of the two boxes corresponding to the two clauses. See Figure~\ref{fig:main:log-resolution}.}

\rev{
The following example explains the opposite connection: It explains the logical interpretation of geometric resolution.
\begin{example}[Geometric resolution is a special case of logical resolution] Consider the geometric resolution depicted in Figure~\ref{fig:main:resolution} between the two dyadic rectangles
$\mv w_1 = \dbox{\lambda,00}$ and $\mv w_2 = \dbox{10,01}$. The $X$-dimension (i.e. the horizontal one) is encoded using two bits.
Let $x_1, x_2$ be the truth values corresponding to those two bits (i.e. $\true$ if the bit is $1$ and $\false$ otherwise).
Similarly, the (vertical) $Y$-dimension is encoded using two bits, corresponding to $y_1$, $y_2$.
The rectangle $\mv w_1$ corresponds to the conjunctive clause
\[C_1 = (\bar y_1 \wedge \bar y_2),\]
whose negation is the  (disjunctive) clause
\[D_1 = (y_1 \vee y_2).\]
Similarly, $\mv w_2$ corresponds to the conjunctive clause
\[C_2=(x_1 \wedge \bar x_2 \wedge \bar y_1 \wedge y_2),\]
whose negation is the clause
\[D_2=(\bar x_1 \vee x_2 \vee y_1 \vee \bar y_2).\]
The resolvent of $D_1$ and $D_2$ is the following clause
\[D=(\bar x_1 \vee x_2 \vee y_1),\]
whose negation is the conjunctive clause
\[C=(x_1 \wedge \bar x_2 \wedge \bar y_1),\]
which corresponds to the dyadic rectangle $\mv w=\dbox{10,0}$,
which is exactly the result of the geometric resolution of $\mv w_1$ and $\mv w_2$.
\label{ex:geo-to-logic}
\end{example}}

\rev{Appendix~\ref{sec:cnf-dnf} contains more details on the connection between logical resolution and geometric resolution. The following proposition is based on this connection: It follows from the completeness of logical resolution.}

\bprop[Completeness of geometric resolution]
\label{prop:geo-complete}
Given a set of boxes $\st A$ such that the union of all the boxes in $\st A$ covers some box $\mv b$, there exists a sequence of geometric resolutions on $\st A$ that results in a box $\mv b'$ that contains $\mv b$.
\eprop
The crux of this paper is to show that one can efficiently find a small sequence of geometric resolutions that solves \bcp. 
%However, we note that geometric resolution is strictly less powerful than general resolution in the sense that resolution proofs can be much shorter than geoemtric resolution proofs. (See the appendix for more details.)
%\ar{I need to add this stuff.}
%\hqn{It's not clear if the above claim is true -- given the same binary
%encodings; if it's not binary encoding then geometric resolution doesn't apply.
%Why do we even need to mention general resolution here?}
The inputs to most of the resolutions made by our algorithms will have an even more restricted structure: %(see e.g. Lemma~\ref{lmm ordered res} for a formal statement):
\begin{eqnarray}
\mv w_1 &=& \dbox{y_1\wc \dots\wc y_{\ell-1}\wc  x_\ell0\wc  \lambda\wc  \dots\wc \lambda}\label{eqn:main w1 format}\\
\mv w_2 &=& \dbox{z_1\wc \dots\wc z_{\ell-1}\wc  x_\ell1\wc  \lambda\wc  \dots\wc \lambda}\label{eqn:main w2 format},
\end{eqnarray}
where for every $i<\ell$, %both $y_i$ and $z_i$ are prefixes of $x_i$. This means that
 either $y_i$ or $z_i$ is a prefix
of the other. 
%We use $y_i \cap z_i$ to denote the {\em longer} of the two
%strings $y_i$, $z_i$. The geometric meaning is clear: we are taking the intersection
%of those two dyadic segments which are contained in one another.

\bdefn[Ordered geometric resolution]
Given two dyadic boxes $\mv w_1$ and $\mv w_2$ of the format shown in
\eqref{eqn:main w1 format} and \eqref{eqn:main w2 format},
the {\em ordered geometric resolution} of $\mv w_1$ and $\mv w_2$ is the dyadic box
\begin{equation}
\mv w = \dbox{y_1\cap z_1\wc\dots\wc y_{\ell-1}\cap z_{\ell-1}\wc  x_\ell\wc  \lambda\wc  \dots\wc \lambda} \label{eqn:main w format}
\end{equation}
We say that $\mv w$ is the result of resolving $\mv w_1$ and
$\mv w_2$ {\em on attribute $A_\ell$}. (Note that $x_\ell$ might be $\lambda$.)
\label{defn:main:ord-res}
\edefn

%Our \Resolve routine uses ordered resolution, and clearly $\mv w$ is a positive
%witness for $\mv b$ as desired.

\subsection{The Algorithm}
\label{sec:the-algo}

Our algorithm for \bcp at its core solves essentially the boolean version of
the \bcp using a sub-routine called\\ \tetrisskeleton. The sub-routine
is then repeatedly invoked by the {\em outer} algorithm -- \tetris\ -- to compute 
the output of the \bcp instance.

\subsubsection{The Core Algorithm}
\label{subsec:the-algo:the-core}

The Boolean version of \bcp is the following problem:
given a set of dyadic boxes $\calA$ and a target box $\mv b$, determine if 
$\mv b$ is covered by the (union of) boxes in $\calA$. 
\tetrisskeleton solves this problem by not only answering YES or NO,
but also generating an {\em evidence} for its answer:
\bi
 \item If $\mv b$ is covered by $\calA$,  then output
a box $\mv w$ that covers $\mv b$ such that
$\mv w$ is covered by the union of boxes in $\calA$.
 \item If $\mv b$ is not covered by $\calA$, then output a
point/tuple in $\mv b$ that is
not covered by any box in $\calA$.
\ei

%More concretely, the output is a pair $(v, \mv w)$, where $v$
%is a boolean value indicating whether $\mv w$ is a ve witness ($v$ is
%{\sc true}) or a negative witness ($v$ is {\sc false}).

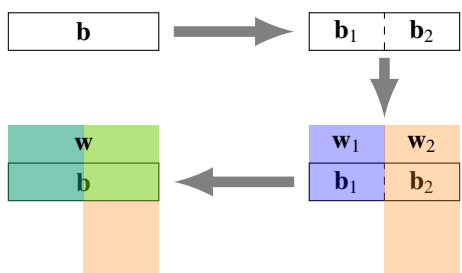
\begin{figure}[th]
\begin{center}
\begin{tikzpicture}[yscale=.5]

%%Box b
\draw (0,0) rectangle (2,1);
\node at (1,.5) {$\mv b$};

%%Split the box
\draw (4,0) rectangle (6,1);
\draw[dashed] (5,0) -- (5,1);
\node at (4.5,.5) {$\mv b_1$};
\node at (5.5,.5) {$\mv b_2$};

%%% Arrows

\draw[line width=4, gray, -latex] (2.2,.5) -- (3.8,.5);
\draw[line width=4, gray, -latex] (5,-.2) -- (5,-1.8);

%%% Witnesses
\begin{scope}[shift={(0,-4)}]
\draw (4,0) rectangle (6,1);
\draw[dashed] (5,0) -- (5,1);
\node at (4.5,.5) {$\mv b_1$};
\node at (5.5,.5) {$\mv b_2$};

\fill[blue,opacity=.3] (4,0) rectangle (5,2);
\fill[orange,opacity=.3] (5,-2) rectangle (6,2);
\node at (4.5,1.5) {$\mv w_1$};
\node at (5.5, 1.5) {$\mv w_2$};

\draw[line width=4, gray, -latex] (3.8,.5) -- (2.2,.5);

%%% Resolution
\draw (0,0) rectangle (2,1);
\node at (1,.5) {$\mv b$};
\fill[blue,opacity=.3] (0,0) rectangle (1,2);
\fill[orange,opacity=.3] (1,-2) rectangle (2,2);
\fill[green, opacity=.3] (0,0) rectangle (2,2);
\node at (1, 1.5) {$\mv w$};

\end{scope}
\end{tikzpicture}
\end{center}
\caption{Illustration of main resolution step.} % \ar{Needs to be TikZed.}}
\label{fig:main:covered}
\end{figure}

\tetrisskeleton has a very natural recursive structure.
We fix a \rao\ (\RAO) of the query,
say $(A_1,\dots,A_n)$. Following this order, we find the first dimension
on which $\mv b$ is {\em thick} \rev{(i.e. the length of the projection of $\mv b$ onto this dimension is $\geq 2$)} and thus can be split into two halves $\mv b_1$ 
and $\mv b_2$.
If we can find an uncovered point in either half, then we can immediately
return. Otherwise, we have recursively found two boxes $\mv w_1$ and $\mv w_2$, 
each of which covers one half of $\mv b$.
Each box may not cover $\mv b$ as a whole.
Hence, we resolve the two boxes $\mv w_1$ and $\mv w_2$ by
creating a maximal box $\mv w \subseteq \mv w_1 \cup \mv w_2$, making sure 
that $\mv w$ covers both $\mv b_1$ and $\mv b_2$; hence, $\mv w$ covers $\mv b$.
Figure~\ref{fig:main:covered} illustrates the main idea.

%\ar{\kb is now a macro under \textbackslash kb: feel free to change to whatever term you prefer.}

\begin{algorithm}[th]
\caption{\tetrisskeleton$(\mv b)$}
\label{alg:main:tetris-skeleton}
\begin{algorithmic}[1]
\renewcommand{\algorithmicrequire}{\textbf{\rev{Global parameter:}}}
\Require{A global set of boxes $\calA$}\Comment{Our \kb}
\renewcommand{\algorithmicrequire}{\textbf{Precondition:}}
\Require{Pick a \rao\ (\RAO) $(A_1,\dots,A_n)$}
\renewcommand{\algorithmicrequire}{\textbf{Input:}}
\Require{Target box $\mv b$}
\Ensure{A pair $(v,\mv w)$, where $\mv w$ is a cover box for $\mv b$ if $v$
is {\sc true}, and an uncovered point if $v$ is {\sc false}}
\If {there is a box $\mv a \in \calA$ such that $\mv a \supseteq \mv b$}\label{line:main:containment check 1}
  \State \Return $(\text{\sc true}, \mv a)$\label{line:main:pretent}
\ElsIf {$\mv b$ is a unit box} \label{line:main:unitbox}
  \State \Return $(\text{\sc false}, \mv b)$%
%\Comment{Where the negative
%  witness comes from}
\label{line:main:root-neg-witness}
\Else
  \State $\pair{\mv b_1}{\mv b_2} \wgets \Call{\firstsplit}{\mv b}$ \label{ln:main covered split} 
  \State \Comment{Cut $\mv b$ into two equal halves}
%  \State Split $\mv b$ into two halves $\mv b_1$ and $\mv b_2$ along the first
%         splittable dimension.
  \State $(v_1,\mv w_1) \la \tetrisskeleton(\mv b_1)$
  \If {$v_1$ is {\sc false}} \label{ln covered if false1}
    \State \Return $(\text{\sc false}, \mv w_1)$
  \ElsIf {$\mv w_1 \supseteq \mv b$} \label{line:main:w1-contains-b}
    \State \Return $\pair{\true}{\mv w_1}$ \label{ln covered true w1}
  \EndIf
  \State $(v_2,\mv w_2) \la \tetrisskeleton(\mv b_2)$\Comment{Backtracking}
  \If {$v_2$ is {\sc false}} \label{ln covered if false2}
    \State \Return $(\text{\sc false}, \mv w_2)$
  \ElsIf {$\mv w_2 \supseteq \mv b$} \label{line:main:w2-contains-b}
    \State \Return $\pair{\true}{\mv w_2}$ \label{ln covered true w2}
  \EndIf
  \State $\mv w \la $ \Call{\Resolve}{$\mv w_1,\mv w_2$} \Comment{Geometric resolution of $\mv w_1, \mv w_2$}\label{line:main:bcp-resolve}
  \State $\calA \la \calA \cup \{\mv w\}$\Comment{Cache the resolution} \label{line:main:add-to-kb}
  \State \Return $(\text{\sc true}, \mv w)$\label{line:main:return-resolvent}
\EndIf
\end{algorithmic}
\end{algorithm}

\tetrisskeleton is presented in Algorithm~\ref{alg:main:tetris-skeleton}.
There are three extra things that Algorithm~\ref{alg:main:tetris-skeleton} 
does over the basic outline above. First, we handle the base cases
when $\mv b$ is already covered by a box 
in $\calA$ itself in lines~\ref{line:main:containment check 1} 
and~\ref{line:main:pretent} (see Appendix~\ref{sec:dyadic-DS} on how we can implement this step in $\tO(1)$ time using a {\em multi-level dyadic tree} data structure) and when $\mv b$ is a unit box that is not 
covered by any box in $\calA$ (and hence cannot be covered by any boxes 
derived from $\calA$ either) in lines~\ref{line:main:unitbox} 
and~\ref{line:main:root-neg-witness}. 
Second, we check boundary conditions in lines~\ref{line:main:w1-contains-b} 
and~\ref{line:main:w2-contains-b}. 
Finally, in line~\ref{line:main:add-to-kb}, we add back the result of 
resolution from line~\ref{line:main:bcp-resolve} to $\calA$. (The last step 
is crucial in proving most of our results.) We defer a more detailed 
discussion on $\Resolve$ and $\firstsplit$ to the end of this section.

%(We will explain how $\firstsplit$ and $\Resolve$ work in Section~{sec:dyadic-DS}.)
%The small change was in line~\ref{line:root-neg-witness} of
%Algorithm~\ref{alg:main:tetris-skeleton}: instead of returning a negative
%witness that corresponds to an output tuple, we report it as an output and return
%$(\text{\sc true}, \mv b)$ {\em exactly as if}
%$\mv b$ was part of the \kb $\calA$.

\subsubsection{The Outer Algorithm: \tetris}
\label{sec:outer}

The \tetrisskeleton algorithm was designed for the boolean \bcp. We now
present the simple idea that allows us to use \tetrisskeleton as a sub-routine
and solve the general \bcp problem. 
The input to the general \bcp problem is a set of boxes $\calB$ to which we have 
oracle access. The {\em oracle} represents the pre-built database indices
of input relations from a join query. In particular, given a unit box $\mv w$, the oracle
can return the set of boxes in $\calB$ containing $\mv w$ in $\tO(1)$-time. (See Appendix~\ref{sec:dyadic-DS} for more.)

Algorithm~\ref{alg:main:tetris}, named \tetris, solves \bcp by continuously 
calling \tetrisskeleton on input $\calA$, called the \kb, with the target 
box being the universal box $\mv b = \dbox{\lambda,\dots,\lambda}$.
We will explain how different initializations of $\calA$ lead to different
guarantees in later sections.

After each invocation of \tetrisskeleton, $\calA$ is amended with a few 
more boxes and the next invocation of\\ \tetrisskeleton is
on the enlarged \kb\ $\calA$. Apart from resolvents that are cached by \tetrisskeleton in $\st A$, \tetris amends $\calA$ with two types of boxes: output (unit) boxes and boxes from $\st B$.
To be more specific, if \tetrisskeleton returns $(\textsc{true},\mv w)$, then 
we know there are no tuples to output and we can stop. However, if \tetrisskeleton 
returns $(\textsc{false},\mv w)$, then we check if $\mv w$ is {\em not} covered 
by any box in $\st B$. If so, we know $\mv b$ is an output point and we can 
output that point and amend $\calA$ with $\mv b$. 
Otherwise we know that $\calA$ was not properly initialized 
in which case we amend $\calA$ with boxes in $\st B$ that 
cover $\mv b$ and repeat. \rev{See Example~\ref{ex:tetris}.}

%We now describe how we can use \tetris to solve the join problem.
%The (very mild) difference between the general \bcp problem and the
%join problem is that we have to list the output tuples in the join problem.

\begin{algorithm}[ht]
\caption{\tetris$(\st B)$}
\label{alg:main:tetris}
\begin{algorithmic}[1]
\Require{Oracle access to a set of boxes $\st B$ \rev{(i.e. $\st B$ is the input of a \bcp instance)}}
\Ensure{All tuples not covered by any box in $\st B$ \rev{(i.e. the output of the \bcp instance)}}
\State $\mathsf{Initialize}(\st A)$\label{ln:main:initialize}
\State $\pair{v}{\mv w}\wgets\Call{\tetrisskeleton}{\UB}$
\While {$v$ = \false}
        \State $\calB' \wgets \{\mv b\in \st B \wsuchthat \mv b\supseteq \mv
        w\}$ \label{ln:main online B'} \Comment{from the oracle}
        \If {$\st B'=\emptyset$}
              \State {\bf Report} $\mv w$ as an output tuple\label{line:main:report-output}
              \State $\calB'\wgets \{\mv w\}$ \label{line:main:add-output}
        \EndIf
        \State $\st A \wgets \st A\cup \st B'$ \Comment{Amend the \kb}\label{line:main:add-to-kb-outer}
        \State $\pair{v}{\mv w}\wgets\Call{\tetrisskeleton}{\UB}$
\EndWhile
\end{algorithmic}
\end{algorithm}

\subsubsection{Recursion and Resolution}

We next flesh out the two key operations that were not specified
in the description of Algorithm~\ref{alg:main:tetris-skeleton}:
how to split a box $\mv b$ into two halves in line~\ref{ln:main covered split}, 
and how to resolve two
{\em witnesses} $\mv w_1$ and $\mv w_2$ in line~\ref{line:main:bcp-resolve} of
\tetrisskeleton (Algorithm~\ref{alg:main:tetris-skeleton}).

We first explain what the \firstsplit routine does.
Consider a dyadic box $\mv b = \dbox{x_1,x_2,\dots,x_n}.$
If $|x_i|=d$, then $x_i$ represents a unit dyadic segment, which corresponds to
a flat slice through the $A_i$-dimension. The box $\mv b$ is flat and is not
splittable along such dimension.
The first thick dimension is the smallest value $\ell\in[n]$ for which
$|x_\ell|<d$.
Because $\mv b$ is not a unit box, there must exist such an $\ell$.
In that case, the call
\[ \pair{\mv b_1}{\mv b_2} \wgets \firstsplit(\mv b) \]
in line~\ref{ln:main covered split} of Algorithm~\ref{alg:main:tetris-skeleton} 
returns the following pair:
\begin{eqnarray*}
\mv b_1 &=& \dbox{x_1\wc \dots\wc x_{\ell-1}\wc  x_\ell0\wc  x_{\ell+1}\wc \dots\wc x_n}\\
\mv b_2 &=& \dbox{x_1\wc \dots\wc x_{\ell-1}\wc  x_\ell1\wc  x_{\ell+1}\wc \dots\wc x_n}.
\end{eqnarray*}
Note again that by definition $|x_i|=d$ for all $i<\ell$.
%Using the {\em multi-level dyadic tree} data structure discussed
%Appendix~\ref{sec:dyadic-DS}, we can easily find the first splittable dimension 
%of a dyadic box in $\tilde O(1)$ time.
It is easy to implement the above bitstring operation in $O(dn)$ time, which by our convention is $\tO(1)$ time.

Next, we explain the resolution step.
$\Resolve$ is geometric resolution as defined in Section~\ref{sec:geo-res}.
Note that by the time \Resolve is called in line~\ref{line:main:bcp-resolve} 
we know none of $\mv w_1$ and $\mv w_2$ covers $\mv b$.
There are a lot of boxes we can infer from $\mv w_1$ and $\mv w_2$ if those 
two boxes are general dyadic boxes that can overlap in peculiar ways.
However, \tetrisskeleton forces $\mv w_1$ and $\mv w_2$ to be somewhat 
special, making resolution much more intuitive and clean. 
In Lemma~\ref{lmm ordered res}, we show that all the resolutions in 
line~\ref{line:main:bcp-resolve} are ordered geometric resolutions 
(see Definition~\ref{defn:main:ord-res}).

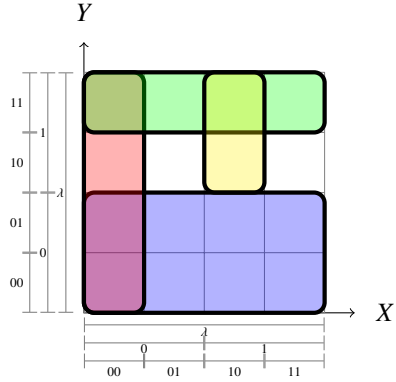
\begin{figure}
\begin{center}
    \begin{tikzpicture}[scale=.8]
    \draw [<->] (0, 4.5) -- (0, 0) -- (4.5, 0);
    \draw[help lines] (0,0) grid (4,4);
    \node [black] at (5,0) {$X$};
    \node [black] at (0, 5) {$Y$};
    \draw[ultra thick, rounded corners, fill=blue, fill opacity=.3] (0,0) rectangle (4,2);
    \draw[ultra thick, rounded corners, fill=red, fill opacity=.3] (0,0) rectangle (1,4);
   \draw[ultra thick, rounded corners, fill=green, fill opacity=.3] (0,3) rectangle (4,4);
   \draw[ultra thick, rounded corners, fill=yellow, fill opacity=.3] (2,2) rectangle (3,4);

\draw [gray, |-|] (0,-0.2) -- (4,-0.2); 
\node [below] at (2,-.08) {\tiny{$\lambda$}};
\draw [gray, |-|] (0,-0.5) -- (2,-0.5); 
\node [below] at (1,-.37) {\tiny{$0$}};
\draw [gray, |-|] (2,-0.5) -- (4,-0.5); 
\node [below] at (3,-.37) {\tiny{$1$}};
\draw [gray, |-|] (0,-0.8) -- (1,-0.8); 
\node [below] at (0.5,-.75) {\tiny{$00$}};
\draw [gray, |-|] (1,-0.8) -- (2,-0.8); 
\node [below] at (1.5,-.75) {\tiny{$01$}};
\draw [gray, |-|] (2,-0.8) -- (3,-0.8); 
\node [below] at (2.5,-.75) {\tiny{$10$}};
\draw [gray, |-|] (3,-0.8) -- (4,-0.8); 
\node [below] at (3.5,-.75) {\tiny{$11$}};

\begin{scope}[rotate=-90, shift={(-4, -.1)}]
\draw [gray, |-|] (0,-0.2) -- (4,-0.2); 
\node [left] at (2,-.08) {\tiny{$\lambda$}};
\draw [gray, |-|] (0,-0.5) -- (2,-0.5); 
\node [left] at (1,-.37) {\tiny{$1$}};
\draw [gray, |-|] (2,-0.5) -- (4,-0.5); 
\node [left] at (3,-.37) {\tiny{$0$}};
\draw [gray, |-|] (0,-0.8) -- (1,-0.8); 
\node [left] at (0.5,-.75) {\tiny{$11$}};
\draw [gray, |-|] (1,-0.8) -- (2,-0.8); 
\node [left] at (1.5,-.75) {\tiny{$10$}};
\draw [gray, |-|] (2,-0.8) -- (3,-0.8); 
\node [left] at (2.5,-.75) {\tiny{$01$}};
\draw [gray, |-|] (3,-0.8) -- (4,-0.8); 
\node [left] at (3.5,-.75) {\tiny{$00$}};
\end{scope}
    \end{tikzpicture}
\end{center}
\caption{\rev{A \bcp instance with two dimensions/attributes $(X, Y)$ and a box set 
$\calB:=\left\{\dbox{\lambda, 0}, \dbox{00, \lambda}, \dbox{\lambda, 11}, \dbox{10, 1}\right\}.$
The output tuples are $\dbox{01, 10}$ and $\dbox{11, 10}$.}}
\label{fig:tetris-example}
\end{figure}

\rev{
\begin{example}
\label{ex:tetris}
Consider the following set of boxes in two dimensions/attributes $(X, Y)$:
\[\calB:=\left\{\dbox{\lambda, 0}, \dbox{00, \lambda}, \dbox{\lambda, 11}, \dbox{10, 1}\right\}.\]
$\calB$ is depicted in Figure~\ref{fig:tetris-example}. Suppose that we apply \tetris (Algorithm~\ref{alg:main:tetris}) to solve 
the \bcp instance with box set $\calB$ (Recall Definition~\ref{defn:bcp-boxcert}).
Suppose that we initialize $\calA$ (in Line~\ref{ln:main:initialize} of Algorithm~\ref{alg:main:tetris}) to be the following subset of $\calB$:
\[\calA=\left\{\dbox{\lambda, 0}, \dbox{00, \lambda}, \dbox{\lambda, 11}\right\}.\]
\tetris now invokes $\tetrisskeleton(\dbox{\lambda,\lambda})$. Let $\sigma=(X,Y)$ be the chosen \rao.  Since no box in $\calA$ covers $\dbox{\lambda,\lambda}$, 
\tetrisskeleton splits $\dbox{\lambda,\lambda}$ into $\dbox{0,\lambda}$ and $\dbox{1,\lambda}$ and recurses.
Similarly, $\tetrisskeleton(\dbox{0,\lambda})$ will split $\dbox{0,\lambda}$ into $\dbox{00,\lambda}$ and 
$\dbox{01,\lambda}$ and recurse. $\tetrisskeleton(\dbox{00,\lambda})$ will find a box in $\calA$ that covers $\dbox{00,\lambda}$, which is $\dbox{00,\lambda}$, and will return $\pair{\true}{\dbox{00,\lambda}}$.
$\tetrisskeleton(\dbox{01,\lambda})$ will split into $\dbox{01, 0}$ and $\dbox{01, 1}$. $\tetrisskeleton(\dbox{01, 0})$ will find a box in $\calA$ that covers $\dbox{01, 0}$ and will return $\pair{\true}{\dbox{\lambda, 0}}$.
$\tetrisskeleton(\dbox{01, 1})$ will split into $\dbox{01,10}$ and $\dbox{01, 11}$. $\tetrisskeleton(\dbox{01,10})$ will not find 
any box in $\calA$ covering $\dbox{01,10}$ and will return $\pair{\false}{\dbox{01,10}}$, which will go all the way up the recursion. Since no boxes in $\calB$ cover $\dbox{01,10}$, \tetris will report $\dbox{01,10}$ as an output tuple, and will add
$\dbox{01,10}$ to $\calA$.

\tetris will now invoke $\tetrisskeleton(\dbox{\lambda,\lambda})$ again (but now $\calA$ has been amended with $\dbox{01,10}$).
The recursion will go on as before except that $\tetrisskeleton(\dbox{01,10})$ will now return $\pair{\true}{\dbox{01,10}}$.
$\tetrisskeleton(\dbox{01, 1})$ will now resolve $\dbox{01, 10}$ with $\dbox{\lambda, 11}$ (that was returned by $\tetrisskeleton(\dbox{01, 11})$) into the box $\dbox{01,1}$, will add this new box to $\calA$, and return $\pair{\true}{\dbox{01,1}}$.
$\tetrisskeleton(\dbox{01,\lambda})$ will resolve $\dbox{\lambda, 0}$ with $\dbox{01, 1}$ into $\dbox{01, \lambda}$, and add it to $\calA$. $\tetrisskeleton(\dbox{0,\lambda})$ will resolve $\dbox{00, \lambda}$ with $\dbox{01, \lambda}$ into $\dbox{0, \lambda}$.

$\tetrisskeleton(\dbox{1,\lambda})$ will recursively discover that $\dbox{10, 10}$ is not covered by any box in $\calA$ and will return $\pair{\false}{\dbox{10, 10}}$. \tetris will look in $\calB$ for boxes that cover $\dbox{10, 10}$, will find $\dbox{10,1}$, add it to $\calA$, and call $\tetrisskeleton(\dbox{\lambda,\lambda})$ again.

$\tetrisskeleton$ now will recursively resolve $\dbox{\lambda,0}$ and $\dbox{10, 1}$ into $\dbox{10, \lambda}$, and then discover that $\dbox{11, 10}$ is not covered by any box in $\calA$. \tetris will report $\dbox{11, 10}$ as an output tuple, and add it to $\calA$. Finally, \tetrisskeleton will resolve $\dbox{11, 10}$ and $\dbox{\lambda, 11}$ into $\dbox{11, 1}$, and then resolve $\dbox{11, 1}$ and $\dbox{\lambda, 0}$ into $\dbox{11, \lambda}$, which in turn resolves with $\dbox{10, \lambda}$ into $\dbox{1, \lambda}$. Finally, the latter resolves with $\dbox{0,\lambda}$ into $\dbox{\lambda,\lambda}$.
\end{example}}

We now state the key analytical tool that will be used throughout this paper
to bound the runtime of our algorithm in different settings. The tool is a 
very simple but important combinatorial lemma that says the following:
hiding behind the potential poly-log factor in $\tilde O$, we can bound the 
runtime of \tetris by the number of resolutions it performs.
The main observation is that in \emph{most} cases when the algorithm 
backtracks, it does one resolution. The amount of work it does modulo the 
recursive calls is $\tilde O(1)$: inserting a new box, querying for 
boxes containing a box, and resolving. Finally, line~\ref{line:main:add-to-kb} and line~\ref{line:main:containment check 1} make sure that we are not repeating any 
resolution more than once.

\blmm[Runtime is bounded by \#resolutions]
\label{lmm:main time<=res}
Let $M$ denote the total number of resolutions performed by 
Algorithm~\ref{alg:main:tetris}.
Then, the total runtime of Algorithm~\ref{alg:main:tetris} is $\tO(M)$.
\elmm

\subsubsection{\tetris as DPLL with clause learning}
\label{sec:dpll-tetris}

We briefly explain how \tetris can be viewed as a form
of DPLL with clause learning.
\rev{(See Appendix~\ref{sec:cnf-dnf} for more details.)}
A tuple in the output space is an $n$-dimensional dyadic box each of whose
components is a string of length $d$. When viewed as a bit-string, this tuple
is a truth assignment. A dyadic gap box $\mv w$ under this view can be encoded
with a clause, containing all tuples {\em not} belonging to $\mv w$.
Under this encoding, geometric resolution becomes a particular form of 
propositional logic resolution.
\rev{(Recall Example~\ref{ex:geo-to-logic}.)}
The resolvent of a geometric resolution is a new clause that was inferred
and cached in the computation.
Hence, \tetris can be cast as a DPLL algorithm for \#SAT 
with a fixed variable ordering
and with a particular way of learning new clauses.
(It is for \#SAT because the algorithm keeps running even after 
a satisfying assignment is found. See Appendix~\ref{sec:cnf-dnf}.)

\rev{Alternatively, when viewed from a geometric perspective, DPLL (with clauses learning) can be viewed as \tetris:
As was shown in Figure~\ref{fig:main:log-resolution}, the negation of each clause corresponds to a box in the Boolean cube.
Assigning a truth value to some literal in DPLL corresponds to splitting the target box $\mv b$ in \tetris in half and considering only one half. Resolving two clauses in DPLL corresponds to applying a geometric resolution between the corresponding boxes. Caching in DPLL corresponds to storing a resolvent $\mv w$ in the knowledge base $\calA$ of \tetris.}

%Finally, we point out that \tetris\ does the following extra step over 
%the algorithm idea mentioned in Section~\ref{sec:outer}. When we hit an 
%uncovered point  $\mv b$ on line~\ref{line:main:root-neg-witness} of
%Algorithm~\ref{alg:main:tetris-skeleton}, if $\mv b$ turns out to be an output
%tuple then \tetris will
%(1) report it (line~\ref{line:main:report-output}), and
%(2) add the output (unit) box $\mv w$ to the \kb $\calA$ 
%(line~\ref{line:main:add-output}).
%Step (2) essentially allows us to think of the output points as being part 
%of the input boxes, which in turn allows us to essentially only consider 
%the empty output case in our analysis.

%and
%(3) {\em continue operating as if $\mv w$ was a box in $\calA$ all along}.
%In other words, line~\ref{line:main:root-neg-witness} becomes
%line~\ref{line:main:pretent}.
%The above idea will be the key analytical device that allows us to essentially only consider the empty output case and will be
%used throughout this paper; hence, we make it stand out as a definition.
%
%\bdefn[Shadow box idea]
%We will refer to the above idea as the {\em shadow box idea}, where
%for analytical purposes the algorithm works as if the already identified output boxes were
%part of the input \kb $\calA$ from the beginning.
%\label{defn:main:shadow-box}
%\edefn

\subsection{Worst-case Results}
\label{SEC:WORST-CASE}

The initialization of the \kb $\calA$ has a crucial implication in terms of
the kind of runtime result \tetris is able to attain.
In this section, we discuss one extreme where we can load the \kb 
$\calA$ with {\em all} boxes from the input set of boxes $\calB$.
For notational convenience, we call \tetris\ with this specific 
instantiation of $\Initialize$ to be\\ \tetrispreloaded.

It turns out that \tetrispreloaded achieves the
following type of runtime guarantee: given a join query $Q$, 
under some assumption about the type of boxes in $\calB(Q)$,
\tetrispreloaded runs in time at most the maximum \agm-bound on a bag of any 
tree decomposition of $Q$.
\rev{(See Appendix~\ref{sec:tree-decomp} for background about tree decompositions.)}
And we can construct $\calB(Q)$ satisfying the assumption in time linear
in the input relations' sizes.

Since the \rev{above} result requires some lengthy definitions, we state below
a slightly weaker result, in terms of the fractional hypertree width of
the query $Q$. We prove our full (stronger) result in Appendix~\ref{sec offline}.

\bthm[\tetrispreloaded achieves fractional hypertree width bound]
\label{thm:main fhtw}
Let $Q$ be a join query, $N$ the total number of input tuples,
$\fhtw$ the fractional hypertree width of the query, and 
$Z$ the total number of output tuples.
Then, there exists a global attribute order (GAO) $\sigma$ such that the 
following holds.
Suppose for all $R\in\atoms(Q)$, $\calB(R)$ is $\sigma$-consistent.
Then, by setting $\RAO$ to be $\sigma$, 
\tetrispreloaded on input $\calB(Q)$ runs in time \rev{$\tO(N^{\fhtw}+Z)$}.
\ethm

%The above along with Proposition~\ref{prop:bcp=join} implies the following:
%\bcor
%\tetris\ can compute any join query $Q$ in time $O(N+N^{fhtw(Q)}+Z)$, where $N$ is the total number of input tuples and $Z$ is the total number of output tuples.
%\ecor

Recall that \tetris uses ordered geometric resolution. It turns out that 
\nocache\ is enough to recover the \agm\ bound 
(see Theorem~\ref{thm:app agm}). However, \nocache\ is not powerful enough 
to recover Theorem~\ref{thm:main fhtw} (see Theorem~\ref{thm:Y-tree}).

%\ar{HUNG: Please put in something about the adaptive width stuff.}

\subsection{Beyond Worst-case Results}
\label{SEC:BEYOND-WC}

Our algorithm \tetris\ not only can recover some existing results as we have
seen, but also leads to new results, as presented in this section. 
In particular, we show that \tetris\ can extend the bounded treewidth 
results of~\cite{nnrr}, which only hold for GAO-consistent input indices,
to handle cases of arbitrary input indices, including sophisticated indices 
such as dyadic trees (and multiple indices per relation). 

%\subsubsection{Acyclic and Bounded Treewidth Queries}
%\label{sec:acyclic}

The crux of {\em beyond worst-case} guarantee is for the runtime of the algorithm
to be measured in the finer notion of (box) certificate size $|\boxcert|$
of the \bcp instance, instead of input size. 
It is easy to construct arbitrarily large input instances for which the 
certificate size is $\tO(1)$.  (See Appendix~\ref{sec:cert}.)
Consequently, preloading the \kb $\calA$ with all boxes from 
$\calB$ as we did with \tetrispreloaded is no longer an option.

To obtain certificate-based results, we only load the boxes from 
$\st B$ into $\calA$ that are absolutely needed.  
In particular, we go the other extreme and set $\calA\gets\emptyset$ 
in $\Initialize(\calA)$ (and let lines~\ref{ln:main online B'} 
and~\ref{line:main:add-to-kb-outer} in Algorithm~\ref{alg:main:tetris} load 
the required boxes from $\st B$ into $\calA$). For notational convenience, 
we use\\ \tetrisreloaded to refer to \tetris\ with this specific instantiation 
of $\Initialize$.

In the following results, we use the well-known fact that if a hypergraph
(or a query) has treewidth $w$, then there is a vertex ordering (or an attribute
ordering) that has an {\em elimination width} $w$; and, this 
ordering can be computed in $\tO(1)$-time in data complexity.
We get a near-optimal result for treewidth-$1$ queries:
\bthm[$\tO(|\boxcert|+Z)$-runtime for treewidth $1$]
\label{thm:main acyclic arity<=2}
For any set of boxes $\st B$ with $\tw(\calH(\calB)) = 1$,
by setting \RAO to be the attribute ordering with elimination width $1$, 
\tetrisreloaded solves \bcp\ on input $\st B$ in time $\tO(|\boxcert|+Z)$.
%Here, $\boxcert$ is any optimal box certificate for the instance, 
%and $Z$ is the output size.
\ethm

Along with Propositions~\ref{prop:tw box vs tw Q} and \ref{prop:bcp=join}, 
the above result implies the following:
\bcor
\tetrisreloaded\ evaluates any join query $Q$ with treewidth $1$ in time 
$\tO(|\boxcert|+Z)$.
%by setting the \RAO to be the attribute ordering with elimination
%width $1$. 
%Here, $\boxcert$ is
%an optimal box certificate for the join instance, and $Z$ is the output size.
\ecor

Note that a treewidth of $1$ implies that all relations are binary. 
In Proposition~\ref{prop omega 4/3} in Appendix~\ref{sec:tightness},
we show that as soon as there is
a relation of arity $\geq 3$, a runtime of $\tO(|\boxcert|+Z)$ is not possible
modulo the hardness of $\mathsf{3SUM}$.

For general treewidths, we prove a slightly weaker result.
\bthm[$\tO(|\boxcert|^{w+1}+Z)$-runtime for treewidth $w$]
For any set of boxes $\st B$ with $\tw(\calH(\calB)) = w$,
by setting \RAO to be the attribute ordering with elimination width $w$, 
\tetrisreloaded solves \bcp\ on input $\st B$ in time $\tO(|\boxcert|^{w+1}+Z)$.
%Here, $\boxcert$ is any optimal box certificate for the instance, 
%and $Z$ is the output size.
\label{thm:main:C^{w+1}+Z}
\ethm

Along with Propositions~\ref{prop:tw box vs tw Q} and \ref{prop:bcp=join}, 
the above result implies the following:
\bcor
\tetrisreloaded\ evaluates any join query $Q$ with treewidth $w$ in time 
$\tO(|\boxcert|^{w+1}+Z)$. %, by setting the \RAO to be the attribute ordering with 
%elimination width $w$. 
%Here, $\boxcert$ is an optimal box certificate 
%for the join instance, and $Z$ is the output size.
\ecor

%\ar{Put in lower bound for \ordered. It seems that Theorem~\ref{thm:C^{w+1}-lowerbound} is only stated for \tetrisreloaded. So maybe the general lower bound still needs to be proved?}

\subsection{Arbitrary queries}
\label{SEC:CERT-N/2}

We now show that an enhancement of \tetris\ gives an improved beyond worst-case result for arbitrary join queries.
In particular, we show the following result (see Corollary~\ref{cor:block-size} in Appendix~\ref{app:block-size} for a more general result):
\bthm
\label{thm offline as}
For any integer $n\ge 2$, the problem \bcp\ on $n$ dimensions can be solved in
time $\tO(\abs{\boxcert}^{n/2}+Z)$.
\ethm
Theorem~\ref{thm:cert^n-1-lb} shows that this result cannot be
achieved by an algorithm that only performs ordered geometric resolution (like \tetris),
no matter which \RAO it chooses.
The main reason is that it might
get stuck in resolving boxes along a fixed dimension due to the fixed
\RAO, while it could have covered the entire space faster by dynamically
switching to resolutions in other dimensions.

We get around this bottleneck by transforming the input boxes into 
boxes in a higher-dimensional space, then applying \tetris.
The idea is to carefully construct this map so that the amount of work
per dimension is balanced out.
It is worth noting that we are still using the same algorithm \tetris, under a
transformed input.
Since the analysis of the algorithm is quite involved, we sketch in this section
some of the key ideas by making some assumptions about the input.
In particular, in this section we assume that the algorithm is 
given as its input the box certificate of the instance of the \bcp, which we 
will denote by $\cert$. We will also refer to this version of the \bcp\ as
the {\em offline} case of the problem. At the end of the section, we outline how
we can remove this restriction, leaving the full description to 
Appendix~\ref{sec online as}.

\subsubsection{Divide and conquer}
\label{sec:balanced-case}

To build intuition, we start off with a very special case. 
Call the input box set $\cert$ {\em balanced} if there exists an attribute 
$X$ and a partition $P_X$ of the domain $\D(X)$ into $\tO(\sqrtC)$ many 
(disjoint) dyadic intervals such that 
$(i)$ for each box $\mv b\in\cert$, the interval $\pi_X(\mv b)$ is contained 
in one of the intervals in $P_X$ and 
$(ii)$ for each interval $x\in P_X$, the number of boxes $\mv b\in\cert$ such that
 $\pi_X(\mv b)\subseteq x$ is also bounded by $\tO(\sqrtC)$. 
 The idea is that, when $\cert$ is balanced we can solve 
 $\tO(\sqrtC)$ independent sub-instances of \bcp, one for each {\em layer} $x\in
 P_X$; each sub-instance has an input box set of size $\tO(\sqrtC)$. 
 This divide and conquer strategy is useful because 
 \tetris\ can solve \bcp\ on $\calB$ in time
 $\tO(\abs{\calB}^{n-1}+Z)$ (Theorem~\ref{thm:C^{n-1}+Z-upperbound} in
 Appendix~\ref{sec online GB}).
 This means if we apply \tetris to each of the $\tO(\sqrtC)$ independent
 subproblems, and then output the union of the results, then we have an
 overall run time of $\tO(\sqrtC\cdot(\sqrtC)^{n-1}+Z)=\tO(|\cert|^{n/2}+Z)$, 
 as desired.

In general, the above is too strong a condition.
%\ar{We need to refer to the odds and evens example for the hard instance for triangle query for a specific example.}
%As mentioned already, we will show later in Theorem~\ref{thm:cert^n-1-lb} that no
%algorithm that just makes ordered resolutions (as \tetris does) can beat the
%$\Omega(|\cert|^{n-1})$ bound for \bcp on $n$ dimensions in general. This in turn
%implies that there exist certificates $\cert$ that cannot be balanced for
%any dimension. (
Appendix~\ref{app:unbalanced} gives an explicit certificate
$\cert$ that is not balanced.
%\ar{MAHMOUD: I could not make the hard odds and even instance for the triangle lower bound to show that the corresponding certificate is not balanced: I think those are balanced. Anyhow, put in a similar example in Appendix~\ref{app:unbalanced}-- please check to make sure I did not mess up something there.}

To rectify this situation, we perform a conceptually simple pre-processing
step. We design a procedure called $\balance$ that takes as input the
certificate $\cert$ (recall that we are in the offline case). It outputs
a box set $\cert'$ of the same size. $\cert'$ has a specific \RAO such that if one runs \tetris on
$\cert'$ with this \RAO, then one ends up with the desired $\tO(|\cert|^{n/2}+Z)$ runtime. Thus, armed with the balancing procedure, our final algorithm has a very simple structure as illustrated in Algorithm~\ref{alg offline as}.

\begin{algorithm}[th]
\caption{\tetrispreloadedas}%\\(i.e. \tetrispreloaded with load balancing)}
\label{alg offline as}
\begin{algorithmic}[1]
\Require{A set of boxes $\st C$}
\Ensure{Output tuples for the \bcp\ on $\cert$}
\State $\st B\wgets \balance(\cert)$ \label{line:main call balance}
\State \Return \Call{\tetrispreloaded}{$\st B$} \label{line:main call tetris}
\end{algorithmic}
\end{algorithm}

\subsubsection{Load-balancing with \texorpdfstring{$\balance$}{Balance}}

To sketch out how load-balancing works, we make one further simplifying
assumption that we are trying to solve only the Boolean version of \bcp: 
given the set of boxes $\cert$, does the union of the boxes in
$\cert$ cover the entire output space? 
We begin by formalizing certain notions that we used in defining a balanced
$\cert$ above.

\bdefn[Dimension partition]
A {\em partition} $P$ of $\D=\{0,1\}^d$ is a collection of disjoint dyadic intervals
whose union is exactly $\D$. 
Given a dimension $X$ of the \bcp,
an {\em $X$-partition} is a partition of the domain $\D(X)$.
We will typically use $P_X$ to denote a partition along dimension $X$.
\edefn

Geometrically, a partition along dimension $X$ divides the output space
into $|P_X|$ layers, one for each interval $x$ in $P_X$. 
%In particular, the layer defined by a fixed interval $x\in P_X$, called the
%{$x$-layer}, is the dyadic box $\dbox{x, \lambda,\lambda}$.
%To verify that the output space is completely covered by the input gap
%boxes, it is sufficient to verify that every $x$-layer is covered, for each
%$x\in P_X$.
An input gap box whose $X$-component is disjoint from $x$ will not affect
whether the $x$-layer is covered. 
Hence, to verify whether the $x$-layer is covered, we can ignore all gap boxes
that do not intersect $x$. If the remaining set of gap boxes is small, then this
verification is fast. At the same time, we do not want too many layers because
that certainly increases the total amount of verification work. This balancing
act leads to our first idea: we find a dimension partition that is somehow
balanced.

More concretely, given a set of boxes $\st C$ and a dyadic interval
$x$ on the domain $\D(X)$, define two sets:
\begin{eqnarray}
\st C_{\subset x}(X) &=& \left\{ \mv b \in \st C \suchthat \pi_X(\mv b)
\subsetneq x \right\} \label{eqn C_subset}\\
\st C_{\supseteq x}(X) &=& \left\{ \mv b \in \st C \suchthat \pi_X(\mv b)
\supseteq x \right\}. \label{eqn C_supset}
\end{eqnarray}
%In other words, the first set $\st C_{\subset x}(X)$ consists of all dyadic boxes in
%$\st C$ that are strictly contained in the $x$-layer.
%The second set $\st C_{\supseteq x}(X)$ is the set of boxes in $\st C$ each of whose
%$X$-component completely covers the interval $x$.
Note that, for every box $\mv b \in \st C \setminus (\st C_{\subset x}(X) \cup \st
C_{\supseteq x}(X))$, the dyadic interval $\pi_X(\mv b)$ is completely disjoint
from the interval $x$.
%It should also be emphasized that $\pi_X(\mv b) \subsetneq x$ means $x$ is a
%strict prefix of the binary string $\pi_X(\mv b)$.

\bdefn[Balanced dimension partition]
Let $\st C$ be the set of input gap boxes, and $X$ be any attribute.
A {\em balanced $X$-partition} is an $X$-partition $P_{X}$ such that
\begin{eqnarray*}
|P_X|                  &=    & \tO(\sqrtC)\\
|\st C_{\subset x}(X)| &\leq & \sqrtC, \text{ for every } x \in P_X.
\end{eqnarray*}
\edefn

Given a set $\st C$ of input gap boxes and an arbitrary attribute $X$,
we can show that a balanced $X$-partition can be computed in time $\tO(|\st C|)$.
(Proposition~\ref{prop:app:construct_P} in Appendix~\ref{app:cert-n/2}.)
Furthermore, if there exists a balanced $X$-partition (for some
dimension $X$) such that for every $x\in P_X$, we also have
$|\cert_{\supseteq x}(X)|\le \tO(\sqrtC)$ and 
$\cert_{\supseteq x}(X) \cap
\cert_{\supseteq x'}(X)=\emptyset$ for every $x\neq x'\in P_X$, then
$\cert$ is balanced.
So, to have a balanced partition, two fairly strong conditions are required.
We next introduce two ideas to relax these requirements.

First, the requirement that $\cert_{\supseteq x}(X)$ be disjoint
for distinct $x\in P_X$ is not strictly required for the divide and
conquer strategy to go through. In particular, for each $x \in P_X$ we 
can create the sub-instance of \bcp by copying every box $\mv
b\in \cert_{\supseteq x}(X)$ and replacing $\pi_X(\mv b)$ by $x$ (for each
copy). These new boxes along with $\cert_{\subset x}(X)$ form a sub-instance.
Now we can solve each sub-instance separately as before. 
Now, if $|\cert_{\supseteq x}(X)|\le \tO(\sqrtC)$ for every $x\in P_X$, then 
we would be done.

%\bprop[Efficient constructibility of balanced partitions]
%\label{prop:construct_P}
%\eprop
%\vspace*{1cm}

Second, unfortunately in general $\abs{\cert_{\supseteq x}(X)}$ could be as 
large as $\Omega(|\cert|)$. The copying trick to divide the \bcp instance 
into disjoint sub-instances for each $x\in P_X$ is too expensive when $n>3$. 
However, when $n=3$, note that we have $|P_X|$ disjoint \bcp instances on
dimension $2$. 
In two dimensions, thanks to Theorem~\ref{thm:C^{n-1}+Z-upperbound}, we know
\tetris can solve the sub-instance of \bcp in time $\tO(|\cert|)$, which 
would lead to the desired $\tO(|\cert|^{3/2})$ time 
(since $|P_X|\le \tO(\sqrtC)$). Applying bruteforcely,
this trick will lead to an overall runtime of $\tO(|\cert|^{n-3/2})$, which
matches our desired bound of $\tO(|\cert|^{n/2})$ only for $n=3$.

For general $n$ we use the following natural recursive strategy. 
We divide up the original \bcp on $n$ dimensions to $|P_X|$ many disjoint 
\bcp problems (with the copying trick as above) on $n-1$ dimensions 
(i.e. on all dimensions except $X$). If a sub-problem has size $\tO(\sqrtC)$ 
then we can run \tetris directly. Otherwise we recurse on each of the 
sub-problems. However, there are two technical issues we need to solve to 
properly implement this recursive strategy.

First, in the discussion above, we did not talk about boxes in
$\cert_{\subset x}(X)$ for $x\in P_X$. In particular, to define the disjoint $(n-1)$-dimensional
sub-problems, we have to perform all possible resolutions of boxes in
$\cert_{\subset x}(X)$ and only retain those boxes $\mv b$ such that
$\pi_X(\mv b)=x$ for the sub-problem corresponding to $x\in P_X$. Theorem~\ref{thm:C^{n-1}+Z-upperbound} implies that the
number of such boxes could potentially be as large as $\tO(|\cert|^{(n-1)/2})$.
When $n=3$, this is still not a problem.
However when $n>3$, we will need to reason about such boxes carefully. (In particular, we
cannot consider a sub-problem on $\tO(|\cert|^{(n-1)/2})$ many boxes.)

Second, we also need to be careful about the number of times we apply the
copying trick above. Some of the boxes that need to be copied might themselves be previous copies, in which case the copying effect would be accumulated. In particular, it is possible for the outcome of a
resolution (call it $\mv b$) among boxes in $\cert_{\subset x}(X)$ for $x\in
P_X$ to become a box in $\cert_{\supseteq y}(Y)$ for $y\in P_Y$, 
where $Y$ is some dimension that is encountered later in the recursion. 
Now in this case, we have to be careful when we copy $\mv b$ while 
applying the copying trick for dimension $Y$.

To tackle the issues above, our final implementation of $\balance$ ends up
mapping the \bcp\ on $n$ dimension to a \bcp on $2n-2$ dimensions. Note that in
this case we cannot use the simple analysis we used for balanced certificates earlier:
since even if we can get $\sqrtC$ disjoint
problems with certificate size $\sqrtC$, each sub-problem will be in $2n-3$ dimensions.
\tetris on each sub-problem will take time $\tO(|\cert|^{n-2})$, which is generally too 
costly. Thus, for our $\balance$ function, we have to do a more careful analysis. 
We would like to stress  that Algorithm~\ref{alg offline as} is still valid: just 
that the earlier intuitive analysis needs to be 
tweaked a fair bit.

We present the full analysis of Algorithm~\ref{alg offline as} along with the complete definition of $\balance$ in Appendix~\ref{app:cert-n/2}. The analysis is a bit involved since we have to carefully analyze the number of input boxes each witness depends on. As alluded to earlier, resolvents from earlier levels of recursive calls can interact with resolutions at lower levels of recursions, which foils a straightforward recursive analysis. However, we prove a recursive structural lemma on how resolvents are supported by appropriate number of boxes from previous resolutions, which is enough to appropriately bound the total number of resolutions.

For the {\em online} version of the problem where the certificate is not given
as input, we use the same strategy as \tetrisreloaded: load boxes from the input
$\calB$ only when necessary. The number of boxes loaded is $\tO(|\cert|)$. Since
we are now loading boxes on the go, the notion of a balanced set of boxes
changes over time as new boxes are added. For example, a dyadic interval $x \in
P_X$ might define a {\em good} layer whose sub-problem can be solved efficiently for a while, but as new boxes come the layer might 
eventually become overloaded. Furthermore, the notion of balancedness depends on
the total number of boxes. Thus, a {\em bad} layer might also become good after
some time if new boxes do not intersect this layer. We take care of the above
key issues by periodically re-adjusting the partitions and show that the total
amount of readjustments is not too high.

\section{Lower Bounds and Extensions} 
\label{SEC:LOWERBOUNDS}

%!TEX root = main.tex
%\ar{Still needs to be written. Below is just a dump from earlier sections for now}

%\input{fig}

%We have seen so far that ordered geometric resolution (what \tetris\ does) is
%enough to generalize the results of~\cite{nnrr} to general input indices. 
%However, to obtain the stronger result on general queries in 
%Section~\ref{SEC:CERT-N/2}, we had to
%move beyond ordered geometric resolution (technically ordered geometric resolution in the
%transformed space). In this section, we will see that this switch was necessary.
In this section, we clarify the classes of geometric resolution that are needed to
compute various classes of \bcp\ (and hence joins) and prove their limitations.
Then, we prove some conditional lower bounds showing that the restrictions in 
our beyond worst-case results are necessary. Finally, we present some extensions 
where we prove sharper upper bounds that depend more on the query structure 
(but only hold for weaker forms of certificates).

\subsection{Limitations of Resolution Strategies}

So far in this paper, we have seen the class of geometric resolution in
Section~\ref{SEC:CERT-N/2}. This is the most general class of resolution we will
use in this paper. Recall from Section~\ref{sec:intro} that we denote this class
of resolution by \geo. In Sections~\ref{SEC:WORST-CASE} and~\ref{SEC:BEYOND-WC}, we saw a subclass of \geo: ordered geometric resolutions. Recall from Section~\ref{sec:intro} that we denote this class of resolutions by \ordered. It turns out that another subclass of \ordered, which we call \nocache\ is also an interesting class. \nocache\ (as mentioned in Section~\ref{sec:intro}) is the subclass of \ordered\ that only re-uses the input gap boxes: in other words, if an intermediate box has to be used more than once, then all the set of resolutions leading up to the intermediate box has to be repeated.\footnote{The qualifier \textsc{Tree} comes from the following fact. We can consider any set of resolutions in \geo\ (and hence \ordered) as a DAG-- each box is a node and the inputs to a resolution point towards the output of the resolution. \nocache\ is the subset of \ordered, where the resolution DAG is a tree.}

Figure~\ref{fig:res} summarizes the power and limitations of the three classes of resolution above. All of our lower bounds (which we present next) follow by constructing explicit hard examples with an empty output for various classes of resolution that we consider in this paper. (Note that for these hard examples any resolution strategy will have to generate the box $\dbox{\lambda,\dots,\lambda}$.)

We begin with the power of \nocache. We show that one can modify \tetris\ so that no resolution results are ever cached (this essentially corresponds to running Algorithm~\ref{alg:main:tetris} but without line~\ref{line:main:add-to-kb} in Algorithm~\ref{alg:main:tetris-skeleton}) so that one can achieve the \agm\ bound. Note that this change implies that the modified algorithm falls under \nocache. This implies that (see Corollary~\ref{cor agm tree}):
\bthm[\nocache\ achieves \agm bound]
\label{thm:app agm}
Let $Q$ be a join query, $N$ the total number of input tuples,
%$Z$ total number of output tuples,
and $\agm(Q)$ the best \agm bound for this instance.
%Let $\st B$ be set of boxes with \agm\ bound $\rho$.
Then there exists a scheme in \nocache\ that computes $Q$ with
$\tO(\agm(Q))$ many resolutions.
%, provided that the following conditions are met:
%\bi
%\item The GAO used by the algorithm is arbitrary, but fixed.
%\item The set of gap boxes $\calB$ provided to the algorithm contains
%all GAO-consistent gap boxes from each input relation.\footnote{The theorem still
%holds even if some relations have additional indices that are incompatible with
%the GAO chosen by the algorithm and the algorithm is using gap boxes from those
%indices as well.} (This is the fixed but arbitrary GAO chosen by \tetrispreloaded.)
%\ei
\ethm
%\ar{We need to add in something in the appendix that proves the above.}
%\mak{Corollay~\ref{cor agm tree}.}

Now recall that \tetris uses \ordered\ and in particular, by Theorem~\ref{thm:main fhtw} \tetris is powerful enough to recover the fractional hypertreewidth bound. In turn, this implies that \ordered\ is enough to compute the \bcp\ on boxes with treewidth $1$. Next, we argue that \nocache\ is not powerful enough to recover such a result. (See Theorem~\ref{thm:N^{n/2}-lowerbound} and its proof in the appendix.)

\bthm
\label{thm:Y-tree}
There exists a query $Q$ with treewidth $1$ such that every \nocache\ algorithm on input $\calB(Q)$ needs to make $\Omega(N^{n/2})$ many resolutions, where $N$ is the number of input tuples.
\ethm
%\ar{MAHMOUD: Please put in a proof of the above.}
%\mak{Theorem~\ref{thm:N^{n/2}-lowerbound}.}

We now move to \ordered. Since \tetris\ only uses \ordered, Theorem~\ref{thm:main:C^{w+1}+Z} immediately implies that there exists an \ordered\ algorithm that can solve the \bcp\ on boxes with treewidth $w$ with $\tO(|\boxcert|^{w+1}+Z)$ many resolutions. %We note next that the exponent cannot be improved by more than an {\em additive} factor of one if one sticks with \ordered.
 Next, we show that this in general is the best possible (see Theorem~\ref{thm:C^{w+1}-lowerbound} and its proof).
\bthm
\label{thm:cert^w+1-lb}
There exists a set $\st B$ of boxes with $1<\tw(\calH(\calB))<n-1$ such that any \ordered\ algorithm that solves the \bcp on $\st B$ needs to make $\Omega(|\boxcert(\calB)|^{w+1})$ many resolutions.
\ethm
We have already seen that we can prove bounds of the form $\tO(\abs{\boxcert}^w+Z)$ for the special case of $w=1$ (Theorem~\ref{thm:main acyclic arity<=2}) and $w=n-1$ (Theorem~\ref{thm:C^{n-1}+Z-upperbound}).
%\ar{The correct result in the appendix needs to be referred to here.}
  Next, we show the upper bound for $w=n-1$ is the best possible. (See Theorem~\ref{thm:C^{n-1}-lowerbound} and its proof.)
%\mak{Theorem~\ref{thm:C^{n-1}+Z-upperbound}.}
%\ar{MAHMOUD: Please prove the above.}
%\mak{Theorem~\ref{thm:C^{w+1}-lowerbound}}
%\ar{ALL: There are two lower bounds for \ordered: one is the above and the other is that for $n$-dimensions we have a $|\boxcert|^{n-1}$ lower bound. We probably should put only one in here: which one?}
\bthm
\label{thm:cert^n-1-lb}
There exists a set $\st B$ of boxes on $n$ dimensions such that any \ordered\ algorithm that solves the \bcp on $\st B$ needs to make $\Omega(|\boxcert(\calB)|^{n-1})$ many resolutions.
\ethm
%\mak{$\Omega(\abs{\st C}^{n-1})$ lower bound is Theorem~\ref{thm:C^{n-1}-lowerbound}.}

Note that the above implies that our move to \geo\ to obtain the bound of $\tO(|\boxcert|^{n/2}+Z)$ for the \bcp\ problem on dimension $n$ was necessary. It is natural to wonder if this bound can be further improved. We show that this is not possible with \geo. (See Theorem~\ref{thm:C^{n/2}-lowerbound} and its proof.)
\bthm
\label{thm:geo-lb-n-clique}
For every $n\ge 3$, there exists an instance for the \bcp\ on $n$ dimensions on which every \geo\ algorithm needs to make $\Omega(|\boxcert|^{n/2})$ many resolutions.
\ethm
%\ar{need to import in the half-n-half lower bound in the appendix to prove the above.}
%\mak{Theorem~\ref{thm:C^{n/2}-lowerbound}.}
The proof of Theorem~\ref{thm:geo-lb-n-clique} follows by a volume argument. We construct the boxes so that resolving any two of them results in a box with a small volume in the output tuples space. Thus, to cover the box $\dbox{\lambda,\dots,\lambda}$ one needs to perform a lot of geometric resolutions.

Another natural question is whether \geo\ is strictly less powerful than general resolution. In Appendix~\ref{app:paul}, we show that this indeed is the case by showing that general resolution can solve the hard instance for the proof of Theorem~\ref{thm:geo-lb-n-clique} with $\tO(\abs{\boxcert})$ many general resolutions. However, we do not know if general resolution can solve all \bcp\ instances on $n$ dimensions with $o(|\boxcert|^{n/2})$ resolutions.
%\ar{Put Paul's example in the appendix. We should also explicitly get Paul's permission to use his example.}

\subsection{Other Results}

A natural question we have not addressed so far is whether we can extend Theorem~\ref{thm:main acyclic arity<=2} to all $\beta$-acyclic queries. These queries do admit linear time algorithms but for the weaker comparison based certificate~\cite{nnrr}. We show that under the \textsf{3SUM} conjecture, one cannot hope for such a result for box certificates for $\beta$-acyclic queries if relations are allowed arities of $3$: see Proposition~\ref{prop omega 4/3} in the appendix.

Finally, we are able to prove upper bounds with better dependence on the query than the result in Section~\ref{SEC:CERT-N/2} {\em if} we work with weaker notions of certificates. See Appendix~\ref{sec gao consistent results}.

%I forgot all the things in this section, please fill
%  in with paragraphs.

\section{Conclusion}
%We presented the \tetris algorithm that showed a simple geometric
%resolution system allowed us to derive algorithms that match the
%efficiency of several of the best known algorithms for worst-case
%analysis and to derive new results for beyond worst-case analysis. 

We presented a simple geometric
resolution system that allowed us to derive algorithms that match the
efficiency of several of the best known algorithms for worst-case
analysis and to derive new results for beyond worst-case analysis. Of
purely conceptual interest, these rederivations in our simple
framework unify and -we argue- simplify their presentation. More
technically, our notion of certificate supports a wide range of
indexing schemes, compared to previous work that essentially focused
on Btrees with a total attribute order. We are excited about further
opportunities to more carefully study the impact of indexing on query
performance. In addition, we made a connection to proof complexity via
geometric resolution that we believe may further strengthen
the connection between constraint satisfaction and database join
processing. We conclude with two technical questions. First, as observed by~\cite{nnrr} it is not possible to obtain a certificate based result with the \fhtw\ in the exponent. It is a very interesting question to figure out the `correct' notion of fractional cover for certificate-based results. Second, it would be interesting to extend the results of~\cite{OZ14} to the certificate setting.

\section*{Acknowledgments}

We thank Paul Beame for clarifying the relation of our notion of geometric 
resolution with general resolution and we thank Javiel Rojas-Ledesma for bringing Klee's measure problem to our attention.

MAK's research is supported in part by NSF grant CCF-1161196. HQN's research
is supported in part by NSF grants CNF-1409551 and CCF-1319402.
CR gratefully acknowledges the support of DARPA's XDATA Program under
No. FA8750-12-2-0335, DEFT Program under No. FA8750-13-2-0039,
DARPA's MEMEX program under No. FA8750-14-2-0240, NSF CAREER Award
No. IIS-1353606, CCF-1356918 and EarthCube Award under No. ACI-1343760, the
ONR N000141210041 and N000141310129, the Sloan Research Fellowship, the
Moore Foundation Data Driven Investigator award, and gifts from American
Family Insurance, Google, Lightspeed Ventures, and Toshiba.
AR's research is supported in by part by NSF grants CCF-0844796 and CCF-1319402.

\bibliographystyle{acm}
\bibliography{main}

\def\cprime{$'$} \def\shortbib{0}
\begin{thebibliography}{10}

\bibitem{DBLP:books/aw/AbiteboulHV95}
{\sc Abiteboul, S., Hull, R., and Vianu, V.}
\newblock {\em Foundations of Databases}.
\newblock Addison-Wesley, 1995.

\bibitem{geometric-io}
{\sc Afshani, P., Barbay, J., and Chan, T.~M.}
\newblock Instance-optimal geometric algorithms.
\newblock In {\em FOCS\/} (2009), pp.~129--138.

\bibitem{DBLP:conf/coco/AllenderHMPS06}
{\sc Allender, E., Hellerstein, L., McCabe, P., Pitassi, T., and Saks, M.~E.}
\newblock Minimizing {DNF} formulas and {AC$^0_d$} circuits given a truth
  table.
\newblock In {\em IEEE Conference on Computational Complexity\/} (2006), IEEE
  Computer Society, pp.~237--251.

\bibitem{MR599482}
{\sc Alon, N.}
\newblock On the number of subgraphs of prescribed type of graphs with a given
  number of edges.
\newblock {\em Israel J. Math. 38}, 1-2 (1981), 116--130.

\bibitem{MR985145}
{\sc Arnborg, S., and Proskurowski, A.}
\newblock Linear time algorithms for {NP}-hard problems restricted to partial
  {$k$}-trees.
\newblock {\em Discrete Appl. Math. 23}, 1 (1989), 11--24.

\bibitem{AGM08}
{\sc Atserias, A., Grohe, M., and Marx, D.}
\newblock Size bounds and query plans for relational joins.
\newblock In {\em FOCS\/} (2008), IEEE Computer Society, pp.~739--748.

\bibitem{DBLP:conf/soda/BarbayK02}
{\sc Barbay, J., and Kenyon, C.}
\newblock Adaptive intersection and t-threshold problems.
\newblock In {\em SODA\/} (2002), pp.~390--399.

\bibitem{DBLP:journals/talg/BarbayK08}
{\sc Barbay, J., and Kenyon, C.}
\newblock Alternation and redundancy analysis of the intersection problem.
\newblock {\em ACM Transactions on Algorithms 4}, 1 (2008).

\bibitem{Bertino:2012:ITA:2480898}
{\sc Bertino, E., Ooi, B.~C., Sacks-Davis, R., Tan, K.-L., Zobel, J.,
  Shidlovsky, B., and Andronico, D.}
\newblock {\em Indexing Techniques for Advanced Database Systems}.
\newblock Springer Publishing Company, Incorporated, 2012.

\bibitem{BGL13}
{\sc Beyersdorff, O., Galesi, N., and Lauria, M.}
\newblock Parameterized complexity of {DPLL} search procedures.
\newblock {\em {ACM} Trans. Comput. Log. 14}, 3 (2013), 20.

\bibitem{DBLP:conf/uai/BidyukD04}
{\sc Bidyuk, B., and Dechter, R.}
\newblock On finding minimal w-cutset.
\newblock In {\em UAI\/} (2004), D.~M. Chickering and J.~Y. Halpern, Eds., AUAI
  Press, pp.~43--50.

\bibitem{Blanas:2011:DEM:1989323.1989328}
{\sc Blanas, S., Li, Y., and Patel, J.~M.}
\newblock Design and evaluation of main memory hash join algorithms for
  multi-core {CPUs}.
\newblock In {\em SIGMOD\/} (2011), ACM, pp.~37--48.

\bibitem{6686177}
{\sc Chan, T.}
\newblock Klee's measure problem made easy.
\newblock In {\em Foundations of Computer Science (FOCS), 2013 IEEE 54th Annual
  Symposium on\/} (Oct 2013), pp.~410--419.

\bibitem{Chaudhuri:1998:OQO:275487.275492}
{\sc Chaudhuri, S.}
\newblock An overview of query optimization in relational systems.
\newblock In {\em PODS\/} (1998), ACM, pp.~34--43.

\bibitem{DBLP:journals/tcs/ChekuriR00}
{\sc Chekuri, C., and Rajaraman, A.}
\newblock Conjunctive query containment revisited.
\newblock {\em Theor. Comput. Sci. 239}, 2 (2000), 211--229.

\bibitem{DBLP:conf/soda/ChenLSZ07}
{\sc Chen, J., Lu, S., Sze, S.-H., and Zhang, F.}
\newblock Improved algorithms for path, matching, and packing problems.
\newblock In {\em SODA\/} (2007), N.~Bansal, K.~Pruhs, and C.~Stein, Eds.,
  SIAM, pp.~298--307.

\bibitem{MR0149690}
{\sc Davis, M., Logemann, G., and Loveland, D.}
\newblock A machine program for theorem-proving.
\newblock {\em Comm. ACM 5\/} (1962), 394--397.

\bibitem{MR0134439}
{\sc Davis, M., and Putnam, H.}
\newblock A computing procedure for quantification theory.
\newblock {\em J. Assoc. Comput. Mach. 7\/} (1960), 201--215.

\bibitem{MR0043115}
{\sc de~Bruijn, N.~G., van Ebbenhorst~Tengbergen, C., and Kruyswijk, D.}
\newblock On the set of divisors of a number.
\newblock {\em Nieuw Arch. Wiskunde (2) 23\/} (1951), 191--193.

\bibitem{DBLP:journals/ai/Dechter90}
{\sc Dechter, R.}
\newblock Enhancement schemes for constraint processing: Backjumping, learning,
  and cutset decomposition.
\newblock {\em Artif. Intell. 41}, 3 (1990), 273--312.

\bibitem{Dechter:2003:CP:861888}
{\sc Dechter, R.}
\newblock {\em Constraint Processing}.
\newblock Morgan Kaufmann Publishers Inc., San Francisco, CA, USA, 2003.

\bibitem{DBLP:conf/aaai/DechterP88}
{\sc Dechter, R., and Pearl, J.}
\newblock Tree-clustering schemes for constraint-processing.
\newblock In {\em AAAI\/} (1988), H.~E. Shrobe, T.~M. Mitchell, and R.~G.
  Smith, Eds., AAAI Press / The MIT Press, pp.~150--154.

\bibitem{journals/ai/DechterP89}
{\sc Dechter, R., and Pearl, J.}
\newblock Tree clustering for constraint networks.
\newblock {\em Artificial Intelligence 38}, 3 (1989), 353--366.

\bibitem{DBLP:conf/kr/DechterR94}
{\sc Dechter, R., and Rish, I.}
\newblock Directional resolution: The {D}avis-{P}utnam procedure, revisited.
\newblock In {\em KR\/} (1994), J.~Doyle, E.~Sandewall, and P.~Torasso, Eds.,
  Morgan Kaufmann, pp.~134--145.

\bibitem{DBLP:conf/soda/DemaineLM00}
{\sc Demaine, E.~D., L{\'o}pez-Ortiz, A., and Munro, J.~I.}
\newblock Adaptive set intersections, unions, and differences.
\newblock In {\em SODA\/} (2000), pp.~743--752.

\bibitem{DBLP:conf/sat/EenS03}
{\sc E{\'e}n, N., and S{\"o}rensson, N.}
\newblock An extensible sat-solver.
\newblock In {\em SAT\/} (2003), E.~Giunchiglia and A.~Tacchella, Eds.,
  vol.~2919 of {\em Lecture Notes in Computer Science}, Springer, pp.~502--518.

\bibitem{DBLP:journals/jacm/Fagin83}
{\sc Fagin, R.}
\newblock Degrees of acyclicity for hypergraphs and relational database
  schemes.
\newblock {\em J. ACM 30}, 3 (1983), 514--550.

\bibitem{Fagin:2001:OAA:375551.375567}
{\sc Fagin, R., Lotem, A., and Naor, M.}
\newblock Optimal aggregation algorithms for middleware.
\newblock In {\em Proceedings of the Twentieth ACM SIGMOD-SIGACT-SIGART
  Symposium on Principles of Database Systems\/} (New York, NY, USA, 2001),
  PODS '01, ACM, pp.~102--113.

\bibitem{Fan:2015:QBD:2745754.2745771}
{\sc Fan, W., Geerts, F., Cao, Y., Deng, T., and Lu, P.}
\newblock Querying big data by accessing small data.
\newblock In {\em Proceedings of the 34th ACM SIGMOD-SIGACT-SIGAI Symposium on
  Principles of Database Systems\/} (New York, NY, USA, 2015), PODS '15, ACM,
  pp.~173--184.

\bibitem{MR1639767}
{\sc Friedgut, E., and Kahn, J.}
\newblock On the number of copies of one hypergraph in another.
\newblock {\em Israel J. Math. 105\/} (1998), 251--256.

\bibitem{MR1279424}
{\sc Goerdt, A.}
\newblock Davis-{P}utnam resolution versus unrestricted resolution.
\newblock {\em Ann. Math. Artificial Intelligence 6}, 1-3 (1992), 169--184.

\bibitem{DBLP:journals/jcss/GottlobLS03}
{\sc Gottlob, G., Leone, N., and Scarcello, F.}
\newblock Robbers, marshals, and guards: game theoretic and logical
  characterizations of hypertree width.
\newblock {\em J. Comput. Syst. Sci. 66}, 4 (2003), 775--808.

\bibitem{graefe93}
{\sc Graefe, G.}
\newblock Query evaluation techniques for large databases.
\newblock {\em ACM Computing Surveys 25}, 2 (June 1993), 73--170.

\bibitem{MR745586}
{\sc Griggs, J.~R.}
\newblock Maximum antichains in the product of chains.
\newblock {\em Order 1}, 1 (1984), 21--28.

\bibitem{grohebounds}
{\sc Grohe, M.}
\newblock Bounds and algorithms for joins via fractional edge covers.
\newblock In {\em In Search of Elegance in the Theory and Practice of
  Computation\/} (2013), V.~Tannen, L.~Wong, L.~Libkin, W.~Fan, W.-C. Tan, and
  M.~P. Fourman, Eds., vol.~8000 of {\em Lecture Notes in Computer Science},
  Springer, pp.~321--338.

\bibitem{GM06}
{\sc Grohe, M., and Marx, D.}
\newblock Constraint solving via fractional edge covers.
\newblock In {\em SODA\/} (2006), ACM Press, pp.~289--298.

\bibitem{DBLP:journals/ai/GyssensJC94}
{\sc Gyssens, M., Jeavons, P., and Cohen, D.~A.}
\newblock Decomposing constraint satisfaction problems using database
  techniques.
\newblock {\em Artif. Intell. 66}, 1 (1994), 57--89.

\bibitem{DBLP:conf/adbt/GyssensP82}
{\sc Gyssens, M., and Paredaens, J.}
\newblock A decomposition methodology for cyclic databases.
\newblock In {\em Advances in Data Base Theory\/} (1982), pp.~85--122.

\bibitem{Kim:2009:SVH:1687553.1687564}
{\sc Kim, C., Kaldewey, T., Lee, V.~W., Sedlar, E., Nguyen, A.~D., Satish, N.,
  Chhugani, J., Di~Blas, A., and Dubey, P.}
\newblock Sort vs. hash revisited: fast join implementation on modern
  multi-core {CPUs}.
\newblock {\em Proc. VLDB Endow. 2}, 2 (Aug. 2009), 1378--1389.

\bibitem{DBLP:journals/jcss/KolaitisV00}
{\sc Kolaitis, P.~G., and Vardi, M.~Y.}
\newblock Conjunctive-query containment and constraint satisfaction.
\newblock {\em J. Comput. Syst. Sci. 61}, 2 (2000), 302--332.

\bibitem{DBLP:books/cs/Maier83}
{\sc Maier, D.}
\newblock {\em The Theory of Relational Databases}.
\newblock Computer Science Press, 1983.

\bibitem{marques2009conflict}
{\sc Marques-Silva, J., Lynce, I., and Malik, S.}
\newblock Conflict-driven clause learning {SAT} solvers.
\newblock {\em Handbook of satisfiability 185\/} (2009), 131--153.

\bibitem{DBLP:journals/talg/Marx10}
{\sc Marx, D.}
\newblock Approximating fractional hypertree width.
\newblock {\em ACM Transactions on Algorithms 6}, 2 (2010).

\bibitem{DBLP:conf/stoc/Marx10}
{\sc Marx, D.}
\newblock Tractable hypergraph properties for constraint satisfaction and
  conjunctive queries.
\newblock In {\em STOC\/} (2010), pp.~735--744.

\bibitem{DBLP:journals/mst/Marx11}
{\sc Marx, D.}
\newblock Tractable structures for constraint satisfaction with truth tables.
\newblock {\em Theory Comput. Syst. 48}, 3 (2011), 444--464.

\bibitem{DBLP:journals/jacm/Marx13}
{\sc Marx, D.}
\newblock Tractable hypergraph properties for constraint satisfaction and
  conjunctive queries.
\newblock {\em J. {ACM} 60}, 6 (2013), 42.

\bibitem{masek1979}
{\sc Masek, W.~J.}
\newblock Some {NP}-complete set covering problems.
\newblock unpublished manuscript.

\bibitem{alon-network-motifs}
{\sc Milo, R., Shen-Orr, S., Itzkovitz, S., Kashtan, N., Chklovskii, D., and
  Alon, U.}
\newblock Network motifs: simple building blocks of complex networks.
\newblock {\em Science 298}, 5594 (October 2002), 824--827.

\bibitem{DBLP:conf/dac/MoskewiczMZZM01}
{\sc Moskewicz, M.~W., Madigan, C.~F., Zhao, Y., Zhang, L., and Malik, S.}
\newblock Chaff: Engineering an efficient sat solver.
\newblock In {\em DAC\/} (2001), ACM, pp.~530--535.

\bibitem{nnrr}
{\sc Ngo, H.~Q., Nguyen, D.~T., R\'e, C., and Rudra, A.}
\newblock Beyond worst-case analysis for joins with {Minesweeper}.
\newblock In {\em PODS\/} (2014), pp.~234--245.

\bibitem{NPRR}
{\sc Ngo, H.~Q., Porat, E., R{\'e}, C., and Rudra, A.}
\newblock Worst-case optimal join algorithms: [extended abstract].
\newblock In {\em PODS\/} (2012), pp.~37--48.

\bibitem{skew}
{\sc Ngo, H.~Q., R{\'e}, C., and Rudra, A.}
\newblock Skew strikes back: New developments in the theory of join algorithms.
\newblock In {\em SIGMOD RECORD\/} (2013), pp.~5--16.

\bibitem{Nguyen:2015:JPG:2764947.2764948}
{\sc Nguyen, D., Aref, M., Bravenboer, M., Kollias, G., Ngo, H.~Q., R{\'e}, C.,
  and Rudra, A.}
\newblock Join processing for graph patterns: An old dog with new tricks.
\newblock In {\em Proceedings of the GRADES'15\/} (New York, NY, USA, 2015),
  GRADES'15, ACM, pp.~2:1--2:8.

\bibitem{OZ14}
{\sc Olteanu, D., and Zavodny, J.}
\newblock Size bounds for factorised representations of query results.
\newblock {\em ACM Transactions on Database Systems\/} (2014).
\newblock To appear.

\bibitem{DBLP:journals/sigmod/ONeilG95}
{\sc O'Neil, P.~E., and Graefe, G.}
\newblock Multi-table joins through bitmapped join indices.
\newblock {\em SIGMOD Record 24}, 3 (1995), 8--11.

\bibitem{DBLP:conf/sigmod/ONeilQ97}
{\sc O'Neil, P.~E., and Quass, D.}
\newblock Improved query performance with variant indexes.
\newblock In {\em SIGMOD Conference\/} (1997), J.~Peckham, Ed., ACM Press,
  pp.~38--49.

\bibitem{doi:10.1137/0220065}
{\sc Overmars, M.~H., and Yap, C.-K.}
\newblock New upper bounds in klee’s measure problem.
\newblock {\em SIAM Journal on Computing 20}, 6 (1991), 1034--1045.

\bibitem{DBLP:conf/pods/PapadimitriouY97}
{\sc Papadimitriou, C.~H., and Yannakakis, M.}
\newblock On the complexity of database queries.
\newblock In {\em PODS\/} (1997), pp.~12--19.

\bibitem{3sum}
{\sc P{\v a}tra{\c s}cu, M.}
\newblock Towards polynomial lower bounds for dynamic problems.
\newblock In {\em Proc. 42nd ACM Symposium on Theory of Computing (STOC)\/}
  (2010), pp.~603--610.

\bibitem{DBLP:books/daglib/0066829}
{\sc Pearl, J.}
\newblock {\em Probabilistic reasoning in intelligent systems - networks of
  plausible inference}.
\newblock Morgan Kaufmann series in representation and reasoning. Morgan
  Kaufmann, 1989.

\bibitem{DBLP:journals/bioinformatics/PrzuljCJ04}
{\sc Przulj, N., Corneil, D.~G., and Jurisica, I.}
\newblock Modeling interactome: scale-free or geometric?
\newblock {\em Bioinformatics 20}, 18 (2004), 3508--3515.

\bibitem{Ramakrishnan:2002:DMS:560733}
{\sc Ramakrishnan, R., and Gehrke, J.}
\newblock {\em Database Management Systems}, 3~ed.
\newblock McGraw-Hill, Inc., New York, NY, USA, 2003.

\bibitem{MR855559}
{\sc Robertson, N., and Seymour, P.~D.}
\newblock Graph minors. {II}. {A}lgorithmic aspects of tree-width.
\newblock {\em J. Algorithms 7}, 3 (1986), 309--322.

\bibitem{RoughgardenCS264}
{\sc Roughgarden, T.}
\newblock Lecture notes on beyond worst-case analysis (cs264).

\bibitem{DBLP:journals/sigmod/Scarcello05}
{\sc Scarcello, F.}
\newblock Query answering exploiting structural properties.
\newblock {\em SIGMOD Record 34}, 3 (2005), 91--99.

\bibitem{DBLP:conf/iccad/SilvaS96}
{\sc Silva, J. P.~M., and Sakallah, K.~A.}
\newblock Grasp - a new search algorithm for satisfiability.
\newblock In {\em ICCAD\/} (1996), pp.~220--227.

\bibitem{DBLP:journals/tc/Marques-SilvaS99}
{\sc Silva, J. P.~M., and Sakallah, K.~A.}
\newblock Grasp: A search algorithm for propositional satisfiability.
\newblock {\em IEEE Trans. Computers 48}, 5 (1999), 506--521.

\bibitem{DBLP:journals/ipm/SpieglerM85}
{\sc Spiegler, I., and Maayan, R.}
\newblock Storage and retrieval considerations of binary data bases.
\newblock {\em Inf. Process. Manage. 21}, 3 (1985), 233--254.

\bibitem{DBLP:conf/www/SuriV11}
{\sc Suri, S., and Vassilvitskii, S.}
\newblock Counting triangles and the curse of the last reducer.
\newblock In {\em WWW\/} (2011), pp.~607--614.

\bibitem{Tsourakakis:2008:FCT:1510528.1511415}
{\sc Tsourakakis, C.~E.}
\newblock Fast counting of triangles in large real networks without counting:
  Algorithms and laws.
\newblock In {\em ICDM\/} (2008), IEEE Computer Society, pp.~608--617.

\bibitem{DBLP:books/cs/Ullman89}
{\sc Ullman, J.~D.}
\newblock {\em Principles of Database and Knowledge-Base Systems, Volume II}.
\newblock Computer Science Press, 1989.

\bibitem{leapfrog}
{\sc Veldhuizen, T.~L.}
\newblock Triejoin: A simple, worst-case optimal join algorithm.
\newblock In {\em ICDT\/} (2014), pp.~96--106.

\bibitem{DBLP:conf/vldb/Yannakakis81}
{\sc Yannakakis, M.}
\newblock Algorithms for acyclic database schemes.
\newblock In {\em VLDB\/} (1981), pp.~82--94.

\bibitem{DBLP:conf/cade/Zhang97}
{\sc Zhang, H.}
\newblock Sato: An efficient propositional prover.
\newblock In {\em CADE\/} (1997), W.~McCune, Ed., vol.~1249 of {\em Lecture
  Notes in Computer Science}, Springer, pp.~272--275.

\end{thebibliography}
%\end{bibunit}

\newpage
%\onecolumn

\appendix
\section{Background}
\label{sec prelim}
% ----------------------------------------------------------------------------

\subsection{\agm bound}
\label{sec:agm-bound}

The structure of a join query $Q$ can be represented by a hypergraph
$\calH(Q)$, or simply $\calH$. The hypergraph has vertex set
$\calV = \vars(Q)$, and edge set
\[ \calE = \left\{ \vars(R) \suchthat R \in \atoms(Q) \right\}. \]
We often index the relations using edges from this hypergraph.
Hence, instead of writing $R(\vars(R))$, we can write $R_F$, for $F\in\calE$.

A {\em fractional edge cover} of a hypergraph $\calH$ is a point
$\mv x = (x_F)_{F\in\calE}$ in the following polyhedron: 
\[ \left\{ \mv x \suchthat \sum_{F: v \in F} x_F \geq 1, \forall v \in \calV,
\mv x \geq \mv 0 \right\}. \]

Atserias-Grohe-Marx \cite{AGM08} and Grohe-Marx \cite{GM06}
proved the following remarkable inequality, which shall be 
referred to as the {\em \agm's inequality} henceforth.
For any fractional edge cover $\mv x$ of the query's hypergraph, 
\begin{equation}
\label{eqn:AGM}
 |Q| = |\Join_{F\in\calE}R_F| \leq \prod_{F\in\calE} |R_F|^{x_F}. 
\end{equation}
Here, $|Q|$ is the number of tuples in the (output) relation $Q$.

The optimal edge cover for the \agm bound depends on the relation sizes.
To minimize the right hand side of \eqref{eqn:AGM}, we can solve the
following linear program:
\begin{eqnarray*}
\min && \sum_{F\in \calE} (\log_2 |R_F|) \cdot x_F\\
\text{s.t.} && \sum_{F: v \in F} x_F \geq 1, v \in \calV\\
&&\mv x \geq \mv 0
\end{eqnarray*}
Implicitly, the objective function above depends on the database instance
$\calD$ on which the query is applied. 
Let $\rho^*(Q, \calD)$ denote the 
optimal objective value to the above linear program. 
We refer to $\rho^*(Q,\calD)$ as the {\em fractional edge cover number} of the 
query $Q$ with respect to the database instance $\calD$,
following Grohe \cite{grohebounds}.
The \agm's inequality can be summarized simply by
$|Q| \leq 2^{\rho^*(Q, \calD)}$.

\bdefn[$\agm(Q)$]
We also define $\agm(Q, \calD) = 2^{\rho^*(Q,\calD)}$; in particular
$\agm(Q, \calD)$ is the best \agm-bound for query $Q$ on database instance
$\calD$.
\edefn

When $\calD$ is clear from context, we will simply use $\rho^*(Q)$ to denote
$\rho^*(Q, \calD)$, and $\agm(Q)$ to denote $\agm(Q, \calD)$.

\bdefn[Fractional edge cover number]
Sometimes it is convenient to replace individual relation sizes by a single
number $N$ representing the entire input size. In that case, in the linear
program above we can drop the $(\log_2|R_F|)$ factors. 
The optimal objective value for this simplified program is called the
{\em fractional edge cover number} of the hypergraph $\calH$,
denoted by $\rho^*(\calH)$.
\edefn

\subsection{Tree decomposition, acyclicity, and various notions of widths}\label{sec:tree-decomp}

\bdefn[Acyclicity and GYO elimination]\label{defn:acyclic}
There are many definitions of acyclic hypergraphs. A hypergraph $(\cal
V, \cal E)$ is $\alpha$-acyclic if the GYO procedure returns
an empty hypergraph~\cite[p.~128]{DBLP:books/aw/AbiteboulHV95}. Essentially, in GYO
one iterates two steps: (1) remove any edge that is empty or contained
in another hyperedge, or (2) remove vertices that appear in at most
one hyperedge. If the result is empty, then the hypergraph is
$\alpha$-acyclic. A hypergraph $\calH$
is $\beta$-acyclic if the hypergraph formed by
any subset of hyperedges of $\calH$ is $\alpha$-acyclic.
\edefn

\bdefn[Tree decomposition]\label{defn:tree-dec}
Let $\calH = (\calV, \calE)$ be a hypergraph.
A {\em tree decomposition} of $\calH$ is a pair $(T, \chi)$
where $T = (V(T), E(T))$ is a tree and $\chi : V(T) \to 2^{\calV}$ assigns to
each node of the tree $T$ a set of vertices of $\calH$.
The sets $\chi(t)$, $t\in V(T)$, are called the {\em bags} of the 
tree decomposition.  There are two properties the bags must satisfy:
\bi
 \item[(a)] For every hyperedge $F \in \calE$, there is a bag $\chi(T)$ 
 such that $F\subseteq \chi(t)$.
 \item[(b)] For every vertex $v \in \calV$, the set 
 $\{ t \suchthat t \in T, v \in \chi(t) \}$ is not empty and forms a 
 connected subtree of $T$.
\ei
\edefn

The rest of this section roughly follows the definitions given in 
\cite{DBLP:journals/talg/Marx10}.
The {\em width} of a tree decomposition is the quantity 
\[ \max_{t \in V(T)} |\chi(t)|-1. \]
The {\em treewidth} of a hypergraph $\calH$, denoted by $\tw(\calH)$,
is the minimum width over all tree decompositions of the hypergraph.
Let $\rho^*(t)$ denote the fractional edge cover number of the hypergraph
$\calH[\chi(t)]$: the hypergraph $\calH$ restricted to the bag $\chi(t)$.
Then the quantify
\[ \max_{t \in V(T)} |\rho^*(t)| \]
is called the {\em fractional hypertree width} (fhtw) of this tree decomposition of the hypergraph $\calH$.
The fractional hypertree width of $\calH$, denoted by fhtw$(\calH)$, is the minimum one over all tree decompositions of $\calH$.
It should be clear that fhtw$(\calH) \leq$ tw$(\calH)$.

\bdefn[Treewidth of a query]
The treewidth of a query $Q$, denoted by $\tw(Q)$, is the treewidth of its
hypergraph $\calH(Q)$.
\edefn

% ----------------------------------------------------------------------------
\section{Indices as Gap Collections, Geometric Certificates}
\label{sec:cert}
% ----------------------------------------------------------------------------

In this section, we attempt to make the gap box abstraction more concrete.
We will explain what the ``boxes'' are and why the idea that every database
index is a collection of boxes holds for ordered input index data structures 
such as a sorted list, B-tree, B+-tree, or a trie. Note again
that a hash table can be simulated by a search tree within a $\log$-factor;
hence this observation holds for hash-based indices as far as the theoretical
bounds in this paper are concerned.

% ----------------------------------------------------------------------------
\subsection{\ms and GAO-consistent certificate}
% ----------------------------------------------------------------------------

\begin{figure}                                                                  
\centerline{\includegraphics[width=4in]{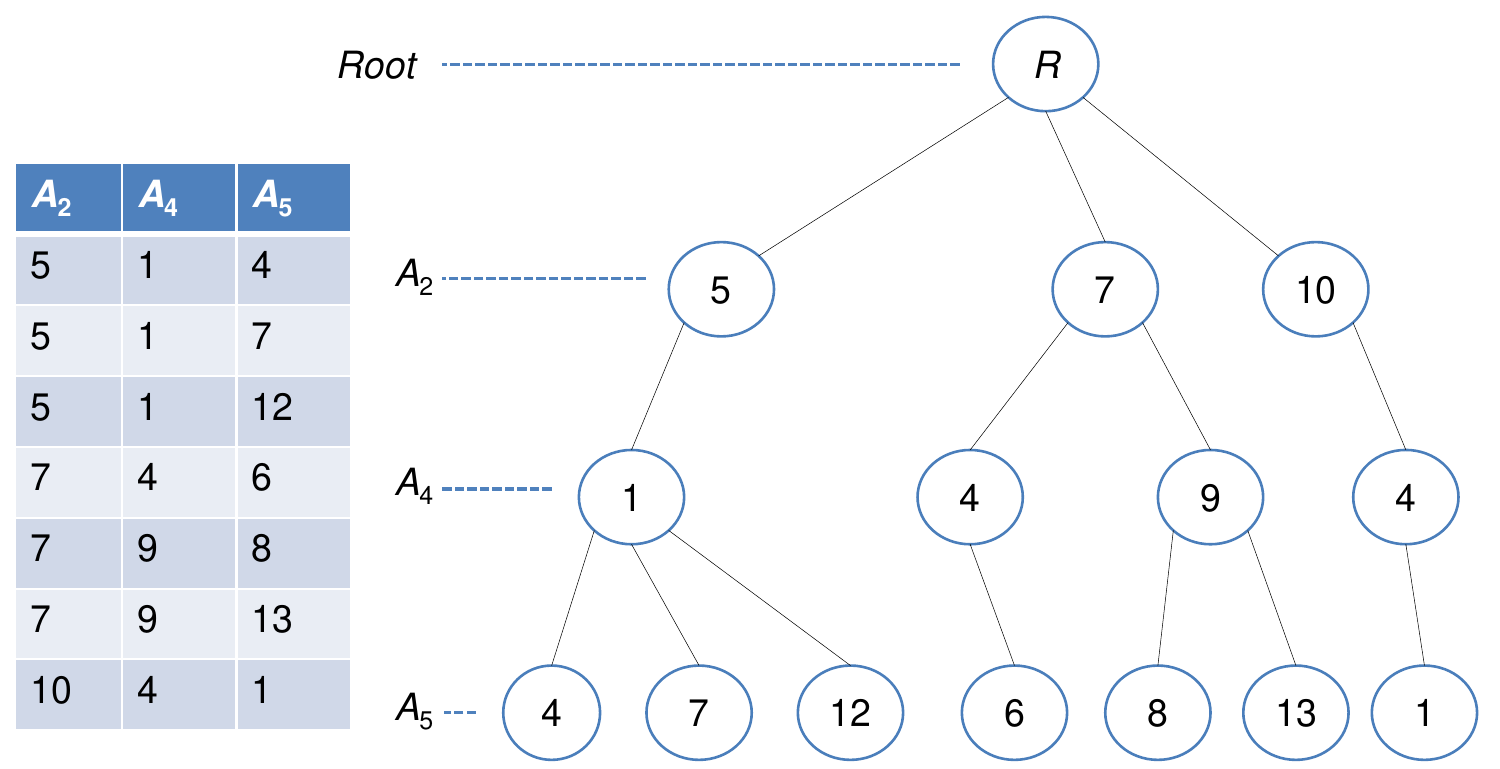}}                           
\caption{The (unbounded fanout) $\btree$ data structure used by \ms}
\label{fig:R-trie}                                                              
\end{figure}

We recall key concepts and ideas from \cite{nnrr} from which a new notion of
{\em geometric certificate} called {\em GAO-consistent certificate} arises
naturally.
The \ms algorithm assumes that the input relations are already indexed
using a search-tree data structure such as a traditional B-tree 
which is widely used in commercial relational database 
systems~\cite[Ch.10]{Ramakrishnan:2002:DMS:560733}          
or a Trie~\cite{leapfrog}.
For example, Figure~\ref{fig:R-trie} shows the index for a relation
$R$ on attribute set $\vars(R) = \{A_2, A_4, A_5\}$.
This index for $R$ is in the order $A_2,A_4,A_5$.
Furthermore, there is an ordering of all the attributes 
in $\vars(Q)$ -- called the {\em global attribute order} (GAO) --
such that all input relations are indexed consistent with this GAO.
This assumption shall be referred to as the {\em GAO-consistency assumption}.
For example, for the triangle query $Q = R(A,B) \Join S(B, C) \Join T(A, C)$,
if the GAO is $B, A, C$, then
$R$ is indexed in the $(B, A)$ order,
$S$ in the $(B, C)$, and $T$ in the $(A, C)$.

\ms views the set of potential output tuples of a join
query as ``grid points'' in a high-dimensional space called the {\em output
space}. The output space is the cross-product of all domains of input
attributes.
\ms maintains a data structure called the {\em constraint data structure} (\cds) 
which holds a collection of
``gaps'' in the output space whose union contains all points which are either
non-output tuples or output tuples already reported.
When the union of gaps in the $\cds$ covers the entire output space,
the algorithm terminates.

More concretely, \ms starts off by obtaining an arbitrary ``{\em probe point}'' 
$\mv t$ from the output space. 
Given the current probe point, \ms performs the following steps:

\begin{enumerate}                                                                   
\item \ms queries into the indices of the input relations to verify whether
the current probe point $\mv t$ is indeed an output tuple, in which case $\mv              
t$ is output and an appropriate gap is inserted into the \cds signifying
that the output $\mv t$ has already been reported.
If $\mv t$ is not an output tuple, the input index structures              
return some gaps ``around'' $\mv t$. These gaps are inserted into the
\cds.

\item Then, \ms queries into the \cds to obtain the next probe point, which is
a point not covered by the union of the gaps inserted thus far.
If no probe point exists, the algorithm terminates, because all inserted gaps
cover the output space.
\end{enumerate}   

To be more concrete, suppose $\vars(Q) = \{A_1,\dots,A_6\}$ with 
$(A_1,A_2,\dots,A_6)$ being the GAO. Suppose the relation $R$ shown in Figure~\ref{fig:R-trie}
is an input relation. Consider the following probe point
\[ \mv t = (t_1,t_2,t_3,t_4,t_5,t_6) = (6, 6, 1, 3, 7, 9). \]
We first project this probe point down to the coordinate subspace spanned by the
attributes of $R$: $(t_2, t_4, t_5) = (6, 3, 7)$.
From the index structure for $R$, we see that $t_2 = 6$ falls between
the two $A_2$-values $5$ and $7$ in the relation. Thus, this index returns a
gap consisting of all points lying between the two hyperplanes $A_2=5$ and
$A_2=7$. 
This gap is encoded with the constraint 
\[ \langle *, (5, 7), *, *, *, * \rangle, \]
where $*$ is the wildcard character matching any value in the corresponding
domain, and $(5,7)$ is an open interval on the $A_2$-axis.
On the other hand, suppose the probe point is
\[ \mv t = (t_1,t_2,t_3,t_4,t_5,t_6) = (6, 7, 1, 5, 8, 9). \]
Then, a gap returned might be the band in the hyperplane $A_2=7$,\;
$4 < A_4 < 9$. The encoding of this gap is
\[ \langle *, 7, *, (4, 9), *, * \rangle. \]
The number $7$ indicates that this gap is inside the hyperplane $A_2=7$,
and the open interval encodes all points inside this hyperplane where
$4 < A_4 < 9$.

Due to the GAO-consistency assumption, all the constraints returned by the input
indices have the property that for each constraint there is only one interval 
component, after that there are only wildcard component. Henceforth, these
constraints will be called {\em GAO-consistent constraints}.
The key result from \cite{nnrr} is the following: 
If we consider the class of join algorithms that
only perform comparisons between input elements, then the number of comparisons 
necessary to certify the correctness of the output is a lowerbound on the run 
time of this class of (non-deterministic) algorithms.
This class of comparison-based algorithms models a very wide class of join 
algorithms, including index-nested-loop join, block-nested-loop join,
hash-join (up to a $\log$-factor), grace join.

Note that every constraint inserted into the $\cds$ is a (multi-dimensional)
rectangle inside the output space. We will call these rectangles {\em boxes}.
The analysis from \cite{nnrr} shows the following:
\bi
 \item Let $\gaocert$ denote the minimum set of comparisons sufficient to
certify the output, then the number of probe points $\mv t$ issued by \ms is 
$\tilde O(|\gaocert|+Z)$, where $Z$ is the output size,
and $\tilde O$ hides a query-dependent factor.
 \item Every probe point $\mv t$ is a point that is not covered by existing boxes stored
in the \cds.
 \item If a new probe point $\mv t$ is not an output tuple, then \ms will
insert at least one new box containing the probe point $\mv t$. We will refer
to these boxes as {\em gap boxes}. Note the
important fact that \ms often also inserts boxes that {\em do
not} contain $\mv t$; this is because the analysis needs to show that we
can ``pay'' for this iteration by a fresh comparison in $\gaocert$.
 \item If a new probe point $\mv t$ {\em is} an output tuple, then \ms will
also insert a new box containing $\mv t$ to rule it out; but we will refer
to these boxes as {\em output boxes}.
\ei

Now, let's forget about $\gaocert$ and examine what the \cds sees and processes.
The \cds has a set of {\em output
boxes}, and a collection of {\em gap boxes} which do not contain any output point. 
When the \cds cannot find a probe point anymore, 
the union of output boxes and gap boxes is the entire output space $\outspace$.
In other words, every point in the output space 
is either an output point, or is covered by a gap box. 
We will call the collection of gap boxes satisfying this property 
{\em box certificate} (to be defined more precisely below): It certifies that the reported output points are all the output points of the join query.
A box certificate is a purely geometric notion, and on the surface does not
seem to have anything to do with comparisons. Yet from the results in
\cite{nnrr}, we now know that a box certificate of minimum size
is a lowerbound on the number of comparisons issued by {\em any}
comparison-based join algorithm (with the GAO-consistency assumption).

With the GAO-consistency assumption, the gap boxes can only come from the gaps 
issued by the input relations following their GAO-consistent index structures. 
We make this notion more precise here.

\bdefn[GAO-consistent box certificate]
Let $Q$ be a natural join query whose input relations have already been indexed
consistent with a GAO $(A_1, \dots, A_n)$. 
The gap boxes owned by $R \in \atoms(Q)$ are formed by the gaps 
between adjacent sibling nodes in the search trees for $R$.
These boxes can be of dimension $n$, $n-1$, down to
$1$. (A $1$-dimensional box is a segment on a line.)
Let $\calB(R)$ denote the set of gap boxes owned by $R$.

A {\em GAO-consistent box certificate} is a subset of gap boxes from 
$\bigcup_{R\in\atoms(Q)} \calB(R)$ that cover every point not in the output. 
We use $\gaoboxcert$ to also denote an optimal GAO-consistent box
certificate, which means it has the minimum number of gap boxes.
\edefn

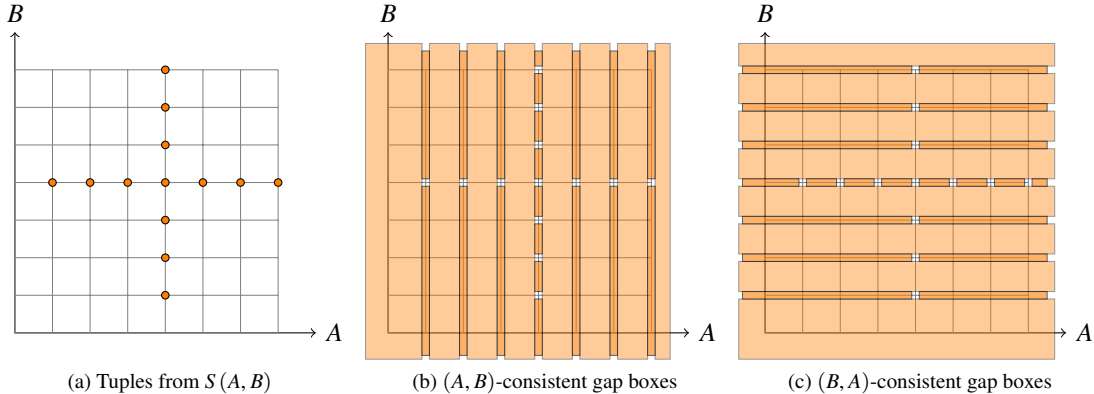
\begin{figure}[!htp]
\centering
\subfloat[Tuples from $S(A,B)$]{
    \begin{tikzpicture}[scale=0.5]
    \draw [<->] (0, 8) -- (0, 0) -- (8, 0);
    \draw[help lines] (0,0) grid (7,7);
    \draw [fill=white,opacity=0] (-0.7,-0.7) rectangle (7.7,0.9);
    \draw [fill=orange] (1,4) circle [radius=0.1];
    \draw [fill=orange] (2,4) circle [radius=0.1];
    \draw [fill=orange] (3,4) circle [radius=0.1];
    \draw [fill=orange] (5,4) circle [radius=0.1];
    \draw [fill=orange] (6,4) circle [radius=0.1];
    \draw [fill=orange] (7,4) circle [radius=0.1];
    \draw [fill=orange] (4,1) circle [radius=0.1];
    \draw [fill=orange] (4,2) circle [radius=0.1];
    \draw [fill=orange] (4,3) circle [radius=0.1];
    \draw [fill=orange] (4,4) circle [radius=0.1];
    \draw [fill=orange] (4,5) circle [radius=0.1];
    \draw [fill=orange] (4,6) circle [radius=0.1];
    \draw [fill=orange] (4,7) circle [radius=0.1];
    \node [black] at (8.5,0) {$A$};
    \node [black] at (0, 8.5) {$B$};
    \end{tikzpicture}
}
\subfloat[$(A,B)$-consistent gap boxes]{
    \begin{tikzpicture}[scale=0.5]
    \draw [<->] (0, 8) -- (0, 0) -- (8, 0);
    \draw[help lines] (0,0) grid (7,7);
    \draw [fill=orange,opacity=0.4] (-0.6,-0.7) rectangle (0.9,7.7);
    \draw [fill=orange,opacity=0.4] (1.1,-0.7) rectangle (1.9,7.7);
    \draw [fill=orange,opacity=0.4] (2.1,-0.7) rectangle (2.9,7.7);
    \draw [fill=orange,opacity=0.4] (3.1,-0.7) rectangle (3.9,7.7);
    \draw [fill=orange,opacity=0.4] (4.1,-0.7) rectangle (4.9,7.7);
    \draw [fill=orange,opacity=0.4] (5.1,-0.7) rectangle (5.9,7.7);
    \draw [fill=orange,opacity=0.4] (6.1,-0.7) rectangle (6.9,7.7);
    \draw [fill=orange,opacity=0.4] (7.1,-0.7) rectangle (7.5,7.7);
    \draw [fill=orange,opacity=0.6] (0.9,3.9) rectangle (1.1,-0.6);
    \draw [fill=orange,opacity=0.6] (0.9,4.1) rectangle (1.1,7.5);
    \draw [fill=orange,opacity=0.6] (1.9,3.9) rectangle (2.1,-0.6);
    \draw [fill=orange,opacity=0.6] (1.9,4.1) rectangle (2.1,7.5);
    \draw [fill=orange,opacity=0.6] (2.9,3.9) rectangle (3.1,-0.6);
    \draw [fill=orange,opacity=0.6] (2.9,4.1) rectangle (3.1,7.5);
    \draw [fill=orange,opacity=0.6] (4.9,3.9) rectangle (5.1,-0.6);
    \draw [fill=orange,opacity=0.6] (4.9,4.1) rectangle (5.1,7.5);
    \draw [fill=orange,opacity=0.6] (5.9,3.9) rectangle (6.1,-0.6);
    \draw [fill=orange,opacity=0.6] (6.9,4.1) rectangle (7.1,7.5);
    \draw [fill=orange,opacity=0.6] (6.9,3.9) rectangle (7.1,-0.6);
    \draw [fill=orange,opacity=0.6] (5.9,4.1) rectangle (6.1,7.5);
    \draw [fill=orange,opacity=0.6] (3.9,0.9) rectangle (4.1,-0.6);
    \draw [fill=orange,opacity=0.6] (3.9,1.9) rectangle (4.1,1.1);
    \draw [fill=orange,opacity=0.6] (3.9,2.9) rectangle (4.1,2.1);
    \draw [fill=orange,opacity=0.6] (3.9,3.9) rectangle (4.1,3.1);
    \draw [fill=orange,opacity=0.6] (3.9,4.9) rectangle (4.1,4.1);
    \draw [fill=orange,opacity=0.6] (3.9,5.9) rectangle (4.1,5.1);
    \draw [fill=orange,opacity=0.6] (3.9,6.9) rectangle (4.1,6.1);
    \draw [fill=orange,opacity=0.6] (3.9,7.5) rectangle (4.1,7.1);
    \node [black] at (8.5,0) {$A$};
    \node [black] at (0, 8.5) {$B$};
    \end{tikzpicture}
}
\subfloat[$(B,A)$-consistent gap boxes]{
    \begin{tikzpicture}[scale=0.5]
    \draw [<->] (0, 8) -- (0, 0) -- (8, 0);
    \draw[help lines] (0,0) grid (7,7);
    \draw [fill=orange,opacity=0.4] (-0.7,-0.7) rectangle (7.7,0.9);
    \draw [fill=orange,opacity=0.4] (-0.7,1.1) rectangle (7.7,1.9);
    \draw [fill=orange,opacity=0.4] (-0.7,2.1) rectangle (7.7,2.9);
    \draw [fill=orange,opacity=0.4] (-0.7,3.1) rectangle (7.7,3.9);
    \draw [fill=orange,opacity=0.4] (-0.7,4.1) rectangle (7.7,4.9);
    \draw [fill=orange,opacity=0.4] (-0.7,5.1) rectangle (7.7,5.9);
    \draw [fill=orange,opacity=0.4] (-0.7,6.1) rectangle (7.7,6.9);
    \draw [fill=orange,opacity=0.4] (-0.7,7.1) rectangle (7.7,7.7);
    \draw [fill=orange,opacity=0.6] (3.9,0.9) rectangle (-0.6,1.1);
    \draw [fill=orange,opacity=0.6] (4.1,0.9) rectangle (7.5,1.1);
    \draw [fill=orange,opacity=0.6] (3.9,1.9) rectangle (-0.6,2.1);
    \draw [fill=orange,opacity=0.6] (4.1,1.9) rectangle (7.5,2.1);
    \draw [fill=orange,opacity=0.6] (3.9,2.9) rectangle (-0.6,3.1);
    \draw [fill=orange,opacity=0.6] (4.1,2.9) rectangle (7.5,3.1);
    \draw [fill=orange,opacity=0.6] (3.9,4.9) rectangle (-0.6,5.1);
    \draw [fill=orange,opacity=0.6] (4.1,4.9) rectangle (7.5,5.1);
    \draw [fill=orange,opacity=0.6] (3.9,5.9) rectangle (-0.6,6.1);
    \draw [fill=orange,opacity=0.6] (4.1,5.9) rectangle (7.5,6.1);
    \draw [fill=orange,opacity=0.6] (4.1,6.9) rectangle (7.5,7.1);
    \draw [fill=orange,opacity=0.6] (3.9,6.9) rectangle (-0.6,7.1);
    \draw [fill=orange,opacity=0.6] (0.9,3.9) rectangle (-0.6,4.1);
    \draw [fill=orange,opacity=0.6] (1.9,3.9) rectangle (1.1,4.1);
    \draw [fill=orange,opacity=0.6] (2.9,3.9) rectangle (2.1,4.1);
    \draw [fill=orange,opacity=0.6] (3.9,3.9) rectangle (3.1,4.1);
    \draw [fill=orange,opacity=0.6] (4.9,3.9) rectangle (4.1,4.1);
    \draw [fill=orange,opacity=0.6] (5.9,3.9) rectangle (5.1,4.1);
    \draw [fill=orange,opacity=0.6] (6.9,3.9) rectangle (6.1,4.1);
    \draw [fill=orange,opacity=0.6] (7.5,3.9) rectangle (7.1,4.1);
    \node [black] at (8.5,0) {$A$};
    \node [black] at (0, 8.5) {$B$};
    \end{tikzpicture}
}
\caption{A relation and its GAO-consistent gap boxes}
\label{fig:gap-boxes-from-S}
\end{figure}

\begin{example}[GAO-consistent gap boxes]
Consider a relation $S(A,B)$ on two attributes $A$ and $B$, represented by
points shown in Figure~\ref{fig:gap-boxes-from-S}(a).
If $S$ was indexed in the $A,B$ order, then the gap boxes from $S$
(i.e. the set $\calB(S)$) are shown in Figure~\ref{fig:gap-boxes-from-S}(b).
If $S$ was indexed in the $B,A$ order, then the gap boxes from $S$
are shown in Figure~\ref{fig:gap-boxes-from-S}(c).
\end{example}

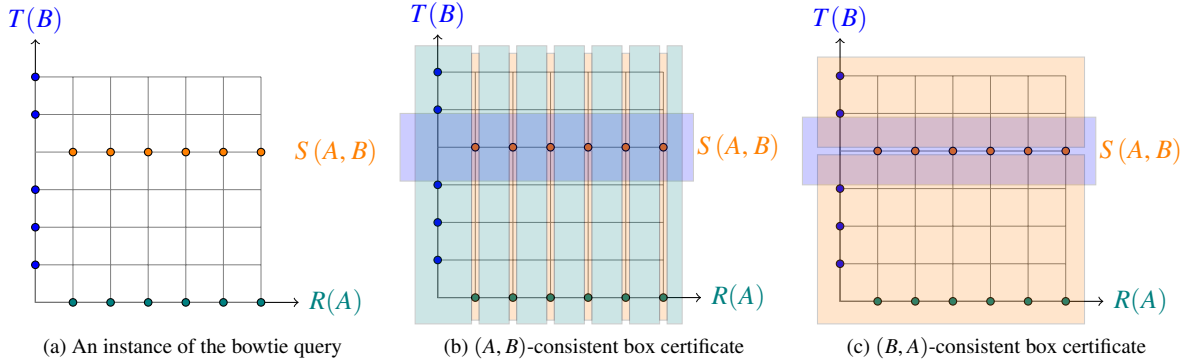
\begin{figure}[!thp]
\centering
\subfloat[An instance of the bowtie query]{
    \begin{tikzpicture}[scale=0.5]
    \draw [<->] (0, 7) -- (0, 0) -- (7, 0);
    \draw[help lines] (0,0) grid (6,6);
    \draw [fill=blue] (0,1) circle [radius=0.1];
    \draw [fill=blue] (0,2) circle [radius=0.1];
    \draw [fill=blue] (0,3) circle [radius=0.1];
    \draw [fill=blue] (0,5) circle [radius=0.1];
    \draw [fill=blue] (0,6) circle [radius=0.1];
    \draw [fill=teal] (1,0) circle [radius=0.1];
    \draw [fill=teal] (2,0) circle [radius=0.1];
    \draw [fill=teal] (3,0) circle [radius=0.1];
    \draw [fill=teal] (4,0) circle [radius=0.1];
    \draw [fill=teal] (5,0) circle [radius=0.1];
    \draw [fill=teal] (6,0) circle [radius=0.1];
    \draw [fill=orange] (1,4) circle [radius=0.1];
    \draw [fill=orange] (2,4) circle [radius=0.1];
    \draw [fill=orange] (3,4) circle [radius=0.1];
    \draw [fill=orange] (4,4) circle [radius=0.1];
    \draw [fill=orange] (5,4) circle [radius=0.1];
    \draw [fill=orange] (6,4) circle [radius=0.1];
    \node [teal] at (8,0) {$R(A)$};
    \node [blue] at (0, 7.5) {$T(B)$};
    \node [orange] at (8, 4) {$S(A,B)$};
    \end{tikzpicture}
}
\subfloat[$(A,B)$-consistent box certificate] {
    \begin{tikzpicture}[scale=0.5]
    \draw [<->] (0, 7) -- (0, 0) -- (7, 0);
    \draw[help lines] (0,0) grid (6,6);
    \draw [fill=blue] (0,1) circle [radius=0.1];
    \draw [fill=blue] (0,2) circle [radius=0.1];
    \draw [fill=blue] (0,3) circle [radius=0.1];
    \draw [fill=blue] (0,5) circle [radius=0.1];
    \draw [fill=blue] (0,6) circle [radius=0.1];
    \draw [fill=teal] (1,0) circle [radius=0.1];
    \draw [fill=teal] (2,0) circle [radius=0.1];
    \draw [fill=teal] (3,0) circle [radius=0.1];
    \draw [fill=teal] (4,0) circle [radius=0.1];
    \draw [fill=teal] (5,0) circle [radius=0.1];
    \draw [fill=teal] (6,0) circle [radius=0.1];
    \draw [fill=orange] (1,4) circle [radius=0.1];
    \draw [fill=orange] (2,4) circle [radius=0.1];
    \draw [fill=orange] (3,4) circle [radius=0.1];
    \draw [fill=orange] (4,4) circle [radius=0.1];
    \draw [fill=orange] (5,4) circle [radius=0.1];
    \draw [fill=orange] (6,4) circle [radius=0.1];
    \node [teal] at (8,0) {$R(A)$};
    \node [blue] at (0, 7.5) {$T(B)$};
    \node [orange] at (8, 4) {$S(A,B)$};
    \draw [fill=blue,opacity=0.2] (-1,3.1) rectangle (6.8,4.9);
    \draw [fill=orange,opacity=0.2] (0.9,3.9) rectangle (1.1,-0.6);
    \draw [fill=orange,opacity=0.2] (0.9,4.1) rectangle (1.1,6.5);
    \draw [fill=orange,opacity=0.2] (1.9,3.9) rectangle (2.1,-0.6);
    \draw [fill=orange,opacity=0.2] (1.9,4.1) rectangle (2.1,6.5);
    \draw [fill=orange,opacity=0.2] (2.9,3.9) rectangle (3.1,-0.6);
    \draw [fill=orange,opacity=0.2] (2.9,4.1) rectangle (3.1,6.5);
    \draw [fill=orange,opacity=0.2] (3.9,3.9) rectangle (4.1,-0.6);
    \draw [fill=orange,opacity=0.2] (3.9,4.1) rectangle (4.1,6.5);
    \draw [fill=orange,opacity=0.2] (4.9,3.9) rectangle (5.1,-0.6);
    \draw [fill=orange,opacity=0.2] (4.9,4.1) rectangle (5.1,6.5);
    \draw [fill=orange,opacity=0.2] (5.9,3.9) rectangle (6.1,-0.6);
    \draw [fill=orange,opacity=0.2] (5.9,4.1) rectangle (6.1,6.5);
    \draw [fill=teal,opacity=0.2] (-0.6,-0.7) rectangle (0.9,6.7);
    \draw [fill=teal,opacity=0.2] (1.1,-0.7) rectangle (1.9,6.7);
    \draw [fill=teal,opacity=0.2] (2.1,-0.7) rectangle (2.9,6.7);
    \draw [fill=teal,opacity=0.2] (3.1,-0.7) rectangle (3.9,6.7);
    \draw [fill=teal,opacity=0.2] (4.1,-0.7) rectangle (4.9,6.7);
    \draw [fill=teal,opacity=0.2] (5.1,-0.7) rectangle (5.9,6.7);
    \draw [fill=teal,opacity=0.2] (6.1,-0.7) rectangle (6.5,6.7);
    \end{tikzpicture}
}
\subfloat[$(B,A)$-consistent box certificate]{
    \begin{tikzpicture}[scale=0.5]
    \draw [<->] (0, 7) -- (0, 0) -- (7, 0);
    \draw[help lines] (0,0) grid (6,6);
    \draw [fill=blue] (0,1) circle [radius=0.1];
    \draw [fill=blue] (0,2) circle [radius=0.1];
    \draw [fill=blue] (0,3) circle [radius=0.1];
    \draw [fill=blue] (0,5) circle [radius=0.1];
    \draw [fill=blue] (0,6) circle [radius=0.1];
    \draw [fill=teal] (1,0) circle [radius=0.1];
    \draw [fill=teal] (2,0) circle [radius=0.1];
    \draw [fill=teal] (3,0) circle [radius=0.1];
    \draw [fill=teal] (4,0) circle [radius=0.1];
    \draw [fill=teal] (5,0) circle [radius=0.1];
    \draw [fill=teal] (6,0) circle [radius=0.1];
    \draw [fill=orange] (1,4) circle [radius=0.1];
    \draw [fill=orange] (2,4) circle [radius=0.1];
    \draw [fill=orange] (3,4) circle [radius=0.1];
    \draw [fill=orange] (4,4) circle [radius=0.1];
    \draw [fill=orange] (5,4) circle [radius=0.1];
    \draw [fill=orange] (6,4) circle [radius=0.1];
    \node [teal] at (8,0) {$R(A)$};
    \node [blue] at (0, 7.5) {$T(B)$};
    \node [orange] at (8, 4) {$S(A,B)$};
    \draw [fill=blue,opacity=0.2] (-1,3.1) rectangle (6.8,4.9);
    \draw [fill=orange,opacity=0.2] (-0.6,3.9) rectangle (6.5,-0.6);
    \draw [fill=orange,opacity=0.2] (-0.6,4.1) rectangle (6.5,6.5);
    \end{tikzpicture}
}
\caption{A GAO-consistent box certificate for the bowtie query}
\label{fig:gao-consistent certificate}
\end{figure}

\begin{example}[GAO-consistent box certificate]
    Consider the bowtie query 
    \[ Q = R(A) \Join S(A, B) \Join T(B) \]
    where the GAO is $(A, B)$. An input instance to this problem is shown
    in Figure~\ref{fig:gao-consistent certificate}(a).
    A GAO-consistent box certificate where the GAO is $(A, B)$ is shown 
    in Figure~\ref{fig:gao-consistent certificate}(b).
    A GAO-consistent box certificate where the GAO is $(B, A)$ is shown 
    in Figure~\ref{fig:gao-consistent certificate}(c).
\label{ex:gao-consistent cert}
\end{example}

The following proposition follows from the above discussion on the main
results of \cite{nnrr}, leading the way to a more general geometric notion of
certificates.

\bprop[Implicit from \cite{nnrr}]
For any input instance to the join evaluation problem where all input relations
are indexed consistent with a fixed GAO,
we have $|\gaoboxcert| = O(|\gaocert|)$.
\label{defn:cert-box=<}
\eprop

% ----------------------------------------------------------------------------
\subsection{General box certificates}
\label{subsec:general box certificates}
% ----------------------------------------------------------------------------

It can be seen from Example~\ref{ex:gao-consistent cert} that a GAO-consistent 
certificate is highly sensitive to the GAO for the same input data. 
In the $(A, B)$ GAO, all gap boxes from $R$ are vertical strips,
all the gap boxes from $S$ are the same vertical strips as $R$ and the 
two segments above and below each orange point.
The gap boxes from $T$ are horizontal strips between the blue points.
To cover the entire space, we will need $\Omega(N)$ strips.
In the $(B, A)$ GAO, $S$ has two big boxes above and below the orange
line. So the space can be covered with just three boxes: two orange boxes and
one blue box.

A fact often seen in practice \cite{Bertino:2012:ITA:2480898} is that relations
are indexed with multiple search keys. In such case, there is no reason to stick
to a particular GAO. For example, if the relation $S$ from
Example~\ref{ex:gao-consistent cert} was already indexed in {\em both} the $(A,
B)$ order and the $(B, A)$ order, we should make use of the information form
both indices to speed up join processing. From this view point, the notion of 
geometric certificate starts to gain traction: if we stick to a comparison-based
algorithm it is not clear how we can make use of the availability of multiple
indices per relation. From the geometric view point, we can simply think of the
set of indices for every relation as a collection of boxes whose union is
exactly the set of tuples {\em not} in the relation.

Abstracting away, henceforth we will model the input relations as a collection
of boxes, denoted by $\calB(R)$ for each input relation $R$. This way, it does
not matter anymore how many indices are available per input relation.
The notion of {\em box certificate} comes naturally.

%Marked app
\bdefn[Box certificate]
Let $Q$ be a natural join query whose input relations have already been indexed.
There can be multiple indices per relation.
Let $R$ be a $k$-ary relation on attributes, say, $(A_1,\dots,A_k)$.
The gap boxes owned by $R \in \atoms(Q)$ are a set of rectangles whose union
contains precisely the set of tuples on $\D(A_1)\times \cdots \times \D(A_k)$
which {\em do not} belong to $R$.
Let $\calB(R)$ denote the set of gap boxes owned by $R$.

A {\em box certificate} is a subset of gap boxes from 
$\bigcup_{R\in\atoms(Q)} \calB(R)$ that cover every point not in the output. 
We use $\boxcert$ to denote a box certificate of minimum size for the instance.
\label{defn:boxcert}
\edefn

The following proposition is straightforward.

\bprop
For any input instance to the join problem, and for any fixed GAO, we have
$|\boxcert| = O(|\gaoboxcert|)$ (and thus $|\boxcert| = O|\gaocert|$).
And, for some input instances $|\boxcert| = O(1)$ while 
$|\gaoboxcert| = \Omega(N)$.
\label{prop:C<<Cgao}
\eprop
%Marked app
\bp
The first statement is obvious but needs a bit of clarification. 
Suppose the input instances
are indexed with multiple search key orderings. For example, $S(A, B, C)$ can be
indexed with $6$ different B-tree key orderings. Then, if we fixed a GAO we are
only able to use one of the $6$ available indices from $S$.
To see the second statement, we can refer back to Example~\ref{ex:gao-consistent
cert}. If the GAO was $(A, B)$, then obviously $|\gaocert| = \Omega(N)$, while
the general box certificate can take the boxes in 
Figure~\ref{fig:gao-consistent certificate}(c).
\ep

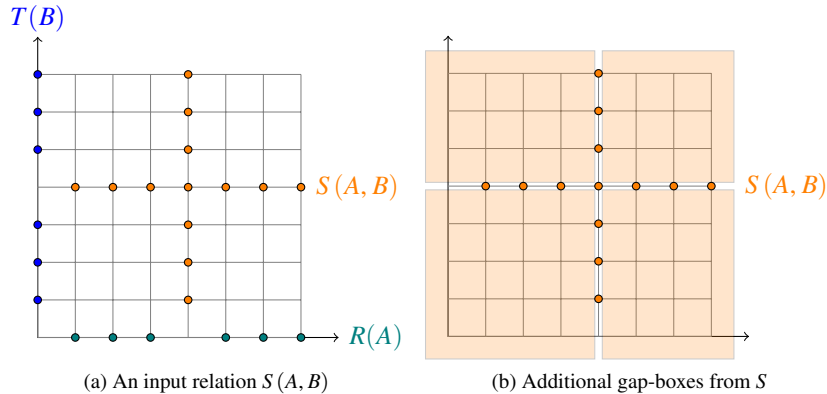
\begin{figure}[htbp]
\centering
\subfloat[An input relation $S(A, B)$] {
    \begin{tikzpicture}[scale=0.5]
    \draw [<->] (0, 8) -- (0, 0) -- (8, 0);
    \draw[help lines] (0,0) grid (7,7);
    \draw [fill=blue] (0,1) circle [radius=0.1];
    \draw [fill=blue] (0,2) circle [radius=0.1];
    \draw [fill=blue] (0,3) circle [radius=0.1];
    \draw [fill=blue] (0,5) circle [radius=0.1];
    \draw [fill=blue] (0,6) circle [radius=0.1];
    \draw [fill=blue] (0,7) circle [radius=0.1];
    \draw [fill=teal] (1,0) circle [radius=0.1];
    \draw [fill=teal] (2,0) circle [radius=0.1];
    \draw [fill=teal] (3,0) circle [radius=0.1];
    \draw [fill=teal] (5,0) circle [radius=0.1];
    \draw [fill=teal] (6,0) circle [radius=0.1];
    \draw [fill=teal] (7,0) circle [radius=0.1];
    \draw [fill=orange] (1,4) circle [radius=0.1];
    \draw [fill=orange] (2,4) circle [radius=0.1];
    \draw [fill=orange] (3,4) circle [radius=0.1];
    \draw [fill=orange] (5,4) circle [radius=0.1];
    \draw [fill=orange] (6,4) circle [radius=0.1];
    \draw [fill=orange] (7,4) circle [radius=0.1];
    \draw [fill=orange] (4,1) circle [radius=0.1];
    \draw [fill=orange] (4,2) circle [radius=0.1];
    \draw [fill=orange] (4,3) circle [radius=0.1];
    \draw [fill=orange] (4,4) circle [radius=0.1];
    \draw [fill=orange] (4,5) circle [radius=0.1];
    \draw [fill=orange] (4,6) circle [radius=0.1];
    \draw [fill=orange] (4,7) circle [radius=0.1];
    \node [teal] at (9,0) {$R(A)$};
    \node [blue] at (0, 8.5) {$T(B)$};
    \node [orange] at (8.5, 4) {$S(A,B)$};
    \end{tikzpicture}
}
\subfloat[Additional gap-boxes from $S$] {
    \begin{tikzpicture}[scale=0.5]
    \draw [<->] (0, 8) -- (0, 0) -- (8, 0);
    \draw[help lines] (0,0) grid (7,7);
    \draw [fill=orange] (1,4) circle [radius=0.1];
    \draw [fill=orange] (2,4) circle [radius=0.1];
    \draw [fill=orange] (3,4) circle [radius=0.1];
    \draw [fill=orange] (5,4) circle [radius=0.1];
    \draw [fill=orange] (6,4) circle [radius=0.1];
    \draw [fill=orange] (7,4) circle [radius=0.1];
    \draw [fill=orange] (4,1) circle [radius=0.1];
    \draw [fill=orange] (4,2) circle [radius=0.1];
    \draw [fill=orange] (4,3) circle [radius=0.1];
    \draw [fill=orange] (4,4) circle [radius=0.1];
    \draw [fill=orange] (4,5) circle [radius=0.1];
    \draw [fill=orange] (4,6) circle [radius=0.1];
    \draw [fill=orange] (4,7) circle [radius=0.1];
    \draw [fill=orange,opacity=0.2] (-0.6,-0.6) rectangle (3.9,3.9);
    \draw [fill=orange,opacity=0.2] (-0.6,4.1) rectangle (3.9,7.6);
    \draw [fill=orange,opacity=0.2] (4.1,-0.6) rectangle (7.6,3.9);
    \draw [fill=orange,opacity=0.2] (4.1,4.1) rectangle (7.6,7.6);
    \node [orange] at (9, 4) {$S(A,B)$};
    \end{tikzpicture}
}
\caption{B-tree gap boxes can still be bad, and more general boxes help}
\label{fig:boxcert-still-bad}
\end{figure}

\begin{example}[Gaps from B-tree indices can be bad]
\label{ex Btree can be bad}
A box certificate is a function of the input gap boxes.
Perhaps the most natural gap boxes are the gap boxes coming from search
tree indices such as B-tree or trie as shown in 
Figure~\ref{fig:gap-boxes-from-S}.
In this example, we show that even when the input relations are indexed using
all possible GAOs, the box certificate can still be much larger than necessary
for some input instances.

Consider again the bowtie query $Q = R(A) \Join S(A, B) \Join T(B)$. 
The relation $S(A,B)$ is indexed in both directions $(A,B)$ and $(B,A)$.
For the input instance shown in Figure~\ref{fig:boxcert-still-bad}(a),
$\Omega(N)$ rectangles are still required to cover the output space.

To see this, note that all the B-tree-style gap boxes from $R$ and $T$
cannot cover any grid point $(i,j)$ for $i, j\in [7] - \{4\}$.
And the gap boxes from $S$ from the $S(A,B)$-index and from the
$S(B,A)$-index are shown in Figure~\ref{fig:gap-boxes-from-S}.
The big (light orange) boxes can't cover the grid points. And we will need
$\Omega(N)$ thin boxes to cover all of them, in spite of the fact that two
gap boxes from $R$ and $T$ already cover all points in $S$.
\end{example}

\begin{example}[Non-B-tree gap boxes can help]
Suppose the indices for relation $S(A,B)$ store four additional
gap-boxes as shown in Figure~\ref{fig:boxcert-still-bad}(b). Then, we'd
have a constant-sized certificate for the bad instance from
Figure~\ref{fig:boxcert-still-bad}(a).
Unfortunately, the commonly used B-tree indices do not allow for returning
these types of gap boxes. One of the key contributions of this paper is the
observation that the kind of gap boxes shown in 
Figure~\ref{fig:boxcert-still-bad}(b) can be {\em inferred} from the 
B-tree gap boxes. In fact, they can be inferred from {\em any} collection
of gap boxes encoding the same relation. The inference framework developed
in this paper can be used to analyze the complexity of such inference algorithms,
which helps build better input indices.
\label{ex general dyadic box can help}
\end{example}

% ----------------------------------------------------------------------------
\subsection{On discovering an optimal box certificate and the case for dyadic 
box certificate}
\label{app:subsec:the case for dyadic}

Algorithm~\ref{algo:outer-shell} is essentially the skeleton of the \ms algorithm.
There are a couple of subtle differences which do not affect our discussion here.
If for every input relation $R$ and every probe point $\mv t$, there are 
only $\tilde O(1)$ maximal gap boxes
in $\calB(R)$ containing $\pi_{\vars(R)}(\mv t)$, then it is easy to see that
Algorithm~\ref{algo:outer-shell} discovers any box certificate
$\boxcert$ with $\tilde O(m|\boxcert| + Z)$ many insertions into \cds, where $m$ is the
number of input relations.

\begin{algorithm}[t]
\caption{Algorithm for discovering a box certificate}
\label{algo:outer-shell}
\begin{algorithmic}[1]
\State $\cds \gets \emptyset$ \Comment{No box discovered yet}
\While {$\cds$ can find $\mv t$ not in any stored box}
  \If {$\pi_{\vars(R)}(\mv t) \in R$ for every $R\in \atoms(Q)$}
    \State Report $\mv t$ and insert $\mv t$ as an output box back to the \cds
  \Else
    \State Query all $R \in \atoms(Q)$ for all maximal gap boxes containing
          $\pi_{\vars(R)}(\mv t)$
    \State Insert those gap boxes into \cds
  \EndIf
\EndWhile
\end{algorithmic}
\end{algorithm}

For a GAO-consistent index, it certainly is true that there is only {\em one} 
maximal GAO-consistent gap box in $\calB(R)$ containing $\pi_{\vars(R)}(\mv t)$.
However, this is not true in general, when we allow for arbitrary gap boxes.
This is because the number of maximal gap-boxes containing a given probe point
can be linear in the input size, and it is possible that only one of them is
used in the optimal box certificate. 

\begin{example}[There can be $\Omega(N)$ maximal gap-boxes]
Consider, for example, the situation shown in
Figure~\ref{fig:many-maximal-gap-boxes}.
\begin{figure}[htbp]
\centering
    \begin{tikzpicture}[scale=0.5]
    \draw[help lines] (0,0) grid (7,7);
    \draw [fill=orange] (1,7) circle [radius=0.1];
    \draw [fill=orange] (2,6) circle [radius=0.1];
    \draw [fill=orange] (3,5) circle [radius=0.1];
    \draw [fill=orange] (4,4) circle [radius=0.1];
    \draw [fill=orange] (5,3) circle [radius=0.1];
    \draw [fill=orange] (6,2) circle [radius=0.1];
    \draw [fill=orange] (7,1) circle [radius=0.1];
    \draw [fill=orange,opacity=0.1] (-0.6,-0.6) rectangle (0.9,7.5);
    \draw [fill=orange,opacity=0.2] (-0.6,-0.6) rectangle (1.9,6.9);
    \draw [fill=orange,opacity=0.3] (-0.6,-0.6) rectangle (2.9,5.9);
    \draw [fill=orange,opacity=0.4] (-0.6,-0.6) rectangle (3.9,4.9);
    \draw [fill=orange,opacity=0.5] (-0.6,-0.6) rectangle (4.9,3.9);
    \draw [fill=orange,opacity=0.6] (-0.6,-0.6) rectangle (5.9,2.9);
    \draw [fill=orange,opacity=0.7] (-0.6,-0.6) rectangle (6.9,1.9);
    \draw [fill=orange,opacity=0.8] (-0.6,-0.6) rectangle (7.5,0.9);
    \node [orange] at (9, 4) {$S(A,B)$};
    \node [black] at (-4, 1) (x) {probe point};
    \draw [fill=black] (0.5,0.5) circle [radius=0.1];
    \draw [->,black] (x) -- (0.4,0.5);
    \draw [<->] (0, 8) -- (0, 0) -- (8, 0);
    \end{tikzpicture}
\caption{There can be $\Omega(N)$ maximal gap-boxes}
\label{fig:many-maximal-gap-boxes}
\end{figure}
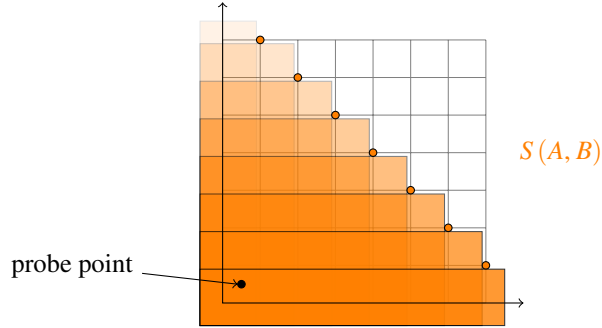
\end{example}

In this section, we make the case for a particular kind of gap boxes called
{\em dyadic boxes} to overcome the above problem. We make the following 
observations (to be proved below):
\bi
 \item for every probe point
$\mv t$, the number of maximal dyadic gap boxes
containing $\pi_{\vars(R)}(\mv t)$ is $\tilde O(1)$.
 \item any gap box can be broken down into $\tilde O(1)$ dyadic gap boxes.
 \item for any box certificate $\boxcert$, there is a dyadic box certificate of 
 size $\tilde O(|\boxcert|)$.
\ei
From these observations, we will then assume (up to a $\polylog$-factor)
in the rest of the paper that
the sets $\calB(R)$ contain only dyadic gap boxes.

%Marked app
Consider any join query $Q$ with $\vars(Q) = \{A_1,\dots,A_n\}$.
For simplicity, but without any loss of generality, let us assume
the domain of each attribute is the set of all binary strings of length $d$,
i.e.
\[ \D(A) = \{0,1\}^d, \ \forall A \in \vars(Q). \]
This is equivalent to saying that the domain of each attribute is the set of all 
integers from $0$ to $2^d-1$.
Since $d$ is the number of bits needed to encode a data value of the input, 
$d$ is logarithmic in the input size.

\bdefn[Dyadic interval]
A {\em dyadic interval} is a binary string $x$ of length $\leq d$.
We use $|x|$ to denote its length. This interval represents all the 
binary strings $y$ such that $|y| = d$ and $x$ is a prefix of 
$y$. \rev{Translating to the integral domain, let $i$ be the integer corresponding to the string $x$. The dyadic interval represents 
all integers in the range
\[[i2^{d-|x|}, (i+1)2^{d-|x|}-1].\]}
The empty string $x=\lambda$ is a dyadic interval consisting
of all possible values in the domain.
If $x$ has length $d$, then it is called a {\em unit dyadic interval}
or a {\em full-length dyadic interval}.
In this case, $x$ represents a point in the domain.
\edefn

\bdefn[Dyadic box]
Let $\vars(Q) = \{A_1,\dots,A_n\}$. A {\em dyadic box} is an $n$-tuple of 
dyadic intervals:
\[ \mv b = \dbox{x_1, \dots, x_n}, \]
If all components of $\mv b$ are unit dyadic intervals, then $\mv b$
represents a {\em point} in the output space.
The dyadic box {\em is} the set of all tuples 
$$ \mv t = (t_1,\dots,t_n) \in \prod_{i=1}^n\mv D(A_i)$$
such that $t_i$ belongs to the dyadic interval $x_i$, for all $i\in[n]$.
\edefn

Note again that some dyadic intervals can be $\lambda$, matching 
arbitrary domain values; also,
a dyadic box $\mv b$ contains a dyadic box $\mv b'$ if each of 
$\mv b$'s components is a prefix of the corresponding component in $\mv b'$. 
The set of all dyadic boxes form a partially ordered set (poset) under 
this containment relation.

%Marked app
\bprop[Number of (maximal) dyadic boxes containing a point is $\tilde O(1)$]
Let $R$ be an arbitrary input relation of arity $k$.
Then, given a probe point $\mv t$, the number of maximal dyadic gap boxes
from $R$ that contain $\pi_{\vars(R)}(\mv t)$ is bounded by
\[ \sum_{i=0}^k (-1)^i \binom{k}{i} \binom{\lfloor kd/2 \rfloor+k-1-id-i}{k-1} \leq
\binom{\lfloor kd/2 \rfloor + k-1}{k-1} \leq d^k.
\]
The number of dyadic boxes containing $\mv t$ is at most $d^k$.
(Note that $d^k = \tilde O(1)$, because in this paper $\tilde O$ hides factors 
that are query dependent and poly-log dependent on the data.)
\label{prop:number-of-boxes-containing-t}
\eprop
\bp
Consider the poset $\calP_d = \{0,1\}^{\leq d}$ of binary strings of
length $\leq d$ under the reversed prefix order, i.e. $x \preceq y$ in this 
poset if $y$ is a prefix of $x$. 
(This is just a binary tree in reverse.) 

Let $\calP_d^k$ denote the $k$th {\em Cartesian power} of $\calP_d$
(under the product order), i.e. $\calP_d^k = \calP_d \times \calP_d \times \cdots
\times \calP_d$, $k$ times.
Then, $\calP_d^k$ is the poset of all dyadic boxes under the containment order.
The set of all dyadic boxes {\em containing} a point $\mv p$ is a
principal filter of $\calP_d^k$ at $\mv p$, denoted by 
$\calP_d^k[\uparrow \mv p]$.
Recall that a point is nothing but a dyadic box all of whose components are
unit dyadic intervals.

The principal filter $\calP_d^k[\uparrow \mv p]$ is isomorphic to the $k$th
Cartesian power of a chain, each chain is of length $d$. It is well-known
from poset theory that any Cartesian power of a chain is graded, 
rank symmetric, rank unimodal, and satisfies the {\em Sperner property} 
\cite{MR0043115,MR745586}. 
In particular, the middle-rank set forms a maximum antichain. 

For $0 \leq m \leq kd$, the 
$m$th-rank of the chain product $\calP_d^k[\uparrow \mv p]$
has size equal to the number of integral solutions to the following equation
\[ x_1 + \dots + x_k = m, \ \ 0 \leq x_i \leq d, i \in [k]. \]
By inclusion-exclusion, the number of such solutions is
\[ \sum_{i=0}^k (-1)^i \binom{k}{i} \binom{m+k-1-id-i}{k-1}. \]
We should pick $m=kd/2$ because of rank-unimodality. 
The sum does not have a closed-form solution, even with the help of
hypergeometric series. But the first term is bounded by
$\binom{kd/2 + k-1}{k-1}.$
(Of course, one can also take the much simpler but weaker bound of $d^k$.)
This is an upperbound on the number of maximal dyadic gap-boxes which contain a
given probe point. 
The number is exponential in the query size, and up to poly-log in the input
size, which we hide in the $\tO$-notation for brevity.\footnote{In fact, most 
of our algorithms has only a $\log$-dependence on the input
size instead of poly-log; but we will not make this distinction precise for the
sake of clarity.}
\ep

\brmk
For completeness, let us also discuss the case when the domain sizes are not
uniform. If $|\mv D(A_i)| = D_i \leq 2^{d_i}$, then the set of all dyadic boxes
is the product $\calP = \calP_{d_1} \times \cdots \times \calP_{d_k}$.
Then, the principal filter of $\calP$ at $\mv p$ is a product of chains:
\[ \calP[\uparrow \mv p] = \calP_{d_1}[\uparrow \mv p] \times \cdots \times
\calP_{d_k}[\uparrow \mv p]. 
\]
The Cartesian product of $k$ chains also satisfies the Sperner property; and, it
is graded, rank-symmetric and rank-unimodal. 
The minimum rank is $0$, and the maximum rank is $\sum_{i=1}^k d_i$.
For any $m$ such that
\[ 0 \leq m \leq d_1 + d_2 + \cdots + d_k, \]
the $m$th-rank of the poset $\calP[\uparrow \mv p]$ has size equal to the number
of integral solutions to the following equation
\[ x_1+\cdots+x_k = m, \ \ 0 \leq x_i \leq d_i, i \in [k]. \]
By inclusion-exclusion, the number of such solutions is
\[ \sum_{i=0}^k (-1)^i \sum_{S\in \binom{[k]}{i}} \binom{m+k-1-\sum_{j\in S}d_j-i}{k-1}. \]
This is a nasty sum. But it can always be upperbounded by the first term (or the
first odd number of terms):
\[ \binom{m+k-1}{k-1}. \]
To get the largest-sized rank, we pick
\[ m = \left\lfloor \frac{d_1+\cdots+d_k}{2} \right\rfloor. \]
\ermk

\bprop
\label{prop box = O(1) dyadic}
Every (not necessarily dyadic) box in $n$ dimensions can be decomposed into a 
disjoint union of at most $(2d)^n = \tilde O(1)$ dyadic boxes.
\eprop
\bp
It is sufficient to show that every closed interval (i.e. a $1$-dimensional box) 
can be written as a disjoint union of at most $2d$ dyadic segments.
Consider the complete binary tree where each node is labeled with the $01$ path
from the root down to the node; a left branch is labeled $0$ and the right is
labeled $1$. Then, the set of all leaves represent the domain $\mv D(A)$ for
an attribute $A$.
Each closed interval is a set of consecutive leaves. Each node in the tree
corresponds to a dyadic segment. Now, when we have a set of consecutive nodes at
some level, we can merge every two sibling nodes into their parent, which is a 
dyadic segment containing the two siblings. Continue this merging process until
no more merging is possible, then we get a collection of dyadic segments
covering the original set of leaves. There are at most two nodes left per depth
of this tree, because if there were $3$ then some $2$ of them can be merged.
\ep

The following proposition follows immediately.

\bprop
For every box certificate, there is a dyadic box certificate of size at
most a factor of $(2d)^n = \tilde O(1)$ larger.
\eprop

% ----------------------------------------------------------------------------
\section{Main Algorithm, Data Structure, and Analytical Idea}
\label{sec:mainidea}
\subsection{Dyadic data structure and dyadic resolution}
\label{sec:dyadic-DS}

There are two key operations that \tetris performs repeatedly: (1) querying for
all boxes containing a given box, %(2) splitting a box into two equal halves,
and (2) ``resolving'' two boxes to get a new box.
We explain how a dyadic data structure can be used to perform those operations
efficiently.

\paragraph{Data structure for storing dyadic boxes}

The operations in Line~\ref{line:main:containment check 1} of 
Algorithm~\ref{alg:main:tetris-skeleton},
%Lines~\ref{ln covered j1} and \ref{ln covered if1} of Algorithm~\ref{alg covered},
and Line~\ref{ln:main online B'} of Algorithm~\ref{alg:main:tetris}
are essentially the same operation: given a box $\mv b$, return the set of all 
dyadic boxes containing $\mv b$ from a storage of dyadic boxes.
This operation can easily be supported in $\tilde O(1)$ time, as shown by Proposition \ref{prop:number-of-boxes-containing-t}.

To store a collection of dyadic boxes, there are many options (including a
hash table). We briefly describe here a natural implementation using a multi-level
dyadic tree.

A {\em dyadic tree} is a binary tree storing dyadic segments, i.e. binary strings of
length at most $d$. At each node, the left branch corresponds to bit $0$
and the right corresponds to bit $1$.
Each time we insert a new dyadic segment $x$ into the tree,
we follow the bits of $x$ down the tree. If we end up with a node already created
in the tree, then we mark that node as a storage node. If we end up at a leaf
while there are still bits left in $x$, then we create new nodes according
to the bits of $x$ that still need to be visited.

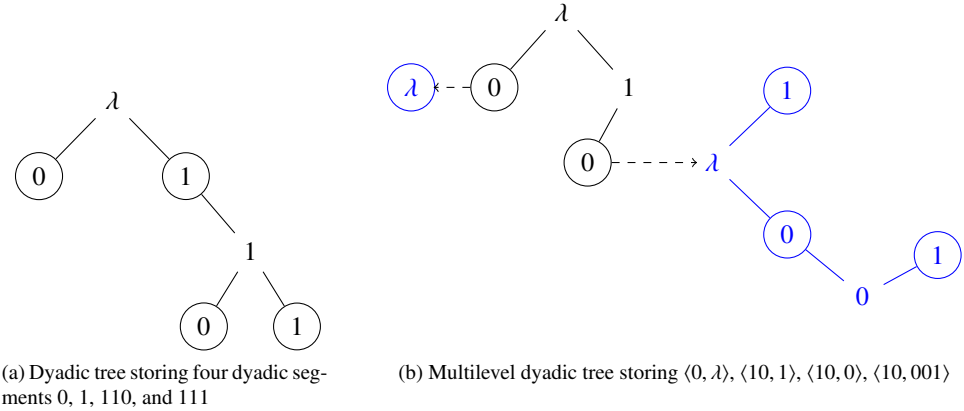
\begin{figure}[!htp]
\centering
\subfloat[Dyadic tree storing four dyadic segments $0$, $1$, $110$, and $111$]{
\tikzset{every tree node/.style={minimum width=1em,draw,circle},
         blank/.style={draw=none},
         edge from parent/.style=
         {draw, edge from parent path={(\tikzparentnode) -- (\tikzchildnode)}},
         level distance=1cm,
         sibling distance=0.6cm}
    \begin{tikzpicture}%[scale=0.5]
    \Tree [.\node[draw=none]{$\lambda$};
           [.0 ]
           [.1  
            \edge[blank]; \node[blank]{};
            \edge[]; [.\node[draw=none]{1};
                          \edge[]; {0}
                          1
                     ]
           ]
          ]
    \end{tikzpicture}
}
\qquad
\subfloat[Multilevel dyadic tree storing $\dbox{0, \lambda}$, $\dbox{10, 1}$,
$\dbox{10, 0}$, $\dbox{10, 001}$]{
    \begin{tikzpicture}%[scale=0.5]
    \tikzset{every tree node/.style={minimum width=1em,draw,circle},
         blank/.style={draw=none},
         edge from parent/.style=
         {draw, edge from parent path={(\tikzparentnode) -- (\tikzchildnode)}},
         level distance=1cm,
         sibling distance=0.6cm}
    \Tree [.\node[draw=none]{$\lambda$};
           [.\node(a){0}; ]
           [.\node[draw=none]{1};
            \edge[]; [.\node(b){0};]
            \edge[blank]; \node[blank]{};
           ]
          ]
    \node [blue,draw,circle] at (-2,-1) {$\lambda$};
    \begin{scope}[shift={(2, -2)}]
    \tikzset{grow'=right,draw=blue,text=blue,every path/.style={draw=blue}}
    \Tree [.\node[draw=none]{$\lambda$};
           [.1 ]
           [.0  
            \edge[blank]; \node[blank]{};
            \edge[]; [.\node[draw=none]{0};
                          \edge[]; {1}
                          \edge[blank]; \node[blank]{};
                     ]
           ]
          ]
    \end{scope}
    \draw [->,dashed] (a) -- (-1.7,-1);
    \draw [->,dashed] (b) -- (1.8,-2);
    \end{tikzpicture}
}
\caption{A simple data structure for storing dyadic boxes}
\label{fig:dyadic tree}
\end{figure}

A dyadic tree can be used to store a set of dyadic segments.
Recall that each dyadic box is an $n$-tuple of dyadic segments, i.e. an
$n$-tuple of strings of the form
$\mv{b} = \dbox{x_1, x_2, \dots, x_n}$, where $x_i \in \{0,1\}^{\leq d}$ for all $i\in[n]$.
Fix some global attribute order $(A_1, A_2, \dots, A_n)$. 
In a {\em multi-level dyadic tree}, we store a dyadic box $\dbox{x_1,\dots,x_n}$ 
by using a dyadic tree to store $x_1$, then at the storage node of $x_1$, there is a pointer
to the root of a second dyadic tree storing $x_2$, etc.
If there are two dyadic boxes \rev{having} the same dyadic segment $x_1$, then
their second dyadic segments are stored in the same tree, naturally.
Figure~\ref{fig:dyadic tree} illustrates this simple data structure.
\rev{More generally, a storage node for $x_k$ has a pointer to the root of a dyadic tree for storing $x_{k+1}$.}

From Proposition~\ref{prop:number-of-boxes-containing-t}, the number
of dyadic boxes containing $\mv b$ is at most $d^n = \tilde O(1)$.
It is straightforward to see that the multilevel dyadic tree can return the
set of all boxes containing a given dyadic box in $\tilde O(1)$ time.

\paragraph{Resolution}

\iffalse
%Marked app
We next need to explain what the \firstsplit routine does.
Consider a dyadic box $\mv b = \dbox{x_1,x_2,\dots,x_n}.$
If $|x_i|=d$, then $x_i$ represents a unit dyadic segment, which corresponds to
a flat slice through the $A_i$-dimension. The box $\mv b$ is flat and is not
splittable along such dimension.
The first splittable dimension is the smallest value $\ell\in[n]$ for which
$|x_\ell|<d$. 
Because $\mv b$ is not a unit box, there must exist such an $\ell$.
In that case, the call 
\[ \pair{\mv b_1}{\mv b_2} \wgets \firstsplit(\mv b) \]
on Line~\ref{ln covered split} of Algorithm~\ref{alg covered} returns the following pair:
\begin{eqnarray*}
\mv b_1 &=& \dbox{x_1\wc \dots\wc x_{\ell-1}\wc  x_\ell0\wc  x_{\ell+1}\wc \dots\wc x_n}\\
\mv b_2 &=& \dbox{x_1\wc \dots\wc x_{\ell-1}\wc  x_\ell1\wc  x_{\ell+1}\wc \dots\wc x_n}.
\end{eqnarray*}
Note again that by definition $|x_i|=d$ for all $i<\ell$. 
Using the multilevel dyadic tree data structure discussed in the previous section,
we can easily find the first splittable dimension of a dyadic box in $\tilde O(1)$ time.
%Marked app
\fi

%Marked app
%The idea of resolution, as was shown in Figure~\ref{fig:main:covered}, is to ``infer''
%from the two positive witnesses $\mv w_1$ and $\mv w_2$ for $\mv b_1$ and
%$\mv b_2$ a new box $\mv w$, the positive witness for $\mv b$.
Note that by the time \Resolve is called in Line~\ref{line:main:bcp-resolve} of Algorithm~\ref{alg:main:tetris-skeleton}, we know
none of $\mv w_1$ and $\mv w_2$ covers $\mv b$.
There are a lot of boxes we can infer from $\mv w_1$ and $\mv w_2$ if those two
boxes are general dyadic boxes which can overlap in peculiar ways.
However, \tetris forces $\mv w_1$ and $\mv w_2$ to be somewhat special.

%, making
%resolution much more intuitive and clean.
%
%We will prove in the following lemma 
%that $\mv w_1$ and $\mv w_2$ must have the following form:
%\begin{eqnarray}
%\mv w_1 &=& \dbox{y_1\wc \dots\wc y_{\ell-1}\wc  x_\ell0\wc  \lambda\wc  \dots\wc \lambda}\label{eqn w1 format}\\
%\mv w_2 &=& \dbox{z_1\wc \dots\wc z_{\ell-1}\wc  x_\ell1\wc  \lambda\wc  \dots\wc \lambda}\label{eqn w2 format}.
%\end{eqnarray}
%where for every $i<\ell$, both $y_i$ and $z_i$ are prefixes of $x_i$. This means that either $y_i$ or $z_i$ is a prefix
%of the other. We use $y_i \cap z_i$ to denote the {\em longer} of the two
%strings $y_i$, $z_i$. The geometric meaning is clear: we are taking the intersection
%of those two dyadic segments which are contained in one another.
%
%\bdefn[Ordered resolution]
%Given two dyadic boxes $\mv w_1$ and $\mv w_2$ of the format shown in
%\eqref{eqn w1 format} and \eqref{eqn w2 format},
%the {\em ordered resolution} of $\mv w_1$ and $\mv w_2$ is the dyadic box
%\begin{equation}
%\mv w = \dbox{y_1\cap z_1\wc\dots\wc y_{\ell-1}\cap z_{\ell-1}\wc  x_\ell\wc  \lambda\wc  \dots\wc \lambda} \label{eqn w format}
%\end{equation}
%In this case, we say that $\mv w$ is the result of resolving $\mv w_1$ and 
%$\mv w_2$ {\em on attribute $A_\ell$}. (Note that $x_\ell$ might be $\lambda$.)
%\label{defn:ord-res}
%\edefn

Our \Resolve routine uses ordered geometric resolution (recall Definition~\ref{defn:main:ord-res}), and clearly $\mv w=\Resolve(\mv w_1,\mv w_2)$ 
 covers $\mv b$ as desired.
%Marked app
We next prove the claimed property of $\mv w_1$ and $\mv w_2$.

\blmm[All resolutions are ordered]
\label{lmm ordered res}
If the initial call to \tetrisskeleton is with the universal box $\mv b = \UB$, then the
following hold:
\bi
 \item[(i)] At any point in time, the box $\mv b$ to be split must have the form
\begin{equation}
\mv{b} \weq
   \dbox{x_1\wc\ldots\wc x_{\ell-1}\wc x_\ell\wc \lambda\wc \ldots\wc\lambda}
   \label{eqn b format}
\end{equation}
where $\ell \in [n]$ and $x_1, \ldots, x_{\ell-1}$ are strings of length $d$
(maximal-length strings), and $|x_\ell|<d$. 
 \item[(ii)] every time we are
calling \Resolve$(\mv w_1, \mv w_2)$, the two witnesses have the
format shown in \eqref{eqn:main w1 format} and \eqref{eqn:main w2 format}.
\ei
In particular, it is sufficient for \Resolve to apply ordered resolution.
\elmm
\bp
We prove $(i)$ by induction. The universal box certainly has the claimed format.
When we split a box $\mv b$ (in line \ref{ln:main covered split} of Algorithm \ref{alg:main:tetris-skeleton})
that has the format \eqref{eqn b format}, $\mv b_1$ and $\mv b_2$ will be:
\begin{eqnarray*}
\mv b_1&=&\dbox{x_1\wc\ldots\wc x_{\ell-1}\wc x_\ell0\wc \lambda\wc \ldots\wc\lambda}\\
\mv b_2&=&\dbox{x_1\wc\ldots\wc x_{\ell-1}\wc x_\ell1\wc \lambda\wc \ldots\wc\lambda}
\end{eqnarray*}
This completes the proof of the invariant.
To see $(ii)$, note that when we call 
$\Resolve(\mv w_1, \mv w_2)$, we know $\mv w_1$ covers $\mv b_1$ but not 
$\mv b$ (due to line~\ref{line:main:w1-contains-b} in Algorithm~\ref{alg:main:tetris-skeleton}), and $\mv w_2$ covers $\mv b_2$ but not $\mv b$ (due to line~\ref{line:main:w2-contains-b} in Algorithm~\ref{alg:main:tetris-skeleton}). 
It follows that $\mv w_1$ and $\mv w_2$ must be of the forms
\eqref{eqn:main w1 format} and \eqref{eqn:main w2 format}, respectively.
\ep

%Marked app
\paragraph{Runtime is linear in the number of resolutions}
%We finally prove a very simple but important combinatorial lemma that will be used
%throughout this paper to bound the runtime of our algorithm in 
%various different settings.
%The main observation is that in \emph{most} cases when the algorithm backtracks, it does one resolution. The amount of work it does modulo the recursive calls is 
%$\tilde O(1)$: inserting a new box, querying for boxes containing a box, and
%resolving. And we are not repeating any resolution more than once. 
%Consequently, hiding behind the potential poly-log factor in $\tilde O$
%(which can easily be shown to not explode), we can bound the runtime
%of \tetris by the number of resolutions it performs.

We first prove Lemma~\ref{lmm:main time<=res}.

%\blmm[Runtime is bounded by \#resolutions]
%\label{lmm time<=res}
%Let $M$ denote the total number of resolutions performed by Algorithm~\ref{alg covered}.
%Then, the total runtime of Algorithm~\ref{alg covered} is 
%$\tO(M)$.
%\elmm
%\ar{Since you talked about ordered resolutions above, it might be good to clarify if the above result needs it: from what I can understand it does not.}
Remember that whenever Algorithm \ref{alg:main:tetris-skeleton} does a resolution, it adds the result back to $\st A$ (in Line \ref{line:main:add-to-kb}). Therefore, $M$ can be thought of as the total increase in the size of $\st A$ since the initial call to \tetrisskeleton. %Notice also here that we are not requiring the initial call to be with the universal box $\mv b=\UB$. Hence, this lemma does not depend on the one above.
\bp[Proof of Lemma~\ref{lmm:main time<=res}]

We begin with the simplifying assumption that \tetris calls \tetrisskeleton only once. In this case if $M$ is the number of resolutions made, we will prove that \tetrisskeleton (and hence \tetris) runs in time $\tO(M+1)$. Later on in the proof, we will see how to get rid of this assumption.

\tetrisskeleton can be thought of as a depth-first traversal of a binary tree whose nodes are dyadic boxes. Calling \mbox{\tetrisskeleton($\mv b$)} corresponds to visiting node $\mv b$ in this binary tree. As soon as \tetrisskeleton visits a node $\mv b$, it checks whether $\mv b$ is covered by some dyadic box $\mv a$ in $\st A$ and whether $\mv b$ is a point (lines \ref{line:main:containment check 1} and \ref{line:main:unitbox} of Algorithm \ref{alg:main:tetris-skeleton}): If $\mv b$ is either covered or a point, then $\tetrisskeleton$ backtracks directly without visiting any children of $\mv b$, in which case $\mv b$ is a \emph{leaf node} of the visited binary tree. If $\mv b$ is neither covered nor a point, then \tetrisskeleton recursively visits its children $\mv b_1$ and $\mv b_2$, in which case $\mv b$ is an \emph{internal node}. \tetrisskeleton spends $\tO(1)$ at each node it visits. In any full binary tree, the number of leaf nodes is equal to one plus the number of internal nodes. Therefore, to bound the runtime of \tetrisskeleton, we only need to bound the number of internal nodes it visits (i.e.\ the number of recursive calls of \tetrisskeleton in which the execution reaches line \ref{ln:main covered split}).

Moreover, whenever \tetrisskeleton returns from visiting a leaf with a \false\, (in line \ref{line:main:root-neg-witness}), then (thanks to lines \ref{ln covered if false1} and \ref{ln covered if false2}) it will keep backtracking while passing this \false\, upwards the tree until it reaches the root. Because the depth of any node in the tree is $\tO(1)$, the total number of internal nodes that return \false\, is $\tO(1)$. (This term contributes to the $+1$ term in our final bound of $\tO(M+1)$.) Therefore, to bound the runtime of \tetrisskeleton, we only need to bound the number of internal nodes that return \true\, (i.e.\ the number of recursive calls that return in lines \ref{ln covered true w1}, \ref{ln covered true w2} or \ref{line:main:return-resolvent}). We will be referring to those nodes as \emph{\true\, internal nodes}.

While \tetrisskeleton is traversing the tree, new boxes are being created by resolution and added to $\st A$ (in lines \ref{line:main:bcp-resolve} and \ref{line:main:add-to-kb}). We will be referring to those boxes as \emph{resolution boxes}. We will show that the total number of \true\, internal nodes is within a $\tO(1)$ factor from the total number of resolution boxes. And to do that, we will establish a mapping between \true\, internal nodes and resolution boxes such that this mapping satisfies the following two conditions:
\bi
\item No resolution box is mapped to more than $\tO(1)$ \true\, internal nodes.
\item Every \true\, internal node is mapped to at least one resolution box.
\ei
During the traversal, \tetrisskeleton makes sure not to visit any internal node $\mv b$ that is already covered by some box $\mv a$ in $\st A$ (\tetrisskeleton might still visit $\mv b$ as a leaf node, but thanks to line \ref{line:main:containment check 1}, the execution will never make it to line \ref{ln:main covered split} in order for $\mv b$ to become an internal node). However, when a new resolution box $\mv w$ is added to $\st A$, the current internal node $\mv b$ that is being visited might already be covered by $\mv w$. If this happens, then (thanks to line \ref{line:main:containment check 1}) \tetrisskeleton will keep backtracking upwards the tree until it reaches a node that is not covered by $\mv w$, and it will never visit any internal node that is covered by $\mv w$ ever after. The depth of the tree is $\tO(1)$. Therefore, from the moment a new resolution box $\mv w$ is added to $\st A$, \tetrisskeleton will traverse no more than $\tO(1)$ \true\, internal nodes that are covered by $\mv w$. We define the mapping between resolution boxes and \true\, internal nodes as follows: Every resolution box $\mv w$ is mapped to all \true\, internal nodes that are covered by $\mv w$ and that are traversed by \tetrisskeleton after $\mv w$ is added to $\st A$. From this definition, we can see that no resolution box is mapped to more than $\tO(1)$ \true\, internal nodes.

Before \tetrisskeleton returns from visiting any internal node $\mv b$ with \true, $\mv b$ must be covered by some box in $\st A$ (in lines \ref{ln covered true w1}, \ref{ln covered true w2} and \ref{line:main:return-resolvent}, $\mv b$ is covered by $\mv w_1, \mv w_2$ and $\mv w$, which all have been added to $\st A$). Moreover, this box could not have existed in $\st A$ from the very beginning because earlier in line \ref{line:main:containment check 1} \tetrisskeleton could not find any box in $\st A$ that covers $\mv b$. In other words, this box must be a resolution box that $\mv b$ can be mapped into. Therefore, every \true\, internal node is mapped to at least one resolution box.

\rev{Finally, we consider the case when \tetris calls \tetrisskeleton more than once. 
Since a single invocation of  \tetrisskeleton takes time $\tO(M+1)$, all invocations of \tetrisskeleton combined take time $\tO(M+I)$ where $M$ is the total number of resolutions performed during all invocations, and $I$ is the total number of invocations. To prove the lemma, we need to prove that $I=\tO(M)$.

First we prove that $I\leq\abs{\boxcert}+Z+1$ where $\boxcert$ is a minimal box certificate (See Definition~\ref{defn:bcp-boxcert}). Each invocation of \tetrisskeleton that returns $\false$ returns also a unit box $\mv w$ that is not covered by any box in $\calA$. If $\mv w$ is not an output tuple, then at least one box from $\boxcert$ that covers $\mv w$ is added to $\calA$. Otherwise, one output tuple is added to $\calA$ (out of $Z$ tuples). Once a box (either from $\boxcert$ or from the $Z$ output tuples) is added to $\calA$, it is not going to be added again.

The last invocation of \tetrisskeleton returns $(\true,\UB)$, indicating that $\UB$ is covered. At that time, \tetris would have implicitly built a resolution proof of $\UB$ using the $Z$ output unit boxes along with at least $\abs{\boxcert}$ input boxes. Hence, $M\geq \abs{\boxcert}+Z-1$.}

%Note that the last call to \tetrisskeleton returns $(\textsc{true},\mv w)$ and we can just absorb the $+1$ factor in the term $\tO(M+1)$ for this last invocation in our final $\tO(M)$ bound. So let us now consider an invocation of \tetrisskeleton that returns $(\textsc{false},\mv b)$. Then line~\ref{line:main:add-to-kb-outer} in Algorithm~\ref{alg:main:tetris} adds box(es) $\st B'$ that cover $\mv b$ to $\st A$ before the next invocation of \tetrisskeleton. Notice that at least one of box in $\st B'$ is necessary to cover the point $\mv b$
%Further, in the next invocation of \tetrisskeleton it will either (i) redo all of the work from the previous invocation to the point where $\mv b$ was discovered as an uncovered point (and in the current invocation, we will find a covering box) or (ii) we will stop at an ancestor of $\mv b$ that is covered by one of the newly added boxes. Note that in both cases in the next invocation of \tetrisskeleton we get a new \textsc{true} leaf node. By making our bound three times larger, we can charge the work that is redone (in case (i) above) to the previous invocation of \tetrisskeleton and to charge the $+1$ term in the $\tO(M+1)$ bound above for the current invocation of \tetrisskeleton to the newly created \textsc{true} leaf node that be created due to $\mv b$ in the next invocation of \tetrisskeleton.
\ep
%Marked app

The following definition comes up naturally in the runtime analysis of \tetris in Section~\ref{SEC:WORST-CASE} and later.
It is just a generalization 
of the concept of prefix of a string.

\bdefn[Prefix of a box]
\label{defn box prefix}
Given any box $\mv{b}$:
$$\mv b \weq \dbox{x_1\wc\ldots\wc x_n}$$
(where $x_1, \ldots, x_n$ are strings of arbitrary length), we define a \emph{prefix} of $\mv b$ to be any box $\mv{b'}$ that has the form:
$$\mv{b'}\weq\dbox{x_1\wc \ldots \wc x_{l-1}\wc \prefix{x_l}\wc \lambda\wc\ldots\wc\lambda}$$
where $l \in [n]$.
\edefn
Notice that $\mv{b'}$ contains $\mv b$.
It results from $\mv b$ by removing bits from the end. If we ignore the 
commas and consider all strings of each box to be a single string, then 
$\mv{b'}$ is indeed a prefix of $\mv b$.

In the context of joins (and \bcp), when we resolve two boxes $\mv w_1$ and $\mv w_2$, each one 
of them could be 
either an input gap box or an output box (the latter is the box $\mv b$ added due to lines~\ref{line:main:add-output} and~\ref{line:main:add-to-kb-outer} in Algorithm~\ref{alg:main:tetris}) or a result of a previous resolution.

\bdefn[\gbresolution]
A \emph{\gbresolution} is any resolution where each 
one of the two boxes to be resolved is either an input gap box or a result of a 
previous \gbresolution (i.e., a \gbresolution does not involve any output box, 
neither directly nor indirectly). The result of a \gbresolution is called a \emph{\gbresolvent}, as defined in Section~\ref{SEC:BEYOND-WC}.
\edefn

\bdefn[\obresolution]
An \emph{\obresolution} is any resolution that is not a \gbresolution (i.e., at least one of the two boxes to be resolved is either an output box or a result of a previous \obresolution). The result of an \obresolution is called an \emph{\obresolvent}, as defined in Section~\ref{SEC:BEYOND-WC}.
\edefn

The following theorem can be thought of as a template for almost all 
runtime bounds that are presented in this paper. 
We apply this theorem by bounding the number of \gbresolutions that \tetris
performs depending on the input query's structural information.
%and whether
%\tetrisreloaded or \tetrispreloaded was called.

\bthm[The gap box resolution bound]
\label{thm X+Z}
%When \tetrispreloaded (Algorithm \ref{alg offline}) or
%\tetrisreloaded (Algorithm \ref{alg online}) is used on a join query, it runs in 
An invocation of \tetris runs in
time $\tO(X+Z)$, where $X$ is the total number of \gbresolutions that have been performed, and $Z$ is the total number of output tuples of the \bcp.
\ethm
%\ar{Again would be good to clarify the part about ordered resolutions here.}

\bp
We will use Lemma \ref{lmm:main time<=res}, and bound the runtime by bounding the 
number of resolutions. The number of \gbresolutions is $X$. Next, we 
bound the number of \obresolutions.

We will prove the following claim: For every witness $\mv w$ that is either an output box or an \obresolvent, there is some output box $\mv z$ such that $\mv w$ is a prefix box of $\mv z$. Because every box has $\tO(1)$ possible prefix boxes, proving the claim will bound the number of \obresolutions by $\tO(Z)$, as desired.

This claim can be proved by induction. The base case clearly holds. For the inductive step, suppose that $\mv w=\Resolve(\mv w_1, \mv w_2)$ where the claim holds for at least one of $\mv w_1$ or $\mv w_2$. Without loss of generality, let the claim hold for $\mv w_1$.\; $\mv w_1$ must be a prefix box of some output box $\mv z$.\; $\mv z$ is a unit box (i.e. a point) and hence has the form
$$\mv z\weq\dbox{t_1\wc \ldots \wc t_n}$$
where $t_1, \ldots, t_n$ are strings of maximal-length $d$. In Algorithm \ref{alg:main:tetris}, the calls to
$\tetrisskeleton$ are with the universal box $\mv b=\UB$. According to Lemma \ref{lmm ordered res}, all 
performed resolutions are going to be ordered. In particular, $\mv w_1$ and $\mv w_2$ must have the form shown in \eqref{eqn:main w1 format} and \eqref{eqn:main w2 format}. Since $\mv w_1$ is a prefix box of $\mv z$, $\mv w_1$ and $\mv w_2$ will have the form
$$\begin{array}{ccccccc}
\mv w_1\weq\dbox{&          t_1 &\wc\ldots\wc&           t_{l-1} &\wc x_l 0\wc\lambda\wc\ldots\wc\lambda&}\\
\mv w_2\weq\dbox{&\prefix{t_1}&\wc\ldots\wc&\prefix{t_{l-1}}&\wc x_l 1\wc\lambda\wc\ldots\wc\lambda&}
\end{array}$$
%\ar{Might not hurt to argue why the above claim holds on $\mv w_1$ and $\mv w_2$.}
where $l\in[n]$. The output resolvent $\mv w$ will be
$$\mv w\weq\dbox{t_1\wc\ldots\wc t_{l-1}\wc x_l\wc\lambda\wc\ldots\wc\lambda},$$
which is a prefix of $\mv w_1$ and hence $\mv z$. Therefore, the claim holds for $\mv w$.
\ep

% ----------------------------------------------------------------------------
\section{Omitted details from Section~\ref{SEC:WORST-CASE}}
% ----------------------------------------------------------------------------
\label{sec offline}

In this section, we show that by calling \tetrispreloaded 
with an appropriate collection of gap boxes $\calB(Q)$, we can achieve 
a runtime of
the form \rev{$\tilde O(N^{\fhtw} + Z)$}, where $N$ is the total input size, $\fhtw$
is the fractional hypertree width for the instance, and $Z$ is the output size.
When the query is $\alpha$-acyclic, $\fhtw=1$, and thus \tetrispreloaded runs in
time linear in the input plus the output size. This is the celebrated Yannakakis
result.

We will prove the above result by breaking it up into three steps.
First, we show that \tetrispreloaded achieves a run time of
$\tO(N + \agm(Q))$, where $\agm(Q)$ is the best (i.e.\ tightest) \agm bound for the
input instance. (See Section~\ref{sec:agm-bound}.)
Then, we show that \tetrispreloaded runs in time
$\tO(N + Z)$ if the query is $\alpha$-acyclic.
Finally, we prove the \rev{$\tilde O(N^{\fhtw} + Z)$} result by using the first two 
results.

These results are conditioned on two key aspects of \tetris:
\bi
 \item The \RAO used internally by \tetrisskeleton, which decides the dimension order 
 in which the algorithm splits a target box $\mv b$.
 \item The type of gap boxes that we chose to pre-load \tetrispreloaded with (i.e., the set $\calB$ in Algorithm~\ref{alg:main:tetris}.)
\ei

% ----------------------------------------------------------------------------
\subsection{\tetrispreloaded achieves \agm bound}

\bdefn[$\DT{\st A}$: a multilevel dyadic tree pruned by a set of boxes $\st A$]
\label{defn:tree-pruned-by-A}
Let $\st A$ be a set of dyadic boxes. The multilevel dyadic tree that is pruned by $\st A$, denoted by $\DT{\st A}$, is the set of all boxes $\mv b$ that have the format described by \eqref{eqn b format} and that satisfy the following condition:
\bi
\item For every box $\mv b'$ that is a prefix box
\footnote{Check Definition~\ref{defn box prefix}.}
of $\mv b$ such that $\mv b'\neq \mv b$
and for every box $\mv a\in\st A$, we have $\mv b'\nsubseteq \mv a$ (i.e., $\mv b'$ is not covered by $\mv a$).
\ei
\edefn

%Marked aop
\bthm[\tetrispreloaded achieves \agm bound]
\label{thm agm}
Let $Q$ be a join query, $N$ the total number of input tuples,
%$Z$ total number of output tuples,
$\agm(Q)$ the best \agm bound for this instance,
and $\st B(Q)=\bigcup_{R\in\atoms(Q)}\calB(R)$ the set of all input gap boxes.
Then, \tetrispreloaded runs in time 
$\tO(N + \agm(Q))$ and $\abs{\DT{\st B(Q)}}$ is bounded by $\tO(\agm(Q))$, provided that the following conditions are met:
\bi
\item The \RAO $\sigma$ used by the algorithm is arbitrary, but fixed.
\item For each input relation $R\in\atoms(Q)$, 
     $\calB(R)$ contains only $\sigma$-consistent gap boxes.
        (This condition can easily be met if the search tree structure for $R$
        is consistent with $\sigma$.)
% HQN: I removed this footnot because it will confuse people
%     \footnote{The 
%        theorem still holds even if some relations have additional indices 
%        that are not $\sigma$-consistent and the algorithm is using gap boxes 
%        from those indices as well.}
\ei

\ethm
%Marked aop
\bp
The term $\tO(N)$ is the time needed to initialize $\st A$ with all boxes of $\st B$ (in line~\ref{ln:main:initialize} of Algorithm~\ref{alg:main:tetris}). While explaining the proof, we will assume that the data structure that is being used to store boxes of $\st A$ is a multilevel dyadic tree (Section~\ref{sec:dyadic-DS}).

We will consider a slightly different version of \tetrisskeleton, denoted by $\tetrisnewskeleton$. The difference is that whenever $\tetrisnewskeleton$ encounters a unit box $\mv b$ that is not covered by any box in $\st A$, instead of returning $(\false, \mv b)$ as \tetrisskeleton would do, $\tetrisnewskeleton$ reports $\mv b$ as an output, adds $\mv b$ to $\st A$, and returns $(\true, \mv b)$. In particular, instead of line~\ref{line:main:root-neg-witness} of \tetrisskeleton (Algorithm~\ref{alg:main:tetris-skeleton}), \tetrisnewskeleton has the following three lines:
\begin{algorithmic}
\State {\bf Report} $\mv b$ as an output tuple
\State $\calA \la \calA \cup \{\mv b\}$
\State \Return $(\text{\true}, \mv b)$
\end{algorithmic}
Unlike \tetrisskeleton which returns one output point per call, a single call to \tetrisnewskeleton reports all output points. First, we will explain the proof under the assumption that we are using \tetrisnewskeleton. Later on, we will see how the proof holds for \tetrisskeleton as well.

We use the proof strategy from \cite{skew}, implicit in \cite{NPRR}:
by induction on the total number of attributes, then apply H\"older inequality
to the inductive step.
To bound the total runtime, it is sufficient to bound the number of boxes
$\mv b$ that the algorithm considers.
Each box $\mv b$ is nothing but a node in a multilevel dyadic tree,
one level for each component of $\mv b$. 
Hence, in the proof of this theorem, we will often speak of \tetrisnewskeleton ``visiting''
a node in the dyadic tree. ``Visiting a node'' means the current box $\mv b$ in \tetrisnewskeleton
corresponds to that node.

The base case is when there is $n=1$ attribute $A$ and $m$ unary relations,
i.e. $|\atoms(Q)| = m$.
The join problem reduces to the problem of computing the 
intersection of $m$ sets of strings of length $d$. The gaps are dyadic segments,
i.e. strings of length at most $d$. When we insert gaps (from $\calB(Q)$)
into the internal dyadic data structure of \tetrisnewskeleton, each gap corresponds to
a node in the dyadic tree. We call nodes storing gaps {\em storage nodes}.
All leaves of the dyadic tree are storage nodes (while internal nodes of the dyadic tree \emph{could} be storage nodes).
\tetrisnewskeleton starts from the root of the tree, visiting the left and right branches
of each node recursively. If a storage node is hit, then the algorithm backtracks.
%When two sibling nodes are storage nodes, they are {\em resolved} to make the
%parent of the two sibling nodes also a storage node.

%\ar{The term 'tuple box' has not been defined and is used in the para below. Also in para below, when you say you backtrack from a tuple box you also want to say that you never go back to that box later.}
Tuples of any relation are unit boxes; we will be referring to them as 
\emph{tuple boxes}.
Now, consider the smallest relation $R$ from the input. 
The gap boxes
and tuple boxes of $R$ correspond to storage nodes and depth-$d$ leaves of the dyadic
tree. (We do not store those depth-$d$ leaves.)
When \tetrisnewskeleton visits a storage node corresponding to a gap from $R$, it will backtrack.
It might have backtracked earlier because it might have hit a storage node
corresponding to a gap from a different relation. In either case, each gap box from $R$ is
visited at most once. If \tetrisnewskeleton hits a depth-$d$ leaf corresponding to a tuple
from $R$, it might or might not report this tuple as an output depending on whether it corresponds to a gap from a different relation or not. But either way, it will backtrack and never visit this leaf again. If \tetrisnewskeleton does not hit
this leaf in the first place, it must have backtracked earlier at a parent node
of this leaf. There are at most $d$ such parent nodes.
In summary, the total runtime of \tetrisnewskeleton is at most linear in the number of tuples
from $R$ plus the number of gap boxes from $R$. Thanks to
Proposition~\ref{prop:number-of-boxes-containing-t}, the total runtime is
bounded by $\tO(|R|)$.

%\ar{Hypergrpah corresponding to a query/fraction edge cover would need to be defined in th currently non-existent section that defined joins etc.}
Next, let $\{x_S\}_{S\in\atoms(Q)}$ be a fractional edge cover of the hypergraph of
this query, then $\sum_{S\in \atoms(Q)} x_S \geq 1$ and $x_S\geq 0$ for all 
$S \in \atoms(Q)$. The overall runtime is $\tO$ of
\[ \min_{S\in\atoms(Q)} |S|  = |R| \leq
         |R|^{\sum_{S\in\atoms(Q)} x_S}
         \leq \prod_{S\in\atoms(Q)} |S|^{x_S}. 
\]
The right-hand side is the \agm bound that corresponds to the chosen 
fractional edge cover. (See Equation~(\ref{eqn:AGM}).)  The base case is thus proved.

Next, consider the general case when $n>1$.
Without loss of generality, assume $\sigma = (A_1,\dots,A_n)$.
Recall that a $\sigma$-consistent gap box $\mv b = \dbox{x_1,\dots,x_n}$
has the following property: there is one position $i \in [n]$ for which
$|x_j| = 0$ (i.e. $x_j = \lambda$) for every $j>i$, and
$|x_j| \in \{0, d\}$ for every $j<i$. (Definition~\ref{defn:GAO-consistent-box}.)

We bound the number of boxes $\mv b = \dbox{x_1,\dots,x_n}$ 
visited by \tetrisnewskeleton by considering
two types of $\mv b$: 
(type 1) the boxes $\mv b$ for which $x_2=\lambda$, and 
(type 2) the boxes $\mv b$ for which
$x_2 \neq \lambda$.
Note that by Lemma~\ref{lmm ordered res}, if $x_2=\lambda$ then
$x_i=\lambda$ for all $i>2$.

To bound the number of type-1 boxes \tetrisnewskeleton visits, we only have
to note that \tetrisnewskeleton visits these nodes exactly as if it was 
computing $\Join_{R : A_1 \in \vars(R)} \pi_{A_1}(R)$.
To see this, consider any value 
$t_1 \notin \Join_{R : A_1 \in \vars(R)} \pi_{A_1}(R)$.
Then, $t_1 \notin \pi_{A_1}(R)$ for some $R$. In that case, out of the
$\sigma$-consistent gap boxes from $R$, there must be a gap box of the form
$\mv a = \dbox{a_1,\lambda, \dots,\lambda}$ where
$a_1$ is a prefix of $t_1$.
And thus, when the algorithm hits the storage node corresponding to $a_1$
at the first-level dyadic tree, it will have to backtrack.
Hence, as shown in the base case, the total number of type-1 boxes
\tetrisnewskeleton visits is bounded by (within a $\tO(1)$ factor)
\[ \min_{R : A_1 \in \vars(R)} |R| \leq \prod_{R : A_1\in\vars(R)} |R|^{x_R}
\leq \agm(Q). \]

%\ar{Maybe explain a bit more why each type-2 box must have the claimed form.}
To bound the number of type-2 boxes \tetrisnewskeleton visits, we note that each such box
$\mv b = \dbox{x_1,x_2,\dots,x_n}$ must have $|x_1|=d$
(Lemma \ref{lmm ordered res})
and $x_1 \in \ \Join_{R : A_1 \in \vars(R)} \pi_{A_1}(R)$.
For every value $x_1 \in \ \Join_{R : A_1 \in \vars(R)} \pi_{A_1}(R)$, \tetrisnewskeleton will be visiting type-2 boxes $\mv b$ that start with this $x_1$ (i.e.\ $\mv b = \dbox{x_1,x_2,\dots,x_n}$) exactly as if it was computing the join
\[ \left( \Join_{R : A_1\in\vars(R)} \pi_{A_2,\dots,A_n}\sigma_{A_1=x_1} R \right)
 \Join \left( \Join_{R : A_1\notin\vars(R)} R \right).
\]
The reason is that relations that contain $A_1$ provide gap boxes of the form $\mv a=\dbox{x_1,a_2,\dots,a_n}$, while other relations provide gap boxes of the form $\mv a=\dbox{\lambda,a_2,\dots,a_n}$. All those gap boxes are pre-loaded in $\st A$. Once \tetrisnewskeleton visits a box $\mv b$ that is covered by any box $\mv a$ in $\st A$, it will backtrack (thanks to Line \ref{line:main:containment check 1} of Algorithm~\ref{alg:main:tetris-skeleton}).

To sum up, \tetrisnewskeleton computes the intersection of the 
projections of all relations on $A_1$. And for each value $x_1$ in this intersection, 
\tetrisnewskeleton applies itself recursively on the selection of $x_1$ from all relations. 
This behavior is identical to that of the generic join algorithm 
(Algorithm 3 in \cite{skew}).
The inductive step follows from the query decomposition lemma (i.e. Lemma 3.1 from \cite{skew}
with $|I|=1$). Hence, we have proved that the number of boxes $\mv b$ visited by \tetrisnewskeleton is $\tO(\agm(Q))$.

Every box $\mv b$ \tetrisnewskeleton visits must belong to $\DT{\st B(Q)}$, but the converse is not necessarily true. However, the above proof not only bounds the number of boxes $\mv b$ visited by \tetrisnewskeleton, but also the superset $\DT{\st B(Q)}$.

Finally, we remove our assumption about \tetrisnewskeleton. In particular, we show that using \tetrisskeleton instead is not going to slow down the outer algorithm \tetrispreloaded by more a $\tO(1)$ factor. Before \tetrisskeleton returns from visiting a node $\mv b$ with \true, it adds a box $\mv w$ containing $\mv b$ to $\st A$. This way, \tetrisskeleton makes sure it never visits any children of $\mv b$ ever after. Suppose that \tetrisskeleton encounters an output point $\mv o$. Then, it returns directly to the root with \false. The next call of \tetrisskeleton will be heading straight towards $\mv o$, thanks to the boxes $\mv w$ that were added to $\st A$ by the previous call. Hence, within a $\tO(1)$ time, the next call will be resuming the work of the previous call, as if the two were a single continues call. As a result, the multiple calls to \tetrisskeleton are equivalent to a single call to \tetrisnewskeleton.
\ep

If we prevent \tetris from caching resolution results (i.e.\ if we drop
line~\ref{line:main:add-to-kb} of Algorithm~\ref{alg:main:tetris-skeleton}), then \tetris will
be performing what is known as \emph{tree} resolution: Every box that results from a resolution
will become an input to maximally one resolution. Hence, the resolution proof
generated by \tetris will have a tree structure. Because $\DT{\st
B(Q)}=\tO(\agm(Q))$, \tetrispreloaded without caching can still achieve the bound of $\tO(N + \agm(Q))$.
\footnote{For that to hold, the step of reporting output points has to be moved from the outer \tetrispreloaded into \tetrisskeleton, the same way we did in the proof of Theorem~\ref{thm agm} when we defined \tetrisnewskeleton.}

\bcor[\nocache\ achieves \agm bound]
\label{cor agm tree}
Let $Q$ be a join query, $N$ the total number of input tuples,
%$Z$ total number of output tuples,
$\agm(Q)$ be the best \agm bound for this instance, and let the conditions of Theorem~\ref{thm agm} be met.
Then, $Q$ can be solved using $\tO(\agm(Q))$ tree ordered geometric resolutions.
\ecor

% ----------------------------------------------------------------------------
\subsection{\tetrispreloaded matches Yannakakis algorithm on \texorpdfstring{$\alpha$}{alpha}-acyclic queries}

To show the main result of this section, we need a simple auxiliary lemma.
Recall that a GYO-elimination order \cite{DBLP:books/aw/AbiteboulHV95} is 
obtained by repeating the following two operations on the hypergraph of an 
$\alpha$-acyclic query:
(a) remove a vertex that belongs to only one hyperedge (i.e.\ a \emph{private} vertex),
(b) remove a hyperedge which is a subset of another hyperedge.
As mentioned earlier in Definition~\ref{defn:acyclic}, a hypergraph is $\alpha$-acyclic if and only if repeating the above two operations
arbitrarily results in an empty hypergraph. The order in which vertices
are removed is called a {\em GYO elimination order}. 
Notice that there can be many GYO elimination orders for a given $\alpha$-acyclic query.
For example, if the query has two relations $R(A_1,\dots,A_n)$ and
$S(A_1,\dots,A_n)$, then there are $n!$ GYO elimination orders.

\blmm
Let $Q$ be an $\alpha$-acyclic query.
Let $\sigma=(A_1,\dots,A_n)$ be the reverse of any GYO elimination order for $Q$.
For any $k \in [n]$, define $\calR_k = \{ R \in\atoms(Q) \suchthat A_k \in \vars(R)\}$.
Then, there exists a relation $R \in \calR_k$ satisfying the following property:
\begin{equation}
\forall S\in\calR_k\quad\quad
\left[\vars(S) \cap \{A_1,\dots,A_k\} \wsubseteq \vars(R) \cap \{A_1,\dots,A_k\}\right]
\label{eqn R_k support}
\end{equation}
\label{lmm bottom relation alpha acyclic}
\elmm
\bp
When $A_n$ is first removed by GYO-elimination, it must be contained in only one hyperedge $e$ of the hypergraph of $Q$. 
All other hyperedges containing $A_n$ must have been removed
earlier. Therefore, all those hyperedges must have been contained in other hyperedges that contain $A_n$. Hence, the relation $R$
with $\vars(R) = e$ satisfies property (\ref{eqn R_k support}).
After $A_n$ is removed, the residual graph is $\alpha$-acyclic. Induction
completes the proof.
\ep

\bdefn[Bottom relation and $\support(A_k)$]
\label{defn support A_k}
Fix a GAO that is any reversed GYO elimination order.
We will refer to a relation $R\in\calR_k$ satisfying condition
(\ref{eqn R_k support}) as a {\em bottom relation for $A_k$} with respect to the
GAO.
The set $\vars(R) \cap \{A_1\dots,A_k\}$ is called the {\em support for $A_k$},
denoted by $\support(A_k)$.
Another way to state condition (\ref{eqn R_k support}) is to say that
the union of all the sets $\vars(S) \cap \{A_1,\dots,A_k\}$ is a subset
of $\support(A_k)$.
\edefn

Note that there can be multiple bottom relations for $A_k$.
For example, in the query $Q = R(A,B,C) \Join S(A, B, C) \Join T(A, B, C)$,
all the relations are bottom relations for $A$, for $B$, and for $C$.

\bdefn[Bottom boxes for $A_k$]
Fix the GAO $\sigma$ to be any reversed GYO elimination order, and consider only
$\sigma$-consistent gap boxes from all relations.
%An input gap box $\mv a$ is called a bottom gap box for $A_k$
%if $\support(\mv a)$ is exactly the support for $A_k$ with respect to
%the reversed GYO elimination order.
Let $\calN_k$ denote the set of all input gap boxes {\em and}
input tuple boxes $\mv b$ whose supports are precisely $\support(A_k)$.
\edefn

Obviously, the boxes from $\calN_k$ can only come from the bottom relations for $A_k$;
and only the gap boxes (and not the tuple boxes)
from the bottom relations are part of the input to
\tetrispreloaded.
The following proposition is straightforward.

\bprop
Fix the GAO $\sigma$ to be any reversed GYO elimination order, and consider only
$\sigma$-consistent gap boxes from all relations.
For any $k\in [n]$, the number of bottom boxes for $A_k$ is linear (within a $\tO(1)$ factor) in the total 
number of tuples from all the bottom relations for $A_k$.
In particular,
\[ \sum_{k=1}^n |\calN_k| = \tO(N). \]
\label{prop total bottom box size}
\eprop

\bthm[\tetrispreloaded matches Yannakakis algorithm]
\label{thm yan}
Let $Q$ be an $\alpha$-acyclic join query, $N$ the total number of input tuples,
and $Z$ total number of output tuples.
Then, \tetrispreloaded runs in time 
$\tO(N + Z)$, provided that the following conditions are met:
\bi
\item The \RAO $\sigma$ used by the algorithm is the reverse of some 
    GYO elimination order.\footnote{i.e., the vertex that is eliminated first 
        in GYO should be the last in the \RAO $\sigma$.}
    \item For each $R\in \atoms(Q)$, the boxes in $\calB(R)$ are
        $\sigma$-consistent. 
\ei
\ethm
\bp
WLOG, assume $\sigma = (A_1,\dots,A_n)$.
We prove this theorem by applying Theorem \ref{thm X+Z} and bounding the number
of \gbresolutions to be $\tO(N)$. Thanks to Proposition~\ref{prop total bottom box size},
it is sufficient to show the following claim:

{\bf Claim 1.} For any $k\in [n]$, the number of \gbresolutions on $A_k$ is $\tO(|\calN_k|)$.

We first show that the claim holds for $k=n$,
i.e. the number of \gbresolutions on $A_n$ is $\tO(|\calN_n|)$.
Our plan is to show that every \gbresolution on $A_n$ produces a box $\mv w$
which {\em covers} some box $\mv a \in \calN_n$. 
From Proposition~\ref{prop:number-of-boxes-containing-t}, the number of such
boxes $\mv w$ is thus bounded by $\tO(|\calN_n|)$.

Resolution occurs only in line~\ref{line:main:bcp-resolve} of Algorithm~\ref{alg:main:tetris-skeleton}. 
For completeness, we repeat some of the arguments leading to Lemma~\ref{lmm ordered res}.
For the witnesses $\mv w_1$ and $\mv w_2$ to resolve on the last attribute $A_n$, their $n$-th 
components must be $x_n0$ and $x_n1$ respectively for some string $x_n$. 
We only need to resolve $\mv w_1$ and $\mv w_2$ when each one of them covers 
one half of $\mv b$ but neither one of them covers $\mv b$ as a whole. 
This only happens when $\mv b, \mv b_1, \mv b_2$ have the forms
\begin{eqnarray*}
\mv b   &=& \dbox{x_1\wc\ldots\wc x_{n-1}\wc x_n}\\
\mv b_1 &=& \dbox{x_1\wc\ldots\wc x_{n-1}\wc x_n0}\\
\mv b_2 &=& \dbox{x_1\wc\ldots\wc x_{n-1}\wc x_n1}.
\end{eqnarray*}
where $|x_i|=d$ for all $i<n$. 
When this is the case, then $\mv w_1, \mv w_2$ and $\mv w$ must have the forms
\begin{eqnarray*}
\mv w_1&\weq&\dbox{\prefix{x_1}\wc\ldots\wc\prefix{x_{n-1}}\wc x_n0}\\
\mv w_2&\weq&\dbox{\prefix{x_1}\wc\ldots\wc\prefix{x_{n-1}}\wc x_n1}\\
\mv{w}&\weq&\dbox{\prefix{x_1}\wc\ldots\wc\prefix{x_{n-1}}\wc x_n}.
\end{eqnarray*}
where one prefix in $\mv w_1$ can have a different length than the corresponding prefix 
in $\mv w_2$, and $\mv w$ will inherit the longer prefix between the two
(i.e.\ the one that corresponds to the intersection of the two dyadic segments).

Positive witnesses
\footnote{$\mv w$ is a positive witness iff it covers $\mv b$. Remember that in \tetrispreloaded, input gap boxes and output tuples cover the whole space.}
can be obtained in three ways: 
either from input gap boxes, or from output tuples, or by the (ordered) 
resolution of other witnesses (The first two cases can occur in Line 
\ref{line:main:pretent} while the third one occurs in Line \ref{line:main:bcp-resolve} 
of Algorithm \ref{alg:main:tetris-skeleton}). 
Since we are only considering \gbresolutions, the witnesses
$\mv w_1$ and $\mv w_2$ do not contain any output tuple; in particular, each of
them is either an input gap box, or is a result of previous resolutions on input
gap boxes.

To prove Claim 1 (for $k=n$), we will show that $\mv w$ covers some
box in the set $\calN_n$. 
Note that, by definition, $\calN_n$ contains either gap boxes or tuple boxes
from the bottom relations of $A_n$.
To show that $\mv w$ covers some box in $\calN_n$, we show by induction
that every witness involved in a resolution on $A_n$ must cover some box in
$\calN_n$.
For the base case, suppose $\mv w_1$ and $\mv w_2$ are input gap boxes.
We want to show that each one of them covers a box in $\calN_n$.
Note that because they are input gap boxes,
$\support(\mv w_1) \subseteq \support(A_n)$ and
$\support(\mv w_2) \subseteq \support(A_n)$.

Let $\overline{\mv b}$ denote the projection
\footnote{See Definition~\ref{defn box proj} for the notion of projection of a box onto some support.}
 of $\mv b$ onto the support of
$A_n$, i.e.  $\overline{\mv b} = \pi_{\support(A_n)}(\mv b)$.
Since no input gap boxes cover $\mv b$ (otherwise we would not have
called $\firstsplit(\mv b)$), no input gap boxes cover $\overline{\mv b}$
either. 
In particular, each one of the boxes $\overline{\mv b}$,\; $\overline{\mv b_1} = \pi_{\support(A_n)}(\mv b_1)$ 
and $\overline{\mv b_2} = \pi_{\support(A_n)}(\mv b_2)$ must be a prefix box of some
box in $\calN_n$. (See Definition~\ref{defn box prefix}.)
Because $\support(\mv w_1) \subseteq \support(A_n)$, $\mv w_1$ (which covers $\mv b_1$) covers $\overline{\mv b_1}$ as well. As a result, $\mv w_1$ covers some box in $\calN_n$. And the same holds for $\mv w_2$.

Now, to complete the induction, we show that $\mv w$ (which is the resolution of $\mv w_1$ and $\mv w_2$) must cover some
box in $\calN_n$. Notice that
$$\support(\mv w)\subseteq \support(\mv w_1)\cup\support(\mv w_2)\subseteq\support(A_n)$$
Therefore, $\mv w$ (which covers $\mv b$) covers $\overline{\mv b}$ as well. But $\overline{\mv b}$ is a prefix of some box in $\calN_n$. As a result, $\mv w$ covers some box in $\calN_n$.

Next, we prove Claim 1 above for some $k<n$. The proof is almost the same as
the $k=n$ case. However, we will need one additional claim, which can be proved by induction.

{\bf Claim 2.} Every witness $\mv w$ that is involved in a $\gbresolution$
on attribute $A_k$ must satisfy $\support(\mv w) \subseteq \support(A_k)$.

Before proving Claim 2, let us see how it helps complete the proof of Claim 1.
According to Lemma \ref{lmm ordered res}, when we resolve $\mv w_1$ and $\mv w_2$ 
on $A_{k}$, all the attributes $A_{k+1}, \ldots, A_n$ will be $\lambda$'s. 
Each one of the boxes $\mv b, \mv b_1, \mv b_2,  \mv w_1, \mv w_2, \mv w$ will 
have $(n-k)$ trailing $\lambda$s. For example,
\begin{eqnarray*}
\mv b   & \weq&\dbox{x_1\wc\ldots\wc x_{k-1}\wc x_{k}\wc\lambda\wc\ldots\wc\lambda}\\
\mv b_1 & \weq&\dbox{x_1\wc\ldots\wc x_{k-1}\wc x_{k}0\wc\lambda\wc\ldots\wc\lambda}\\
\mv b_2 & \weq&\dbox{x_1\wc\ldots\wc x_{k-1}\wc x_{k}1\wc\lambda\wc\ldots\wc\lambda}.
\end{eqnarray*}

Now, let $\overline{\mv b'} = \pi_{\support(A_k)}(\mv b)$.
Then no input gap boxes cover $\overline{\mv b'}$, otherwise we would not
have split $\mv b$. In particular, each one of the boxes $\overline{\mv b'}$,\; $\overline{\mv b_1'} = \pi_{\support(A_k)}(\mv b_1)$ 
and $\overline{\mv b_2'} = \pi_{\support(A_k)}(\mv b_2)$ must be a prefix box of some
box in $\calN_k$.
From Claim 2, $\support(\mv w_1) \subseteq \support(A_k)$ and
$\support(\mv w_2) \subseteq \support(A_k)$.
Consequently, both $\mv w_1$ and $\mv w_2$ cover boxes in $\calN_k$. However, 
\begin{equation}
\support(\mv w)\subseteq \support(\mv w_1)\cup\support(\mv w_2)\subseteq\support(A_k)
\label{eq claim2 inductive step}
\end{equation}
Consequently, $\mv w$ covers some box in $\st N_k$ as well. (\ref{eq claim2 inductive step}) proves the inductive step of Claim 2 within the same value of $k$. For different values of $k$, Claim 2 can be proved inductively from $k=n$ down to $1$.
\ep

% ----------------------------------------------------------------------------
\subsection{\tetrispreloaded matches the fractional hypertree width bound}

To obtain a runtime of \rev{$\tO(N^{\fhtw} + Z)$}, a typical strategy is as follows.
We first compute a tree decomposition (Definition~\ref{defn:tree-dec}) of the query for which the the maximum
\agm bound over the bags is minimized. Then, we compute a set of intermediate
relations, one for each bag, using an \agm-bound matching algorithm such
as \cite{NPRR} or \cite{leapfrog}. Finally, we run Yannakakis algorithm on
the resulting ``bag relations'', because those relations form an $\alpha$-acyclic
query. Since each intermediate relation is of
size at most $N^{\fhtw}$, we have the claimed runtime.

The above strategy is essentially to ``hide'' the non $\alpha$-acyclic parts inside bags.
The strategy is somewhat unsatisfactory as we have to run the algorithm in two stages. 
We will show that \tetrispreloaded can achieve the objective in ``one shot.''

We have already seen (in Theorems \ref{thm agm} and \ref{thm yan}) that 
\tetrispreloaded achieves \agm bound and is capable of playing the 
role of Yannakakis algorithm. Next, we will show that it is even more ``polymorphic'' 
than that: When applied on a join query given some tree decomposition, it will 
achieve \agm bound on each bag and simulate Yannakakis on the bags. 
In other words, we don't have to apply \tetrispreloaded on each 
bag individually, and then one more time on all the bags together. 
The algorithm is ``smart'' enough to produce all this behavior when it is 
applied directly on the original query.

Given a bag $G$ of some tree decomposition of a join query $Q$, we will use $\vars(G)$ to denote the set of all attributes in $G$.

\bthm[\tetrispreloaded achieves \agm and Yannakakis together]
\label{thm agm+yan}
Let $Q$ be a join query, $N$ the total number of input tuples, 
$Z$ the total number of output tuples, $T\!D$ some tree decomposition of $Q$,
and $\overline{\agm_{T\!D}(Q)}$ the maximum over all $T\!D$ bags of the best \agm bound for this bag.
Then, \tetrispreloaded runs in time
$\tO(N + \overline{\agm_{T\!D}(Q)}+Z)$, provided that the following conditions are met:
\bi
\item The \RAO $\sigma$ used by the algorithm is the reverse of some GYO elimination order on the bags of $T\!D$.\footnote{The bags of any tree decomposition form an $\alpha$-acyclic hypergraph. Hence, we can apply GYO elimination on this hypergraph by repeating those two steps in any order: (1) remove any bag that is contained in another bag, (2) remove any vertex/attribute that appears in only one bag.}
\item For each relation $R\in \atoms(Q)$, the boxes in $\calB(R)$ are
    $\sigma$-consistent.
\item For each bag $G$ of $T\!D$, there exists a set of relations $\atoms(G)\subseteq\atoms(Q)$ that achieves the best \agm bound for $G$ and that satisfies $\vars(R)\subseteq\vars(G)$ for all $R\in\atoms(G)$.
\footnote{This condition can always be enforced as follows: If relation $R$ is necessary to achieve \agm bound on bag $G$ while $\vars(R)\nsubseteq\vars(G)$, then $\pi_{\vars(G)}R$ has to be precomputed, added to $\atoms(Q)$, and has to replace $R$ in \agm bound of $G$.}
\ei
\ethm
\bp
WLOG, assume $\sigma = (A_1,\dots,A_n)$.
We will use the same outline of the proof of Theorem \ref{thm yan}. We 
start with bounding the total number of \mbox{\gbresolutions} on the last 
attribute $A_n$. Bags form an $\alpha$-acyclic hypergraph. $A_n$ is the 
first vertex/attribute that is eliminated in GYO. Consider all bags that contain 
$A_n$. At least one of them (let's call it $G_n$) must contain all the 
attributes that are contained in any of them (Lemma \ref{lmm bottom relation
alpha acyclic}). Hence, $G_n$ is a \emph{bottom bag} for $A_n$ (in the sense of
Definition \ref{defn support A_k}). Let $\agm_n$ be the best \agm bound 
for $G_n$. We define a set $\calN_n$ of size $\tO(\agm_n)$. 
The plan is to show that every \gbresolution on $A_n$ produces a box that
covers some box in $\calN_n$.

Let $\calB(G_n)$ denote the set of input gap boxes from all relations in $\atoms(G_n)$ projected onto $\vars(G_n)$. The set $\calN_n$ is chosen to be
$\DT{\calB(G_n)}$. (See Definition~\ref{defn:tree-pruned-by-A}.) By
Theorem~\ref{thm agm}, $\abs{\calN_n}=\tO(\agm_n)$.

While running \tetrispreloaded on $Q$, consider the parameter $\mv b$ of 
\tetrisskeleton. Let $\overline{\mv b}$ be the projection of $\mv b$ onto $\vars(G_n)$. It is not hard to see that $\overline{\mv b}$ must belong to 
$\calN_n$.

To prove that every \gbresolution on $A_n$ produces a box that covers some 
box in $\calN_n$, we can prove two things: First, every input gap box that is 
used as a witness on $A_n$ covers some box in $\calN_n$ (and that 
box is $\overline{\mv b}$). Second, every resolution on $A_n$ between 
two boxes that cover boxes in $\calN_n$ produces another box that covers some box 
in $\calN_n$ (and that box is also $\overline{\mv b}$).

Inductively, we can extend this to any attribute $A_{k}$ for $k<n$.
\ep

The following corollary is immediate.

%Marked aop
\bcor[\tetrispreloaded achieves fractional hypertree width bound]
\label{cor fhtw}
Let $Q$ be a join query, $N$ the total number of input tuples,
$\fhtw$ be the fractional hypertree width of the query, and $Z$ the total number of output tuples.
Then, \tetrispreloaded runs in time 
\rev{$\tO(N^{\fhtw}+Z)$}, provided that the following conditions are met:
\bi
\item The \RAO $\sigma$ used by the algorithm is the reverse of some GYO elimination order on the bags
of some tree decomposition $T\!D$ whose fractional hypertree width is $\fhtw$.
\item For each relation $R\in\atoms(Q)$, the box set $\calB(R)$ is
    $\sigma$-consistent.
\item For each bag $G$ of $T\!D$, there exists a set of relations $\atoms(G)\subseteq\atoms(Q)$ that forms an optimal fractional edge cover of $G$ and that satisfies $\vars(R)\subseteq\vars(G)$ for all $R\in\atoms(G)$.
\footnote{As before, precomputing $\pi_{\vars(G)}R$ can always enforce this condition when needed.}
\ei
\ecor

% ----------------------------------------------------------------------------
\section{Omitted details from Section~\ref{SEC:BEYOND-WC}}
% ----------------------------------------------------------------------------
\label{sec online GB}

In this section, we consider join queries (and \bcp instances) whose treewidth 
\footnote{\rev{See Appendix~\ref{sec:tree-decomp} or Definition~\ref{defn:induced-width}. There are no known results that achieve the fractional hypertree width in the 
certificate size. Instead, the treewidth is more natural for certificate-based results.
While the treewidth is no smaller the fractional hypertree width,
the certificate size is no larger than input size.
Hence, we have a tradeoff between worst-case results and certificate-based results.}}
is $w$. We show that \tetrisreloaded can be used to solve those queries in time $\tO(\abs{\boxcert}^{w+1} +Z)$. Moreover, we show that \tetrisreloaded runs in time $\tO(\abs{\boxcert}^w +Z)$ in two special cases: when $w=1$ and $w=n-1$. However, when $1<w<n-1$, Theorem~\ref{thm:cert^w+1-lb} shows a lowerbound of $\Omega(\abs{\boxcert}^{w+1} +Z)$ on \ordered\ proof sizes, and hence a similar lower bound on the runtime of \tetrisreloaded. Moreover, when $w=n-1$, Theorem~\ref{thm:cert^n-1-lb} shows a lower bound of $\Omega(\abs{\boxcert}^{n-1}+Z)$.

To analyze the runtime of \tetrisreloaded, 
we first argue that the number of accesses it makes to $\calB$ is of the correct order:
\blmm%[\tetris loads only $\tO(|\boxcert|)$ boxes from $\calB$]
\label{lmm sum B's = cert}
The total number of boxes that \tetrisreloaded loads from $\st B$ into $\st A$ (in line~\ref{ln:main online B'}) is $\tO(|\boxcert|)$.
%$\boxcert$ is a minimum-sized box certificate for $\st B$.
\elmm
\bp
In each iteration of \tetrisreloaded, if $\mv w$ is an output point, then no box from 
$\st B$ is loaded into $\calA$.
If $\mv w$ is not an output point, there must be at least one
box from $\boxcert$ that covers $\mv w$. Moreover, no box from $\boxcert$ is
loaded twice.
Hence, the total number of iterations that load boxes from $\st B$ is at most $|\boxcert|$.
The number of boxes that are loaded from $\st B$ in each iteration
(i.e.\ the size of $\st B'$) is $\tO(1)$, 
thanks to Proposition~\ref{prop:number-of-boxes-containing-t} in
Appendix~\ref{app:subsec:the case for dyadic}, completing the proof.
\ep

From Lemma~\ref{lmm:main time<=res}, the runtime of $\tetris$ is in the
order of the number of resolutions $\mv w = \Resolve(\mv w_1,\mv w_2)$ it
performs. We call the box $\mv w$ a {\em resolvent} (box).
It is not hard to see that a box cannot be a resolvent twice. Hence, it is
sufficient to bound the total number of resolvents.
We develop a simple technique for bounding the total number of 
resolvents \tetris encounters. 
This technique will be used many times to prove other runtime bounds in the
paper.

\subsection{The integral cover support lemma}
\label{subsec:integral cover support}

In the context of \tetrisreloaded, a box $\mv a$ is called an 
\emph{input gap box} if it was loaded from $\st B$ into $\st A$ 
(i.e. if $\mv a\in \st B'$) at some point in time during the execution of 
the algorithm.
Note that we load into the \kb $\calA$ either gap boxes from $\calB$, or 
an output (unit) box.
We distinguish between two types of resolvents.
We call a resolvent $\mv w = \Resolve(\mv w_1,\mv w_2)$ an {\em \obresolvent} if either $\mv w_1$ or $\mv w_2$ is an output box, or
(recursively) if either $\mv w_1$ or $\mv w_2$ is an \obresolvent.
Other resolvents are called {\em \gbresolvents}.

\bdefn[Projection of a box onto some support]
Let $\mv b = \dbox{x_1, \ldots, x_n}$ be any dyadic box, and 
$V$ be some subset of attributes. Then, the projection of $\mv b$ onto $V$, 
denoted by $\pi_V(\mv b)$, is the box 
$\overline{\mv b} = \dbox{y_1,\dots,y_n}$ where
\[ y_i = \begin{cases} x_i & \mbox{if } i \in V\\ \lambda& \mbox{if } i \notin V.  \end{cases} \]
\label{defn box proj}
\edefn

\bdefn[Resolvent supported on an integral cover]
Let $\mv w$ be a \gbresolvent.
Let $S$ be a subset of $\support(\mv w)$, and $\mv a$ be an input gap box.
Then, $\mv w$ is said to be {\em supported by  $\mv a$ on $S$} if
$\pi_S(\mv a) \subseteq \pi_S(\mv w).$
(Geometrically, the shadow of $\mv w$ on the coordinate subspace of
the variables in $S$ contains the shadow of $\mv a$ on the same subspace.)
An {\em integral cover} of $\support(\mv w)$ is a collection of subsets
of $\support(\mv w)$, say $S_1, \dots, S_c$, such that
\[ S_1 \cup \cdots \cup S_c = \support(\mv w). \]
The resolvent $\mv w$ is said to be {\em supported on an integral cover}
$S_1, \dots, S_c$ if for each $i\in [c]$, $\mv w$ is supported by some
input gap box on $S_i$.
\edefn

Note that the collection $\{S_1,\dots,S_c\}$ viewed as a hypergraph
forms an integral (edge) cover of the ground set $\support(\mv w)$.

\blmm[The integral cover support lemma]
Suppose there is a positive integer $c \in [n]$ such that
every \gbresolvent $\mv w$ is supported by an integral cover of size at most $c$.
Then, \tetrisreloaded runs in time $\tO(|\boxcert|^c + Z)$.
\label{lmm IC support}
\elmm
\bp
Noting Theorem~\ref{thm X+Z}, we only need to show that the number of \gbresolvents is at most $\tO(|\boxcert|^c)$.
The total number of boxes that \tetrisreloaded loads from $\st B$ into $\st A$ is $\sum_{\calB'} |\calB'|$.
For a given subset $K$ of the $n$ attributes of $\st B$, there are only $\tO(1)$ 
possible integral covers of size $c$.
For each such integral cover $\{S_1, \dots, S_c\}$, there are at most 
$\tO((\sum_{\calB'} |\calB'|)^c)$
witnesses $\mv w$ supported on this integral cover. (This is because given an integral cover of size $c$ along with $c$ input gap boxes , there are maximally $\tO(1)$ witnesses supported on this integral cover by those $c$ input gap boxes, thanks to Proposition \ref{prop:number-of-boxes-containing-t}.)
By Lemma \ref{lmm sum B's = cert}, $(\sum_{\calB'} |\calB'|)^c=\tO(|\boxcert|^c)$.
%\ar{Maybe explain why the bound of $(\sum_{\calB'} |\calB'|)^c$ is valid?}
The number of possible choices of $K$ is also $\tO(1)$, which completes the proof. % The lemma is thus proved.
\ep

%Marked app

% ----------------------------------------------------------------------------
\subsection{The \texorpdfstring{$\tO(|\boxcert|^{w+1}+Z)$}{O(|C|\{w+1\}+Z)} runtime for queries with treewidth \texorpdfstring{$w$}{w}}
\label{sec:C^{w+1}+Z}

In this section, we will prove Theorem~\ref{thm:main:C^{w+1}+Z}. We begin with some background.

\bdefn[$\support(A_k)$ and the induced width of a GAO]
\label{defn:induced-width}
Let $\calH_n=(\calV,\calE)$ be a hypergraph whose vertex set is $\st V=\{A_1, \ldots, A_n\}$.
Let $\sigma=(A_1, \ldots, A_n)$ be a GAO for the vertices (attributes) in $\st V$.
The support of $A_n$ (denoted by $\support(A_n)$) is the union of
all hyperedges in $\calH_n$ that contain $A_n$.
Construct $\calH_{n-1}$ from $\calH_n$ by inserting $\support(A_n)$
as a new hyperedge, and then removing $A_n$ from the vertex set and
from all the hyperedges of $\calH_n$.
Then, define $\support(A_{n-1})$ as the union of all hyperedges
containing $A_{n-1}$ in $\calH_{n-1}$.
We keep constructing hypergraphs $\calH_k$ and defining the supports
of $A_k$, for $k=n-2,\dots,1$ in the same way:
\bi
 \item $\support(A_k)$ is the union of all hyperedges of $\calH_k = (\calV_k,
 \calE_k)$ that contain $A_k$.
 \item $\calH_{k-1}$ is constructed from $\calH_k$ by adding a new
 hyperedge $\support(A_k)$ to $\calH_k$ and removing the vertex
 $A_k$ from $\calH_k$.
\ei

The quantity
\begin{equation}
w = \max_{k\in [n]} |\support(A_k)| - 1 
\label{eqn induced width}
\end{equation}
is called the {\em induced width} of $\sigma=(A_1, \ldots, A_n)$ (with respect
to the hypergraph $\st H$).
Furthermore, if $\st H$ has treewidth $w$, then there exists a GAO
with induced width $w$. 
This follows from the well-known fact that the smallest induced treewidth 
(over all elimination orders) of a hypergraph is the same as the treewidth of
the hypergraph (see, e.g., \cite{journals/ai/DechterP89, MR985145}). 
Such a GAO with optimal induced width
can be computed in time exponential in the size of $\st H$.
\footnote{If $\st H$ is a hypergraph representing a join query, then the size of $\st H$ is data-independent. We say that $\st H$ has size $\tO(1)$ in data complexity.}
\edefn
Notice that if $\st H$ is $\alpha$-acyclic and $\sigma$ is a reversed GYO elimination order, then the above definition reduces back to Definition~\ref{defn support A_k}.

\rev{
\begin{example}
Suppose that $\calH=(\calV,\calE)$ where $\calV=\{A_1, A_2, A_3, A_4\}$ and 
$\calE=\left\{\{A_1, A_2\},\{A_1, A_3\},\{A_2,A_4\},\{A_3,A_4\}\right\}$. Let the GAO be $\sigma=(A_1,A_2,A_3,A_4)$.
$\support(A_4)=\{A_2, A_3, A_4\}$ and $\calH_{3}=(\calV_{3},\calE_{3})$ where $\calV_{3}=\{A_1, A_2, A_3\}$ and $\calE_{3}=\left\{\{A_1,A_2\},\{A_1,A_3\},\{A_2,A_3\}\right\}$.
$\support(A_3)=\{A_1,A_2,A_3\}$ and $\calH_2=(\calV_2,\calE_2)$ where $\calV_2=\{A_1,A_2\}$ and
$\calE_2=\left\{\{A_1,A_2\}\right\}$. Similarly, $\support(A_2)=\{A_1, A_2\}$ and $\support(A_1)=\{A_1\}$.
By \eqref{eqn induced width}, the induced width is 2.
\end{example}
}

\blmm
Let $\st B$ be a set of boxes over the attributes $\{A_1, \ldots, A_n\}$ and suppose that we run \tetrisreloaded on $\st B$ with the \RAO $\sigma=(A_1, \ldots, A_n)$.
Let $\mv w$ be an input gap box or a \gbresolvent,
and suppose that the last non-$\lambda$ component of $\mv w$ is on attribute $A_k$ (i.e.\ $k=\max \left\{i\suchthat A_i\in\support(\mv w)\right\}$). Then
\[ \support(\mv w) \subseteq \support(A_k). \]
\label{lmm support(w)}
\elmm
\bp
We prove this lemma by induction. If $\mv w$ is an input gap box,
then clearly $\support(\mv w) \in \calE_k$. For the inductive step, suppose
$\mv w$ is the resolution of $\mv w_1$ and $\mv w_2$ on attribute $A_k$,
where the induction hypothesis is $\support(\mv w_1) \subseteq \support(A_k)$
and $\support(\mv w_2) \subseteq \support(A_k)$.

If the last non-$\lambda$ component of $\mv w$ is on $A_k$, then 
by Lemma~\ref{lmm ordered res}
\[ \support(\mv w) \subseteq \support(\mv w_1) \cup \support(\mv w_2) 
                   \subseteq \support(A_k).
\]
If the resolution turns the $k$th component of $\mv w$ into a $\lambda$,
then $\support(\mv w) \subseteq \support(A_k) - \{A_k\}$.
Suppose the last non-$\lambda$ component of $\mv w$ is on $A_{k'}$, then
$A_{k'} \in \support(A_k) - \{A_k\}$, and hence $\support(\mv w)$ is a
hyperedge of the graph $\calH_{k'}$. This means
$\support(\mv w) \subseteq \support(A_{k'})$ as desired.
\ep

%Marked app
For the sake of completeness, we re-state Theorem~\ref{thm:main:C^{w+1}+Z} 
and then prove it.
\bthm[Theorem~\ref{thm:main:C^{w+1}+Z} re-stated]
%\tetrisreloaded runs in time $\tO(|\boxcert|^{w+1}+Z)$]
For any set of boxes $\st B$ with $\tw(\calH(\calB)) = w$,
by setting \RAO to be the attribute ordering with elimination width $w$, 
\tetrisreloaded solves \bcp\ on input $\st B$ in time $\tO(|\boxcert|^{w+1}+Z)$.
Here, $\boxcert$ is any optimal box certificate for the instance, 
and $Z$ is the output size.
\label{thm:C^{w+1}+Z}
\ethm
%Marked app
\bp
We apply Lemma~\ref{lmm IC support}. We will show that for every witness
$\mv w$ which is a either an input gap box or a \gbresolvent,
its support $\support(\mv w)$ is the union of $w+1$ singleton sets
$V_1, \dots, V_{w+1}$ such that for each set $V_i$,
$\pi_{V_i}(\mv a) \subseteq \pi_{V_i}(\mv w)$ for some input gap box 
$\mv a$.

From Lemma~\ref{lmm support(w)} and equation~\eqref{eqn induced width},
it follows that $\support(\mv w) \leq w+1$. Therefore, it is sufficient to prove the following claim:
every non-$\lambda$ component of $\mv w$ is a prefix of some 
component of an input gap box $\mv a$.

The claim is proved easily by induction.
For the base case,
if $\mv w$ is an input gap box, then clearly the claim holds.
For the inductive step, suppose $\mv w = \Resolve(\mv w_1,\mv w_2)$ where
the claim holds for $\mv w_1$ and $\mv w_2$, and the resolution is on attribute $A_k$. 
The claim holds for $\mv w$ because every non-$\lambda$ 
component of $\mv w$ is either the same as that of $\mv w_1$ or of $\mv w_2$, 
except for the component corresponding to $A_k$ which is a prefix of the component
from $\mv w_1$ (and $\mv w_2$).
\ep

% ----------------------------------------------------------------------------
\subsection{The \texorpdfstring{$\tO(|\boxcert|+Z)$}{O(|C|+Z)} runtime for queries with treewidth \texorpdfstring{$1$}{1}}
\label{subsec:treewidth 1 results}

%Marked app
Given a hypergraph $\st H$ (or a query $Q$), there exists a GAO with induced width $1$ if and only if 
$\st H$ is a forest (all relations in $Q$
have arity at most $2$, and $Q$ is $\alpha$-acyclic
\footnote{When all relations have arity $\leq 2$, $\alpha$-acyclicity and 
$\beta$-acyclicity coincide.}).
When the induced width is $1$, by definition $\support(A_k) \leq 2$
for all $k\in [n]$.

The essence of the proof of Theorem~\ref{thm:main acyclic arity<=2} 
is to make use of the fact that when resolving $\mv w_1$ and $\mv w_2$
with support of size at most $2$, we end up with a resolvent $\mv w$
which contains either $\mv w_1$ or $\mv w_2$ or both.
(See Figure~\ref{fig:main:resolution} for an illustration.)

\blmm[2D-resolution expands] %2D resolution expands]
Suppose \tetris performs $\mv w=\Resolve(\mv w_1, \mv w_2)$
such that $|\support(\mv w_1) \cup \support(\mv w_2)| \leq 2$, 
then the resulting box $\mv w$ is a prefix box
\footnote{See Definition \ref{defn box prefix} for the notion of ``prefix box''. Note that in this situation $\mv w$ being a prefix box of $\mv w_i$ is the same as $\mv w$ containing $\mv w_i$.}
of either $\mv w_1$ or $\mv w_2$
or both.
\label{lmm 2D res expands}
\elmm
\bp
Given two strings $x_1, x_2$ and two boxes
$\mv{w_1}, \mv{w_2}$ WLOG of the form:
\begin{eqnarray*}
\mv w_1 \weq \dbox{\lambda\wc &x_1         &\wc \lambda\wc \lambda\wc x_2 0\wc \lambda}\\
\mv w_2 \weq \dbox{\lambda\wc &\prefix{x_1}&\wc \lambda\wc \lambda\wc x_2 1\wc \lambda}
\end{eqnarray*}
The resolvent of $\mv w_1$ and $\mv w_2$ is
\[ \mv w \weq \dbox{\lambda\wc x_1 \wc \lambda\wc \lambda\wc x_2\wc \lambda}.
\]
By definition~\ref{defn box prefix}, $\mv w$ is a prefix box of $\mv w_1$.
\ep

\bthm[Theorem~\ref{thm:main acyclic arity<=2} re-stated]
%\tetrisreloaded runs in time $\tO(|\boxcert|^{w+1}+Z)$]
For any set of boxes $\st B$ with $\tw(\calH(\calB)) = 1$,
by setting \RAO to be the attribute ordering with elimination width $1$, 
\tetrisreloaded solves \bcp\ on input $\st B$ in time $\tO(|\boxcert|+Z)$.
Here, $\boxcert$ is any optimal box certificate for the instance, 
and $Z$ is the output size.
\label{thm acyclic arity<=2}
\ethm
\bp
In light of Lemma~\ref{lmm IC support}, we show that every 
\gbresolvent $\mv w$ is supported by {\em one} input gap box on $\support(\mv w)$.
In particular, we use induction to show that
$\mv a \subseteq \mv w$ for some input gap box $\mv a$.
If $\mv w$ was an input gap box, then the above obviously holds.
We can use that as a base case.

\tetris selects a \RAO\ with elimination width $1$. 
From Lemma~\ref{lmm support(w)} and Equation~\ref{eqn induced width},
any boxes $\mv w_1$ and $\mv w_2$ that are resolved (on any dimension) 
satisfy $|\support(\mv w_1) \cup \support(\mv w_2)| \leq 2$. 
From Lemma~\ref{lmm 2D res expands}, every resolution
$\mv w = \Resolve(\mv w_1,\mv w_2)$ results in a  box $\mv w$ containing either
$\mv w_1$ or $\mv w_2$. Hence, by induction every \gbresolvent
contains an input gap box.
\ep

\subsection{The \texorpdfstring{$\tO(|\boxcert|^{n-1}+Z)$}{O(|C|\{n-1\}+Z)} runtime for any query over \texorpdfstring{$n$}{n} attributes}

In this section, we will prove the following result.
(An $n$-clique is an example of a hypergraph whose treewidth $w=n-1$.)
\bthm[\tetrisreloaded runs in time $\tO(|\boxcert|^{n-1}+Z)$]
For any set of boxes $\st B$ in $n$ dimensions, 
\tetrisreloaded solves \bcp\ on input $\st B$ in time $\tO(|\boxcert|^{n-1}+Z)$.
Here, $\boxcert$ is any optimal box certificate for the instance, 
and $Z$ is the output size.
\label{thm:C^{n-1}+Z-upperbound}
\ethm

Along with Proposition~\ref{prop:bcp=join}, 
the above result implies the following:
\bcor
\tetrisreloaded\ evaluates any join query $Q$ over $n$ attributes in time 
$\tO(|\boxcert|^{n-1}+Z)$, where $\boxcert$ is an optimal box certificate 
for the join instance, and $Z$ is the output size.
\ecor
\bp[Proof of Theorem~\ref{thm:C^{n-1}+Z-upperbound}]
We inductively prove the following claim: Every witness $\mv w$ which is involved in a \gbresolution on $A_n$ is supported by one input gap box on $\{A_i, A_n\}$ for each $i\in[n-1]$. In the base case, $\mv w$ is an input gap box in which case the claim is obviously true. For the inductive step, $\mv w=\Resolve(\mv w_1, \mv w_2)$. For each $i\in[n-1]$, either $\pi_{\{A_i, A_n\}}(\mv w_1)\subseteq \pi_{\{A_i, A_n\}}(\mv w)$ or $\pi_{\{A_i, A_n\}}(\mv w_2)\subseteq \pi_{\{A_i, A_n\}}(\mv w)$. Assuming the claim holds for $\mv w_1$ and $\mv w_2$, it holds for $\mv w$.

For each $i \in [n-1]$, every witness $\mv w$ which is involved in a \gbresolution on $A_i$ is supported on its support set by $\leq i$ input gap boxes. Lemma~\ref{lmm IC support} completes the proof.
\ep

%Marked app

\section{Omitted details from Section~\ref{SEC:CERT-N/2}}
\label{app:cert-n/2}

\subsection{The high-level ideas}

While Theorem~\ref{thm:C^{n-1}+Z-upperbound} shows that \tetrisreloaded runs in time $\tO(\abs{\boxcert}^{n-1}+Z)$ for $n$-attributes/dimensions, Theorem~\ref{thm:cert^n-1-lb} shows a lower bound of $\Omega(\abs{\boxcert}^{n-1}+Z)$ for all algorithms that use only \ordered.

In this section, we highlight the key ideas that lead to a new enhancement of \tetris
that enables it to bypass the above lower bound and solve join queries (and more generally \bcp instances) in time $\tO(\abs{\boxcert}^{n/2} + Z)$.
We take two specific values of $n$ as examples: $n=3$ and $n=4$, attempting to introduce one new idea at a time. Note that this result is independent of the
input query structure.

In these examples, we assume that the algorithm is 
given as its input the box certificate of the instance of \bcp, which we 
will denote by $\cert$. We will also refer to this version of \bcp\ as
the {\em offline} case of the problem.
Moreover, we will only consider the \emph{Boolean} version of \bcp where the
objective is to determine whether the output is empty. Boolean \bcp is
equivalent to determining whether the set $\st C$ of input gap boxes covers the
entire output space. (See Definition~\ref{defn:boolean-bcp}.)

Later on in Section~\ref{sec offline as}, we will be using Theorem~\ref{thm X+Z} to convert
the offline algorithm solving \emph{Boolean} \bcp in time $\tO(\abs{\cert}^{n/2})$
into an offline algorithm solving \bcp in time $\tO(\abs{\cert}^{n/2}+Z)$.
In Section~\ref{sec online as}, we will show how to convert the \emph{offline} algorithm into an \emph{online} one. Theorem~\ref{thm:geo-lb-n-clique} shows a lower bound of $\Omega(\abs{\st C}^{n/2}+Z)$ for all \geo\ algorithms.

\subsection{Idea 1: load-balancing}
We start with a motivating example.
\begin{example}
\label{ex:Omega(sqr(|C|))}
Consider an input query $Q$ over $n=3$ attributes $X, Y$, and $W$,
and the set $\st C:=\st C_1 \cup \st C_2 \cup \st C_3$ of input gap
boxes, where
\begin{eqnarray*}
    \st C_1 &=&
\left\{\dbox{0x, \lambda, 0} \wsuchthat x \in \{0,1\}^{d-2}\right\}
\quad\cup\quad
\left\{\dbox{0, y, 1} \wsuchthat y \in \{0,1\}^{d-2}\right\}\\
\st C_2 &=&
\left\{\dbox{10x, 0, \lambda} \wsuchthat x \in \{0,1\}^{d-2}\right\}
\quad\cup\quad
\left\{\dbox{10, 1, z} \wsuchthat z \in \{0,1\}^{d-2}\right\}\\
\st C_3 &=&
\left\{\dbox{110, y, \lambda} \wsuchthat y \in
\{0,1\}^{d-2}\right\} \quad\cup\quad
\left\{\dbox{111, \lambda, z} \wsuchthat z \in \{0,1\}^{d-2}\right\}.
\end{eqnarray*}
Note that $\st C_1, \st C_2, \st C_3$, and $\st C$ cover $\dbox{0, \lambda, \lambda}$, $\dbox{10, \lambda, \lambda}$, $\dbox{11, \lambda, \lambda}$, and $\dbox{\lambda, \lambda, \lambda}$ respectively. 
The output is empty and only $O(|\st C|)$ out-of-order geometric resolutions are 
sufficient to prove it.\footnote{Indeed note that by resolving on the $X$-attribute of the first subset of $\st C_1$, we can get $\dbox{0,\lambda,0}$. Similarly by resolving on the $Y$-attribute of the second subset of $\st C_1$, we get the box $\dbox{0,\lambda,1}$. With one more resolution we get the box $\dbox{0,\lambda,\lambda}$ from $\st C_1$. Similarly by doing resolutions on $X$, $W$, then $Y$-attributes of $\st C_2$, one can get the box $\dbox{10,\lambda,\lambda}$. Finally, by resolving on $Y$, $W$, then $X$ attributes of $\st C_3$, we can get the box $\dbox{11,\lambda,\lambda}$. Then with two more resolutions on the three boxes from $\st C_1, \st C_2,\st C_3$, we get $\dbox{\lambda,\lambda,\lambda}$, as desired.} However, $\Omega(\abs{\st C}^2)$ ordered geometric resolutions 
are necessary for the proof of $\dbox{\lambda, \lambda, \lambda}$, no matter what the \RAO is. This is because if $W$ is the last attribute in the \RAO, then we will need $\Omega(\abs{\st C}^2)$ resolutions just to infer $\dbox{0, \lambda, \lambda}$ from $\st C_1$. If $Y$ (or $X$) is the last attribute, then $\st C_2$ (or $\st C_3$) is going to create the same problem.
\end{example}

The main reason \tetrispreloaded (and any \ordered\ algorithm)
is slow in the kind of inputs shown above is
because it got ``stuck'' in performing two many resolutions on one particular
attribute, creating many witnesses along a particular dimension. At a very high
level, our first idea is to cut the output space along a particular dimension 
into a collection of ``layers,'' each of which has relatively ``few'' input 
gap boxes. Then, our algorithm explores whether each of
these layers is completely covered by the input gap boxes. Since those
layers form a partition of a given dimension, the output space is covered iff each
layer is covered. The layers are chosen so that the number of layers is
``small'' and the number of boxes that can contribute to covering each layer
is also ``small.'' From there, verifying that a layer is covered takes little time.

To be more concrete, we first define the notion of ``dimension partition.''

\bdefn[Dimension partition]
Recall that the domain of each attribute is assumed to be $\D = \{0,1\}^d$,
i.e. the set of all binary strings of length $d$. Each binary string $x$ of
length $\leq d$ is a dyadic interval.
A {\em partition} $P$ of $\D=\{0,1\}^d$ is a collection of disjoint dyadic intervals
whose union is exactly $\D$. In particular, $P$ is a partition of $\D$ iff it
satisfies the following two properties:
\bi
 \item Strings in $P$ are prefix-free, i.e. no string is a prefix of another.
 \item For every string $s$ of length $d$, there is some string $x \in P$ so
 that $x$ is a prefix of $s$.
\ei
Given an attribute $X$ of the input query,
an {\em $X$-partition} is a partition of the domain $\D(X)$.
We will typically use $P_X$ to denote a partition along dimension $X$.
\edefn

\begin{example*}[\continued{ex:Omega(sqr(|C|))}]
$P_X=\left\{0, 10, 110, 111\right\}$ is one possible $X$-partition. Another $X$-partition could be $P_X = \left\{0, 100, 101, 110, 111\right\}$.
\end{example*}

Geometrically, a partition along dimension $X$ divides the output space
into $|P_X|$ ``layers'', one for each interval $x$ in $P_X$. 
In particular, the layer defined by a fixed interval $x\in P_X$, called the
{$x$-layer}, is the dyadic box $\dbox{x, \lambda,\lambda}$.
To verify that the output space is completely covered by the input gap
boxes, it is sufficient to verify that every $x$-layer is covered, for each
$x\in P_X$.
An input gap box whose $X$-component is disjoint from $x$ will not affect
whether the $x$-layer is covered. 
Hence, to verify whether the $x$-layer is covered, we can ignore all gap boxes
that do not intersect $x$. If the remaining set of gap boxes is small, then this
verification is fast. At the same time, we do not want too many layers because
that certainly increases the total amount of verification work. This balancing
act leads to our first idea: we find a dimension partition that is somehow
``balanced.''

For notational convenience, given a set of boxes $\st C$ and a dyadic interval
$x$ on the domain $\D(X)$, define two sets:
\begin{eqnarray}
\st C_{\subset x}(X) &=& \left\{ \mv b \in \st C \suchthat \pi_X(\mv b)
\subsetneq x \right\} \label{eqn:app C_subset}\\
\st C_{\supseteq x}(X) &=& \left\{ \mv b \in \st C \suchthat \pi_X(\mv b)
\supseteq x \right\}. \label{eqn:app C_supset}
\end{eqnarray}
In other words, the first set $\st C_{\subset x}(X)$ consists of all dyadic boxes in
$\st C$ that are strictly contained in the $x$-layer.
The second set $\st C_{\supseteq x}(X)$ is the set of boxes in $\st C$ each of whose
$X$-component completely covers the interval $x$.
Note that, for every box $\mv b \in \st C - (\st C_{\subset x}(X) \cup \st
C_{\supseteq x}(X))$, the dyadic interval $\pi_X(\mv b)$ is completely disjoint
from the interval $x$.
It should also be emphasized that $\pi_X(\mv b) \subsetneq x$ means $x$ (as a string) is a
strict prefix of the binary string $\pi_X(\mv b)$. (Alternatively, $x$ as a dyadic segment strictly contains the dyadic segment $\pi_X(\mv b)$.)

\bdefn[Balanced dimension partition]
\label{def:balanced-partition}
Let $\st C$ be the set of input gap boxes, and $X$ be any attribute.
A {\em balanced $X$-partition} is an $X$-partition $P_{X}$ such that
\begin{eqnarray*}
|P_X|                  &=    & \tO(\sqrtC)\\
|\st C_{\subset x}(X)| &\leq & \sqrtC, \text{ for every } x \in P_X.
\end{eqnarray*}
\edefn

Geometrically, the first condition states that the number of layers is small.
The second condition states that the number of input gap boxes completely
contained in every $x$-layer is small.
Since there are $|\st C|$ input gap boxes, the quantity $\sqrtC$ is exactly the
mid-point, balancing the number of layers and the number of boxes contained
within each layer.

\begin{example*}[\continued{ex:Omega(sqr(|C|))}]
The following is a balanced $X$-partition: (Recall that $\abs{\st C}=6\cdot2^{d-2}$ and that Definition~\ref{def:balanced-partition} does not put any restrictions on $|\st C_{\supseteq x}(X)|$.)
\begin{equation}
P_X=\left\{0x'\suchthat x'\in\{0, 1\}^{\left\lceil (d-2)/2\right\rceil}\right\}\cup\left\{10x'\suchthat x'\in\{0, 1\}^{\left\lceil (d-2)/2\right\rceil}\right\}\cup\Bigl\{110, 111\Bigr\}.
\label{eq:ex:Omega(sqr(|C|)):P_X}
\end{equation}
\end{example*}

A balanced partition can be constructed easily, as the following proposition
shows.

\bprop[Efficient \rev{construction} of balanced partitions]
\label{prop:app:construct_P}
Given a set $\st C$ of input gap boxes and an arbitrary attribute $X$,
a balanced $X$-partition can be computed in time $\tO(|\st C|)$.
\eprop
\bp
We start with the trivial partition $P = \{ \lambda \}$
that has only one layer, and keep revising it until it becomes balanced.
An interval $x \in P$ is said to be {\em heavy} if 
\begin{equation}
\label{eq:app:heavy-interval}
\abs{\st C_{\subset x}(X)} > \sqrtC.
\end{equation}
While there is still a heavy interval $x$ in $P$, replace
$x$ by the two sub-intervals $x0,x1$. These sub-intervals are called {\em
children} of $x$, and $x$ is a \emph{parent} of both $x0$ and $x1$.
This process certainly terminates when no interval in $P$ is heavy anymore. (Note that by definition a unit interval $x$ is not heavy because $\abs{\st C_{\subset x}(X)}$=0.) It 
remains to show that $|P| = \tO(\sqrtC)$ in the end.
To see this, consider the set $H$ of all heavy intervals. If an interval is
heavy then all its prefixes are heavy. 
Let $H'$ be the maximal set of prefix-free intervals in $H$. Then the intervals in
$H'$ are disjoint. Because they are all heavy, $|H'| \leq \sqrtC$. 
Furthermore, because the intervals in $H-H'$ are prefixes of intervals in $H'$, we conclude that
$|H| = \tO(\sqrtC)$. Finally, since the intervals in $P$ are all children of some
intervals in $H$, we conclude that $|P| \leq 2|H| = \tO(\sqrtC)$ as well.
\ep

Next, we explain how a balanced $X$-partition can be used to solve the offline
version of the problem in time $\tO(|\st C|^{3/2})$ when $n=3$.
Consider an input query $Q$ over $n=3$ attributes $X, Y$, and $W$.
Let $P_X$ be a balanced $X$-partition.
For every interval $x\in P_X$, we would like to verify that the $x$-layer, i.e.
the box $\dbox{x, \lambda,\lambda}$ is covered by all gap boxes in $\st C$.
The boxes in $\st C_{\subset x}(X) \cup \st C_{\supseteq x}(X)$ are the only boxes
in $\st C$ that intersect the $x$-layer. Hence, it is sufficient to verify that
boxes in this union cover the $x$-layer. 

At this point, we introduce another simple idea: we reduce the $x$-layer
coverage problem above to a slightly different 3-dimensional coverage problem
and run \tetrispreloaded on it.
Define
\begin{eqnarray}
\st C[x] &=& \left\{ \dbox{\lambda, y, w} \suchthat \dbox{x',y,w} \in \st
C_{\supseteq x}(X) \text{ for some } x'\right\}\label{eq:C[x]-lb}\\
\st F[x] &=& \left\{ \dbox{x', y, w} \suchthat \dbox{x x',y,w} \in \st
C_{\subset x}(X) \right\}.\label{eq:F[x]-lb}
\end{eqnarray}
Then, verifying whether 
the boxes in $\st C_{\subset x}(X) \cup \st C_{\supseteq x}(X)$ cover
the $x$-layer is exactly identical to verifying whether $\st C[x] \cup
\st F[x]$ covers the entire space $\dbox{\lambda,\lambda,\lambda}$.
In other words, because the $X$-components of
the boxes in $\st C_{\supseteq x}(X)$ already contain the entire $x$-interval, we 
might as well truncate all but the $x$-interval part of those boxes.
And, we convert the $X$-component of boxes in $\st C_{\subset x}(X)$ to
``relative'' values in the coordinate system restricted to the $x$-layer.

Now, we run \tetrispreloaded on the $\st C[x] \cup \st F[x]$ input with
the \RAO being $(Y, W, X)$. While this \RAO might seem a bit unnatural, we will
show in the next sub-section why this \RAO embodies a new idea.
We bound the runtime of \tetrispreloaded using Lemma~\ref{lmm IC support}.
%\yell{Lemma 6.5 from FOCS submission}. 

\bi
 \item If a witness $\mv w$ is a result of a \gbresolution on $X$, then
its support is (integrally) covered by two boxes from $\st F[x]$ (as all boxes
from $\st C[x]$ \rev{have $X$-components} already equal $\lambda$). Thus, the number of
such witnesses is at most $\tO(|\st F[x]|^2) = \tO(\abs{\st C_{\subset x}(X)}^2) = \tO(|\st C|)$.

 \item The number of witnesses $\mv w$ resulting from \gbresolutions on $Y$ or $W$
is also at most $|\st C|$. For example, consider a witness 
$\mv w = \Resolve(\mv w_1,\mv w_2)$ which is resolved on $Y$ or $W$.  
By Lemma~\ref{lmm 2D res expands},
$\mv w$ is a prefix box of either $\mv w_1$ or $\mv w_2$.
The witnesses $\mv w_i$ ($i\in \{1,2\}$) fall into three classes:
(1) $\mv w_i$ is the result of a \gbresolution on $Y$ or $W$, 
(2) $\mv w_i \in \st C[x]$,
(3) $\mv w_i$ is the result of \gbresolution on $X$ (the resolution
turns $\mv w_i$'s $X$-component into a $\lambda$).
The number of witnesses of types (2) and (3) is at most $\tO(|\st C|)$.
Hence, by induction the number of \gbresolutions on $Y$ or $W$ is also
$\tO(|\st C|)$.
\ei
(We will make the above argument more formal in the next section.)
Hence, the overall $x$-layer verification process takes $\tO(|\st C|)$ time;
and, since there are $\tO(\sqrtC)$ intervals in $P_X$, the entire algorithm takes
$\tO(|\st C|^{3/2})$-time as desired.

\begin{example*}\continued{ex:Omega(sqr(|C|))}
We apply \eqref{eq:F[x]-lb} and \eqref{eq:C[x]-lb} for every $x$ in the $X$-balanced partition $P_X$ given earlier in \eqref{eq:ex:Omega(sqr(|C|)):P_X}. For every $x'\in\{0, 1\}^{\left\lceil (d-2)/2\right\rceil}$, we get (assuming $d\geq 4$):
\begin{eqnarray*}
\st F[0x'] &=& \left\{\dbox{x'', \lambda, 0} \wsuchthat x'' \in \{0,1\}^{\left\lfloor (d-2)/2\right\rfloor}\right\}\\
\st C[0x'] &=& \left\{\dbox{\lambda, y, 1} \wsuchthat y \in \{0,1\}^{d-2}\right\}\\
\st F[10x'] &=& \left\{\dbox{x'', 0, \lambda} \wsuchthat x'' \in \{0,1\}^{\left\lfloor (d-2)/2\right\rfloor}\right\}\\
\st C[10x'] &=& \left\{\dbox{\lambda, 1, z} \wsuchthat z \in \{0,1\}^{d-2}\right\}.
\end{eqnarray*}
Moreover
\begin{eqnarray*}
\st C[110] &=& \left\{\dbox{\lambda, y, \lambda} \wsuchthat y \in
\{0,1\}^{d-2}\right\}\\
\st C[111] &=& \left\{\dbox{\lambda, \lambda, z} \wsuchthat z \in \{0,1\}^{d-2}\right\}\\
\st F[110] &=& \st F[111] = \emptyset.
\end{eqnarray*}
Now for every $x'\in\{0, 1\}^{\left\lceil (d-2)/2\right\rceil}$, we verify that the $0x'$-layer is covered by verifying that $\st C[0x']\cup \st F[0x']$ covers $\dbox{\lambda,\lambda,\lambda}$. While running \tetrispreloaded on $\st C[0x']\cup \st F[0x']$ using the \RAO of $(Y, W, X)$, we need $\tO(\sqrtC)$ resolutions on $X$ to infer $\dbox{\lambda, \lambda, 0}$ from $\st F[0x']$. After that, we need $\tO(\abs{\st C})$ resolutions on $Y$ and $W$ to infer $\dbox{\lambda, \lambda, \lambda}$ from $\st C[0x']\cup\left\{\dbox{\lambda, \lambda, 0}\right\}$. (Recall that we were looking at the $0x'$-layer.)

In a similar way, we can handle the remaining $10x',110,111$-layers.
\end{example*}

\subsection{Idea 2: taking the high road (to higher dimension)}

The algorithm described above can be expressed more cleanly as follows.
Strategies based on ordered geometric resolutions do not meet the runtime
target (of $\tO(|\st C|^{n/2})$) for the original problem. 
However, it will meet the runtime target if we map
each input gap box to a higher dimensional gap box, {\em then} run
\tetrispreloaded on these new gap boxes. This idea is similar in spirit to the
kernel method in machine learning: data in the original dimensions are not
linearly separable, but they become linearly separable in higher dimensions
after a kernel map.

More concretely, we will re-cast the above algorithm by explicitly constructing 
the map that transforms each input gap box to a gap box in a higher-dimensional
space. This set of new gap boxes are constructed so that they are
``load-balanced'', making \tetris work efficiently on them.

Let $s$ be any dyadic interval, and let $\prefixes{s}$ denote the set of all binary
strings that are prefixes of $s$, including $s$ itself.
For any set $S$ of dyadic intervals. Define
\[ \prefixes{S} = \bigcup_{s\in S} \prefixes{s}. \]

Let $P_X$ denote a fixed balanced $X$-partition of $\st C$.
Let $X'$ and $X''$ be two new attribute names.
The following map, called the {\em $X$-load balancing map},
\[ \lbm_X : \D(X) \times \D(Y) \times \D(W) \to 
            \D(X') \times \D(Y) \times \D(W) \times \D(X''), \]
is defined as follows.
Let $\mv b = \dbox{x,y,w}$ be any box in $\D(X) \times \D(Y) \times \D(W)$.
Then,
\[ \lbm_X(\mv b) = \lbm_X(\dbox{x, y, w}) :=
   \begin{cases}
     \dbox{x', y, w, x''} & \text{ if } x = x' x'' \text{ where } x' \in
     P_X \text{ and } x'' \neq \lambda\\
     \dbox{x, y, w, \lambda} & \text{ otherwise. (i.e., $x \in \prefixes{P_X}$)}
   \end{cases}.
\]
Let $\lbm_X(\st C) = \bigl\{ \lbm_X(\mv b) \suchthat \mv b \in \st C\bigr\}$. Then the
algorithm described in the previous section is simply to run \tetrispreloaded
on input $\lbm_X(\st C)$.
This input has gap boxes in $4$ dimensions $(X', Y, W, X'')$, which is also the
\RAO the algorithm adopts.

For each $x \in P_X$, define the following two sets:
\begin{eqnarray}
\st C[x] &=& \left\{ \dbox{x', y, w, \lambda} \in \lbm_X(\st C) \suchthat x' \in \prefixes{x} \right\} \label{eq:C[x]-hr}\\
\st F[x] &=& \left\{ \dbox{x, y, w, x''} \in \lbm_X(\st C) \suchthat x'' \neq
\lambda \right\}. \label{eq:F[x]-hr}
\end{eqnarray}
Note that
\[ \lbm_X(\st C) = \bigcup_{x\in P_X} \left( \st C[x] \cup \st F[x] \right), \]
and that the sets $\st F[x]$ are disjoint for $x\in P_X$.
Because $P_X$ is a balanced $X$-partition, we know $|\st F[x]| \leq \sqrtC$
for all $x \in P_X$ and, obviously, $|\st C[x]| \leq |\st C|$.

\bdefn[$A$-witness]
Let $A$ be an attribute from the set $\{X',Y,W,X''\}$.
A witness $\mv w$ is called an {\em $A$-witness} if it participates in a
\gbresolution on attribute $A$ during the execution of the algorithm.
(If $\mv w = \Resolve(\mv w_1,\mv w_2)$, then all three boxes
$\mv w, \mv w_1,\mv w_2$ participate in the resolution.)
\edefn

We analyze the above algorithm by counting the number of $A$-witnesses for
each $A\in \{X',Y,W,X''\}$.
\bi
 \item First, we bound the number of $X''$-witnesses.
 When \tetris resolves two boxes $\mv w_i = \dbox{x'_i,y_i,w_i,x''_i}, i\in
 \{1,2\}$ on $X''$, the strings $x'_1$ and $x'_2$ must be a prefix of one another.
 Furthermore, $x''_i \neq \lambda$ for $i\in \{1,2\}$. Hence, $x'_1, x'_2 \in
 P_X$.
 But because strings in $P_X$ are prefix-free, $x'_1=x'_2$. Consequently, 
 the \gbresolutions on $X''$ can be grouped into disjoint groups, one for each
 $x\in P_X$. For each such $x\in P_X$, the number of $X''$-witnesses $\mv w =
 \dbox{x,y,w,x''}$ is at most $\tO(|\st F[x]|^2)$ because $\mv w$ must be 
 supported on $\{Y, X''\}$ and $\{W, X''\}$ by boxes from $\st F[x]$. 
 Consequently, the total number of $X''$-witnesses is at most 
 $\sum_{x\in P_X}\tO(|\st F[x]|^2) = \tO(|\st C|^{3/2})$.

 \item Similarly, we count the number of $W$-witnesses by fixing an $x \in
 P_X$ and counting the number of $W$-witnesses of the form $\mv w =
 \dbox{x',y,w,\lambda}$ for some string $x'\in\prefixes{x}$.
 Then, by summing this count over all $x\in P_X$, we obtain an upperbound on the
 number of $W$-witnesses. We sketch the counting argument below, leaving the
 completely rigorous description to the next sections when we describe the
 general algorithm and its formal analysis.

 Suppose $\mv w = \Resolve(\mv w_1,\mv w_2)$ is a $W$-witness. 
 Note that since \tetris  only performs ordered resolutions, $\pi_{X''}(\mv w_1) = \pi_{X''}(\mv w_2) = \lambda$,
 and $\pi_{W}(\mv w_1) \neq \lambda, \pi_{W}(\mv w_2) \neq \lambda$.
 By Lemma~\ref{lmm 2D res expands},
 $\pi_{Y, W}(\mv w)$ is a prefix box of either $\pi_{Y,W}(\mv w_1)$ or 
 $\pi_{Y,W}(\mv w_2)$.
 Each witness $\mv w_i$ ($i\in \{1,2\}$) belongs to one of three classes:
 (1) $\mv w_i$ is the result of a \gbresolution on $W$, 
 (2) $\mv w_i \in \st C[x]$,
 (3) $\mv w_i$ is the result of \gbresolution on $X''$ (this resolution makes
 $\pi_{X''}(\mv w_i) = \lambda$).
 The number of witnesses of types (2) and (3) is at most $\tO(|\st C|)$.
 (For (3), this is because $\mv w_i$ is supported on $\{Y,X''\}$ and $\{W,X''\}$
 by two boxes from $\st F[x]$.)
 Hence, for each $x\in P_X$, by induction the number of $W$-witnesses of the form
 $\dbox{x', y, w, \lambda}$ where $x' \in \prefixes{x}$ is $\tO(|\st C|)$.
 Overall, the total number of $W$-witnesses is $\tO(|\st C|^{3/2})$
 as desired.
 \item The cases for $Y$ and $X'$ are much simpler and thus omitted.
\ei

\begin{example*}\continued{ex:Omega(sqr(|C|))} We have $\lbm_X(\st C)=\lbm_X(\st C_1)\cup\lbm_X(\st C_2)\cup\lbm_X(\st C_3)$, where
\begin{eqnarray*}
\lbm_X(\st C_1) &=&
\left\{\dbox{0x', \lambda, 0, x''} \wsuchthat x'\in\{0, 1\}^{\left\lceil (d-2)/2\right\rceil}, x''\in\{0, 1\}^{\left\lfloor (d-2)/2\right\rfloor}\right\}
\quad\cup\quad\\
& &\left\{\dbox{0, y, 1, \lambda} \wsuchthat y \in \{0,1\}^{d-2}\right\}\\
\lbm_X(\st C_2) &=&
\left\{\dbox{10x', 0, \lambda, x''} \wsuchthat x'\in\{0, 1\}^{\left\lceil (d-2)/2\right\rceil}, x''\in\{0, 1\}^{\left\lfloor (d-2)/2\right\rfloor}\right\}
\quad\cup\quad\\
& &\left\{\dbox{10, 1, z, \lambda} \wsuchthat z \in \{0,1\}^{d-2}\right\}\\
\lbm_X(\st C_3) &=&
\left\{\dbox{110, y, \lambda, \lambda} \wsuchthat y \in
\{0,1\}^{d-2}\right\} \quad\cup\quad\\
& &\left\{\dbox{111, \lambda, z, \lambda} \wsuchthat z \in \{0,1\}^{d-2}\right\}.
\end{eqnarray*}
Applying \eqref{eq:F[x]-hr} and \eqref{eq:C[x]-hr} on $P_X$ from \eqref{eq:ex:Omega(sqr(|C|)):P_X}, for any $x'\in\{0, 1\}^{\left\lceil (d-2)/2\right\rceil}$, we get (assuming $d\geq 4$)
\begin{eqnarray*}
\st F[0x'] &=& \left\{\dbox{0x', \lambda, 0, x''} \wsuchthat x''\in\{0, 1\}^{\left\lfloor (d-2)/2\right\rfloor}\right\}\\
\st C[0x'] &=& \left\{\dbox{0, y, 1, \lambda} \wsuchthat y \in \{0,1\}^{d-2}\right\}\\
\st F[10x'] &=& \left\{\dbox{10x', 0, \lambda, x''} \wsuchthat x''\in\{0, 1\}^{\left\lfloor (d-2)/2\right\rfloor}\right\}\\
\st C[10x'] &=& \left\{\dbox{10, 1, z, \lambda} \wsuchthat z \in \{0,1\}^{d-2}\right\}.
\end{eqnarray*}
Moreover
\begin{eqnarray*}
\st C[110] &=& \left\{\dbox{110, y, \lambda, \lambda} \wsuchthat y \in
\{0,1\}^{d-2}\right\}\\
\st C[111] &=& \left\{\dbox{111, \lambda, z, \lambda} \wsuchthat z \in \{0,1\}^{d-2}\right\}\\
\st F[110] &=& \st F[111]=\emptyset.
\end{eqnarray*}
\end{example*}

\subsection{Idea 3: global accounting}

\rev{
We next consider a query over $n=4$ attributes. 
We will be using a straightforward generalization of the balanced partition idea from the previous section.
However, as we will see, the analysis is not going to be a straightforward generalization of the analysis for $n=3$.
In particular, while it sufficed to use two simple inductive assumptions to count the number of witnesses for $n=3$,
now we will be needing three (more complicated and seemingly incompatible) inductions to do the counting for $n=4$.
In the next section, we will develop a single quite-involved inductive assumption for any $n$ subsuming all the previous ones for $n=3, 4$.

\iffalse
We explain why a straightforward
generalization of the balanced partition idea above does not quite work for
$n\geq 4$. Then, we introduce a simple `global accounting' idea that makes it
work.
\fi

Consider a query $Q$ on attributes $X,Y,W,T$. Let $\st C$ be the set of input
gap boxes. 
As before, let $P_X$ denote a balanced $X$-partition.
Since now we have an extra dimension, the natural idea is to use an extra balanced partition.
More specifically, we will use a balanced $Y$-partition, and break
the $x$-layer coverage problem into multiple subproblems, one for each interval
$y$ in the balanced $Y$-partition of $\st C[x]$.
}

Before defining a more refined notion of balanced map, we need a few
terminologies.
Let $P$ be a domain partition. Let $s$ be an arbitrary dyadic interval. Define
\begin{equation}\label{eqn:app:s'(P)}
 s'(P) = \begin{cases}
             s & \text{ if } s \in \prefixes{P}\\
             s' & \text{ if } s = s' s'' \text{ for some } s' \in P.
           \end{cases}
\end{equation}
and
\begin{equation}\label{eqn:app:s''(P)}
 s''(P) = \begin{cases}
             \lambda & \text{ if } s \in \prefixes{P}\\
             s'' & \text{ if } s = s' s'' \text{ for some } s' \in P.
           \end{cases}
\end{equation}

Let $P_X$ be a balanced $X$-partition of $\st C$, and
let $P_Y$ be a balanced $Y$-partition of $\st C$.
Define the {\em $X,Y$-load balancing map}
\[ \lbm_{X,Y} : \D(X) \times \D(Y) \times \D(W) \times \D(T) \to 
            \D(X') \times \D(Y') \times \D(W) \times \D(T) \times \D(Y'') \times \D(X''), \]
by setting
\[ \lbm_{X,Y}(\dbox{x, y, w, t}) := \dbox{x'(P_X), y'(P_Y), w, t, y''(P_Y), x''(P_X)}. \]
Let $\lbm_{X,Y}(\st C) = \bigl\{ \lbm_{X,Y}(\mv b) \suchthat \mv b \in \st C\bigr\}$. 
Our algorithm is to run \tetrispreloaded with $\lbm_{X,Y}(\st C)$ as the input.
Note that the problem has now been ``lifted up'' to become a $6$-dimensional
problem!

To analyze the algorithm, for each $x \in P_X$, $y \in P_Y$, define the
following sets:
\begin{eqnarray*}
\st F[x] &=& \left\{ \dbox{x,y',w,t,y'',x''} \in \lbm_{X,Y}(\st C) \suchthat x'' \neq \lambda\right\}\\
\st F[x, y] &=& \left\{ \dbox{x',y,w,t,y'',\lambda} \in \lbm_{X,Y}(\st C)
\suchthat x' \in \prefixes{x} \text{ and } y'' \neq \lambda \right\}\\
\st C[x, y] &=& \left\{ \dbox{x',y',w,t,\lambda,\lambda} \in \lbm_{X,Y}(\st C)
\suchthat x' \in \prefixes{x} \text{ and } y' \in \prefixes{y} \right\}
\end{eqnarray*}
Note that $x' \in \prefixes{x}$ as binary strings
means $x' \supseteq x$ when viewed as dyadic intervals. The reader should keep in
mind that we use these two notations interchangeably, for sometimes one notation
is more succinct than the other. Note also that,
\begin{eqnarray*}
|\st F[x]| &\leq& \sqrtC\\
|\st F[x, y]| &\leq& \sqrtC\\
|\st C[x, y]| &\leq& |\st C|.
\end{eqnarray*}
We will use these facts to bound the number of witnesses of various kinds.
To do so, we need a new notation. 

\bdefn[Conditional witness sets]\label{defn:app conditional witnesses}
Let $A$ be an arbitrary attribute, and $S$ be a set of attributes that does not
contain $A$. Then, we use $\st W(A \suchthat \text{some condition on $S$})$
to denote the set of $A$-witnesses $\mv w$ such that $\pi_S(\mv w)$ satisfies
the condition on $S$.
For example, let $X$, $Y$, and $W$ be three different attributes, then
\begin{eqnarray*}
\st W(X \suchthat Y=y) &=& 
      \left\{ \mv w \suchthat \mv w \text{ is an $X$-witness with } \pi_Y(\mv w)
      = y \right\}\\
\st W(X \suchthat Y \supseteq y \wedge W = w) &=& 
      \left\{ \mv w \suchthat \mv w \text{ is an $X$-witness with } \pi_Y(\mv w)
      \supseteq y \wedge \pi_W(\mv w) = w \right\}.
\end{eqnarray*}
When the condition part is empty, $\st W(A)$ denotes the total set of
$A$-witnesses.
\edefn

We are now ready to bound the number of $A$-witnesses for $A \in \{X'', Y'', W,
T, X', Y'\}$.
 
\bi
 \item The number of $X''$-witnesses is easy to bound. 
 By definition, an $X''$-witness $\mv w$ is a box that is involved in a
 \gbresolution on $X''$; this means $\pi_{X'}(\mv w) \in P_X$. 
 Moreover, $X''$-witnesses that do not share the $X'$-value will not resolve 
 with one another to form another $X''$-witness, because the strings $x\in P_X$ 
 are prefix-free. Consequently,
 \[ \abs{\st W(X'')} = \sum_{x\in P_X} \abs{\st W(X'' \suchthat X'=x)}. \]
 We bound $\abs{\st W(X'' \suchthat X'=x)}$ for each $x \in P_X$ by proving the
 following claim.

 \begin{claim}
\label{clm:app:claim1}
 For any $x\in P_X$, an $X''$-witness 
 $\mv w \in \st W(X'' \suchthat X'=x)$
 is supported on $\{Y',Y'',X''\}$, $\{W,X''\}$, and $\{T, X''\}$ by three boxes
 from $\st F[x]$. 
\end{claim}
 
 From the claim, the total number of $X''$-witnesses is at most
 \[ \sum_{x\in P_X} \tO(|\st F[x]|^3) = \tO( |P_X| \cdot \sqrtC^3) = \tO(|\st C|^2).
 \]
 Claim~\ref{clm:app:claim1} is easily shown by induction, whose proof we omit.

 \item Bounding the number of $Y''$-witnesses requires a new simple but
 subtle idea. We use the following obvious bound,
 \[ \abs{\st W(Y'')} \leq \sum_{x\in P_X} \abs{\st W(Y'' \suchthat X' \supseteq x)}, \]
 and then bound each term on the right hand side by proving the second claim.

 \begin{claim}
\label{clm:app:claim2} For any $x\in P_X$, a witness $\mv w \in \st W(Y'' \suchthat X'
 \supseteq x)$ must satisfy the following two conditions
  \bi \item[(i)] $\mv w$ either is supported on $\{Y', Y''\}$ and $\{W\}$ by two boxes 
 from $\st F[x]$, or is supported on $\{Y', Y'', W\}$ by a box from $\st F[x,\pi_{Y'}(\mv w)]$.
 (Notice that by definition of \lbm, for each $\mv w\in\st W(Y'')$, $\pi_{Y'}(\mv w)\in P_Y$.)
      \item[(ii)] $\mv w$ either is supported on $\{Y', Y''\}$ and $\{T\}$ by two boxes 
 from $\st F[x]$, or is supported on $\{Y', Y'', T\}$ by a box from $\st F[x,\pi_{Y'}(\mv w)]$.
  \ei
\end{claim}
 Assuming Claim~\ref{clm:app:claim2}, then for each witness
 $\mv w \in \st W(Y'' \suchthat X' \supseteq x)$,
 the box
 $\pi_{\{Y',W,T,Y''\}}(\mv w)$ either is (1) supported by three
 boxes from $\st F(x)$, or is (2) supported by two boxes from $\st F[x] \cup \st
 F[x,y]$ for some $y \in P_Y$.
 The number of witnesses satisfying (1) is at most
 $\tO(|\st F[x]|^3)$.
 The number of witnesses satisfying (2) is at most
 $\tO(1)\sum_{y \in P_Y} (|\st F[x]| + |\st F[x,y]|)^2$. Consequently,
 \[
    |\st W(Y'')| 
    \leq 
    \tO(1)\sum_{x\in P_X} \left(|\st F[x]|^3 + \sum_{y \in P_Y} (|\st F[x]| + |\st F[x,y]|)^2 \right)
    = \tO(|\st C|^2).
 \]
 We next prove Claim~\ref{clm:app:claim2} above by induction. Consider a witness 
 $\mv w \in \st W(Y'' \suchthat X' \supseteq x)$. In the base case, 
 $\mv w$ is {\em not} the result of a resolution on $Y''$. In this case, 
 either $\mv w \in \st W(X'' \suchthat X' = x)$, or 
 $\mv w \in \st F[x,\pi_{Y'}(\mv w)]$.
 In both cases, by noting Claim~\ref{clm:app:claim1}, it is easy to see that both (i) and (ii) hold.

 Next, suppose $\mv w = \Resolve(\mv w_1,\mv w_2)$, where $\mv w_1$ and $\mv
 w_2$ satisfy (i) and (ii).
 Because the two witnesses resolve on $Y''$, 
 $\pi_{Y''}(\mv w_1) \neq \lambda$
 and $\pi_{Y''}(\mv w_2) \neq \lambda$. Thus,
 $\pi_{Y'}(\mv w_1) \in P_Y$ and
 $\pi_{Y'}(\mv w_2) \in P_Y$. Because the strings in $P_Y$ are prefix-free, we 
 conclude that $\pi_{Y'}(\mv w_1) = \pi_{Y'}(\mv w_2) = \pi_{Y'}(\mv w)$.
 By definition of resolution, either 
 $\pi_{\{Y', Y'', W\}}(\mv w_1) \subseteq \pi_{\{Y', Y'', W\}}(\mv w)$ or 
 $\pi_{\{Y', Y'', W\}}(\mv w_2) \subseteq \pi_{\{Y', Y'', W\}}(\mv w)$, and
 the same holds for $\{Y', Y'', T\}$. The inductive step follows.

 \item The number of $T$-witnesses can be bounded by
 \[ |\st W(T)| \leq \sum_{x\in P_X} \sum_{y\in P_Y} 
    |\st W(T \suchthat X' \supseteq x, Y' \supseteq y)|.
 \]
 We bound the terms $|\st W(T \suchthat X' \supseteq x, Y' \supseteq y)|$ 
 individually using the following claim.

 \begin{claim}
\label{clm:app:claim3} For a fixed $x\in P_X$ and $y\in P_Y$, a witness $\mv w\in \st W(T \suchthat X' \supseteq x, Y' \supseteq y)$ must either be (1) supported on $\{W\}$ and $\{T\}$ by two boxes
 from $\st F[x] \cup \st F[x,y]$, or be (2) supported on $\{W,T\}$ by a box from
 $\st C[x,y]$. 
\end{claim}
 
 Assuming the claim holds, then the number of $T$-witnesses is bounded by
 \begin{eqnarray*}
  |\st W(T)| &\leq&
  \sum_{x\in P_X, y \in P_Y} 
    |\st W(T \suchthat X' \supseteq x, Y' \supseteq y)|\\
  &\leq&
  \sum_{x\in P_X, y \in P_Y} 
    \tO\left( (|\st F[x]| + |\st F[x,y]|)^2 + |\st C[x,y]| \right)\\
  &\leq&
  \sum_{x\in P_X, y \in P_Y} 
  \tO(|\sqrtC|^2 + |\st C|) = \tO(|\st C|^2). 
 \end{eqnarray*}

 We next prove Claim~\ref{clm:app:claim3}.  For a fixed $x \in P_X$ and $y \in P_Y$, 
 consider a witness $\mv w \in \st W(T \suchthat X' \supseteq x, Y' \supseteq y)$.
 Since our resolutions are ordered, $\pi_{X''}(\mv w) = \pi_{Y''}(\mv w) = \lambda$.
 In the base case, $\mv w$ could be either an input gap box or a result of 
 a \gbresolution on either $X''$ or $Y''$. If it is an input gap box, then 
 $\mv w\in\st C[x,y]$. If it is a result of \gbresolution on $X''$, then 
 $\pi_{X'}(\mv w)=x$ and Claim~\ref{clm:app:claim3} holds because Claim~\ref{clm:app:claim1} holds. 
 If it is a result of a \gbresolution on $Y''$, then $\pi_{Y'}(\mv w)=y$ and 
 the claim holds because Claim~\ref{clm:app:claim2} holds. The inductive step is very similar to 
 that of Claim~\ref{clm:app:claim2}. Suppose that Claim~\ref{clm:app:claim3} holds for $\mv w_1$ and $\mv w_2$. 
 Either $\pi_{\{W, T\}}(\mv w_1) \subseteq \pi_{\{W, T\}}(\mv w)$ or 
 $\pi_{\{W, T\}}(\mv w_2) \subseteq \pi_{\{W, T\}}(\mv w)$, and hence the claim 
 holds for $\mv w$.

 \item Finally, we bound the number of \gbresolutions on $W$, $X'$, and $Y'$. 
 Each witness $\mv w$ is supported by some box in $\st C$ on each one of the 
 attributes $W$, $X'$ and $Y'$. While $W$ can take 
 $\tO(\abs{\st C})$ values, each one of $X'$ and $Y'$ takes only 
 $\tO(\sqrtC)$ values. The total number of $W$, $Y'$ and $X'$-witnesses is 
 bounded by $\tO(\abs{\st C}^2)$, as desired.
\ei

\subsection{\tetrispreloaded with load balancing}
\label{sec offline as}

In this section, we generalize the ideas presented in the previous section to
analyze \tetrispreloadedas with the load balancing map idea incorporated.
Algorithm~\ref{alg:app offline as} has the key steps. 
The load balancing map is computed as a preprocessing step before calling \tetrispreloaded on the mapped boxes,
which have been ``lifted up'' to a higher-dimensional space.

For $i\in [n-2]$, let $P_i$ denote the balanced $A_i$-partition used in the
algorithm. Define the load-balancing map
\[ \lbm_{A_1,\dots,A_{n-2}} : 
      \D(A_1) \times \cdots \times \D(A_n) \to 
      \D(A'_1) \times \cdots \times \D(A'_{n-2}) \times \D(A_{n}) \times
      \D(A_{n-1}) \times \D(A''_{n-2}) \times \cdots \times \D(A''_1),
\]
by setting
\[ \lbm_{A_1,\dots,A_{n-2}}(\dbox{b_1,\dots,b_n}) := 
   \dbox{b'_1(P_1), \dots, b'_{n-2}(P_{n-2}), b_n, b_{n-1}, 
         b''_{n-2}(P_{n-2}), \dots, b''_1(P_1)}.
\]
(Recall the definitions of the functions $s'(P)$ and $s''(P)$ from
\eqref{eqn:app:s'(P)} and \eqref{eqn:app:s''(P)}.)
Also, naturally define
\[ \lbm_{A_1,\dots,A_{n-2}}(\st C) = 
    \left\{ \lbm_{A_1,\dots,A_{n-2}}(\mv b) \suchthat \mv b \in \st C
    \right\}.
\]

\begin{algorithm}[th]
\caption{\tetrispreloadedas, (i.e. \tetrispreloaded with load balancing)}
\label{alg:app offline as}
\begin{algorithmic}[1]
\Require{A set of boxes $\st C$ in the domain $\D(A_1) \times \cdots \times \D(A_n)$}
\Ensure{Output tuples for the \bcp\ on $\cert$}
\For{$i\gets 1$ to $n-2$} \Comment{$\tO(|\st C|)$-time}
   \State $P_i\wgets$ a balanced $A_i$-partition \label{ln:app Pi}
\EndFor
\State $\sigma\wgets (A_1', A_2', \ldots, A_{n-2}', A_{n}, A_{n-1}, A_{n-2}'', A_{n-3}'', \ldots, A_1'')$
\State $\st B \wgets \lbm_{A_1,\dots,A_{n-2}}(\st C)$ \Comment{$\tO(|\st C|)$-time}
\State \Return \Call{\tetrispreloaded}{$\st B$} \Comment{\RAO $=\sigma$}
\end{algorithmic}
\end{algorithm}

The main theorem of this section is the following.

\bthm[\tetrispreloadedas runs in time $\tO(\abs{\st C}^{n/2}+Z)$]
\label{thm:app offline as}
For any set of boxes $\st C$ in $n$ dimensions, 
\tetrispreloadedas solves \bcp\ on input $\st C$ in time $\tO(\abs{\st C}^{n/2}+Z)$,
where $Z$ is the output size.
\ethm

\bp 
For $i\in [n-2]$, let $P_i$ denote the balanced 
$A_i$-partition as computed in line~\ref{ln:app Pi} of the algorithm.
For $m=1$ to $n-2$, define the following sets for every tuple 
$(a_1, \ldots, a_m) \in P_1 \times \cdots \times P_m$:
\begin{eqnarray*}
\st F[a_1, \ldots, a_m] 
&:=& \bigl\{\mv b\in \lbm_{A_1,\dots,A_{n-2}}(\st C)
\wsuchthat \pi_{A'_i}(\mv b) \supseteq a_i \wedge
           \pi_{A''_i}(\mv b) = \lambda, \forall i\in [m-1] \wedge
           \pi_{A_m'}(\mv b) =  a_m \wedge
           \pi_{A_m''}(\mv b) \neq \lambda
            \bigr\}\\
\st C[a_1, \ldots, a_m]
&:=& \bigl\{\mv b\in \lbm_{A_1,\dots,A_{n-2}}(\st C)
 \wsuchthat \pi_{A'_i}(\mv b) \supseteq a_i \wedge
           \pi_{A''_i}(\mv b) = \lambda, \forall i\in [m] \bigr\}
\end{eqnarray*}
For notational convenience, also define the following set:
\[ \st G[a_1, \ldots, a_m]:= \bigcup_{i=1}^{m} \st F[a_1, \ldots, a_i]. \]
It should be noted that
$\lbm_{A_1,\dots,A_{n-2}}(\st C) =\bigcup_{(a_1, \ldots, a_{n-2}) \in P_1 \times \cdots \times P_{n-2}} \st G[a_1,\dots,a_{n-2}] \cup \st
C[a_1,\dots,a_{n-2}].
$
Recalling the notation defined in~\eqref{eqn:app C_subset}, and the definition of
balanced dimension partitions, we have, for every $m\in [n-2]$ and every tuple 
$(a_1, \ldots, a_m) \in P_1 \times \cdots \times P_m$,
\begin{eqnarray*}
\abs{\st F[a_1, \ldots, a_m]} &\leq & \abs{\st C_{\subset
a_m}(A_m)}\leq\sqrtC,\\
\abs{\st G[a_1, \ldots, a_m]} &\leq & m\sqrtC=\tO(\sqrtC),\\
\abs{\st C[a_1, \ldots, a_m]} &\leq & \abs{\st C}.
\end{eqnarray*}

In addition, for each $i\in[n]$, define $S_i$ as the following set of attributes:
\begin{equation}\label{eqn:app:S_i}
 S_i = \begin{cases}
             \{A_i', A_i''\} & \text{ if } i \in [n-2]\\
             \{A_i\} & \text{otherwise}.
           \end{cases}
\end{equation}

Recall that the \RAO used by \tetrispreloadedas is
$(A_1', A_2', \ldots, A_{n-2}', A_{n}, A_{n-1}, A_{n-2}'', A_{n-3}'', \ldots, A_1'')$. 
We will bound the total number of \gbresolutions and then use Theorem~\ref{thm X+Z}.
%%%%%
%\yell{[5.9 in FOCS submission].}
%%%%%
Note that when an output point is found and \tetris inserts it back to $\calA$
(Line~\ref{line:main:report-output}  of Algorithm~\ref{alg:main:tetris}), this output point is now in
the higher dimension of the $\lbm$ map.

Recall the definition of (conditional) witness sets 
(Definition~\ref{defn:app conditional witnesses}). We write
the total number of \gbresolutions as
\[ \sum_{m=1}^{n-2} |\st W(A''_m)| +
   |\st W(A_n)| + |\st W(A_{n-1})| +
   \sum_{m=1}^{n-2} |\st W(A'_m)|.
\]
In what follows, we bound each of the above terms.

\paragraph{Bounding $|\st W(A''_m)|$ for $m\in [n-2]$}
To bound $|\st W(A''_m)|$, we use the following inequality:
\[ |\st W(A''_m)| \leq \sum_{a_1 \in P_1} \cdots \sum_{a_{m-1} \in P_{m-1}}
   |\st W(A''_m \suchthat a_i \subseteq A'_i, \forall i \in [m-1])|.
\]
Notice that in order for a witness $\mv w$ to be involved in an
$A''_m$-resolution, it must be the case that 
$\pi_{A''_i}(\mv w) = \lambda$, for all $i \in [m-1]$.
In particular,
\[ \support(\mv w) \subseteq
    \{A'_1,\dots,A'_{n-2},A_n,A_{n-1},A''_{n-2},\dots,A''_m\}
    = \{A'_1,\dots,A'_{m-1}\} \cup S_m\cup \cdots \cup S_n.
\]
%From $\pi_{A''_i}(\mv w) = \lambda$, we can infer $\pi_{A'_i}(\mv w) \supseteq a_i$ for some
%$a_i \in P_i$.
It must also hold that $\pi_{A'_m}(\mv w) \in P_m$, which is a fact we
will use below.
The idea is to bound each term
$|\st W(A''_m \suchthat a_i \subseteq A'_i, \forall i \in [m-1])|$
individually, by proving the following claims by induction on $m=1, \dots, n-2$.

{\bf Claim $m$.} 
For every tuple $(a_1, \ldots, a_{m-1})\in P_1 \times\cdots\times P_{m-1}$ 
and for every $A_m''$-witness 
$$\mv w \in \st W(A''_m \suchthat a_i \subseteq A'_i, \forall i \in [m-1]),$$ 
the following property holds
\bi
\item For every $i\in\{m+1,\ldots, n\}$, either $\mv w$ is supported on 
    $S_m$ and $S_i$ by two boxes in $\st G[a_1, \ldots, a_{m-1}]$, or 
    $\mv w$ is supported on $S_m \cup  S_i$ by some box in 
    $\st F[a_1, \ldots, a_{m-1}, \pi_{A_m'}(\mv w)]$.
\ei

Assuming Claim $m$ is correct, the term
$|\st W(A''_m \suchthat a_i \subseteq A'_i, \forall i \in [m-1])|$ is bounded
as follows.
For each 
$\mv w \in \st W(A''_m \suchthat a_i \subseteq A'_i, \forall i \in [m-1])$, 
define the projection of $\mv w$ onto the
set $\support(\mv w) - \{A'_1,\dots,A'_{m-1}\}$:
\[ \overline{\mv w} = \pi_{\support(\mv w) - \{A'_1,\dots,A'_{m-1}\}}(\mv w). \]
Then,
\bi
\item Either $\overline{\mv w}$ is supported by an integral cover of size 
$\leq n-m+1$ using the box set $\st G(a_1, \ldots, a_{m-1})$.
\item Or $\overline{\mv w}$ is supported by an integral cover of size $\leq n-m$ 
using the box set $\st G(a_1, \ldots, a_{m-1}, \pi_{A_m'}(\mv w))$.
\ei
This fact helps bound the total number of such witnesses by
$$\tO(1)\sum_{a_1\in P_1}\ldots\sum_{a_{m-1}\in P_{m-1}}
\left[\left(\sqrtC\right)^{n-m+1}+\sum_{a_m\in P_m}\left(\sqrtC\right)^{n-m}\right]=\tO(\abs{\st C}^{n/2}).$$

We next prove Claim $m$ by induction on \gbresolutions and on $m$.

In the base case, $\mv w$ could be either an input gap box or a result of 
a \gbresolution on $A_l''$ for some $l<m$ (assuming $m>1$). 
If $\mv w$ is an input gap box, then 
$\mv w\in\st F[a_1,\ldots, a_{m-1}, \pi_{A_m'}(\mv w)]$ and the claim holds. 
If $m>1$ and $\mv w$ is a result of a \gbresolution on $A_l''$ for some $l<m$, 
then $A_l'(\mv w)=a_l$. By the induction hypothesis, Claim $l$ holds, which 
means $\mv w$ is supported on each 
one of the sets $S_{l+1}, \ldots, S_n$ by some box in 
$\st G(a_1, \ldots, a_l)\subseteq\st G(a_1, \ldots, a_{m-1})$, and hence 
Claim $m$ holds in this case.

For the inductive step, let's assume that Claim $m$ holds for two witnesses 
$\mv w_1$ and $\mv w_2$, and we will prove that it holds for 
$\mv w:=\Resolve(\mv w_1, \mv w_2)$. Because the two witnesses resolve on 
$A_m''$, we have $\pi_{A_m''}(\mv w_1) \neq \lambda$
and $\pi_{A_m''}(\mv w_2)\neq \lambda$.
This means $\pi_{A_m'}(\mv w_1) \in P_m$
and $\pi_{A_m'}(\mv w_2) \in P_m$.
But the strings in $P_m$ are prefix-free, leading to
$\pi_{A_m'}(\mv w_1)=\pi_{A_m'}(\mv w_2)=\pi_{A_m'}(\mv w)$. 
For every $i\in\{m+1, \ldots, n\}$, either 
$\pi_{S_i \cup  S_m}(\mv w_1) \subseteq \pi_{S_i \cup  S_m}(\mv w)$ 
or $\pi_{S_i \cup  S_m}(\mv w_2) \subseteq \pi_{S_i \cup  S_m}(\mv w)$ 
by definition of resolution. The inductive step follows.

\paragraph{Bounding $|\st W(A_{n-1})|$}
We bound the number of $A_{n-1}$-witnesses using the following inequality
\[ |\st W(A_{n-1})| \leq \sum_{a_1 \in P_1} \cdots \sum_{a_{n-2} \in P_{n-2}}
   |\st W(A_{n-1} \suchthat a_i \subseteq A'_i, \forall i \in [n-2])|.
\]
Notice that in order for a witness $\mv w$ to be involved in an
$A_{n-1}$-resolution, it must be the case that 
$\pi_{A''_i}(\mv w) = \lambda$, for all $i \in [n-2]$.
%From $\pi_{A''_i}(\mv w) = \lambda$, we can infer $\pi_{A'_i}(\mv w) \supseteq a_i$ for some
%$a_i \in P_i$.
So, we bound $|\st W(A_{n-1})|$ by bounding the terms
$|\st W(A_{n-1} \suchthat a_i \subseteq A'_i, \forall i \in [n-2])|$, assisted by
the following claim.

{\bf Claim $n-1$.}  For every tuple 
$(a_1, \ldots, a_{n-2})\in P_1\times\cdots\times P_{n-2}$ and for every 
$A_{n-1}$-witness 
\[ \mv w \in \st W(A_{n-1} \suchthat a_i \subseteq A'_i, \forall i \in [n-2]), \]
the following property holds:
\bi
\item Either $\mv w$ is supported on $\{A_{n}\}$ and $\{A_{n-1}\}$ by two boxes in 
$\st G[a_1, \ldots, a_{n-2}]$, or $\mv w$ is supported on $\{A_{n}, A_{n-1}\}$ 
by some box in $\st C[a_1, \ldots, a_{n-2}]$.
\ei
Assuming Claim $n-1$ is correct, the number of $A_{n-1}$-witnesses is bounded by
$$\tO(1)\sum_{a_1\in P_1}\ldots\sum_{a_{n-2}\in P_{n-2}}
\left[\sqrtC^2+\abs{\st C}\right]=\tO(\abs{\st C}^{n/2}).$$
Claim $n-1$ can be proved in a very similar way to Claim $m$ where $m=n-1$ and 
$\st C[a_1, \ldots, a_{n-2}]$ replaces $\st F[a_1,\ldots, a_{m-1},
\pi_{A_m'}(\mv w)]$.

\paragraph{Bounding $|\st W(A'_{m})|$ for $m\in[n-2]$ and $|\st W(A_{n})|$}

Finally, we will bound the total number of $A_1'$, \ldots, $A_{n-2}'$ and 
$A_{n}$-witnesses. Every witness $\mv w$ that is involved in a \gbresolution must 
be supported on each attribute by some box in $\st C$. Attribute $A_{n}$ can 
take up to $\tO(\abs{\st C})$ different values. However, from the definition of
balanced domain partition, each one of the attributes 
$\{A_1', \ldots, A_{n-2}'\}$ takes only $\tO(\sqrtC)$ values. Therefore, the 
total number of $A_1'$, \ldots, $A_{n-2}'$ and $A_{n}$-witnesses is 
bounded by $\tO(\abs{\st C}^{n/2})$, as desired.
\ep

\bcor[Subsumed by \cite{doi:10.1137/0220065,6686177}]
Given a set of boxes $\calB$ in $n$ dimensions, Klee's measure problem over the Boolean semiring can be solved in time $\tO(\abs{\calB}^{n/2})$.
\label{cor:klee-N^{n/2}}
\ecor

\subsection{\tetrisreloaded with load balancing}
\label{sec online as}

In this section, we generalize the ideas presented in the previous section to
analyze \tetrisreloaded with the load balancing map idea incorporated.
The main theorem we will show is the following.

\bthm[\tetrisreloadedas runs in time $\tO(|\boxcert|^{n/2}+Z)$]
\label{thm online as}
For any set of boxes $\st B$ in $n$ dimensions, 
\tetrisreloadedas solves \bcp\ on input $\st B$ in time $\tO(|\boxcert|^{n/2}+Z)$.
Here, $\boxcert$ is any optimal box certificate for the instance, 
and $Z$ is the output size.
\ethm

\bcor
\tetrisreloadedas evaluates any join query $Q$ over $n$ attributes in time 
$\tO(|\boxcert|^{n/2}+Z)$, where $\boxcert$ is an optimal box certificate 
for the join instance, and $Z$ is the output size.
\ecor

In \tetrisreloaded with load balancing (Algorithm~\ref{alg online as}), the dimension partitions are done with respect to the set $\st C$ of input gap boxes that have actually been loaded 
from $\st B$. Once new boxes are loaded from $\st B$ into $\st C$, the 
partitions are updated as follows: If an interval $x$ of any partition $P_i$ becomes \emph{heavy} (i.e., if $|\st C_{\subset x}(A_i)|$ exceeds $\sqrtC$), $x$ gets replaced by $x0$ and $x1$. Later on when new boxes that are not contained in the $x$-layer are loaded into $\st C$, $\abs{\st C}$ is going to increase while $|\st C_{\subset x}(A_i)|$ remains the same. However, even if $\sqrtC$ exceeds $|\st C_{\subset x}(A_i)|$ again, we do NOT return $x$ into $P_i$ instead of $x0$ and $x1$. In other words, we only allow partitions $P_i$ to ``expand''. Lemma~\ref{lmm:|P|=sqrtC-online} below shows that the partition sizes are still going to remain within $\tO(\sqrtC)$. The proof of Theorem~\ref{thm online as} is going to rely on the fact that partitions expand only.

Whenever any partition $P_i$ is updated, boxes in the partitioned space must be updated accordingly to reflect the new $A_i$-partition. In particular, whenever some $x$ in $P_i$ is replaced by $x0$ and $x1$, the knowledge base $\st A$ must be updated by invoking $\updatelbm_{A_i, x}(\st A)$ (in Line~\ref{ln as updatelbm} of Algorithm~\ref{alg online as}). $\updatelbm_{A_i, x}(\st A)$ works as follows:
\bi
\item For every $\mv a\in\st A$ such that $\pi_{A_i'}(\mv a)=x$ and $\pi_{A_i''}(\mv a)=by$ from some $b\in\{0, 1\}, y\in\{0,1\}^*$,\; $\pi_{A_i'}(\mv a)$ becomes $xb$ and $\pi_{A_i''}(\mv a)$ becomes y.
\ei
$\updatelbm_{A_i, x}(\st A)$ does not update any box $\mv a \in \st A$ for which $\pi_{A_i'}(\mv a)\neq x$ or $\pi_{A_i''}(\mv a)=\lambda$. Notice that because partitions expand only, no box $\mv a\in \st A$ can be updated more than $\tO(1)$ times.

\begin{algorithm}[th]
\caption{\tetrisreloadedas, (i.e. \tetrisreloaded with Load Balancing)}
\label{alg online as}
\begin{algorithmic}[1]
\Require{Oracle access to a set of boxes $\st B$ in the domain $\D(A_1) \times \cdots \times \D(A_n)$}
\Ensure{Output tuples for the \bcp\ on $\st B$}
\State $\st C\wgets\emptyset$
\Comment{$\st C$ is a set of boxes in the domain $\D(A_1) \times \cdots \times \D(A_n)$.}
\For{$i\gets 1$ to $n-2$}
	\State $P_i\wgets\{\lambda\}$
\EndFor
\State $\sigma\wgets (A_1', A_2', \ldots, A_{n-2}', A_{n}, A_{n-1}, A_{n-2}'', A_{n-3}'', \ldots, A_1'')$ 
\State $\st A\wgets\emptyset$
\Comment{$\st A$ is a set of boxes in the domain
$\D(A'_1) \times \cdots \times \D(A'_{n-2}) \times \D(A_{n}) \times
      \D(A_{n-1}) \times \D(A''_{n-2}) \times \cdots \times \D(A''_1)$.}
\State $\pair{v}{\mv w}\wgets\Call{\tetrisskeleton}{\UB}$ \Comment{\RAO $=\sigma$}
\While {$v$ = \false}
	\State $\calB' \wgets \{\mv b\in \st B \wsuchthat \lbm_{A_1, \ldots, A_{n-2}}(\mv b)\supseteq \mv w\}$
	\If {$\st B'=\emptyset$}
              \State {\bf Report} $\mv w$ as an output tuple
              \State $\calB'\wgets \{\mv w\}$
     \Else
	         \State $\st C \wgets \st C\cup \st B'$
	         \For{$i\gets 1$ to $n-2$}
	                \While {$\exists x\in P_i \wsuchthat |\st C_{\subset x}(A_i)|>\sqrtC$} \label{ln as while x heavy}
		         	        \State $P_i \wgets P_i -\{x\} \cup \{x0, x1\}$ \label{ln as s++}
		         	        \Comment{Update the partition $P_i$}
		         	        \State $\updatelbm_{A_i, x}(\st A)$ \label{ln as updatelbm}
		         	        \Comment{Re-balance $\st A$ with the new $P_i$}
		         	\EndWhile
	         \EndFor
	\EndIf
	\State $\st A \wgets \st A\cup \balance_{A_1, \ldots, A_{n-2}}(\st B')$
	\State $\pair{v}{\mv w}\wgets\Call{\tetrisskeleton}{\UB}$ \Comment{\RAO $=\sigma$}
\EndWhile
\end{algorithmic}
\end{algorithm}

\blmm
\label{lmm:|P|=sqrtC-online}
At the end of execution of \tetrisreloadedas, $\abs{P_i}=\tO(\sqrtC)$ for each $i\in [n-2]$.
\elmm
\bp
Consider an arbitrary but fixed $i\in[n-2]$.
When \tetrisreloadedas starts, $\abs{P_i}=1$.
Whenever $\abs{P_i}$ increases by one (in Line~\ref{ln as s++} of Algorithm~\ref{alg online as}), there must be some string $x\in\{0, 1\}^*$ such that $|\st C_{\subset x}(A_i)|>\sqrtC$. Consider an arbitrary but fixed integer $k>0$. From the moment $\abs{\st C}$ reaches $k$ until the moment it reaches $4k$, the increase in $\abs{P_i}$ can be bounded as follows: Whenever $\abs{P_i}$ increases by one, there must be some string $x$ that satisfies $|\st C_{\subset x}(A_i)|>\sqrt{k}$. However, for every box $\mv c\in \st C$ (whose size is $\leq 4k$), there are $\tO(1)$ strings $x$ that satisfy $\mv c\in \st C_{\subset x}(A_i)$. Hence, the increase in $\abs{P_i}$ is bounded by $\tO(\sqrt{k})$.

At the end of execution of \tetrisreloadedas, the total size of $P_i$ is bounded by the sum of $\tO(\sqrt{k})$ over all values $k\in\left\{4^0, 4^1, \ldots, 4^{1/2 \lceil\log_2{\abs{\st C}}\rceil}\right\}$:
$$\abs{P_i}\leq\tO(1)\left[2^0+2^1+\cdots+2^{1/2 \lceil\log_2{\abs{\st C}}\rceil}\right]
\leq\tO(1)2^{1/2 \lceil\log_2{\abs{\st C}}\rceil}\left[1+1/2+1/4+\cdots\right]
\leq\tO(\sqrtC).$$
\ep

\bp[Proof of Theorem~\ref{thm online as}]
Because \tetris performs ordered resolutions only, a box cannot be involved in a resolution on some attribute unless all subsequent attributes in the \RAO are $\lambda$'s. If some box $\mv w$ has a $\lambda$ in attribute $A_i''$, then $\pi_{A_i''}(\mv w)$ is going to remain $\lambda$ even after the $A_i$-partition $P_i$ is updated. This is because partitions expand only: any $x\in P_i$ can be replaced by $x0$ and $x1$, but $x0$ and $x1$ cannot be replaced back by $x$. As a result, resolutions remain ordered throughout the whole execution of \tetrisreloadedas.

During the execution of \tetrisreloadedas (Algorithm~\ref{alg online as}), a light interval $x\in P_i$ might get heavy (line~\ref{ln as while x heavy}), in which case $x$ has to be split (line~\ref{ln as s++}).
Given $i\in[n-2]$ and a binary string $x\in\{0, 1\}^*$, let $\st U_i(x)$ refer to the last value of $\st C_{\subset x}(A_i)$ while $x$ was still a light interval (i.e., the last $\st C_{\subset x}(A_i)$ while $|\st C_{\subset x}(A_i)|$ has not yet exceeded $\sqrtC$ at any previous step). By definition, $\abs{\st U_i(x)}\leq \sqrtC$. Define $\st V_i(x)$ as follows:
$$\st V_i(x):=\bigcup_{y\in\prefixes{x}} \st U_i(y)$$
By definition, $\abs{\st V_i(x)}=\tO(\sqrtC)$.

Given a variable $X$, we will use $\hat{X}$ to refer to the \emph{final value} of $X$ at the end of execution of \tetrisreloadedas. In particular, we will be using $\hat{P}_1, \ldots, \hat{P}_{n-2}$, and $\hat{\st C}$ to refer to the final values of $P_1, \ldots, P_{n-2}$, and $\st C$ respectively. Given a box $\mv w\in \st A$, invocations of $\updatelbm_{A_i, x}(\st A)$ might change $\pi_{A_i'}(\mv w)$ and $\pi_{A_i''}(\mv w)$. However, we will be using $\hat{\mv w}$ to refer to the final value of $\mv w$ (i.e. after the last invocation of $\updatelbm$). Similarly, we will be using $\hat{\lbm}_{A_1, \ldots, A_{n-2}}$ to refer to the final load balancing map (i.e. with respect to the final partitions $\hat{P}_1, \ldots, \hat{P}_{n-2}$):
\[ \hat{\lbm}_{A_1,\dots,A_{n-2}}(\dbox{b_1,\dots,b_n}) := 
   \dbox{b'_1(\hat{P}_1), \dots, b'_{n-2}(\hat{P}_{n-2}), b_n, b_{n-1}, 
         b''_{n-2}(\hat{P}_{n-2}), \dots, b''_1(\hat{P}_1)}.
\]
\[ \hat{\lbm}_{A_1,\dots,A_{n-2}}(\st C) = 
    \left\{ \hat{\lbm}_{A_1,\dots,A_{n-2}}(\mv b) \suchthat \mv b \in \st C
    \right\}.
\]

By Lemma~\ref{lmm:|P|=sqrtC-online}, $\abs{\hat{P}_i}=\tO(\sqrthatC)$ for all $i\in [n-2]$.
For $m=1$ to $n-2$, define the following sets for every tuple 
$(a_1, \ldots, a_m) \in \hat{P}_1 \times \cdots \times \hat{P}_m$:
\begin{eqnarray*}
\st F[a_1, \ldots, a_m] 
:=& \bigl\{\mv b\in \hat{\lbm}_{A_1, \ldots, A_{n-2}}\left(\st V_m(a_m)\right)
 &\wsuchthat \pi_{A'_i}(\mv b) \supseteq a_i \wedge
           \pi_{A''_i}(\mv b) = \lambda, \forall i\in [m-1] \bigr\}\\
\st C[a_1, \ldots, a_m]
:=& \bigl\{\mv b\in \hat{\lbm}_{A_1, \ldots, A_{n-2}}\left(\st C\right)
 &\wsuchthat \pi_{A'_i}(\mv b) \supseteq a_i \wedge
           \pi_{A''_i}(\mv b) = \lambda, \forall i\in [m] \bigr\}
\end{eqnarray*}
For notational convenience, also define the following set:
\[ \st G[a_1, \ldots, a_m]:= \bigcup_{i=1}^{m} \st F[a_1, \ldots, a_i]. \]
Notice that $\abs{\st F(a_1, \ldots, a_m)}\leq\abs{\st V_m(a_m)}=\tO(\sqrthatC)$,\;
$\abs{\st G(a_1, \ldots, a_m)}=\tO(\sqrthatC)$,\, and
$\abs{\st C(a_1, \ldots, a_m)}\leq\abs{\hat{\st C}}$.

In addition, for each $i\in[n]$, define $S_i$ as the following set of attributes:
\begin{equation*}
 S_i = \begin{cases}
             \{A_i', A_i''\} & \text{ if } i \in [n-2]\\
             \{A_i\} & \text{otherwise}.
           \end{cases}
\end{equation*}

We prove the following claim by induction on $m=1, \dots, n-2$.

{\bf Claim $m$.} 
For every tuple $(a_1, \ldots, a_{m-1})\in \hat{P}_1 \times\cdots\times \hat{P}_{m-1}$ 
and for every $A_m''$-witness $\mv w$ that satisfies
$$a_i\subseteq \pi_{A_i'}(\hat{\mv w}) \text{ for each } i\in[m-1], $$
the following property holds:
\bi
\item For every $i\in\{m+1,\ldots, n\}$, either $\hat{\mv w}$ is supported on 
    $S_m$ and $S_i$ by two boxes in $\st G[a_1, \ldots, a_{m-1}]$, or 
    $\hat{\mv w}$ is supported on $S_m \cup  S_i$ by some box in 
    $\st F[a_1, \ldots, a_{m-1}, a_m]$ for every $a_m\in\hat{P}_m$ that satisfies $a_m\subseteq \pi_{A_m'}(\hat{\mv w})$.
\ei
The rest of the proof is similar to that of Theorem~\ref{thm offline as}.
\ep

\bcor
\label{cor:klee-boxcert^{n/2}}
Given a set of boxes $\calB$ in $n$ dimensions, Klee's measure problem over the Boolean semiring can be solved in time $\tO(\abs{\boxcert}^{n/2})$, where $\boxcert$ is any optimal box certificate for $\calB$.
\ecor

By Definition~\ref{defn:bcp-boxcert}, we always have $\boxcert\subseteq\calB$ and hence $\abs{\boxcert}\leq\abs{\calB}$. The following proposition shows that $\abs{\boxcert}$ can be unboundedly smaller than $\abs{\calB}$.

\bprop\emph{(There is a class of input instances for which $\abs{\boxcert}$ is unboundedly smaller than $\abs{\calB}$).} For every integer $n\geq1$ and for every integer $b>0$, there exists a set $\calB$ of $n$-dimensional boxes such that $\abs{\calB}=b$ and $\abs{\boxcert}=1$.
\label{prop:|boxcert|<<|calB|}
\eprop
\bp
As long as $\calB$ contains $\UB$, $\boxcert=\{\UB\}$.
\ep

\subsection{Achieving a runtime of \texorpdfstring{$\tO(\abs{\boxcert}^{\frac{B+1}{2}}+Z)$}{O(|C|\{(B+1)/2\}+Z)} for queries with maximum block size \texorpdfstring{$B$}{B}}
\label{app:block-size}
Given a hypergraph, an \emph{articulation point} (alternatively, a \emph{cut vertex}) is a vertex whose removal increases the number of connected components. A \emph{biconnected component} is a component that does not contain any articulation points. A maximal biconnected component is called a \emph{block}. The blocks of any connected hypergraph can be arranged into a tree called the \emph{block tree}.

Let $\st B$ be a \bcp instance whose supporting hypergraph $\calH(\st B)$ has a maximum block size of $B$. We define a load balancing map described by a GAO $\sigma$ for $\st B$ as follows. Initialize $\sigma$ to be empty (i.e., $\sigma\gets()$). Construct a block tree for $\calH(\st B)$ and repeat the following two steps until the block tree is empty.

\bi
\item Let $L$ be a leaf block in the block tree. $L$ can maximally share one attribute with other blocks. Let $A_1, \ldots, A_k$ be the attributes that belong to $L$ and only $L$.
For $i\in [k-2]$, let $P_i$ denote a balanced $A_i$-partition and let $A_i', A_i''$ be the two attributes that replace $A_i$ in the load balancing map. Update $\sigma$ as follows:
$$\sigma\gets (A_1', A_2', \ldots, A_{k-2}', A_{k}, A_{k-1}, A_{k-2}'', A_{k-3}'', \ldots, A_1'', \sigma)$$

\item Remove $L$ from the block tree.
\ei

Notice that if $L$ above shares one attribute with other blocks, then $k\leq B-1$. Every \gbresolvent $\mv w$ is supported on the shared attribute by some input gab box, while the projection of $\mv w$ on the $k$ non-shared attributes can have $\tO(\abs{\boxcert}^{k/2})$ combinations (thanks to load balancing). The total runtime would be $\tO(\abs{\boxcert}^{\frac{k}{2}+1}+Z)$. Alternatively if $L$ does not share any attribute (i.e.\ $L$ is the root of the block tree), then $k\leq B$ and load balancing takes $\tO(\abs{\boxcert}^{\frac{k}{2}}+Z)$. The next corollary follows.

\bcor
\label{cor:block-size}
Let $\st B$ be a \bcp\ instance, $B$ the maximum block size of $\calH(\st B)$, $\boxcert$ be an optimal box certificate for $\st B$, and $Z$ the output size. If \tetrisreloadedas uses the load balancing map and the attribute order $\sigma$ described above, it runs in time $\tO(\abs{\boxcert}^{\frac{B+1}{2}}+Z)$.
\ecor
The above result can obviously be specialized to join queries.

\subsection{An example of an unbalanced certificate}
\label{app:unbalanced}
We present an example of an unbalanced certificate $\st C$ for $n=3$. (For the definition of \emph{balanced}, check Section~\ref{sec:balanced-case}.) In fact, the boxes of $\st C$ will correspond to gap boxes for the triangle query $R(A,B)\Join S(B,C)\Join T(A,C)$. Fix a parameter $N$. We will use $O$ (and $E$) to denote the set of odd (even resp.) values in $[N]$. Let $\st C$ be the union of the following three sets:
\[\st C_1=\left\{ \dbox{o_1, o_2, \lambda} \;|\; o_1, o_2\in O\right\}\cup\left\{\dbox{e_1, e_2, \lambda} \;|\; e_1, e_2\in E\right\},\]
\[\st C_2=\left\{ \dbox{\lambda, o_1, o_2} \;|\; o_1, o_2\in O\right\}\cup\left\{\dbox{\lambda, e_1, e_2} \;|\; e_1, e_2\in E\right\},\]
\[\st C_3=\left\{ \dbox{o_1, \lambda, o_2} \;|\; o_1, o_2\in O\right\}\cup\left\{\dbox{e_1, \lambda, e_2} \;|\; e_1, e_2\in E\right\}.\]

It is not hard to see that the boxes of $\st C=\st C_1 \cup \st C_2 \cup \st C_3$ cover the entire space, and that every box of $\st C$ is necessary to cover some point in the space. 

We cannot partition the domain of $A$ since the $\Theta(\abs{\st C})$ boxes of $\st C_2$ span that entire domain. At the same time, we cannot take the whole domain of $A$ as a single part because that part would contain $\Theta(\abs{\st C})$ boxes. The example is symmetric so that the same problem occurs when we consider $B$ or $C$.

\section{Omitted details from Section~\ref{SEC:LOWERBOUNDS}}
\label{sec:tightness}

In this section, we present the proofs of the lower bounds that were claimed without proof in Section~\ref{SEC:LOWERBOUNDS}.

%\yell{General note: In many places below $\UB$ is used as the universal box but for {\em different} domains. When using $\UB$ below at least the size should be specified somehow. Maybe $\underbrace{\UB}_{n}$? 
%I did not make this change since I did not want to make mistake on the dimensions etc.: please take care of this. --Atri}

%\ar{Another general comment. When you have a display eqaution the usual punctuation rules apply. So e.g. if the displayed equation is the last part of a senetence, then it should end with a period etc. I tried to change this in display equations where the punctuations were missing but I probably missed some. So please double-check.}

\subsection{The \texorpdfstring{$\Omega(N^{n/2}+Z)$}{Omega(N\{n/2\}+Z)} lowerbound for \nocache} %Tree-Ordered-Geometric-Resolution}

We show that \nocache\ is not powerful enough to recover Yannakakis' result on acyclic queries (let alone Theorem~\ref{thm:main fhtw}).

\bthm
[There is a class of acyclic input instances with $w=1$ on which every \nocache\ algorithm runs in time $\Omega(N^{n/2}+Z)$]
\label{thm:N^{n/2}-lowerbound}
For every integer $n>1$ and for every integer $c>0$, there exists an acyclic join query $Q$ with $n$ attributes, size $N=\Theta(c)$ and of treewidth of 1, such that $Q$ satisfies the following conditions:
\bi
\item The output of $Q$ is empty.
\item For every GYO elimination order $\sigma$ of $Q$ that induces treewidth of 1, every \nocache\ proof (with ordering $\sigma$) of $\UB$ consists of $\Omega(N^{n/2})$ resolutions.
\ei
\ethm

%\ar{What about non-GYO order? I can see that would be worse than a GYO but that has to be spelled out.}

\bp
Let $\vars(Q)$ be $\{A_1, \ldots, A_n\}$. Let $\atoms(Q)$ be $\{R_1, \ldots, R_{n+1}\}$. Let $R_1$ be a unary relation over $A_1$ that contains all odd values, and $R_{n+1}$ be a unary relation over $A_n$ that contains all even values. For each $i\in\{2, \ldots, n\}$, let $R_i$ be a binary relation between $A_{i-1}$ and $A_{i}$ that contains all pairs of equal parity. More precisely, the corresponding gap boxes are:
$$\calB(R_1):=\bigl\{\mv c\wsuchthat \pi_{A_1}(\mv c)\in\{0,1\}^{d-1}0 \text{ and } \pi_{A_k}(\mv c)=\lambda \text{ for all }k\in[n]-\{1\}\bigr\},$$
and
$$\calB(R_{n+1}):=\bigl\{\mv c\wsuchthat \pi_{A_n}(\mv c)\in\{0,1\}^{d-1}1 \text{ and } \pi_{A_k}(\mv c)=\lambda \text{ for all }k\in[n-1]\bigr\}.$$
Finally, for $i\in\{2, \ldots, n\}$:
$$\calB(R_i):=\bigl\{\mv c\suchthat \exists b\in\{0,1\},\; \bar{b}=1-b \left[\pi_{A_{i}}(\mv c)\in\{0,1\}^{d-1}b\text{ and }\pi_{A_{i-1}}(\mv c)\in\{0,1\}^{d-1}\bar{b} \text{ and } \pi_{A_k}(\mv c)=\lambda \text{ for all }k\in[n]-\{i-1, i\}\right]\bigr\}.$$
There are two possible GYO elimination orders: $(A_1, A_2, \ldots, A_n)$ and $(A_n, A_{n-1}, \ldots, A_1)$. Without loss of generality, let's consider the first.

We make the following claims.

\begin{claim}
\label{clm:app:I1:Claim 1} $\st C:=\bigcup_{i\in[n+1]}\calB(R_i)$ is a minimal set of boxes that covers $\UB$.
\end{claim}
\bp
$\st C$ covers $\UB$ because $R_2, \ldots, R_n$ constrain $A_1, \ldots, A_n$ to have the same parity while $R_1$ and $R_{n+1}$ constrain $A_1$ and $A_n$ to have opposite parities. Moreover, $\st C$ is a \emph{minimal} set of boxes that covers $\UB$. This is because if any box of $\bigcup_{i\in\{2, \ldots, n\}}\calB(R_i)$ is dropped out of $\st C$, then $A_1$ and $A_n$ are no longer constrained to have the same parity. Alternatively, if any box of $\calB(R_1)$ or $\calB(R_{n+1})$ is dropped, then $A_1$ and $A_n$ are no longer constrained to have opposite parities.
\ep

\begin{claim}
\label{clm:app:I1:Claim 2} 
In every \nocache\ proof of $\UB$ whose leaves are from $\st C$, every box of $\calB(R_{n+1})$ must appear $\Omega(N^{(n-1)/2})$ times.
\end{claim}

Combined with the fact that $\abs{\calB(R_{n+1})}=\Theta(\sqrt{N})$ (we pick $d=\Theta(\log{c})$), proving Claim~\ref{clm:app:I1:Claim 2} above proves the theorem.

To complete the proof, we prove Claim~\ref{clm:app:I1:Claim 2} by induction on $n$. The base case is when $n=2$. By performing resolutions on $A_2$, we infer the following set of boxes whose $A_2$ components are $\lambda$'s:
$$\st D:=\bigl\{\dbox{x1, \lambda}\wsuchthat x \in\{0, 1\}^{d-1}\bigr\}.$$
Note that $\abs{\st D}=\Theta(\sqrt{N})$. Each box of $\calB(R_3)$ is necessary to infer every box of $\st D$. (Further, note that to generate $\dbox{\lambda,\lambda}$ from $\st B(R_1)$ we precisely need the boxes in $\st D$.)

For the inductive step, we will prove Claim~\ref{clm:app:I1:Claim 2} for $n$ assuming it holds for $n-1$. By performing resolutions on $A_n$, we can infer the following set of boxes whose $A_n$ components are $\lambda$'s:
$$\st D:=\bigl\{\mv c\wsuchthat \pi_{A_{n-1}}(\mv c)\in\{0,1\}^{d-1}1 \text{ and } \pi_{A_k}(\mv c)=\lambda \text{ for all }k\in[n]-\{n-1\}\bigr\}.$$
Note that $\abs{\st D}=\Theta(\sqrt{N})$. Because of the resolution order, a box cannot be resolved on any attribute $A_1, \ldots, A_{n-1}$ unless its $A_n$ component is $\lambda$. Let $\st C':=\bigcup_{i\in[n-1]}\calB(R_i)\bigcup \st D$ be the set of given/inferred boxes whose $A_n$ components are $\lambda$'s. Because Claim~\ref{clm:app:I1:Claim 1} holds for $n-1$, $\st C'$ is a minimal set of boxes that covers $\UB$.
%\ar{Argue why any proof has to compute $\st C'$ first.}
Because Claim~\ref{clm:app:I1:Claim 2} holds for $n-1$, every box of $\st D$ must appear $\Omega(N^{(n-2)/2})$ times in any proof of $\UB$ from $\st C'$. Every box of $\calB(R_{n+1})$ must appear in the proof of every box of $\st D$.
\ep

\subsection{A useful sequence of strings}

Before we move ahead we first define a useful collection of sequences of strings.
\bdefn[$B^{(n)}_i$]
Given an integer $n\geq1$, let $B^{(n)}$ be a sequence of $n$ binary strings $B^{(n)}_1, \ldots, B^{(n)}_n$ that are defined as follows:
$$\text{for }i\in [n-1]:\quad B^{(n)}_i=1^{i-1}0,$$
and
$$B^{(n)}_n=1^{n-1}.$$
\edefn
Based on the above definition, $B^{(1)}=(\lambda)$, $B^{(2)}=(0, 1)$, $B^{(3)}=(0, 10, 11)$, $B^{(4)}=(0, 10, 110, 111)$, etc. In other words, $B^{(n)}$ is a binary representation of $n$ values.

We will be interested in these sequences since they have the following nice property:
\blmm
\label{lem:app:string-seq}
For any $n\ge 1$, $B^{(n)}$ partitions the space $\{0,1\}^n$. Further, there is a unique sequence of resolutions amongst strings/intervals in $B^{(n)}$ that leads to $\lambda$.
\elmm
\bp
The claim on $B^{(n)}$ being a partition just follows from its definition (as well as our string encoding of dyadic intervals on $\{0,1\}^n$).

For the claim on the sequence note that for every $i\in [n]$, define $R^{(n)}_i=1^{i-1}$. Note that $B_n^{(n)}(=R^{(n)}_n)$ and $B^{(n)}_{n-1}$ resolve to get $R^{(n)}_{n-1}$. More generally, $R^{(n)}_{i}$ and $B^{(n)}_{i-1}$ resolve to obtain $R^{(n)}_{i-1}$. Note that this sequence results in $\lambda$ since $R^{(n)}_1=\lambda$. Further, it can be checked that no $B^{(n)}_i$ for $i\in [n-2]$ can be resolved with $B^{(n)}_j$ for $j\neq i$. This implies that the sequence above is the unique way to obtain $\lambda$ (this can be formally proved e.g. by induction on $n$).
\ep

%\ar{Added this section. Pls make sure I did not mess up somewhere.}

\subsection{The \texorpdfstring{$\Omega(\abs{\st C}^{n-1}+Z)$}{Omega(|C|\{n-1\}+Z)} lowerbound for \ordered}

Next, we show that the upper bound in Theorem~\ref{thm:C^{n-1}+Z-upperbound} is tight (even if one used an arbitrary \ordered\ algorithm).
\bthm
[There is a class of input instances on which every \ordered\ algorithm runs in time $\Omega(\abs{\boxcert}^{n-1}+Z)$.]
\label{thm:C^{n-1}-lowerbound}
For every integer $n>1$ and for every integer $c>0$, there exists a set $\st C$ of $n$-dimensional $\Theta(c)$ boxes satisfying the following conditions:
\bi
\item For every box $\mv c\in \st C$, $\abs{\support(\mv c)}\leq3$.
\item $\st C$ is a minimal set of boxes that covers $\UB$.
\footnote{This means that (i) the union of all boxes in $\st C$ covers $\UB$,
and (ii) for any box $\mv c \in \st C$, $\st C-\{\mv c\}$ does not cover $\UB$.}
\item For every fixed order of the $n$ dimensions, every \ordered\ proof of $\UB$ consists of $\Omega(\abs{\st C}^{n-1})$ resolutions.
\ei
Hence, there exists a join query $Q$ whose relations have arity $\leq 3$ and whose set of gap boxes $\st B(Q)$ is $\st C$ and whose output is empty, such that every \ordered\ algorithm runs in time $\Omega(\abs{\st C}^{n-1})$ on $Q$.
\ethm
\bp
Let $d':=\lceil\log_2{c}\rceil$. For each $i\in[n]$, define
$$\st C_i:=\bigl\{\mv c \wsuchthat \exists j\in[n]-\{i\}\;
\bigl[\pi_{A_i}(\mv c)=B^{(n-1)}_j \text{ and } \pi_{A_j}(\mv c)\in\{0,1\}^{d'} \text{ and } \pi_{A_k}(\mv c)=\lambda \text{ for all } k \in [n]-\{i, j\}
\bigr]\bigr\}$$
$$\st C'_i:=\bigl\{\dbox{B^{(n)}_i x_1, x_2, x_3, \ldots, x_n}\wsuchthat \dbox{x_1, x_2, \ldots, x_n}\in \st C_i\bigr\}$$
$$\st C:=\bigcup_{i\in[n]}\st C'_i.$$
\footnote{Above, $B^{(n)}_i x_1$ is a \emph{single} string which is the concatenation of $B^{(n)}_i$ and $x_1$.}
By definition, $\abs{\st C}=\Theta(c)$.

We prove the following claim.

\begin{claim}
\label{clm:app:H3:Claim 1} For each $i\in[n]$, $\st C_i$ is a minimal set of boxes that covers $\UB$. Moreover, for every attribute order that ends with $A_i$, every \ordered\ proof of $\UB$ from $\st C_i$ consists of $\Omega(\abs{\st C_i}^{n-1})$ resolutions.
\end{claim}
\bp[Proof]
From Lemma~\ref{lem:app:string-seq}, $\st C_i$ partitions the universal box $\UB$, which makes $\st C_i$ a minimal set of boxes that covers $\UB$.

Starting from $\st C_i$, by performing all possible resolutions on $A_i$, we can infer the following set of boxes whose $A_i$ components are $\lambda$'s
$$\st D_i:=\bigl\{\mv c \wsuchthat \pi_{A_i}(\mv c)=\lambda \text{ and } \pi_{A_j}(\mv c)\in\{0, 1\}^{d'} \text{ for all }j\in[n]-\{i\}\bigr\}.$$
\footnote{Notice that the boxes of $\st D_i$ are still in the same $n$-dimensional space as those of $\st C_i$. The $A_i$ components being $\lambda$'s does not make the boxes of $\st D_i$ $(n-1)$-dimensional; it just makes them span the entire $A_i$ dimension.}
$\st D_i$ is a minimal set of boxes that covers $\UB$. This is because every point in the space is covered by exactly one box of $\st D_i$.
%\ar{This is one place where using $\UB$ without its dimension specified makes things very confusing becase you seem to be claiming that $\st D_i$ and $ st C_i$ cover the same $\UB$.}
$\abs{\st D_i}=\Omega(\abs{\st C_i}^{n-1})$, which proves the second part of the claim.
\ep

By Claim~\ref{clm:app:H3:Claim 1}, for each $i\in [n]$, $\st C'_i$ is a minimal set of boxes that covers $\dbox{B^{(n)}_i, \lambda, \ldots, \lambda}$ (and covers nothing outside $\dbox{B^{(n)}_i, \lambda, \ldots, \lambda}$). Hence, by Lemma~\ref{lem:app:string-seq}, $\st C$ is a minimal set of boxes that covers $\UB$. Let $A_l$ be the last attribute in the arbitrarily-chosen attribute order.
Let's consider the case when $l\neq 1$. (The case of $l=1$ is going to be very similar.)
For any $i\neq j\in[n]$, no box from $\st C_i'$ can be resolved on $A_l$ with any box from $\st C_j'$.
Performing resolutions on $A_l$, we will infer the following boxes from $\st C_l'$
$$\st D'_l:=\bigl\{\dbox{B^{(n)}_l x_1, x_2, x_3, \ldots, x_n}\wsuchthat \dbox{x_1, x_2, \ldots, x_n}\in \st D_l\bigr\}.$$
Following the proof of Claim~\ref{clm:app:H3:Claim 1}, $\st D_l'$ is a minimal set of boxes that covers $\dbox{B^{(n)}_l, \lambda, \ldots, \lambda}$. No box from $\st C-\st C_l'$ overlaps with $\dbox{B^{(n)}_l, \lambda, \ldots, \lambda}$. Hence, each one of the $\Omega(\abs{\st C}^{n-1})$ boxes of $\st D_l'$ is necessary to cover $\UB$.
\ep

\subsection{The \texorpdfstring{$\Omega(\abs{\st C}^{w+1}+Z)$}{Omega(|C|\{w+1\}+Z)} lowerbound for \ordered}

We begin with a technical lemma.
\blmm
\label{lmm:n/2-construction}
Given integers $n>1$ and $d'\geq0$, let $\st C$ be a set that consists of every $n$-dimensional box whose support has size 2 and whose two non-$\lambda$ components share a common suffix after the first $d'$ bits, and this common suffix belongs to $B^{(n-1)}$:
$$\st C:=\bigl\{\mv c \wsuchthat \exists i\neq j\in [n]\; \exists b \in B^{(n-1)}
\bigl[\pi_{A_i}(\mv c), \pi_{A_j}(\mv c) \in\{0, 1\}^{d'}b \text{ and }
\pi_{A_k}(\mv c)=\lambda \textnormal{ for all $k\in [n]-\{i, j\}$}\bigr]\bigr\}.$$
Then, 
\bi
\item $\st C$ is a minimal set of boxes that covers $\UB$.
\item Every \ordered\ proof of $\UB$ from $\st C$ consists of $\Omega(\abs{\st C}^{n/2})$ resolutions.
\ei
\elmm
\bp
First, we prove that $\st C$ is a minimal set of boxes that covers $\UB$.
Consider an arbitrary but fixed box $\mv b=\dbox{x_1, \ldots, x_n}$ where $x_i\in\{0, 1\}^{d'}$ for all $i\in[n]$. Consider the following set of boxes that forms a partition of $\mv b$ (due to Lemma~\ref{lem:app:string-seq}):
$$\st P_{\mv b}=\bigl\{\dbox{x_1 b_1, \ldots, x_n b_n}
\wsuchthat b_i \in B^{(n-1)} \text{ for all } i\in [n]\bigr\}.$$
Let $\st C_{\mv b}$ denote the subset of $\st C$ whose boxes overlap with $\mv b$:
$$\st C_{\mv b}:=\bigl\{\mv c \wsuchthat \exists i\neq j\in [n]\; \exists b \in B^{(n-1)}
\bigl[\pi_{A_i}(\mv c)=x_i b \text{ and } \pi_{A_j}(\mv c)=x_j b \text{ and }
\pi_{A_k}(\mv c)=\lambda \textnormal{ for all $k\in [n]-\{i, j\}$}\bigr]\bigr\}.$$
We have $n$ variables $b_1, \ldots, b_n$ (in the definition of $\st P_{\mv b}$). Each variable can take $n-1$ possible values $B^{(n-1)}$. Each box in $\st C_{\mv b}$ corresponds to a constraint preventing two of the variables to have the same value. (In particular, a tuple is in a box of $\st C_{\mv b}$ if and only if it satisfies the corresponding constraint.) In every assignment of $n$ variables taking $n-1$ values, at least two variables must have the same value. Hence,  by the definition of the constraint corresponding to each box in $\st C_{\mv b}$, the union of all boxes in $\st C_{\mv b}$ covers $\mv b$. Moreover, if any two of the variables are allowed to have the same value, then there is a feasible assignment of the variables. Hence, for any box $\mv c\in \st C_{\mv b}$, $\st C_{\mv b}$ does not cover $\mv b$ without $\mv c$.
%\ar{It might make it a bit easier to follow if you start off by reminder the reader about the equivalence between constraints and boxes covering another box. State this generally then instantiate that general connection with the stuff above.}

Now, we prove that $\Omega(\abs{\st C}^{n/2})$ ordered resolutions are required to infer $\UB$. The $n$ attributes of $\st C$ are symmetric. WLOG let the attribute order be $(A_1, \ldots, A_n)$.
After performing ordered geometric resolutions on $A_n$ eliminating all but the first $d'$ bits, we can infer the following set of boxes:
$$\st D_n:=\bigl\{\mv c \wsuchthat \pi_{A_n}(\mv c)\in\{0, 1\}^{d'} \text{ and } \bigl(\forall i_1\neq i_2\in[n-1]\bigr) \bigl(\exists b_1\neq b_2 \in B^{(n-1)}\bigr)\bigl[\pi_{A_{i_1}}(\mv c)\in\{0, 1\}^{d'}b_1 \;\;\wedge\;\; \pi_{A_{i_2}}(\mv c)\in\{0, 1\}^{d'}b_2\bigr]\bigr\}.$$
Notice that $\abs{\st D_n}=\Theta(\abs{\st C}^{n/2})$. Let $\st C_n$ be the subset of $\st C$ that contains all boxes whose $A_n$-components have length $\leq d'$:
$$\st C_n:=\bigl\{\mv c \wsuchthat \exists i \in [n-1]\; \exists b \in B^{(n-1)}
\bigl[\pi_{A_n}(\mv c), \pi_{A_i}(\mv c) \in\{0, 1\}^{d'}b \text{ and }
\pi_{A_k}(\mv c)=\lambda \textnormal{ for all $k\in [n-1]-\{i\}$}\bigr]\bigr\}.$$
$\st D_n \cup (\st C-\st C_n)$ is a minimal set of boxes that covers $\UB$. This is because boxes of $\st C-\st C_n$ correspond to constraints preventing any two of the first $n-1$ variables to have the same value, while boxes of $\st D_n$ correspond to constraints preventing the first $n-1$ variables to have $n-1$ different values. (The $n$-th variable is not constrained.) The union of both constraint types is a minimal set of constraints that rules out all possible assignments of the first $n-1$ variables.
%\ar{The above was another place where I got confused because $\UB$ here is for $n-1$ dimensions.}
\ep

Next, we show that Theorem~\ref{thm:main:C^{w+1}+Z} is tight.
\bthm[There is a class of input instances on which every \ordered\ algorithm runs in time $\Omega(|\boxcert|^{w+1} + Z)$]
\label{thm:C^{w+1}-lowerbound}
For every integer $w>1$ and for every integer $c>0$, there is a graph $G$ whose 
treewidth is $w$, and there is a set of boxes $\st C$ whose size 
is $\Theta(c)$, such that the following properties hold:
\bi
\item For every box $\mv c\in \st C$,\; $\support(\mv c)$ corresponds to some edge in $G$.
\item $\st C$ is a minimal set of boxes that covers $\UB$.
\item For every elimination order of $G$ that induces a treewidth of $w$, every \ordered\ proof of $\UB$ consists of $\Omega(\abs{\st C}^{w+1})$ resolutions.
\ei
Hence, there is a join query $Q$ whose (hyper)graph is $G$ and whose set of gap boxes $\st B(Q)$ is $\st C$ and whose output is empty, such that every \ordered\ algorithm runs in time $\Omega(|\boxcert|^{w+1})$ on $Q$.
\ethm
\bp
The vertices of $G$ are divided into two subsets: \emph{primary} and \emph{secondary}.
$G$ has $w+1$ primary vertices $A_1, \ldots, A_{w+1}$. For every $i<j\in[w+1]$ and every $k\in[w]$, $G$ has one secondary vertex $B_{i, j, k}$. The edges of $G$ are also divided into primary and secondary. For every $i<j\in[w+1]$, $G$ has one primary edge $\{A_i, A_j\}$. For every $i<j\in[w+1]$ and every $k\in[w]$, $G$ has two secondary edges $\{A_i, B_{i, j, k}\}$ and $\{A_j, B_{i, j, k}\}$.

Let $d':=\lceil\log_2{c}\rceil$. For every $i<j\in[w+1]$ and every $k\in[w]$, define $\st C_{i, j, k}$ as the union of the following two sets
\bi
\item The set of every box whose support is $\{A_i, B_{i, j, k}\}$, whose $B_{i, j, k}$ component is $0$, and whose $A_i$ component ends with the suffix $B^{(w)}_k$ after the first $d'$ bits.
\item The set of every box whose support is $\{A_j, B_{i, j, k}\}$, whose $B_{i, j, k}$ component is $1$, and whose $A_j$ component ends with the suffix $B^{(w)}_k$ after the first $d'$ bits.
\ei
\begin{align*}
\st C_{i, j, k}:=
&\bigl\{\mv c \wsuchthat \pi_{A_i}(\mv c)\in \{0, 1\}^{d'}B^{(w)}_k\text{ and } \pi_{B_{i, j, k}}(\mv c)=0 \text{ and } \pi_X(\mv c)=\lambda\text{ for all }X\in\vars(Q)-\{A_i, B_{i, j, k}\}\bigr\}\\
\bigcup
&\bigl\{\mv c \wsuchthat \pi_{A_j}(\mv c)\in \{0, 1\}^{d'}B^{(w)}_k\text{ and } \pi_{B_{i, j, k}}(\mv c)=1 \text{ and } \pi_X(\mv c)=\lambda\text{ for all }X\in\vars(Q)-\{A_j, B_{i, j, k}\}\bigr\}.
\end{align*}
%\ar{Please state in English what the above means.}
Define $\st C$ as
$$\st C:=\bigcup_{i<j\in[w+1]}\;\bigcup_{k\in[w]}\st C_{i, j, k}.$$
By definition, $\abs{\st C}=\Theta(c)$.

By performing resolutions on some secondary attribute $B_{i, j, k}$, we can infer the following set of boxes:
$$\st D_{i, j, k}:=
\bigl\{\mv c \wsuchthat \pi_{A_i}(\mv c)\in \{0, 1\}^{d'}B^{(w)}_k\text{ and } \pi_{A_j}(\mv c)\in \{0, 1\}^{d'}B^{(w)}_k \text{ and } \pi_X(\mv c)=\lambda\text{ for all }X\in\vars(Q)-\{A_i, A_j\}\bigr\}.$$
Notice that every box of $\st C_{i, j, k}$ is necessary to infer some box of $\st D_{i, j, k}$. In particular, if one box of $\st C_{i, j, k}$ is missing, then we will no longer be able to infer at least one box of $\st D_{i, j, k}$.
Define $\st D$ as
$$\st D:=\bigcup_{i<j\in[w+1]}\;\bigcup_{k\in[w]}\st D_{i, j, k}.$$
\footnote{The boxes of $\st D$ are still in the same domain as those of $\st C$ even though their secondary components $B_{i, j, k}$ are $\lambda$'s.}
By Lemma~\ref{lmm:n/2-construction}, $\st D$ is a minimal set of boxes that covers $\UB$. Every box of $\st C$ is necessary to infer some box of $\st D$, which makes $\st D$ another minimal set of boxes that covers $\UB$.
(We can verify this using the \emph{completeness} of \ordered\ for any fixed attribute order $\sigma$. In particular, let's choose $\sigma$ such that resolution starts with all secondary attributes. From $\st C$, we can infer exactly $\st D$, which -in turn- is exactly what we need to infer $\UB$. But if some box of $\st C$ goes missing, we can no longer infer $\st D$ or $\UB$.)
%\ar{Please also argue the claim on minimality.}

\begin{claim}
\label{clm:w+1}
Consider some arbitrary but fixed elimination order of $G$ that induces a treewidth of $w$, and let $A_e$ be the first primary vertex that is eliminated. Then, before $A_e$ can be eliminated, all of its adjacent secondary vertices must be eliminated first.
\end{claim}
\bp
$A_e$ is connected with edges to $w$ primary vertices and $w\times w$ secondary vertices. Since non of the $w$ primary vertices could have been eliminated before $A_e$, all the secondary vertices must be eliminated before $A_e$, in order to maintain a width of $w$.
\ep

WLOG let $A_1$ be the first primary vertex to-be-eliminated. Before the elimination of $A_1$, all eliminated vertices were secondary. Whenever we eliminated a secondary vertex $B_{i, j, k}$, we inferred $\st D_{i, j, k}$. However, no box from such $\st D_{i, j, k}$ could have been resolved any further with any box. This is because boxes of $\st D$ can only be resolved on primary vertices, and $A_1$ is the first one.

After eliminating all the secondary vertices that are adjacent to $A_1$ (as dictated by Claim~\ref{clm:w+1}), we can infer the following set of boxes
$$\st D_1:=\bigcup_{1<j\leq w+1}\;\bigcup_{k\in[w]}\st D_{1, j, k}.$$
Equivalently, $D_1$ can be written as
$$\st D_1=\bigl\{\mv c \wsuchthat \exists j \in \{2, \ldots, w+1\}\; \exists b \in B^{(w)}
\bigl[\pi_{A_1}(\mv c), \pi_{A_j}(\mv c) \in\{0, 1\}^{d'}b \text{ and }
\pi_{X}(\mv c)=\lambda \textnormal{ for all $X\in \vars(Q)-\{A_1, A_j\}$}\bigr]\bigr\}.$$
By performing resolutions on $A_1$ eliminating all but the first $d'$ bits, we infer the following set
$$\st E_1:=\bigl\{\mv c \suchthat \pi_{A_1}(\mv c)\in\{0, 1\}^{d'} \text{ and } \bigl(\forall j_1\neq j_2\in\{2, \ldots, w+1\}\bigr) \bigl(\exists b_1\neq b_2 \in B^{(w)}\bigr)\bigl[\pi_{A_{j_1}}(\mv c)\in\{0, 1\}^{d'}b_1 \;\;\wedge\;\; \pi_{A_{j_2}}(\mv c)\in\{0, 1\}^{d'}b_2\bigr]\bigr\}.$$
\footnote{The boxes of $\st E_1$ are also assumed to be in the same domain as those of $\st C$ and $\st D$:
Their $B_{i, j, k}$ components are hiding as $\lambda$'s, but they still exist.}
From the proof of Lemma~\ref{lmm:n/2-construction}, we know that $\st E_1 \cup(\st D-\st D_1)$ is a minimal set of boxes that covers $\UB$. Moreover, for every not-yet-eliminated secondary vertex $B_{i, j, k}$, every box of $\st C_{i, j, k}$ is necessary to infer some box of $D_{i, j, k}\subseteq \st D-\st D_1$.
Noting that $\abs{\st E_1}=\Theta(\abs{\st C}^{w+1})$ completes the proof.
\ep

\subsection{The \texorpdfstring{$\Omega(\abs{\st C}^{n/2}+Z)$}{Omega(|C|\{n/2\}+Z)} lowerbound for \geo}
\label{sec:n/2-lowerbound}

We first prove two structural lemmas that will be useful in proving our lower bound for \geo\ proofs.
\blmm
\label{lmm:r'<=r}
Let $\st A$ be a set of dyadic boxes. Let $\mv z$ be a box that can be inferred from $\st A$ using $r$ geometric resolutions. Let $\st A'$ be a set of boxes that cover the boxes of $\st A$ (i.e., for every box $\mv a \in \st A$, there is a box $\mv a' \in \st A'$ such that $\mv a'$ covers $\mv a$). Then, a box $\mv z'$ that covers $\mv z$ can be inferred from $\st A'$ using $r'\leq r$ geometric resolutions.
\elmm
\bp
By induction. In the base case when $r=0$, $\mv z$ must belong to $\st A$. Hence, $\st A'$ must have some box that covers $\mv z$.

For the inductive step, we will assume that the lemma holds for $r-1$ and prove it for $r$. Let $\mv z$ be a box that can be inferred from $\st A$ using $r$ resolutions. Let one of those resolutions be $\mv b\gets\Resolve(\mv a_1, \mv a_2)$ for some $\mv a_1, \mv a_2\in \st A$. As a result, $\mv z$ can be inferred from $\st B:=\st A\cup\{\mv b\}$ using $r-1$ resolutions. Let $\st A'$ be a set of boxes that cover those of $\st A$. Let $\mv a_1'\in \st A'$ cover $\mv a_1$, and $\mv a_2'\in \st A'$ cover $\mv a_2$. We recognize two case:
\bi
\item If $\mv a_1'$ and $\mv a_2'$ can be resolved, then $\mv b$ must be covered by $\mv b'\gets\Resolve(\mv a_1', \mv a_2')$. The boxes of $\st B':=\st A'\cup\{\mv b'\}$ cover those of $\st B$. Because the lemma holds for $r-1$, a box $\mv z'$ that covers $\mv z$ can be inferred from $\st B'$ in $\leq r-1$ resolutions. Therefore, $\mv z'$ can be inferred from $\st A'$ in $\leq r$ resolutions.
\item If $\mv a_1'$ and $\mv a_2'$ cannot be resolved, then one of them must cover $\mv b$. Because the lemma holds for $r-1$, a box $\mv z'$ that covers $\mv z$ can be inferred from $\st A'$ in $\leq r-1$ resolutions.
\ei
\ep

\bdefn
Given a set of boxes $\st A$, let $r(\st A)$ denote the minimum number of geometric resolutions that is sufficient to infer $\UB$ from $\st A$. (If the union of all boxes in $\st A$ does not cover $\UB$, then $r(\st A)=\infty$.)
\edefn

\blmm
\label{lmm:r(AuB)=r(A)}
Let $\st A$ and $\st B$ be two sets of boxes that satisfy the following conditions:
\bi
\item $\UB$ does not belong to $\st B$.
\item Every geometric resolution between two boxes from $\st B$ results in a box that is contained in some box of $\st A$.
\item No box in $\st B$ can be geometrically-resolved with any box in $\st A$ or with any box that can be inferred from $\st A$ through geometric resolution.
\ei
Then, $r(\st A \cup \st B)=r(\st A)$.
\elmm
\bp
By definition, $r(\st A \cup \st B)\leq r(\st A)$. Next, we prove that $r(\st A \cup \st B)\geq r(\st A)$ by induction on the value of $r(\st A \cup \st B)$. In the base case when $r(\st A \cup \st B)=0$, $\UB$ must belong to $\st A \cup \st B$. Hence, $\UB$ must belong to $\st A$ and $r(\st A)=0$.

For the inductive step, we will assume that the lemma holds when $r(\st A\cup\st B)=k$ for some integer $k\geq 0$, and we will prove that it holds when $r(\st A\cup\st B)=k+1$. Let $\st A$ and $\st B$ be two box sets that satisfy the lemma conditions and $r(\st A\cup \st B)=k+1$. Consider some resolution proof of $\UB$ whose facts are boxes from $\st A\cup\st B$ and whose number of resolutions is exactly $r(\st A\cup\st B)$. Let $\mv a\gets\Resolve(\mv a_1, \mv a_2)$ be some resolution in this proof such that both $\mv a_1$ and $\mv a_2$ belong to $\st A \cup \st B$. By definition, $r(\st A \cup \st B)=r(\st A \cup \st B \cup \{\mv a\})+1$. If both $\mv a_1$ and $\mv a_2$ are from $\st B$, then $\mv a$ must be contained in some box of $\st A$. By Lemma~\ref{lmm:r'<=r}, $r(\st A \cup \st B)\leq r(\st A \cup \st B \cup \{\mv a\})$, which is a contradiction. Because no box from $\st B$ resolves with any box from $\st A$, both $\mv a_1$ and $\mv a_2$ must belong to $\st A$. Let $\st A':=\st A \cup \{\mv a\}$.
$$r(\st A\cup \st B)=r(\st A' \cup \st B)+1\geq r(\st A')+1\geq r(\st A).$$
The first inequality above holds because $r(\st A' \cup \st B)=k$ and the lemma is assumed to hold for $k$. The second inequality holds by definition of $r$.
\ep

We are now ready to argue that Theorem~\ref{thm offline as} is tight.
\bthm[There is a class of input instances on which every \geo\ algorithm runs in time $\Omega(\abs{\boxcert}^{n/2}+Z)$.]
\label{thm:C^{n/2}-lowerbound}
For every integer $n>1$ and for every integer $c>0$, there exists a set $\st C$ of $n$-dimensional $\Theta(c)$ boxes satisfying the following conditions:
\bi
\item For every box $\mv c\in \st C$, $\abs{\support(\mv c)}=2$.
\item $\st C$ is a minimal set of boxes that covers $\UB$.
\item Every \geo\ proof of $\UB$ consists of $\Omega(\abs{\st C}^{n/2})$ resolutions, no matter whether they are ordered or out-of-order.
\ei
Hence, there exists a join query $Q$ whose hypergraph is a clique and whose set of gap boxes $\st B(Q)$ is $\st C$ and whose output is empty, such that every \geo\ algorithm takes time $\Omega(\abs{\st C}^{n/2})$ on $Q$.
\ethm
\bp
Let $d':=\lceil 1/2\log_2{c}\rceil$ and $\st C$ be defined as described in Lemma~\ref{lmm:n/2-construction}:
$$\st C:=\bigl\{\mv c \wsuchthat \exists i\neq j\in [n]\; \exists b \in B^{(n-1)}
\bigl[\pi_{A_i}(\mv c), \pi_{A_j}(\mv c) \in\{0, 1\}^{d'}b \text{ and }
\pi_{A_k}(\mv c)=\lambda \textnormal{ for all $k\in [n]-\{i, j\}$}\bigr]\bigr\}.$$
Note that $\abs{\st C}=\Theta(c)$ and $\st C$ is a minimal set of boxes that covers $\UB$.

Lemma~\ref{lmm:n/2-construction} showed that $\Omega(\abs{\st C}^{n/2})$ ordered resolutions are needed to infer $\UB$ from $\st C$.
Now we prove that even if we use out-of-order resolutions, we still need $\Omega(\abs{\st C}^{n/2})$ many of them.
For each $l\in\{2, \ldots, n\}$, 
%\ar{We are using $m$ to denote the number of relations, so please use another symbol for this index.}
we define a box set $\st D_l$ to consist of every box whose support has size 2 and whose two non-$\lambda$ components share a common suffix after the first $d'$ bits. This common suffix belongs to $B^{(n-l+1)}-\left\{1^{n-l}\right\}$:
$$\st D_l:=\bigl\{\mv c \wsuchthat \exists i\neq j\in [n]\; \exists b \in B^{(n-l+1)}-\left\{1^{n-l}\right\}
\bigl[\pi_{A_i}(\mv c), \pi_{A_j}(\mv c) \in\{0, 1\}^{d'}b \text{ and }
\pi_{A_k}(\mv c)=\lambda \textnormal{ for all $k\in [n]-\{i, j\}$}\bigr]\bigr\}$$
Moreover, for each $l\in\{2, \ldots, n\}$, we define a box set $\st E_l$ to consist of every box whose support has size $l$ and whose $l$ non-$\lambda$ components share a common suffix after the first $d'$ bits. This common suffix is $1^{n-l}$:
$$\st E_l:=\bigl\{\mv c \wsuchthat \exists I\subseteq[n], \abs{I}=l\bigl[\pi_{A_i}(\mv c) \in \{0, 1\}^{d'}1^{n-l}\text{ for all } i\in I \text{ and } \pi_{A_i}(\mv c)=\lambda \text{ otherwise}\bigr]\bigr\}$$
From the above definitions, $\st C=\st D_2 \cup \st E_2$.

Now, we make the following claim.

\begin{claim}
\label{clm:app:H5:Claim 1} For each $l\in\{2, \ldots, n-1\}$, $r(\st D_l \cup \st E_l)\geq r(\st D_{l+1} \cup \st E_{l+1})$.
\end{claim}

Note that $\st D_n=\emptyset$ (because $1^0=\lambda$) and
%\ar{Is $\st D_n=\emptyset$ because you are using the convention that $1^0=\lambda?$}
$\st E_n=\left\{\dbox{x_1, \ldots, x_n}\wsuchthat x_i\in\{0,1\}^{d'}\text{ for all }i\in[n]\right\}$.
Every box in $\st E_n$ is necessary to cover $\UB$ (by only using boxes in $\st E_n$) and $\abs{\st E_n}=\Theta(\abs{\st C}^{n/2})$. Hence, $r(\st E_n)=\Omega(\abs{\st C}^{n/2})$. Assuming Claim~\ref{clm:app:H5:Claim 1} is correct, we have
$$r(\st C)=r(\st D_2 \cup \st E_2)\geq r(\st D_n \cup \st E_n)=\Omega(\abs{\st C}^{n/2}),$$
as desired.
%\ar{Briefly argue why the last equality is true.}

To finish the proof, we now  prove Claim~\ref{clm:app:H5:Claim 1}. The proof of Lemma~\ref{lem:app:string-seq} implies the following: In every resolution that occurs between two boxes from $\st D_l \cup \st E_l$, one of the two resolved boxes must belong to $\st E_l$ while the other must belong to the following subset of $\st D_l$:
$$\st D_l':=\bigl\{\mv c \wsuchthat \exists i\neq j\in [n]
\bigl[\pi_{A_i}(\mv c), \pi_{A_j}(\mv c) \in\{0, 1\}^{d'}1^{n-l-1}0 \text{ and }
\pi_{A_k}(\mv c)=\lambda \textnormal{ for all $k\in [n]-\{i, j\}$}\bigr]\bigr\}.$$
%\ar{Please argue the claim above that one of the resolved boxes has to belong to $\st E_l$ and the other to $\st D_l'$.}
This implies that the geometric resolution result must be contained in some box that belongs to $\st E_{l+1}$. From the above definition, we have $\st D_l'=\st D_l - \st D_{l+1}$. Hence,
$$r(\st D_l \cup \st E_l)=r(\st D_{l+1} \cup \st D_l' \cup \st E_l)\geq
r(\st D_{l+1} \cup \st D_l' \cup \st E_l \cup \st E_{l+1}) = r(\st D_{l+1} \cup \st E_{l+1}).$$
The inequality above holds by definition of $r$. The second equality holds because of Lemma~\ref{lmm:r(AuB)=r(A)}.
\ep

\bcor
\label{cor:klee-boxcert^{n/2}-lb}
Let $\calB$ be the input set of boxes for Klee's measure problem, $n$ the number of dimensions, and $\boxcert$ any optimal certificate for $\calB$.
Klee's measure problem over the Boolean semiring has a lower bound of $\Omega(\abs{\boxcert}^{n/2})$ that holds for all algorithms that are based on \geo.
\ecor

\subsection{The \texorpdfstring{$\omega(\abs{\boxcert}^{4/3-\eps}+Z)$}{omega(|C|\{4/3-eps\}+Z)} lowerbound for arity \texorpdfstring{$\geq3$}{>= 3}}
The following proposition aims to show that general dyadic boxes are a bit too
expressive. In particular, this shows that the restriction of the maximum arity being two in Theorem~\ref{thm:main acyclic arity<=2} is necessary. (We make use of a well-known complexity theoretic assumption about
the hardness of the $3${\sf SUM} problem \cite{3sum}.)
%\ar{Maybe re-phrase to stress in the statement that here acyclic means any of the 5 notions and not just $\alpha$-acyclic as the latter is already covered in the PODS 14 paper.}
\bprop[Hardness of acyclic queries with arity-$\geq 3$ relations]
For any $k>2$, there is a class of ($\alpha$, $\beta$, $\gamma$, and Berge)-{\em acyclic} queries $Q$ whose maximum 
relation arity is $k$ satisfying the following.
Unless the $3$\textsf{SUM} problem can be solved in sub-quadratic time,
there does not exist an algorithm that runs in time 
$O(|\boxcert|^{4/3-\eps} + Z)$ for any $\eps>0$ on all input instances.
And, there is a class of acyclic queries $Q$ with maximum arity $k$ satisfying
the following. Unless the exponential time hypothesis is wrong, no algorithm
runs in time $O(|\boxcert|^{o(k)})$ on all instances.
\label{prop omega 4/3}
\eprop
\bp
We first prove the proposition with $k=3$.
We use the $3${\sf SUM} hardness of the triangle query \cite{3sum}.
Given an instance of the triangle query $R(A, B)$, $S(B, C)$, $T(A, C)$,
we create three new three-dimensional relations $W_R(A, B, C)$,
$W_S(A, B, C)$ and $W_T(A, B, C)$.
The gap boxes for $W_R, W_S, W_R$ are the sets of all gap boxes 
from $R$, $S$, and $T$, respectively, with appropriate $\lambda$'s filled 
in the missing coordinates.
For example, if we have a gap box $\dbox{a, b}$ from $R$, we insert 
$\dbox{a, b, \lambda}$ to $\calB(W_R)$. 
Then, the total size of $\calB(W_R), \calB(W_S), \calB(W_T)$ is linear in 
the total sizes of $R$, $S$, and $T$, which is linear in the number of 
tuples from $R$, $S$, and $T$.

Finally, we let $Q$ be any query containing as a sub-query the join 
$W_R(A, B, C)\Join W_S(A, B, C)\Join W_T(A, B, C)$, while the rest of the 
relations from $Q$ can be of arbitrary form.
Note that $Q$ can be $\alpha$, $\beta$, $\gamma$, or even Berge-acyclic.
In the hard instance, we let them contain all possible tuples in their 
respective domains. Thus, for the rest of the relations in $Q$, they do not
contribute any gap box to the problem. The output of this join is exactly
the join of $R\Join S \Join T$, which cannot be computed in 
$O(|\boxcert|^{4/3-\eps} + Z)$-time, following \cite{3sum}.
For $k>3$, we can pad more $\lambda$'s to the above reduction.

Using a reduction from {\sf unique-$k$-clique}
\cite{DBLP:conf/soda/ChenLSZ07} similar to the
reduction above, we can also show that no algorithm can 
run in time $\tilde O(|\boxcert|^{o(k)})$ for all input instances.
The result from \cite{DBLP:conf/soda/ChenLSZ07} states that {\sf
unique-$k$-clique} does not admit an $O(n^{o(k)})$ algorithm unless the
exponential time hypothesis is wrong.
\ep

% ----------------------------------------------------------------------------
\section{\tetrisreloaded in the GAO-consistent certificate world}
\label{sec gao consistent results}
% ----------------------------------------------------------------------------

When we resolve two boxes $\mv w_1$ and $\mv w_2$, the result $\mv w$ ``inherits'' some of its components from $\mv w_1$ and some from $\mv w_2$ (and some from both). In general, $\mv w$ is going to be a ``mixture'' of $\mv w_1$ and $\mv w_2$. It might also happen that $\mv w$ inherits all of its components from $\mv w_1$ only (or $\mv w_2$ only). Given a set of $M$ boxes, if we know that no matter how we resolve them, we can never ``mix up'' more than $c$ of them in a single box, then we cannot perform more than $\tO(M^c)$ resolutions. The integral cover support lemma (Lemma \ref{lmm IC support}) was a formalization of this intuitive idea.

Given a hypergraph $\calH=(\calV, \calE)$ and a GAO $\sigma$, we can define a
measure of how ``badly'' ordered-geometric-resolution can ``mix up'' any set of boxes,
assuming that those boxes are $\sigma$-consistent and their supports correspond to
hyperedges in $\calH$. We will be referring to this measure as the \coverwidth
of $\calH$ that is induced by $\sigma$, denoted by $\cw(\sigma)$. Based on this 
intuition, it is not surprising that \tetrisreloaded runs in time 
$\tO(|\gaoboxcert|^{\cw(\sigma)} + Z)$, which is what we are going to prove 
in this section. This result generalizes earlier ones found in \cite{nnrr}.

We start by defining the
\coverwidth of a hypergraph and stating some of its properties.

% ----------------------------------------------------------------------------
\subsection{\Coverwidth of hypergraphs}
\label{subsec cover width}

Let $\calH = (\calV, \calE)$ be a hypergraph.
Let $\sigma=(v_1,\dots,v_n)$ be an ordering of vertices of this hypergraph.
In the context of this paper, this ordering is called a GAO.
In traditional graph theory, database, and graphical model applications, an ordering of
vertices of a hypergraph is often called an ``elimination order.'' We use GAO
instead of elimination order because sometimes we will need our GAO to be
a particular elimination order or the reverse of an elimination order
in the traditional sense.

For any hypergraph $G$, let $V(G)$ denote its vertex set, and $E(G)$ denote
its edge set.
The {\em union of two hypergraphs} $G_1$ and $G_2$, denoted by $G_1\cup G_2$,
is a hypergraph $G = \bigl(V(G_1)\cup V(G_2), E(G_1) \cup E(G_2)\bigr)$.
If $G$ is a hypergraph and $v$ is one of its vertices, then
$G-v$ denotes the hypergraph obtained from $G$ by removing $v$ from $V(G)$
and from every edge in $E(G)$. If an empty edge results, we remove the empty
edge too.

We define collections $\calG_k$, $k\in [n]$, of hypergraphs with respect to 
$\sigma$, recursively, as follows. 
A hypergraph is identified using the collection of hyperedges that it has;
hence, a ``system of sets'' and a hypergraph are used interchangeably
whichever makes more intuitive sense.

To facilitate the recursive definition, define 
\[ \calG = \left\{ \{F\} \suchthat F \in \calE \right\}. \]
(Here, $\calE$ is the set of hyperedges of the original hypergraph $\calH =
(\calV, \calE)$, the hypergraph of the input query.
In words, initially the collection $\calG$ consists of $|\calE|$ single-edge 
hypergraphs, one for each member of $\calE$, the edge set of $\calH$.

For each $k$ from $n$ down to $1$, we construct the hypergraph collection 
$\calG_k$ in two steps, then perform two extra steps to prepare for 
the construction of $\calG_{k-1}$.
\bi
 \item {\bf Step 1, initialization:} set $\calG_k$ to be
\[ \calG_k = \left\{ G \suchthat G \in \calG \text{ and } v_k \in V(G) \right\}. \]

 \item {\bf Step 2, taking closure:} 
as long as there are two members $G_1, G_2 \in \calG_k$
such that $G_1\cup G_2 \notin \calG_k$,  add the hypergraph
$G_1\cup G_2$ to $\calG_k$.
The construction of $\calG_k$ is completed until we can no longer find
such two members.

  \item {\bf Step 3, vertex elimination:} set $\calG = \calG \cup \calG_k$,
      then for every $G \in \calG$, replace $G$ by $G-v_k$.
\ei

The {\em minimum integral edge cover number} of a hypergraph $G$, denoted by 
$\rho(G)$, is the minimum number of hyperedges in $E(G)$ that can be used
to cover $V(G)$.

The {\em induced \coverwidth} of the GAO $\sigma = (v_1,\dots,v_n)$ with respect
to the original hypergraph $\calH$, abbreviated by $\cw(\sigma)$, is defined to be 
\[ \cw(\sigma) := \max_{k \in [n]} \max_{G \in \calG_k} \rho(G). \]

The {\em \coverwidth} of a hypergraph $\calH$ is the minimum induced \coverwidth
over all possible GAOs:
\[ \cw(\calH) = \min_{\text{GAO } \sigma} \cw(\sigma). \]

The following Propositions show that the notion of \coverwidth is a natural
``interpolation'' width notion between $\beta$-acyclicity and treewidth.
Furthermore, there are classes of hypergraphs for which the \coverwidth is
arbitrarily smaller than the treewidth.

\bprop
The \coverwidth of a $\beta$-acyclic hypergraph $\calH$ is $1$.
Furthermore, a GAO with \coverwidth $1$ for $\calH$ can be computed in
time polynomial in the query complexity.
\label{prop cw of beta}
\eprop
\bp
Let $\calH=(\calV,\calE)$ be the hypergraph. We showed in \cite{nnrr}
that there exists a GAO $\sigma = (v_1,\dots,v_n)$ which is a nested elimination
order. What that means is, for every $k \in [n]$, if we define the following
set system:
\[ \calF_k = \left\{ F \cap \{v_1,\dots,v_k\} \suchthat F \in \calE 
                     \text{ and } v_k  \in F \right\}
\]
then the sets in $\calF_k$ form a chain, i.e. series of sets where one is 
contained in the next.

It is sufficient to show that if $\sigma$ is a nested elimination order, then
$\rho(\calG_k) = 1$ for every $k \in [n]$, where the $\calG_k$ are defined
above. It is not hard to see that $\calG_k \subseteq 2^{\calF_k}$, the
power set of $\calF_k$. Hence, for every hypergraph $G\in \calG_k$, the
hyperedges of $G$ form a chain. To cover $G$, we can simply take the bottom
hyperedge on this chain, i.e. $\rho(G)=1$ for every $G \in \calG_k$.

Verifying whether a hypergraph is $\beta$-acyclic can be done easily in
polytime by an elimination procedure \cite{nnrr}, which also yields a
nested elimination order if one exists.
\ep

\bprop[$\cw(\calH) \leq \tw(\calH)+1$]
If $\calH$ is a hypergraph with treewidth $w$, then its \coverwidth is at
most $w+1$.
\eprop
\bp
Since $\tw(\calH)=w$, there exists a GAO $\sigma= (v_1,\dots,v_n)$ for which
the induced width of $\sigma$ is $w$. In particular, for every hypergraph
$G \in \calG_k$, we know $V(G) \subseteq \support(v_k)$, and hence
$|V(G)| \leq w+1$. (See relation
\eqref{eqn induced width}.) Any minimal integral edge cover of a hypergraph
with at most $w+1$ vertices must have size at most $w+1$.
Thus, $\rho(G) \leq w+1$ for all $G\in\calG_k$ and for all $k\in [n]$.
Consequently, $\cw(\calH) \leq w+1$ as desired.
\ep

\begin{example}[$\cw \ll \tw$]
There are classes of hypergraphs whose \coverwidth is any given positive
integer and whose treewidth is unbounded.

Let $c$ be any positive integer and $\calH$ be any hypergraph with 
$\tw(\calH) = c$. Create a new hypergraph $\calH'$ from $\calH$ by cloning every
vertex of $\calH$ $t$ times. If an edge of $\calH$ contains a vertex $v$,
then the edge will contain $t$ copies of $v$ in $\calH'$.
It is not hard to see that $\cw(\calH') = \cw(\calH)$ but $\tw(\calH')$ is now
unbounded as $t$ is arbitrary.
\end{example}

\begin{example}[$\cw$ of a Loomis-Whitney query]
The LW$(n)$ query is the query whose hypergraph is $\calH = \left( [n],
\binom{[n]}{n-1}\right)$. Note that LW$(3)$ is the triangle query.
It is not hard to see that the \coverwidth of this hypergraph is
$2$, independent of the GAO. This is (very roughly) because, every hyperedge 
is missing a unique vertex; and thus any of two of them cover their union.
\end{example}

\begin{example}[$\cw$ of a cycle]
    Let $C_n$ be the cycle of length $n$. Then, $\cw(C_n) \geq 2$ because
    no matter which GAO we choose, there is a hypergraph in $\calG_n$ that
    has an integral cover number $2$.
    Now, if the cycle's edges are $\{(1,2), (2,3), \cdots,(n-1,n), (n,1)\}$,
    and let $\sigma = (1,2,\dots,n)$, then $\calG_n$ consists of the following
    hypergraphs (recall that we identify a hypergraph by its set of hyperedges,
    so we will not explicitly write down the vertex set of a hypergraph for
    brevity):
    \begin{eqnarray*}
        G_1 &=& \{(1,n)\},\\
        G_2 &=& \{(n-1,n)\},\\
        G_3 &=& \{ (1,n), (n-1,n) \} \text{ (this is $G_1\cup G_2$) }
    \end{eqnarray*}
    As reasoned above, $\rho(G_3) = 2$. Now, after eliminating vertex $n$,
    $\calG_{n-1}$ consists of the following hypergraphs
    \begin{eqnarray*}
        G'_1 &=& \{ (n-2,n-1) \}\\
        G'_2 &=& \{ (n-1) \} \text{ (this is $G_2 - n$) }\\
        G'_3 &=& \{ (1), (n-1) \} \text{ (this is $G_3-n$) } \\
        G'_4 &=& \{ (n-1), (n-2, n-1) \} \text{ (this is $G'_1 \cup G'_2$) }\\
        G'_5 &=& \{ (1), (n-1), (n-2, n-1) \} \text{ (this is $G'_1 \cup G'_3$) }
    \end{eqnarray*}
    We can see that $\rho(G'_i) \leq 2$. Continuing with this process, we can
    prove by induction that $\cw(C_n) = 2$. 
\end{example}

\begin{example}[$\cw$ of a clique]
    The \coverwidth of a $k$-clique is $k-1$. We omit the proof.
\end{example}

% HQN: Had to remove this for now, doesn't add much but too much work
%\paragraph{Relationship to $\beta$-hypertree-width}
%In \cite{DBLP:journals/siamcomp/GottlobP04}, Gottlob and Pichler defined the
%notion of {\em $\beta$-hypertree-width}, which is also meant to capture how
%$\beta$-acyclic a hypergraph is. Recall that a hypergraph is $\beta$-acyclic iff
%every subhypergraph of $H$ is $\alpha$-acyclic. A hypergraph is $\alpha$-acyclic
%iff its (generalized) hypertree width is $1$ \cite{}. Hence, a hypergraph is 
%$\beta$-acyclic iff every sub-hypergraph has hypertree width equal to $1$.
%Analogously, the $\beta$-hypertree-width of a hypergraph is the maximum
%hypertree width over all its subgraphs:
%\[ \bhtw(H) = \max_{H'\subseteq H} \htw(H'). \]
%The notion of $\beta$-hypertree-width can be characterized using vertex
%elimination ordering (or GAO, in this paper's language).
%In the definition of \coverwidth above, for each $\calG_k$, let 
%$G_k =(V_k,E_k) \in \calG_k$ denote the hypergraph all of whose edges 
%contains $v_k$.
%The $H$-covering number of $V_k$, denoted by $\rho_H(V_k)$ is the 
%minimum number of edges in $H$ that can cover the set $V_k$.
%Then, the {\em induced hypertree-width} of $H$ with respect to this
%GAO is the maximum $H$-covering number of $V_k$ over all $k\in [n]$.
%And the hypertree-width of $H$ is the minimum induced hypertree-width
%overall GAOs.

\subsection{GAO-consistent certificate results}

The expressiveness of the input gap boxes certainly help significantly 
reduce the certificate size (see Examples \ref{ex Btree can be bad} and 
\ref{ex general dyadic box can help}).
However, as we have seen from the negative result of 
Proposition~\ref{prop omega 4/3}, there is a great tension between the 
expressiveness of the input gap boxes
and the runtime of {\em any} join algorithm. 

\begin{example}
Consider a query with four attributes, where the GAO is $(A_1, A_2, A_3, A_4)$,
and the input relations are $R(A_1,A_2,A_3, A_4)$ and
$S(A_2, A_4)$.
For two gaps ($\mv w_1$ from $R$ and $\mv w_2$ from $S$) to be resolvable on 
the last attribute, they must have the form:
$$\mv w_1\weq\dbox{t_1\wc t_2\wc t_3\wc x_40}$$
$$\mv w_2\weq\dbox{\lambda\wc t_2\wc\lambda\wc x_41}$$
where $t_1, t_2, t_3$ are maximal-length strings (The `$0$' and `$1$' 
could have appeared the other way around). The resolution result $\mv{w}$ will be:
$$\mv{w}\weq\dbox{t_1\wc t_2\wc t_3\wc x_4}$$
Notice that $\mv w$ is a prefix box of $\mv{w_1}$. (See Definition \ref{defn box prefix}.) In fact, any ordered geometric resolution between two gaps (or prefixes of gaps) from $R$ and $S$ is going to 
produce a prefix of one of them. 
\end{example}

The above example can be made into a formal result.
In this section, we show that if we restrict the input gap boxes to 
be $\sigma$-consistent, where $\sigma$ is the \RAO used by \tetris, 
then there are classes of queries where we can obtain runtimes
that are arbitrarily better than the $\tO(|\boxcert|^{w+1} + Z)$ runtime
proved in Section~\ref{SEC:BEYOND-WC}. Of course, the new runtime will be measured
on the weaker notion of certificate: the GAO-consistent certificate
$\gaoboxcert$.
GAO-consistent certificates, while weaker than general box certificates,
capture a good class of practical algorithms. For example, the Leapfrog Triejoin
algorithm \cite{leapfrog} implemented in LogicBlox database engine uses only indices
from one GAO.

\bthm
\label{thm:tetris-vs-ms}
Let $\st B$ be a set of $\sigma$-consistent boxes for some fixed GAO $\sigma$.
Let $\calH(\st B)$ be the supporting hypergraph of $\st B$, and $\cw(\sigma)$ be the induced cover-width of $\sigma$ with respect to $\calH(\st B)$.
If \tetrisreloaded uses $\sigma$ as a \RAO,
it solves \bcp on input $\st B$ in time $\tO(|\boxcert|^{\cw(\sigma)} + Z)$. (Notice that because $\st B$ is $\sigma$-consistent, $\boxcert$ in here will be $\sigma$-consistent as well.)
\ethm

\bcor[\tetris generalizes \ms]
Let $Q$ be a join query, $\sigma$ be a fixed GAO, and suppose that the set of input gap boxes $\st B(Q)$ is $\sigma$-consistent.
If \tetrisreloaded uses $\sigma$ as a \RAO,
it solves $Q$ in time $\tO(|\gaoboxcert|^{\cw(\sigma)} + Z)$.
\ecor
\bp[Proof of Theorem~\ref{thm:tetris-vs-ms}]
WLOG, assume $\sigma=(A_1, \ldots, A_n)$.
Let $c = \cw(\sigma)$.
The proof strategy is as follows.
We apply Lemma~\ref{lmm IC support} by showing that, for every \gbresolvent
$\mv w$, there exists an integral
cover $F_1,\dots,F_c$ for $\support(\mv w)$ such that $\mv w$ is supported on
$\{F_1,\dots,F_c\}$.
To get the desired integral cover, we equip $\mv w$ with a hypergraph
$G=(V=\support(\mv w),E)$ such that for {\em every} edge $F$ of this hypergraph,
$\mv w$ is supported on $F$ by some input gap box, and that the minimum
integral cover for the hypergraph has size at most $c$.
We obtain such hypergraph by induction. If $\mv w$ was the result of resolving 
$\mv w_1$ with $\mv w_2$, then the hypergraph $G$ will be constructed from the
hypergraphs of $\mv w_1$ and $\mv w_2$, inductively.

In the base case, if $\mv w$ is an input gap box, then the hypergraph for
$\mv w$ is a single-edge hypergraph $G=(V,E)$ with $V = \support(\mv w)$
and $E = \{V\}$. Clearly $\mv w$ is supported on $V$.

Now, suppose $\mv w_1$ and $\mv w_2$ are two gap boxes or \gbresolvents satisfying the following conditions.
There are two hypergraphs $G_1=(V_1, E_1)$ and $G_2=(V_2,E_2)$
for which $V_i = \support(\mv w_i), i \in[2]$, and for every $F \in E_i$,
$\mv w_i$ is supported on $F$ by some input gap box.

Let $\mv w = \Resolve(\mv w_1,\mv w_2)$, and suppose the resolution is on 
attribute $A_k$ for some $k\in [n]$. We consider two cases.

{\bf Case 1.} The $A_k$-component of $\mv w$ is not $\lambda$. 
It is {\em crucial}
that the gap boxes are $\sigma$-consistent, so every component of $\mv w_1$ 
and $\mv w_2$ after $A_k$ is $\lambda$ and
every component before $A_k$ is either $\lambda$ or full-length. (Recall Definition~\ref{defn:GAO-consistent-box}.) Consequently,
for every subset $S \subseteq \support(\mv w_i)$, we have
$\pi_S(\mv w_i) \subseteq \pi_S(\mv w)$.
It follows that the hypergraph $G = G_1 \cup G_2$ satisfies the condition 
that $\support(\mv w) = V(G)$ and $\mv w$ is supported on each hyperedge in 
$E(G)$. 

{\bf Case 2.} The $A_k$-component of $\mv w$ is $\lambda$. In this case,
we equip $\mv w$ with the hypergraph $(G_1\cup G_2) - A_k$.

Finally, by the definition of $\cw(\sigma)$, for each $\mv w$ and each hypergraph
$G$ that $\mv w$ is equipped with, the minimum integral cover of $G$ has size
at most $\cw(\sigma)$. And since $\mv w$ is supported on any integral cover,
the theorem is proved.
\ep

\bcor[\ms's guarantee]
If $Q$ is $\beta$-acyclic, then there exists a \RAO $\sigma$
for which \tetrisreloaded runs in time 
$\tO(|\gaoboxcert| + Z)$, given that the set of input gap boxes $\st B(Q)$ is $\sigma$-consistent.
\ecor
\bp This follows from the above theorem and Proposition~\ref{prop cw of beta}.\ep

\subsection{Cutset \coverwidth}

The notion of \coverwidth was an attempt to measure how $\beta$-acyclic a
hypergraph is. A $\beta$-acyclic hypergraph has \coverwidth exactly $1$.
In probabilistic graphical model (PGM) inference \cite{DBLP:books/daglib/0066829} 
and constraint satisfaction problem (CSP) solving \cite{Dechter:2003:CP:861888}, 
researchers have noticed for a long time that
most inference or CSP problems are easy on $\alpha$-acyclic hypergraphs.
(Note that, as was shown in \cite{nnrr}, in the certificate world the 
boundary has moved from $\alpha$-acyclicity to $\beta$-acyclicity.) 

In PGM inference or CSP, one way to measure the degree of $\alpha$-acyclicity
of a hypergraph is to count the minimum number of vertices one has to remove to
make the hypergraph $\alpha$-acyclic. This idea gives rise to the {\em cycle cutset
conditioning} algorithms \cite{DBLP:books/daglib/0066829,DBLP:journals/ai/Dechter90}. 
The runtime is then multiplied by an
exponential factor in the number of vertices removed.
We do not necessarily need a resulting $\alpha$-acyclic hypergraph. We might
want to stop at a hypergraph of small treewidth, at which point a
treewidth-based search algorithm takes over. This is the idea of
$w$-cutset conditioning algorithms \cite{DBLP:conf/uai/BidyukD04}.

Following the same line, we can define a notion of {\em cutset $c$-\coverwidth},
which is the minimum number $x$ of vertices, say $X\subseteq V$, 
we have to remove so that the resulting hypergraph $H-X$
has \coverwidth equal to $c$.  
In that case, it is easy to see that \tetrisreloaded runs in
time $\tO(|\gaoboxcert|^{x+c} + Z)$ given the correct \RAO.
All we have to do is to put the removed attributes in $X$ in front of the \RAO,
and the best attribute ordering with respect to the \coverwidth
of the residual graph $H-X$ in the end of the \RAO.
Then, we apply Lemma~\ref{lmm IC support} as follows.
For a given witness $\mv w$, we construct an integral cover of its support
set by having a singleton set for each attribute in $\support(\mv w) \cap X$,
and the usual integral cover on $\support(\mv w) \cap (V-X)$.

However, it is easy to prove the following, essentially saying that cutset
conditioning does not help our cause in this problem setting.

\bprop
If a hypergraph $H$ has cutset $c$-\coverwidth $x$, then
$\cw(H) \leq c+x$.
\eprop

\brmk
What does help, however, is that the dyadic segments in the $X$ components
can be general dyadic segments, because in the integral cover we use only
singleton sets as the supports. Hence, this idea of cutset \coverwidth might
still be useful when we know there is a small
subset of attributes on which the input gap boxes store general dyadic 
segments.
\ermk

% ----------------------------------------------------------------------------
\section{From geometry to logic: DNF certificates and connection to DPLL with 
clause learning}
\label{sec:cnf-dnf}
% ----------------------------------------------------------------------------

The bit-string encoding of dyadic intervals leads to a very natural idea
(in hind sight), and opens up a large number of intriguing
questions regarding the nature of database indices and the join operation on them.
In this section, we briefly touch upon some of the research directions and
the questions that \tetris pointed to.

We discuss an alternative formulation, turning the problem from geometry to
logic. We do so by explaining how to store data as DNF-formulas, 
replacing the gap boxes by the more expressive DNF terms.
This encoding/indexing leads to the very natural notion of DNF-certificates, 
which can be {\em a lot} smaller than geometric certificates, at a price.
Then, we explain how syntactically one can view \tetris as doing a special form 
of DPLL with clause learning.

This section is partly speculative.
In some sense we ``close the loop'': {\sf SAT} is a special case of constraint
satisfaction problem, which is equivalent to conjunctive query evaluation,
which \ms showed to be geometric, which \tetris expanded, which led us back
to {\sf SAT}.

\subsection{DNF database indices}

Fix an input relation $R$ of arity $k$. Again for simplicity, assume the 
domain of all attributes are of size $D = 2^d$. We encode each input tuple 
$\mv t = (t_1,\dots,t_k) \in R$ with a DNF term in a very natural way.
There are $dk$ (bit) variables $x_1,\dots, x_{dk}$, of which $d$ variables are 
used to represent each $t_i$. For example, for $k=2$ and $d=3$, and
\begin{eqnarray*}
    t_1&=&2 \text{ ({\tt 010} in binary)}\\
    t_2&=&6 \text{ ({\tt 110} in binary)},
\end{eqnarray*}
we have a truth assignment -- with notation overloading --
\[ \mv t(x_1,\dots,x_6) = ({\tt 010}, {\tt 110}). \]
This is the unique satisfying assignment to the {\em DNF-term} (or {\em conjunctive
clause})
\[ \bar x_1 \wedge
    x_2 \wedge
    \bar x_3 \wedge
    x_4 \wedge
    x_5 \wedge
    \bar x_6.
\]

\bdefn[Tuple DNF-formula]
Any $k$-ary relation $R$ is simply a {\em DNF formula} (a disjunction of
DNF-terms), where each DNF-term has exactly $dk$ variables.
Each term represents a tuple in the relation. 
The set of tuples in $R$ is precisely the set of truth assignments
satisfying the DNF formula. We will refer to this formula as $\varphi(R)$, and
call it the {\em tuple DNF formula} encoding $R$.
\edefn

The above encoding has size $\tO(|R|)$, and one can certainly envision building
an index for the relation that way; though it is not quite clear what
we gain from doing so.
Next, we draw inspiration from \ms and \tetris: we would like the index to
be able to return ``gaps'' representing a region of space where no tuple
from $R$ resides.

The first natural idea is to use the complement of the tuple 
DNF formula to represent the gaps. This complement is a
{\em CNF-formula}, which is a conjunction of clauses.
This representation would represent all possible gaps at once. 
The problem, of course, is that this representation has
size $\Omega(kd|R|)$. Consequently, if we use such a CNF-formula to answer
a probe, it would be the same as transmitting back the entire relation,
defeating the purpose of a probe.
Note that in this setting, a probe is simply a truth assignment of
$kd$ variables. (The projection of the higher dimensional probe point down
to this relation's attributes.)

Recall that the gap boxes cover the non-input-tuples in the
union-sense: they can overlap.
The gap boxes represent DNF-like formulas. 
This observation leads to the next natural idea.
We can design DNF formulas to represent gaps too. 
Each term of a DNF formula has at most $dk$ variables, and thus to respond
to a probe we can return a few terms that the probe satisfies without 
the space explosion.

The dyadic gap boxes are one type of gap DNF terms we are looking for.
Let us start with a couple of examples.

\begin{example}
Consider the following relation $R(A) = \{ 1, 5, 11 \}$
where $\mv D(A) = \{0,1,\dots,15\}$, which means $d=4$.
The tuple DNF formula is
\[ \varphi(R) = 
    (\bar x_1 \wedge \bar x_2 \wedge \bar x_3 \wedge x_4) \vee
    (\bar x_1 \wedge x_2 \wedge \bar x_3 \wedge x_4) \vee
    (x_1 \wedge \bar x_2 \wedge x_3 \wedge x_4).
\]
The truth assignments not in $R$ form gaps, which are as follows.
\bi
\item Gap 1: $\{{\tt 0000}\}$, represented by the DNF term 
    \[ \bar x_1 \wedge \bar x_2 \wedge \bar x_3 \wedge \bar x_4 \]
\item Gap 2: $\{{\tt 0010}, {\tt 0011}, {\tt 0100} \}$, represented by the 
    DNF formula
    \[ (\bar x_1 \wedge \bar x_2 \wedge x_3) \vee
       (\bar x_1 \wedge x_2 \wedge \bar x_3 \wedge \bar x_4).
    \]
\item Gap 3: $\{{\tt 0110}, {\tt 0111}, {\tt 1000}, {\tt 1001}, {\tt 1010} \}$, 
    represented by the 
    DNF formula
    \[ (\bar x_1 \wedge x_2 \wedge x_3) \vee
       (x_1 \wedge \bar x_2 \wedge \bar x_3) \vee
       (x_1 \wedge \bar x_2 \wedge x_3 \wedge \bar x_4).
    \]
\item Gap 4: $\{{\tt 1100}, {\tt 1101}, {\tt 1110}, {\tt 1111} \}$, represented 
    by the DNF formula
    \[ (x_1 \wedge x_2). \]
\ei
\end{example}

It should be clear that the DNF-terms above correspond precisely to the set of
all dyadic boxes. 
And it should also be clear that gap DNF-formulas
are much more powerful than the dyadic boxes. For example,
it is possible to merge the second DNF-term of Gap 2 with the
DNF-term from Gap 1 to form a DNF term $(\bar x_1 \wedge \bar x_3 \wedge \bar
x_4)$. We will get back to this crucial point later.

\begin{example}[The interleaving case]
Consider the hard instance for \ms and \tetris, where we want to compute
the join $R(A) \Join S(A)$ with
$R(A) = \{0, 2, \dots, 2^k-2\}$,
$S(A) = \{1,3,\dots,2^k-1\}$. 
\ms and \tetris work in the same way for this example: they run in time
$\Omega(|R|+|S|)$.

However, if by magic the database index infers more about the relations and
represents the DNF formulas much more succinctly, then join algorithms can
run a lot faster. The complement of $R(A)$ can be represented by the DNF-formula
\[ \bar R = x_d, \]
and the complement of $S(A)$ is
\[ \bar S = \bar x_d. \]
And the certificate $\bar x_d \vee x_d$ certifies that the output is empty.
\end{example}

\bprop
Every dyadic box (in any index order) can be presented by a DNF-term.
The converse does not hold. 
\eprop
\bp The forward direction is obvious.
The interleaving example above shows that the converse does not hold: the
DNF-term $x_d$ cannot be represented by a single dyadic box.
\ep

\bdefn[Gap DNF formula]
Given a relation $R$, a gap DNF formula is a DNF formula $\bar \varphi(R)$ such
that $\mv t \notin R$ iff $\mv t$ satisfies $\bar \varphi(R)$.
A gap DNF formula is {\em non-redundant} if none of its term logically infers
another.
\edefn

\bdefn[DNF index]
A DNF index is a data structure storing tuples in a relation $R$ such that
the storage maintains a gap DNF formula. Given a queried tuple, the index either
returns {\em YES}, the tuple belongs to $R$, or {\em NO} the tuple does not
belong to $R$. In the NO case, a set of gap DNF terms from the gap DNF formula
which the tuple satisfies are returned as evidence.
\edefn

We leave the {\em many} tradeoffs involved in building such a database index
for an (exciting) future work. The above definition is necessarily vague. As
far as we know there is no such index in the database literature.
({\em Bitmap/Variant indices}, widely implemented in database management systems
\cite{DBLP:journals/sigmod/ONeilG95,
DBLP:conf/sigmod/ONeilQ97,
DBLP:journals/ipm/SpieglerM85}, are 
close to the spirit of DNF indices.)

We would like to emphasize, however, that what \ms and \tetris pointed to is the
following: ordered indices such as B-trees, tries, or even hash tables (up to
a $\log$-factor loss) can effectively be viewed as DNF indices. 

To briefly touch upon the huge space of tradeoffs involved in designing such
an index, let us even leave aside the all-important question of how to 
efficiently maintain and update relations built using a DNF index.
The {\em theoretical} question of which gap DNF-formula to maintain
is already interesting and difficult.
There are {\em many} gap DNF-formula for a given relation $R$. 
The gap DNF formula corresponding to the dyadic gap boxes has at least as many
terms as $|R|$, one for each gap.
As the interleaving example shows, this representation might be wildly redundant.
It might make sense to find a DNF-formula with the minimum number of terms
so that the storage is compact.

Finding a minimum DNF-formula representing a given relation $R$ is known to be 
NP-hard \cite{masek1979, DBLP:conf/coco/AllenderHMPS06}.
It is easy to see that {\sc min-DNF} is a special case of {\sc set cover}, and
thus admits a $\log N$-approximation algorithm.
Here, $N$ is the size of the truth table. In other words, we can approximate
{\sc min-DNF} to within a factor of $dk$, where $k$ is the number of attributes
of $R$, and $d$ is the number of bits to represent each attribute value.
This bound is almost tight, since there is a $\gamma>0$ for which $(\log
N)^\gamma$-approximation is not possible (modulo a well-known complexity
theoretic assumption) \cite{DBLP:conf/coco/AllenderHMPS06}.

% ----------------------------------------------------------------------------
\subsection{DNF-certificate and DPLL with Clause Learning}

In the previous section, we have established that a DNF-index can generalize
indices that store dyadic gap boxes; the next natural step is to define the
notion of DNF-certificate for a join query.
As defined above, each relation can abstractly be viewed as a gap DNF-formula,
containing many DNF terms. A tuple does not belong to the relation iff it
satisfies some term of the gap DNF-formula.
We use $\calD(R)$ to denote the set of all DNF terms of the gap DNF formula
for relation $R$.

\bdefn[DNF-certificate]
A {\em DNF-certificate} for an instance of the join query $Q$ is
a collection $\dnfcert$ of DNF-terms, where
\[ \dnfcert \subseteq \bigcup_{R\in \atoms(Q)} \calD(R), \]
such that a tuple $\mv t$ (i.e. a truth assignment) satisfies $\dnfcert$
iff it is {\em not} an output tuple of $Q$.
\edefn

If the DNF-indices are built such that the number of DNF terms in $\calD(R)$
which a probe point satisfies is $\tO(1)$, then we can discover an optimal
$\dnfcert$ using a \ms-like algorithm in the same way that \tetrisreloaded
was designed.
Each time we probe into a relation, it will return either
(1) YES, meaning the probe belongs to the relation, or
(2) NO, along with a set of gap DNF terms, representing a set of gap DNF terms 
that the probe satisfies.

The set of all gap DNF terms discovered so far can be stored in a data structure.
Then, the next probe point (or negative witness in \tetris' sense)
is computed from this data structure. 
It is a tuple that does not satisfy any of the DNF terms stored in the data
structure. In other words, it is a satisfying truth assignment to a 
CNF formula. (The complement of the gap DNF terms.)

Hence, we are entering the realm of SAT solvers. 
A SAT-solver takes a CNF-formula and either provides a {\em refutation proof}
that the formula is not satisfiable, or a satisfying assignment.
Most known SAT algorithms are based on variations of the Davis-Putnam procedures
\cite{MR0134439}, which is resolution-based,
or the Davis-Putnam-Logemann-Loveland (DPLL) algorithm \cite{MR0149690},
which is pure backtracking search.

The original Davis-Putnam procedure \cite{MR0134439} is based on resolution, 
and suffers from the memory explosion problem. 
DP-resolution can be cast as a variable-elimination algorithm 
\cite{DBLP:conf/kr/DechterR94}, which has the identical structure as 
the variable-elimination algorithm for graphical model inference.
If the variable ordering is chosen so that the 
induced treewidth of the sequence is small, then it might be faster 
then DPLL \cite{DBLP:conf/kr/DechterR94}.
DP-resolution can be very bad \cite{MR1279424} in theory, compared to other
proofs, if we aim to minimize the proof size.
The {\em DPLL-algorithm} \cite{MR0149690} is search-based.
The pure form of this algorithm eliminates the exponential memory requirement 
of DP-resolution. DPLL can be viewed as DP with unit resolution.
Most practically efficient SAT-solving algorithms use DPLL with 
more sophisticated forms of {\em clause-learning} 
\cite{DBLP:conf/iccad/SilvaS96,
DBLP:journals/tc/Marques-SilvaS99,
DBLP:conf/dac/MoskewiczMZZM01, 
DBLP:conf/cade/Zhang97,
DBLP:conf/sat/EenS03} 
(see \cite{marques2009conflict} for a nice survey).
The idea is to insert back into the CNF formula a new clause that the algorithm
has learned during the search, in order to ``cache'' some of 
the computation performed thus far. 

In summary, one can envision applying known SAT solving algorithms to compute
joins with runtime proportional to the optimal DNF certificates. 
General DNF certificates are extremely expressive, and thus the general problem
is probably too difficult. As we have mentioned earlier, we leave open the
question of how a DNF index can be built and maintained efficiently. 
We leave this direction of inquiry to a future work.

There are a couple of observations that, in hind sight, relate \tetris to
DPLL with clause learning. 

\paragraph*{\tetris is DPLL with clause learning}
If we view the input box $\mv b$ to \tetris as
a partial truth assignment, then the partition of $\mv b$ into $\mv b_1$
and $\mv b_2$ is simply the assigned value of the {\em next} variable.
The box $\mv w_1$ or $\mv w_2$ can be thought of as encodings of a learned
clause or a base clause, where we found that the current partition 
assignment $\mv b_1$ or $\mv b_2$ is in conflict with the CNF clauses. 
If $\mv w_1$ or $\mv w_2$ contains $\mv b$, i.e. $\mv b$ violates one of those
clauses, then the algorithm backtracks.
Otherwise, the partial assignment $\mv b$ violates the resolved clause
$\mv w$. This resolution is of DP-style. The newly learned clause $\mv w$
is inserted back into the knowledge base. 
Furthermore, since the algorithm continues running after a 
satisfying truth assignment is found, it should properly be thought of
as \#DPLL.

\paragraph*{Worst-case optimal algorithms are pure DPLL with a fixed variable 
ordering.} Another interesting observation is that of Corollary~\ref{cor agm tree}.
\tetris does not need to cache resolvents (i.e. it does not need to insert back
learned clauses) in order to achieve the worst-case \agm bound for the 
input query. In hind sight this was also obvious from observing the algorithms
from \cite{NPRR} and \cite{leapfrog}: those algorithms do not need caching
at all. 

On the other hand, in order to achieve the fractional hypertree width bound or
Yannakakis linear runtime for $\alpha$-acyclic queries, pure backtracking
search is not sufficient.
This holds true for the certificate world too, where caching is a must.

%!TEX root = main.tex

\section{General Resolution can be more powerful than Geometric Resolution}
\label{app:paul}

%\ar{Paul's example goes in here. Am still working on this.}

\subsection{The example}

We now show that the hard examples in the proof of Theorem~\ref{thm:C^{n/2}-lowerbound} can be solved with $\tO(\abs{\cert})$ many general resolutions. Contrast this with the result in Theorem~\ref{thm:C^{n/2}-lowerbound}, which states that for the same examples, any geometric resolution scheme needs to make $\Omega(\abs{\cert}^{n/2})$ resolutions. Thus, this will show that geometric resolution is strictly less powerful than general resolution.

We begin with the case of $n=3$.
For completeness, we restate the hard instance for $n=3$ from the proof of Theorem~\ref{thm:C^{n/2}-lowerbound} here. We need an instance for the triangle query $R(A,B)\Join S(B,C)\Join T(A,C)$. Let $M=2^d$ be an integer for another integer parameter $d\ge 1$ and let $\odd_M$ and $\even_M$ denote the set of odd and even numbers respectively in $\{0,\dots,M-1\}$. Then the following is a hard instance for geometric resolution:

\[R(A,B)= \odd_M\times \even_M \cup \even_M\times \odd_M,\]
\[S(B,C)= \odd_M\times \even_M \cup \even_M\times \odd_M,\]
\[T(A,C)= \odd_M\times \even_M \cup \even_M\times \odd_M.\]

Before we proceed we need to state how we encode the domains so that we can express the gaps as clauses (since we want to use general resolution on them). We will follow the encoding from Section~\ref{sec:cnf-dnf}. In particular, for attributes $A,B$ and $C$ we define $d$ boolean variables each: $a_0,\dots,a_{d-1}$, $b_0,\dots,b_{d-1}$ and $c_0,\dots,c_{d-1}$. Then for every constant $i\in\D(A)=\{0,1\}^d$, we associate a shorthand notation $A_i$ for the conjunctive clause that naturally encodes the binary representation of $i$. In particular, for $i=\sum_{j=0}^{d-1} i_j\cdot 2^j$ (where $i_j\in \{0,1\}$), the corresponding conjunctive clause will have the literal $\neg a_j$ if $i_j=0$ and $a_j$ otherwise. Note that $\neg A_i$ is a valid (disjunctive) clause. We similarly define $B_i$ and $C_j$ for $i\in \D(B)=\{0,1\}^d$ and $j\in \D(C)=\{0,1\}^d$. We would like to stress here that we are just using $A_i,B_i,C_i$ as notational macros and there is not necessarily a semantic meaning to this notation. (See Section~\ref{sec:enc-issues} for why this might matter in general.)

Before we proceed with the encoding of gaps, we record the following simple observation.
\blmm
\label{lem:resolve-even-odd}
Given the clauses $\neg A_i$ for $i\in \odd_M$, one can generate the clause $\neg a_0$ with $O(M)$ general resolutions. Similarly, given clauses $\neg A_i$ for $i\in \even_M$, one can generate the clause $a_0$ with $O(M)$ general resolutions. (Similar results hold for clauses from attributes $B$ and $C$.)
\elmm
\begin{proof}
This follows by noting that in general resolution one can essentially ignore the variable $a_0$ in the clauses and then make $O(M)$ resolutions to generate $\dbox{\lambda,\dots,\lambda}$ (on the variables $a_1,\dots,a_{d-1}$). The proof follows by noting that the values $\neg A_i$ for $i\in \odd_M$ ($i\in\even_M$ resp.) contain the literal $\neg a_0$ ($a_0$ resp.).
\end{proof}

We are now ready to state the gaps from the relations above in terms of the notation defined above. Let us consider the gaps from relation $R$: note that we have a gap rectangle $\dbox{i,j,\lambda}$ for $(i,j)\in \odd_M\times \odd_M\cup\even_M\times\even_M$. In particular, for each such pair $(i,j)$, we have the following conjunctive clause
\[\gap_{R,i,j}=A_i \wedge B_j.\]
Note that the negation of the clause is a normal disjunctive clause:
\[\neg\gap_{R,i,j}=\neg A_i \vee \neg B_j.\]
Similarly we can define the clauses corresponding to gaps in $S$ (denoted by $\gap_{S,i,j}$) and in $T$ (denoted by $\gap_{T,i,j}$). Note that checking if the gaps cover $\dbox{\lambda,\lambda,\lambda}$ is equivalent to checking that the following CNF has no satisfying solutions:
\begin{equation}
\label{eq:triang-bottom}
\left(\bigwedge_{(i,j)\in \odd_M\times \odd_M\cup\even_M\times\even_M} \neg \gap_{R,i,j}\right)\wedge \left(\bigwedge_{(i,j)\in \odd_M\times \odd_M\cup\even_M\times\even_M} \neg \gap_{S,i,j}\right) \wedge
\left(\bigwedge_{(i,j)\in \odd_M\times \odd_M\cup\even_M\times\even_M} \neg \gap_{T,i,j}\right).
\end{equation}

We will now argue that for the set of clauses above, general resolution is strictly more powerful than geometric resolution.

\blmm
\label{lem:gen-res>geo-res}
Starting with \eqref{eq:triang-bottom} one can derive a contradiction with $O(M^2)$ general resolutions, while doing the same takes $\Omega(M^3)$ geometric resolutions.
\elmm

We thank Paul Beame for telling us the proof below and kindly allowing us to use it here.
\begin{proof}
The lower bound follows from the proof of Theorem~\ref{thm:C^{n/2}-lowerbound} so we focus on the upper bound in this proof. Consider the following sequence of (general) resolutions:
\begin{enumerate}
\item Fix an $o\in\odd_M$. Then note that with $O(M)$ resolutions on clauses $\neg A_o\vee \neg B_{o'}$ for $o'\in \odd_M$, we can generate the clause $\neg A_o \vee \neg b_0$. (This essentially follows from Lemma~\ref{lem:resolve-even-odd}.) In particular, with $O(M^2)$ general resolutions we can generate the clauses
\[ \neg A_o \vee \neg b_0\text{ for every }o\in\odd_M.\]
Again applying Lemma~\ref{lem:resolve-even-odd} to the above set of clauses with further $O(M)$ general resolutions, one can generate the clause
\begin{equation}
\label{eq:na0-nb0}
\neg a_0\vee \neg b_0.
\end{equation}
Similarly with $O(M^2)$ general resolutions on the clauses $\neg \gap_{R,i,j}$ for $(i,j)\in \even_M\times \even_M$, we get the following clause:
\begin{equation}
\label{eq:a0-b0}
a_0\vee b_0.
\end{equation}
\item Using analogous argument as above to the gaps from $S$ and $T$ we can with $O(M^2)$ general resolutions generate the clauses:
\begin{equation}
\label{eq:nb0-nc0}
\neg b_0\vee \neg c_0.
\end{equation}
\begin{equation}
\label{eq:b0-c0}
b_0\vee c_0.
\end{equation}
\begin{equation}
\label{eq:na0-nc0}
\neg a_0\vee \neg c_0.
\end{equation}
\begin{equation}
\label{eq:a0-c0}
a_0\vee c_0.
\end{equation}
\item Now with four more resolutions we can generate the clauses $a_0$ and $\neg a_0$, which will generate the required contradiction. (Indeed to generate $a_0$, resolve \eqref{eq:a0-b0} with \eqref{eq:nb0-nc0} and resolving the resulting clause $a_0\vee \neg c_0$ with \eqref{eq:a0-c0}. Similarly, $\neg a_0$ can be generated from the other three clauses.)
\end{enumerate}
The proof is complete by noting that the above steps use $O(M^2)$ general resolutions to generate a contradiction, as required.
\end{proof}

%\ar{MAHMOUD: Could you please check to make sure that the above argument can be extended to the hard instance for $n$-clqiue for $n>3$ to show that those hard instances can be solved with $\tO(\abs{\boxcert})$ many general resolutions? I'm reasonably sure that this statement is correct but it would be good to double-check. If you do not find any surprises just add in a couple of senetences in how one can handle the general $n$ case.}

The previous lemma can be generalized to any $n\geq 3$. In particular, for every $n\geq 3$, the lowerbound example that needs $\Omega(\abs{\st C}^{n/2})$ geometric resolutions in Theorem~\ref{thm:C^{n/2}-lowerbound} can actually be solved within $\tO(\abs{\st C})$ general resolutions.
To see this, recall that a dyadic box is an $n$-tuple of binary strings of length $\leq d$. In the logic framework, each bit corresponds to a Boolean variable and each dyadic box corresponds to a conjunctive clause: variables that appear in this clause are those whose corresponding bits appear in the corresponding dyadic box. (e.g.~If some string has length $<d$, then the variable corresponding to the last bit does not appear in the clause.) The negation of a dyadic box is nothing but a disjunctive clause.

To solve the lowerbound example within $\tO(\abs{\st C})$ general resolutions, all we have to do is start with making all resolutions on the Boolean variables corresponding to the first $d'$ bits of each one of the $n$ strings. (Recall the parameters and example from the proof of Theorem~\ref{thm:C^{n/2}-lowerbound}.) Within $\tO(\abs{\st C})$ such resolutions, we can infer $\tO(1)$ clauses that do not contain any one of those $d'\times n$ variables. Within $\tO(1)$ more resolutions, we can generate a contradiction.

\subsection{Encoding Issues}
\label{sec:enc-issues}

In the previous section, we used the natural binary encoding for encoding each element of the domain as a clause. Now we consider another natural encoding with the goal of pointing out that for our setting the choice of encoding matters. In particular, for every element $i$ of domain $\D(A)$, we have a variable $A_i$. (Note that now unlike the previous section this is not just a syntactic variable.) One also has to add a clause $\vee_{i\in \D(A)} A_i$ to explicitly define the domain. This essentially corresponds to a unary representation of the domain and is a representation in proof complexity that has been studied before (see e.g.~\cite{BGL13}). Below we state the reasons why the binary encoding that we use above is more reasonable than this unary encoding for our purposes.

First, we note that if we assume that all the domains are $\{0,1\}^d$, then since in our setting we have $n$ attributes, any clause can have at most $dn$ literals. This implies that a single general resolution (and obviously geometric resolution) can be implemented in $O(dn)$ time, which is $\tO(1)$ in our setting. On the other hand, clauses in the unary representation can have $\Theta(nN)$ many literals, which is significantly large in many of our settings. In traditional worst-case proof complexity, this is fine since an extra factor of $O(N)$ in the run time is not a big deal. However, in our beyond worst-case results, this extra factor is too prohibitive.

Second, we note the fact that  unary encoding can lead to long clauses, which allows general clauses to encode arbitrary subsets of the domain. By contrast this is not possible in the binary encoding setting (which can only encode polynomial many subsets). In the general attribute case this corresponds to the fact that general resolution can encode combinatorial rectangles/boxes while in the geometric resolution case, we can only encode geometric boxes/resolutions.

Finally, we note that for the case when we have only one attribute (say $A$ and hence we only have the variables $a_0,\dots,a_{d-1}$), general resolution can be lossy in the following sense. Consider the case where we have gaps at $7$ and $11$ (i.e. we have $d=4$). The negations of these gaps can be encoded as
\[a_3\vee \neg a_2\vee \neg a_1\vee \neg a_0 \text{ and } \neg a_3\vee a_2\vee \neg a_1\vee \neg a_0.\]
Note that the resolution of the two clauses above results in $\neg a_1\vee \neg a_0$, which corresponds to all elements in $\{0,1\}^4$ that do not have their least two significant bits as $0$. Note that this set is a {\em strict} subset of values that are not $5$ and not $7$. By contrast, when we perform geometric resolution on two clauses corresponding to negation of dyadic intervals, the resulting clause corresponds exactly to the negation of the union of the two dyadic intervals.\footnote{We note however that for two attributes geometric resolution is also similarly lossy.} Thus, in some sense binary encoding is more suited for geometric resolution than to general resolution.

%\ar{I was not sure exactly what we wanted from this section, so I kind of just rambled on. If you think something specific should go in here, let me know.}

%\bibliographystyle{acm}
%\bibliography{main}
%\end{bibunit}
\end{document}